\documentclass[12pt,oneside,openright]{book}
\usepackage[T1]{fontenc}
\usepackage[utf8]{inputenc}
\usepackage{verbatim}
\usepackage{url}
\usepackage{setspace}
\usepackage{multirow}
\usepackage{graphicx}
\usepackage{float}
\usepackage{array}
\usepackage{subcaption}
\usepackage{latexsym}
\usepackage{bm}
\usepackage{amsmath}
\usepackage{amssymb}
\usepackage{graphicx}
\usepackage{multirow}
\usepackage{longtable}
\usepackage{appendix}
\usepackage{epsfig,amsmath,graphicx}
\usepackage[numbers,sort&compress]{natbib}
\usepackage{mathtools}
\usepackage{amsfonts}
\usepackage{rotating}
\usepackage{tensor}
\usepackage{lscape}
\usepackage{bm}
\usepackage{units}
\usepackage{amsopn, amstext, wasysym}
\usepackage{colonequals}
\usepackage{rotating}
\usepackage{multirow}
\usepackage{xspace}
\usepackage[usenames,dvipsnames]{color}
\usepackage[bookmarksnumbered, bookmarksopen, breaklinks, colorlinks, linkcolor=blue, citecolor=blue,urlcolor=blue]{hyperref}
\usepackage{bibunits}
\usepackage{chngcntr}
\usepackage{etoolbox}
\usepackage{feynmp}	
\usepackage{url}
\usepackage{alltt}
\usepackage{ulem}
\usepackage{epigraph} 
\usepackage{pdfpages}
\DeclareGraphicsExtensions{.eps,.pdf,.png,.jpg}
\usepackage{epstopdf}
\usepackage{enumitem}
\usepackage{dsfont} 
\usepackage{caption} 
\usepackage{subcaption}

\usepackage{bbm}
\usepackage{etoolbox}
\preto\subequations{\ifhmode\unskip\fi}

\makeatletter
\newenvironment{keeppage}{\let\thispagestyle=\@gobble}{}
\makeatother
\AtBeginEnvironment{subappendices}{
	\chapter*{Appendix}
	\addcontentsline{toc}{chapter}{Appendices}
	\counterwithin{figure}{section}
	\counterwithin{table}{section}
}
\usepackage{array}
\definecolor{lightgray}{gray}{0.8}
\newcolumntype{L}{>{\raggedleft}p{0.14\textwidth}}
\newcolumntype{R}{p{0.8\textwidth}}

\newcommand{\bi}{\begin{itemize}}
\newcommand{\ei}{\end{itemize}}
\renewcommand{\tilde}{\widetilde} 

\newcommand{\beq}{\begin{equation}}
\newcommand{\eeq}{\end{equation}}
\def\fr{\frac}

\newcommand{\bea}{\begin{eqnarray}}
\newcommand{\eea}{\end{eqnarray}}
\newcommand{\nn}{\nonumber}

\def\OMIT#1{{}}
\newcommand{\lsim}{\mathrel{\rlap{\lower4pt\hbox{\hskip1pt$\sim$}}
\raise1pt\hbox{$<$}}}         
\newcommand{\gsim}{\mathrel{\rlap{\lower4pt\hbox{\hskip1pt$\sim$}}
\raise1pt\hbox{$>$}}}         

\newcommand{\HRule}{\rule{\linewidth}{0.8mm}}
\newcommand{\HHline}{\rule{\linewidth}{0.1mm}}

\newcommand{\beqn}{\begin{equation}}
	\newcommand{\eeqn}[1]{\label{#1}\end{equation}}
\def\beqa{\begin{eqnarray}}
	\def\eeqa#1{\label{#1}\end{eqnarray}}

\def\nn{\nonumber}

\newcommand{\be}{\begin{equation}}
\newcommand{\ee}{\end{equation}}

\newcommand{\mL}{\mathcal{L}}
\newcommand{\mH}{\mathcal{H}}
\newcommand{\Hmath}{\mathcal{H}}
\newcommand{\mR}{\mathcal{R}}
\newcommand{\nc}{\nu_c}
\newcommand{\PS}{\mathcal{P}_\delta(k)}
\newcommand{\Fbar}{\overline{F}}

\def\stacksymbols #1#2#3#4{\def\theguybelow{#2}
	\def\vp{\lower#3pt}
	\def\sp{\baselineskip0pt\lineskip#4pt}
	\mathrel{\mathpalette\intermediary#1}}

\def\intermediary#1#2{\vp\vbox{\sp
		\everycr={}\tabskip0pt
		\halign{$\mathsurround0pt#1\hfil##\hfil$\crcr#2\crcr
			\theguybelow\crcr}}}

\def\gsim{\stacksymbols{>}{\sim}{2.5}{.2}}
\def\lsim{\stacksymbols{<}{\sim}{2.5}{.2}}
\begin{document}

\frontmatter
\pagenumbering{roman}

\newpage

\thispagestyle{empty}

\newpage
\thispagestyle{empty}
\begin{center}
{\bf\LARGE Cosmological Perturbations in the Early Universe}\\
[3em]
 by\\[1em]
{\large Gizem Şengör}\\[1em]
B.Sc. (Physics) Bo\u{g}azi\c{c}i University, Istanbul, 2011\\
M.Sc. (Physics) Bo\u{g}azi\c{c}i University, Istanbul, 2013\\[7em]
DISSERTATION\\
Submitted in partial fulfillment of the requirements for the degree of\\
Doctor of Philosophy in Physics\\[7em]
Syracuse University\\
{June 2018}\\[5em]
\end{center}%

\newpage
\thispagestyle{empty}
\begin{center}
\vspace*{3in}
Copyright 2018 Gizem Şengör\\[1.5em]
All rights reserved
\end{center}

\newpage
\thispagestyle{empty}
\clearpage
\vspace*{1in}
\begin{center}
{\em To Meral Çakır} \\
	\vspace{2 cm}
	{\em Menna'ya sevgilerle} 
\end{center}
\clearpage

\newpage
\begin{center}
	{\bf\Large Abstract}\\[0.5em]
\end{center}

\begin{quote}
	One essence in capturing the history of primordial fluctuations that arise during inflation and eventually lead to formation of large scale structures in the universe, relies on quantizing general relativity coupled to a scalar field. This is a system that respects diffeomorphism invariance, the freedom of choosing the coordinate system to work with without changing the physics. Hence the way to handle its quantization requires an understanding of how to quantize diffeomorphisms. 
Deciding on suitable coordinate choices and making sure that this gauge fixing, which is unavoidable in any calculation, does not effect the end result, is tricky. In this thesis we make full use of the effects of diffeomorphism invariance of the theory on the primordial fluctuations to pursue two different approaches that focus on treating perturbations after gauge fixing. On one hand we work towards developing our understanding of how to handle quantization in terms of Dirac quantization and Becchi, Rouet, Stora, Tyutin (BRST) quantization. On another, we focus on how to generalize the allowed interactions and understand the scales they bring in the era of preheating that follows inflation, with effective field theory (EFT) methods on cosmological backgrounds.
\end{quote}

\newpage
\begin{center}
	{\bf\Large Foreword}\\[0.5em]
\end{center}

\begin{quote}
	\vspace*{2em}
This manuscript is a revised version of my PhD thesis in Physics, which was presented on May 11th, 2018 with the presence of  William Wylie (chair), Scott Watson (advisor), Cristian Armendariz-Picon, Simon Catterall, Jay Hubisz and Carl Rosenzweig as the defense committee at Syracuse University, NY.
 
We begin with a review of the formalism we will build on, with non expert audience interested in cosmological perturbation theory in mind. Cosmological perturbations, how they are defined, effected by gauge choices, and how their time evolution can be captured are developed in sections \ref{sec:diffeomorphisms}-\ref{sec:spontaneously broken}. Within Chapter \ref{chp:review} itself, everything is defined in terms of gauge invariant variables. The formal development on how to handle gauge fixing, discussed in Chapter \ref{chp:quantization methods}, has been published in its entirety in \cite{Armendariz-Picon:2016dgd}.

Section \ref{sec:powerspectra} starts describing the observables of perturbation theory. We will mainly focus on the two-point function, referred to as the power spectrum, consider the growth of density perturbations and structure formation in the early universe, and focus on primordial black hole constraints for this era and how they constrain models that give rise to alternative histories in section \ref{sec:BPH}. The main results from this section have been published as constraints on possible dark matter scenarios \cite{Georg:2016yxa}.

Lastly, we will move on to study cosmological perturbations in all generality of interactions with the effective field theory (EFT) formalism in Chapter \ref{chp:EFTpreheating}. The standard literature on low energy degrees that arise from spontaneously symmetry breaking are reviewed in section \ref{sec:spontaneously broken}. Preheating within this framework is considered starting from \ref{chp:EFTpreheating} with a focus on two different approaches. The first approach is the formalism that handles the physics of the perturbations without addressing the physics of the background. The second approach discussed in Chapter \ref{chp:towardsEFTbackground} follows the pursuit of capturing the evolution of the full fields. The initial work from these last chapters first appeared in \cite{Ozsoy:2015rna} and the main results were published in \cite{Giblin:2017qjp}.

Sections \ref{subsec:hidden} on Hidden Preheating, \ref{sec:adiabaticmodes} on adiabatic modes through preheating and section \ref{sec:towardshamiltonian} on towards a Hamiltonian construction of EFT of two fields appear for the first time as part of this dissertation.

All in all, we hope that the reader will find that enough room has been reserved, for the old lore just as much as needed to be self contained, and the original research to be inspirational. Let us begin by quoting:
\emph{`` What seems to be most fascinating is the fact that one unique flat spectrum of metric perturbation with one arbitrary constant leads to the calculation of the `content' of the Universe (its entropy, primeaval chemical composition, photon-baryon ratio etc) and to the calculation of the `structure' of the Universe (the mass and density of clusters, the degree of uniformity on greater scale).'' Zeldovich 1972}
\end{quote}

\newpage

\begin{center}
{\bf\Large Acknowledgments}
\end{center}
\vspace*{2em}
I am grateful to Scott Watson for being an encouraging and understanding advisor. As much as he has shown me good techniques to work with, he has also left a door open for all else that is out there and he has always been easy to reach when needed. I acknowledge many conversations both in physics and on a good range of other topics. I especially thank him for allowing me to find my own way and being supportive for me to work with other professors.

It has been a great pleasure to learn from and work with Cristian Armendariz-Picon. I thank him for a lot of skills and knowledge he has imparted. He has been a second advisor to me and I sincerely thank him for taking me on, and working with me through a lot of things which I believe have given me solid grounds to stand on and his approach has been inspirational.

I also thank Professors Kameshwar Wali and Carl Rosenzweig for supporting me at all times. I have had the chance to listen to very nice stories on the world of Physics from them and I thank Carl Rosenzweig for his enlightening advices.

It was a great opportunity to join the University of Amsterdam as a visiting student for a semester. I thank Jan Pieter van der Schaar for being my advisor there and making this visit possible. It was a great joy to work with him. I have had many great discussions with him as well as Dionysios Anninos and other post docs and students there. 

Patty and Yudy have always been welcoming and throughly helpful with any of the administrative issues and I sincerely appreciated running into them at any point in the Physics building. I owe special thanks to Boqian, Leyla and Jerome for being there since the first day I walked onto campus. It would have been hard to survive initially without the logistical support of my neighbors Duygu and Ergin, since then I continue to spend many joyful and memorable times with them. I thank Ogan for the crucial logistics he shared with me even before my arrival and it has been an interesting experience to share an advisor with him.  

I appreciate a lot of technical things I learned from Jayanth and thank him for always finding the time to share them when needed. I have also taken a lot of traveling advice from him, which were very useful. I thank Swetha for joining me on many a journey both within physics and across new countries. I cherish a lot of enjoyable end of the week coffee time and dinners with Lorena and conversations with Suraj. I thank Brandon for the journal club discussions and proofreading sections of the thesis, and to Cem for also lending a second eye on some parts. I dearly thank Cem and Francesco for the hikes across the US. It has been very memorable to perform, dance and play African drums with Biboti Ouikahilo and Wacheva Cultural Arts and to continue on ballet with Karen Mentor and Ruth Arena at Ballet and Dance Center. 

I have also been inspired a lot from my professors in Bogazici University. I would especially like to thank İbrahim Semiz from whom I first learned about gravitation, Yemliha Bilal for grouping a bunch of us together and making this first introduction possible, to my Master's advisor Metin Arık for all he has shown me about research both within cosmology and beyond, and Teoman Turgut, Tonguç Rador, Muhittin Mungan, and Ali Kaya for many technical skills.  I acknowledge the 
presence of three good friends Cem, Emre, Mert from my first week at a university to moving to Syracuse and staying in touch with the world beyond Syracuse.  

I thank all my family, especially my mom and sister for making the separation both in distance and in time manageable, and my dad for drawing my attention to details in a work. I appreciate all that I have learned from Yıldız Alpar, Aylin Kalem, Oya Barbara Karanis about learning, understanding and presenting what lies underneath any concept. Last, but for sure not least, I thank Cynde Sadler who has opened up a lot of doors by teaching me English. I thank for the welcoming hospitality of the Onondaga Nation on whose lands this thesis has been written.

This thesis is dedicated to Meral Çakır who has played an important role for me to develop much of my creativity.



\newpage
    \tableofcontents

\newpage    \listoffigures

\newpage

\begin{center}
	{\bf\Large List of conventions and symbols}
\end{center}

	Throughout this dissertation, we have adopted the following conventions:
	\begin{itemize}
		\item Greek indices $\mu, \nu, \lambda, ...$ label the components of tensors with respect to the coordinate basis and take values $0,1,2,3$.
		\item Latin indices $i, j, k, ...$ run over the spatial coordinates and take the values $1,2,3$. 
		\item Repeated indices are summed over. 
		\item The metric has a $(-,+,+,+)$ signature unless stated otherwise
		\item We work in units such that $\hbar=c=k_B=1$. This results in units where
		
		[Mass] = [Energy] = [Momentum] = [Length$^{-1}$] = [Time$^{-1}$].
%
		
        \item Torsion-free, metric compatible Christoffel Conection is defined as 
        \beq
        \Gamma^{\sigma}_{\mu\nu} = \fr{1}{2}g^{\sigma\rho} \Big(\partial_\mu g_{\nu\rho}+\partial_\nu g_{\mu\rho}-\partial_\rho g_{\mu\nu}\Big)
        \eeq
		\item The Riemann tensor is defined as $R^{\lambda}_{\mu\sigma\nu} \equiv \partial_{\sigma} \Gamma^{\lambda}_{\mu\nu} - \partial_{\nu} \Gamma^{\lambda}_{\mu\sigma} + \Gamma^{\lambda}_{\sigma \rho} \Gamma^{\rho}_{\mu\nu} - \Gamma^{\lambda}_{\nu\rho} \Gamma^{\rho}_{\mu\sigma}$.
		\item The Ricci tensor is defined as $R_{\mu\nu} \equiv R^{\lambda}_{\mu\lambda\nu}$.
		\item The energy momentum tensor $T_{\mu\nu}$ is related to the matter action $S_m$ by\beq 
	\nn	T_{\mu\nu} = - \frac{2}{\sqrt{-g}} \frac{\delta S_m}{\delta g^{\mu\nu}}.
		\eeq
		\item We use the reduced Planck mass $m_{pl} = (8 \pi G)^{-1/2} \approx 2 \times 10^{18} \, \mbox{GeV}$ unless stated otherwise.
		\item We denote cosmic time with $t$ and conformal time with $\eta$. The relation between these two time variables is $d\eta = dt/a(t)$,  and we use the following abbreviations $\dot{f} \equiv \partial_t f$ and $f' \equiv \partial_{\eta} f$.
	\end{itemize}

\newpage

Throughout this dissertation, we have used the following abbreviations:
\begin{itemize}
	\item AdS: Anti-de Sitter
	\item BRST: Becchi, Rouet, Stora, Tyutin
	\item CMB: Cosmic Microwave Background
	\item dS: de Sitter
	\item GW: Gravitational Waves
	\item EFT: Effective Field Theory
	\item FLRW: Friedmann-Lema\^itre-Robertson-Walker
	\item GR: General Relativity
	\item QFT: Quantum field theory
\item PBH: Primordial Black Holes
	\item vev: vacuum expectation value
	\item WKB: Wentzel–Kramers–Brillouin 
\end{itemize}

\mainmatter
\pagenumbering{arabic}

\begin{keeppage}
	
\chapter{Review of Cosmological Perturbation Theory}
\label{chp:review}

\section{Introduction: what we know so far and what we would like to learn next about the universe}
\label{sec:intro}

Within our current understanding, the universe we live in has been through a phase of accelerated expansion, after which it was reheated into radiation domination, then followed by matter taking over, and is again experiencing an epoch of accelerated expansion. We refer to the first era of expansion as the era of inflation and the current one as dark energy. Everything around us has formed throughout this cosmic history; from elements via Big Bang Nucleosynthesis (BBN) during radiation domination, to galaxies during matter domination. A lot of our understanding about these eras have been put together in the framework of general relativity equipped with matter sources. Yet we are aware that we need to develop our understanding of gravity beyond general relativity to address small scales where quantum effects are important. Such improvements will especially be important for understanding the initial stages of inflation where the universe passes from quantum mechanical to classical scales. Moreover it is also an open question if the current dark energy dominated expansion implies that there is something more to gravity at scales larger than the solar system size.  

Through general relativity we have learned that curved spacetime is described by a rank two tensor. Hence, there is always a tensor degree of freedom, a graviton or a gravitational wave, that propagates in curved spacetime. Depending on what type of matter is required for a specific setting of curvature, there can also be scalar and vectorial degrees of freedom. For example, a cosmological constant gives the background for  an era of expansion. A pure cosmological constant is among the vacuum solutions of Einstein Equations and only gives rise to tensor degrees of freedom. However, if there is some time or spatial dependence involved then there will be some matter source responsible for the expansion which can give rise to scalar degrees of freedom. The inflaton is one such matter source. Of course, there are also additional bosons and fermions. In essence, the questions we ask boil down to questions on the type of degrees of freedom, such as the ones we just mentioned, and the kind of interactions they can have.

 Understanding how things fall into place, resembles a bit of a hike during nighttime. As we think on and question how things around us work, we focus on what catches our attention at the moment, like shining a torch straight onto the piece of ground we will take the next step to. Because we do not know the valley ahead of us we do not know which route to follow in looking for answers to our questions. And usually, gaining an understanding of one question gives rise to further questions. Little by little we develop general concepts that can act as guiding marks for us along a trail. In studying how a cosmological spacetime evolves and its possible interactions, our main guide and equipment is \emph{diffeomorphism invariance}, the freedom to choose a coordinate system for calculations without effecting the physical results. The assumptions of spatial homogeneity and isotropy at the level of the background also helps. 

Around 1948, the theory behind the origin of chemical elements was developed \cite{Alpher:1948ve}. It was understood that once an overheated, highly pressurized neutral nuclear fluid is present, a neutron gas, the neutrons involved will decay into protons and electrons as the universe expands and cools. Starting from deuterium, which is present about $3$ minutes later, light elements can form as strong interactions become important at $0.1$ MeV. This is BBN. Hydrogen, Helium and Lithium can form in this way. But to accommodate heavier elements such as Carbon, the universe has to first form stars to be factories for these, and neutron stars for even heavier metals. Long before this, neutrinos decouple from the rest of the matter sources present at $1$ MeV ( at about 1s), and even before $100$ GeV ( $10^{-10}s$ in the early universe) the Electroweak Phase Transition takes place in which the elementary $Z$ and $W^\pm$ bosons gain mass from the spontaneous breaking of Electroweak symmetry. 

In 1929, Hubble's observations of luminosity and distance measurements of supernova \cite{Hubble:1929ig}, showed that we live in an expanding universe. Moreover by 1933, through observations of galactic rotation curves, it was deduced that there is non-matter that interacts only gravitationally, which plays an important role in the structure of galaxy clusters \cite{Zwicky:1937zza}. This type of matter was given the name \emph{dark matter} to account for its insensitivity towards electromagnetic interactions. In the early 1970's, the type of initial spectrum needed for density perturbations to undergo gravitational collapse into protogalaxies in an expanding universe and form the seeds of larger scale structures was worked out \cite{Harrison:1969fb,Zeldovich:1972zz}. These density perturbations, of unknown origin at the time, were later realized to have arised during inflation, the early era of accelerated expansion \cite{Hawking:1982cz}. 

By the early 1980's, the universe was observed to be spatially flat and physicists believed it to have initiated from an adiabatically expanding radiation-dominated state, which would have led to causally disconnected regions. The density perturbations that set the initial conditions for galaxy formation rely on an otherwise homogeneous and flat background. In a universe with such a history, this raises the question of how regions too far apart to be in causal contact, to have communicated with each other even by light since the beginning of the universe, can be in equilibrium? It was proposed that this state of radiation domination should start at some temperature below the Planck Mass, $10^{18}GeV$, and hence at times later then $t=0$, leaving room for quantum gravity effects.  

If there existed a non-adiabatic state with huge entropy before the adiabatic radiation domination, then these problems could be solved. Hence the presence of an early era of accelerated expansion, referred to as \emph{inflation}, was proposed to have taken place before the start of radiation domination \cite{Guth:1980zm}, such that all the regions of the universe were initially in causal contact. In addition, such an era could also solve the monopole problems predicted as a consequence of Grand Unified theories. The original idea of inflation, refined shortly after in \cite{Linde:1981mu,Albrecht:1982wi}, has been enriched in a lot of ways. We now believe that at least a single scalar field with negligible, but not exactly vanishing, time dependence gave rise to such an era of accelerated expansion while it slowly rolled down its potential. As the universe expanded exponentially, inflationary perturbations, froze mode by mode as their physical wavelengths $\lambda\propto e^{Ht}$ became larger than the slowly changing horizon size $R=H^{-1}$. The inflationary perturbations are both scalar (temperature fluctuations) and tensor (gravitational waves) degrees of freedom. These perturbations exited the horizon carrying inflationary information with them. The change in the rate of expansion as the universe evolved was such that the modes that became superhorizon during inflation started reentering into the Hubble radius just before photons decoupled from electrons and nucleons and formed a cosmic background eventually at the microwave range today with a black body spectrum of $2.3K$. This background radiation referred to as the \emph{Cosmic Microwave Background} (CMB) carries the inflationary information untouched up to the time it was formed. It was acknowledged as an observational phenomena by \cite{Dicke:1965zz}, yet was spotted unexpectedly as an isotropic and homogeneous background noise during antenna measurements for purposes of radio astronomy in 1965 \cite{Penziaz65}. Since its discovery, various experiments have been taking measurements of the CMB regularly such as COBE, WMAP and PLANCK. The inflationary information can be extracted from these CMB observations, depending on how well  we understand the later processes that have an effect on CMB photons. 


The passage out of inflation towards radiation domination is the era of \emph{reheating}, which is where the energy stored in the field that drove inflation gets transferred to other fields that eventually make up the rest of the matter content of the universe. Within the picture we presented, the nature of couplings by which the inflaton transfers its energy to other fields, such as dark matter and standard model particles, around the time of reheating is still not fully understood.
As the inflationary modes that contribute to the CMB are outside the horizon at this time, the CMB does not carry information about this stage. At the end of inflation, it is likely that the field that drove inflation underwent an oscillatory phase giving rise to a matter dominated era at first order. This is called the era of \emph{preheating}, during which the conditions are agreeable to obtain resonant growth in perturbations of other fields that had negligible presence during inflation. 
Note that it is not crucial that preheating takes place at all. One major opponent against preheating is the possibility for the inflaton to have perturbatively decayed into other species. At first sight, perturbative decays can only allow for the presence of species lighter than the inflaton itself and hence, lead to radiation domination right away. The main obstacle is that the efficiency for such production may require a long time. The initial proposal for the era of preheating is based on the observation that perturbative decays alone are not efficient enough without a preceding stage of preheating \cite{Kofman:1994rk}. In addition, preheating may result in production of gravitational waves sourced by the preheating products: compact objects such as the primordial black holes, similar to ones we will mention in section \ref{sec:BPH}, and others \cite{Lozanov:2017hjm}. Recently it has also been confirmed that many couplings expected during preheating do not allow the inflaton to decay completely before the end of inflation \cite{Armendariz-Picon:2017llj}.

\section{Diffeomorphisms and Gauge Transformations}
\label{sec:diffeomorphisms}
Here, we study the evolution of linear perturbations around a given background field. That is, we consider fields $\phi(t,\vec{x})=\phi_0(t)+\delta\phi(t,\vec{x})$ and $g_{\mu\nu}(t,\vec{x})=\bar{g}_{\mu\nu}(t)+h_{\mu\nu}(t,\vec{x})$, where in the case of cosmology the evolution of the background fields $\phi_0$ and $\bar{g}_{\mu\nu}$ are determined by Freedman-Lemaitre-Robertson-Walker (FLRW) evolution. This is a homogeneous, isotropic, and dynamic background, where for one temporal and three spatial dimensions the line element is
\be \label{metricfrw} ds^2=-dt^2+a^2(t)\left(\frac{dr^2}{1-kr^2}+r^2d\Omega^2\right)\equiv g_{\mu\nu}dx^\mu dx^\nu.\ee
It is homogeneous because none of the metric components depend on position, yet due to its time dependence, via the factor $a(t)$, it is dynamic. With time, the spatial sections of this metric are scaled by the factor $a(t)$, hence it is referred to as the \emph{scalar factor}. The rate of expansion $H(t)\equiv\frac{\dot{a}}{a}$, which is an observable quantity, is referred to as the Hubble parameter, after Hubble who first measured it through distance measurements of Cepheid variables. The Hubble rate is a time dependent quantity, whose time dependence changes throughout the cosmic evolution, in relation to properties of the matter type that contributes the most to the overall energy density in the universe. The quantity $k$ denotes the overall curvature of spatial sections, it can be $[-1,0,+1]$, depending on the sign of the curvature scalar of the spatial sections, it denotes negatively curved, flat and positively curved spatial sections respectively. We will always consider spatially flat geometries and hence set $k=0$, $d\Omega^n$ denotes the $n-$sphere. This metric is isotropic because all of the spatial sections are rescaled by the same factor, in other words there is not a preference among them.

While the background evolution gives the overall characteristics of the era under question, it is the quantities related to  the expectation values of the perturbations that can be connected with observations.

A diffeomorphism is a spacetime coordinate transformation of the form
\be \label{diff} x^\mu~\to~x'^\mu=x^\mu+\xi^\mu(x)\ee
with a small parameter $\xi$. Under this transformation the fields will transform according to their rank. The functional form of scalars will remain invariant,
\be\label{scalardiff} \phi'(x')=\phi(x'),\ee
and the metric tensor will transform as 
\be \label{metrictr} g'_{\mu\nu}(x')=g_{\alpha\beta}(x)\frac{\partial x^\alpha}{\partial x'^\mu}\frac{\partial x^\beta}{\partial x'^\nu}.\ee 
where the transformation \eqref{metrictr} demonstrates how a rank 2 object transforms under a general diffeomorphism \eqref{diff}. 

By convention we consider the perturbations to be described always around the same background, $\phi'(x')=\phi_0(t)+\delta\phi'(x')$ and $g'_{\mu\nu}(x')=\bar{g}_{\mu\nu}(x)+h'_{\mu\nu}(x')$ \cite{Mukhanov:1990me},\cite{Wbook}. This is called the passive transformation, because the background is fixed and hence passive under the transformation. 
For the scalar field at the background level, a passive transformation implies
\be \label{passivetr}\phi'_0(x')=\phi_0(x').\ee
Our main concern is in the effect of the diffeomorphisms \eqref{diff} on the perturbations. 
We can turn \eqref{scalardiff} around to obtain this change 
\begin{align} \phi'(x')&-\phi(x)=0\\
\phi'_0(x')+\delta\phi'(x')-&\left(\phi_0(x)+\delta\phi(x)\right)=0
\end{align}
Making use of \eqref{passivetr} and noting that $x'=x+\xi$, the change in the scalar perturbation is
\begin{align}\label{phiontheway} \phi_0(x+\xi)+\delta\phi'(x')-\phi_0(x)-\delta\phi(x)=0\\
\delta\phi'(x')-\delta\phi(x)=-\left[\phi_0(x+\xi)-\phi_0(x)\right]\\
\Delta\delta\phi=-\left[\phi_0(x+\xi)-\phi_0(x)\right].\end{align}
Under a passive transformation, the change in the metric perturbations is defined to be \cite{Wbook}
\be \label{metric1} \Delta h_{\mu\nu}(x)\equiv g_{\mu\nu}'(x)-g_{\mu\nu}(x).\ee

Since $\xi$ is a small parameter we can Taylor expand and consider the coordinate dependence of the fields as 
\be\label{phitaylor} \phi'_0(x')=\phi'_0(x+\xi)=\phi_0(x)+\partial_\alpha\phi_0\xi^\alpha,\ee
\be g'_{\mu\nu}(x')=g'_{\mu\nu}(x)+\frac{\partial g'_{\mu\nu}}{\partial x^\alpha}\xi^\alpha.\ee
Using this in \eqref{metric1} we can rewrite the first term as
\be \Delta h_{\mu\nu}(x)=g'_{\mu\nu}(x')-\frac{\partial g'_{\mu\nu}}{\partial x^\alpha}\xi^\alpha.\ee
Now by \eqref{metrictr} the first term gives
\be\Delta h_{\mu\nu}(x)=g_{\beta\lambda}(x)\frac{\partial x^\beta}{\partial x'^\mu}\frac{\partial x^\lambda}{\partial x'^\nu}-\frac{\partial g'_{\mu\nu}}{\partial x^\alpha}\xi^\alpha.\ee
Remember that $x'^\mu=x^\mu+\eta^\mu$, which we can make use of in the form of $x^\mu=x'^\mu-\eta^\mu$ to compute the derivatives. As such, at first order in $\xi^\mu$ we have
\be\label{tensortr} \Delta h_{\mu\nu}(x)=-\left[\bar{g}_{\mu\lambda}\partial_\nu\xi^\lambda+\bar{g}_{\beta\nu}\partial_\mu\xi^\beta+\partial_\lambda\bar{g}_{\mu\nu}\xi^\lambda\right]\ee
And by plugging \eqref{phitaylor} in \eqref{phiontheway} we find that to first order in $\xi^\eta$ the scalar perturbations transform  according to
\be\label{scalartr}\Delta \delta\phi(x)=-\xi^\mu\partial_\mu\phi_0.\ee

The right-hand side of these expressions are minus the Lie derivative \cite{Wald:106274} 
\be\mathcal{L}_\xi{T^{a_1...a_k}}_{b_1...b_l}=\xi^c\nabla_c{T^{a_1...a_k}}_{b_1...b_l}-\sum^{k}_{i=1}{T^{a_1...c...a_k}}_{b_1...b_l}\nabla_c\xi^{a_i}+\sum^{l}_{j=1}{T^{a_1...a_k}}_{b_1...c...b_l}\nabla_{b_j}\xi^c\ee
such that the change in the perturbations of the tensor ${T^{a_1...a_k}}_{b_1...b_l}$ under a diffeomorphism with the parameter $\xi$ are
\be \Delta\delta{T^{a_1...a_k}}_{b_1...b_l}=-\mathcal{L}_{\xi}{T^{a_1...a_k}}_{b_1...b_l}.\ee

The transformations \eqref{tensortr} and \eqref{scalartr}, are gauge transformations. The diffeomorphism parameter $\xi$ can be chosen so as to set some of the cosmological perturbations to zero for convenience. This amounts to picking a set of coordinates where some of the perturbations have been eliminated. Since the physical observables should not be effected by the choice of coordinates, setting some of the cosmological perturbations to zero, is analogous to electromagnetic gauge choices, where some of the components of the electromagnetic potential are set to zero without changing the electric and magnetic fields.

We will end this section with some examples. The choice of a constant shift
\be\label{translation}\xi^{\mu}(x)=a^{\mu}\ee
for the diffeomorphism parameter gives translations, whose generator are the momentum tensor $P_\mu=i\partial_\mu$. Rotations correspond to 
\be \xi^\mu=x^\nu{\omega_\nu}^\mu\ee
where $\omega_{\mu\nu}$ is antisymmetric in its indices, and the corresponding generator is the angular momentum tensor $J_{\mu\nu}=i\left(x_\mu\partial_\nu-x_\nu\partial_\mu\right)$. These first two make up the \emph{Lorentz group}, which is the \emph{Poincare group} and translations. In addition, one can also consider scaling of the spatial coordinates at each point in spacetime \be \label{dilatation}\xi^i=\lambda x^i,~\xi^0=\lambda\eta.\ee
This is called a \emph{dilatation} and is generated by $D=i x^\mu\partial_\mu$. Then, there is also the \emph{special conformal transformation} with the parameter
\be\label{sct}\xi^i=2\vec{x}.\vec{b}x^i-b^i\vec{x}^2\ee
and the generator $K_\mu=i\left(2x_\mu\left(x^\nu\partial_\nu\right)-x^2\partial_\mu\right)$. 
Together, these diffeomorphisms compose the \emph{Conformal Group}. Just as quantum field theory is based on invariance under the Lorentz group, one can also formulate field theories based on invariance under the Conformal group. The later are known as \emph{Conformal Field Theories} (CFT). The diffeomorphisms \eqref{translation}-\eqref{sct} are also the late-time isometries of de Sitter spacetime (dS), which is one of the maximally symmetric (meaning it has the maximum amount of possible killing vectors) solutions to Einstein Equations with a positive cosmological constant, $G_{\mu\nu}+\Lambda g_{\mu\nu}=0$, with positively curved spatial sections who expand in time. Based on its late time symmetries, which are the symmetries of the Conformal group, it has been proposed that at its late time boundary, gravitational physics of de Sitter are equivalent to a conformal field theory \cite{Strominger:2001pn}. This is the basis of the so-called dS/CFT duality.   

Cosmology is built on the basic assumption of homogeneity and isotropy of spatial sections. It is not required for the universe to be static, which allows for the metric to be time dependent. Radiation domination, matter domination and accelerated expansion are all solutions of the Einstein equations with matter, that satisfy these assumptions. From observations we are aware that among these our universe has passed through accelerated expansion in the past (inflation), followed by radiation and matter domination, and is currently undergoing accelerated expansion again (dark energy). Due to homogeneity, all of these eras are invariant under translations in space. Due to isotropy they are all invariant under spatial rotations. The de Sitter solution, which gives eternal expansion of the spatial sections at constant rate, sets the background for the primordial and current eras of expansion. The symmetry group of de Sitter is SO(4,1). Even though the metrics that describe radiation and matter domination do not possess all of these symmetries, in \cite{Hinterbichler:2012nm} it is argued that conformal symmetries have an important role for perturbations in any cosmological era that is dominated by a single scalar field. 

Another maximally symmetric solution is the \emph{Anti de Sitter} spacetime (AdS) which has negative curvature on its spatial sections, also possess conformal symmetries. We have not observed any signs of this geometry dominating at any point within the history of our universe. It arises as the near horizon geometry of black holes.

\section{Cosmological Perturbations}
\label{sec:cosmological perts}
On the whole, we are interested in the components of a rank two tensor, $G_{\mu\nu}$, which is symmetric, and obeys the Bianchi identity
\be \label{bianchi} \nabla^\mu G_{\mu\nu}=0.\ee 
In four dimensions, a symmetric rank two tensor has ten independent entries. The Bianchi identity, which implies four differential equations, will remove four of them. Remember that we also have diffeomorphism invariance, which parametrized by $\xi^\mu$ lets us pick four parameters in four dimensions. Thus by picking the coordinates accordingly we can remove four more degrees of freedom leaving us with two physical degrees of freedom that will propagate.    

What is the nature of these two physical degrees of freedom. A given tensor can be split into two parts, a part composed of its trace and a trace-free part. Notice that in the absence of matter, $T_{\mu\nu}=0$, via the Einstein equations
\be G_{\mu\nu}=\frac{1}{m^2_{pl}}T_{\mu\nu},\ee
where $m^2_{pl}=\frac{\hbar c}{8\pi G}$, the Einstein tensor is traceless, 
\be Tr(G_{\mu\nu})={G^\mu}_\mu={T^\mu}_\mu=0.\ee
The two components of this symmetric, traceless, rank two tensor in four dimensions that satisfies the Bianchi Identity make up two tensor degrees of freedom that propagate in three dimensions. These are the so called gravitational waves, who are the vacuum solutions of General Relativity.

In the presence of matter, $T_{\mu\nu}\neq 0$, the trace becomes important and suggests that there are more modes that propagate depending on the nature of the matter source under consideration. Our main interest in this thesis will be scalar fields as matter sources throughout. In the case of a single scalar field, there is one more component added to the system.

This additional degree of freedom can only come from the eight components which can be set to zero. The Bianchi Identity, which is inherently related to the Jacobi Identity, comes a property of the Einstein Tensor, which we don't want to violate with the matter content. This leads us to think a little bit more on diffeomorphism invariance and time evolution. The hamiltonian of a given system is the generator of time evolution for that system. As we will see below, the hamiltonian of general relativity is just a sum constraints and constraints generate coordinate transformations. Hence  time evolution in general relativity is just a map from one coordinate system to another which can be thought of as ordering spatial surfaces with respect to a parameter understood as time. It is not physical because the Hamiltonian vanishes once the constraints are satisfied. In the physical world we mainly observe quantities related to perturbations, and we build our intuition of physicality, such as our notion of time evolution, based on these observations. While at the level of the Lagrangian we always demand time diffeomorphism invariance, fields that do not respect this symmetry at the background level may be present. It is these type of fields on which our understanding of cosmic evolution is based. Given a time dependent background scalar $\phi_0(t)$, there will be a scalar mode $\delta\phi(\vec{x},t)$ that keeps track of the difference in the time coordinate, or in the rate of expansion depending on how the time coordinate is defined, between two points in space. This fact has first been pointed out in \cite{Hawking:1982cz}, with inflation in mind\footnote{For a recent review see \cite{Senatore:2016aui}, where this is discussed in section 2.3.}. In fact, such a scalar mode is the one observed through temperature fluctuations in the CMB, captured most successfully by inflation. This observation allows for one to focus on the perturbations from the start, by picking the time coordinate to be aligned with the background, such that surfaces of constant time are also surfaces where the value of the scalar field, that sets the nonzero energy momentum density is constant \cite{Creminelli:2006xe}. This is a noncovariant construction of the theory for perturbations based on the remaining spatial diffeomorphism invariance. Under time diffeomorphisms the scalar mode transforms nonlinearly as to make any terms that arise due to the transformation of the background cancel out. This is apparent in equation \eqref{scalartr}, on a time dependent background the scalar perturbation transforms only under a time diffeomorphims, and it is a non-linear transformation. This is complementary to constructing a theory for the full fields, which can determine the background dynamics, and introducing perturbations later. In this latter approach one needs to consider the full diffeomorphism invariance from the beginning. The theories build, in the covariant approach, starting from the full fields will match the noncovariant Lagrangian at the level of perturbations. 

In Chapter 2 we will start from a covariant formalism and work to understand how diffeomorphism invariance is accounted for in the expectation value calculations of operators. In Chapter 3, we will consider the noncovariant approach to focus on perturbations during preheating, yet, we will also comment on how this matches with Lagrangians built in the covariant formalism when one works out their form at the level of perturbations.

Now that we know the nature of our degrees of freedom, let us describe how to include them. The metric can be decomposed in terms of its scalar, vector and tensor degrees of freedom in the following way
\be \label{dec100} g_{00}=\bar{g}_{00}-E,\ee
\be \label{dec1io} g_{i0}=\bar{g}_{ij}+a(t)\left[\partial_i F+G_i\right],\ee
\be \label{dec1ij} g_{ij}=\bar{}g_{ij}+a^2(t)\left[\delta_{ij}A+\partial_i\partial_jB+\partial_iC_j+\partial_jC_i+D_{ij}\right]\ee
with conditions
\be \partial_iC_i=\partial_iG_i=0,~~\partial_iD_{ij}=0,~~D_{ii}=0\ee
where repeated indices are implied to be summed over. Of course, anything that is different from the FLRW form, that is the variables without a bar, are considered to be perturbations. The traceless and transverse tensor $D_{ij}$ makes up the two tensor degrees of freedom. In comparison to our notation with $g_{\mu\nu}=\bar{g}_{\mu\nu}+h_{\mu\nu}$, here $h_{ij}=a^2(t)\left[\delta_{ij}A+\partial_i\partial_jB+\partial_iC_j+\partial_jC_i+D_{ij}\right]$. Which one of the scalar components $[E,A,B]$ describes the scalar mode is gauge dependent. 

There are two gauge invariant scalar quantities $\zeta$ and $\mR$,    
\be\label{eq:zeta} \zeta=\frac{A}{2}-H\frac{\delta\rho}{\dot{\bar{\rho}}},\ee
\be \label{eq:R} \mR=\frac{A}{2}+H\delta u\ee
in this general parametrization \cite{Wbook}. Here $\delta u$ is the velocity perturbation. These two quantities approach each other on super horizon scales. As discussed in \cite{Weinberg:2003sw}, there are always two adiabatic modes at all times. The scalar solutions are \cite{Weinberg:2004kr}
\be \delta\phi_1=-\frac{\mR\dot{\phi}_0}{a(t)}\int_{t_0}^{t}a(t')dt'~~\text{where}~~\mR=constant\ee
and
\be \delta\phi_2=-C\frac{\dot{\phi}_0}{a(t)}~~\text{where}~~C=\text{constant},~~\mR=0.\ee
The theorem demonstrates that on superhorizon scales these modes are physical, i.e. they cannot be set to zero by a gauge choice. These perturbations are called adiabatic because they lie along the inflationary direction, they are curvature perturbations as opposed to perturbations of a secondary matter field and the number of particles they present are fixed. Effectively it is these type of solutions that make up the scalar degree of freedom that is present in the system when there is a time dependent scalar field.

With the general parametrization of the metric above, the energy conservation equation at linear order in perturbations is
\be \label{eq:encon}\partial_0\delta {T^0}_0+\partial_i\delta{T^i}_0+3H\delta{T^0}_0-H\delta{T^i}_i-\frac{\bar\rho+\bar{p}}{2a^2}\left(-\frac{2\dot{a}}{a}h_{ii}+{\dot h}_{ii}\right)=0\ee
where  
\be \label{Tupdown} \delta{T^\mu}_\nu=\bar{g}^{\mu\lambda}\left[\delta T_{\lambda\nu}-h_{\lambda\kappa}{\bar{T}^{\kappa}}_\nu\right]\ee
and repeated indices are summed over.
Through \eqref{eq:zeta} between $\zeta$ and $A$, and 
\be \dot{\bar{\rho}}=-3H\left(\bar{\rho}+\bar{p}\right),\ee
which is the energy conservation at the background level, we can rewrite the energy conservation equation as an equation for change in $\zeta$
\be \label{zetadot}3\left(\bar{\rho}+\bar{p}\right)\dot{\zeta}=\partial_i\delta{T^i}_0-H\delta{T^i}_i+3H\frac{\dot{\bar p}}{\dot{\bar\rho}}\delta\rho-\frac{\bar\rho+\bar{p}}{2}\partial^2_i\dot{B}.\ee
Note that $\delta {T^i}_i=3\delta p$ which turns this equation into
\be\label{dzeta_X}\dot{\zeta}=\frac{\partial_i \delta{T^i}_0}{3(\bar{\rho}+\bar{p})}+\frac{\dot{\bar{\rho}}\delta p-\dot{\bar{p}}\delta\rho}{3(\bar{\rho}+\bar{p})^2}.\ee
The first term will vanish for superhorizon modes and the second term has been argued to decay away leading to $\dot{\zeta}=0$ in agreement with adiabaticity \cite{Weinberg:2003sw}. In terms of the equation of state, defined as $\omega\equiv \frac{\bar{p}}{\bar{\rho}}$ \cite{Finelli:2000ya}
\be\label{zetadoto} (1+ \omega)\dot{\zeta}=0\ee
This suggests that $\dot{\zeta}=0$ and the scalar mode is adiabatic in the presence of cosmological evolution with $\omega\neq -1$. This guarantees the constancy of $\zeta$ during the radiation $\omega_{r}=\frac{1}{3}$ and matter domination $\omega_{m}=0$ that is present in between the two accelerated expansion eras of inflation and dark energy domination with $\omega_a=-1$. Within the post-inflationary era, the equation of state can pass through $\omega_a=-1$. In fact when the energy density is dominated by a single scalar field, if at times when there is no slow roll condition; the background values of this field, $\phi_0$, minimizes its potential, the potential can be Taylor expanded and the leading term will be $V(\phi_0)\sim m_\phi^2\phi_0^2$. With such an approximation for the potential, the background field $\phi_0$ will exhibit an oscillatory behavior as we will see in section \ref{sec:background evolution}. In such cases \eqref{zetadoto} is not applicable. Such eras can indeed arise at the end of inflation, referred to as preheating, or lead to nonthermal cosmologies. In this thesis we will talk about these unconventional epochs and we will return to this point on the constancy of $\zeta$ and the ability of the scalar mode to carry on inflationary information without any effects coming from the post inflationary eras in section \ref{sec:adiabaticmodes}.  

Two common gauge choices in literature are the Newtonian Gauge, also referred to as the Longitudinal gauge, where $F=B=0$; and the Synchronous gauge where $E=F=0$. 


\subsection{ADM Formalism and Constraints}
Previously we introduced scalar, vector, and tensor degrees of freedom based on how one can split a rank two tensor into smaller ranked objects. Of course, this is not the only way. Let us consider how one may parametrize the metric to bring out time evolution. This is the ADM formulation \cite{Arnowitt:1962hi}, which also brings out the constraint nature of diffeomorphisms. Here we will introduce the metric in this formalism following the discussion of \cite{Misner:1974qy}.

The Einstein equations give solutions for the geometry of four dimensional spacetime for given fields that generate an energy density. They inherently take into account coordinate changes. One can consider the four dimensional spacetime as three dimensional hyper-surfaces and one parameter, $\lambda^N(t,x,y,z)$, where the different 3 geometries are ordered along $\lambda^N$, such that the dynamical change from one geometry to the next lies along increasing values of $\lambda^N$. As such, this single parameter, $\lambda^N(t,x,y,z)$ captures the lapse of time and is named the \emph{Lapse function}. 
This stacking up of three geometries along the Lapse function constructs the four dimensional spacetime. In other words the spacetime is sliced into a one parameter family of spacelike hyper-surfases. In this construction of spacetime, one can set up initial value problems by specifying the value of fields and three geometry and their rate of change in time at a certain point in time. In a sense, each three hyper-surface is a coordinate transformed version of another. Time evolution implies a certain ordering of these hyper-surfaces one after the other along the Lapse function. 

To set up the four dimensional spacetime metric out of this one parameter family of slices, one needs to figure out the proper distance on a three dimensional base hypersurface and the proper time in between two consecutive hypersurphases. The lapse of proper time between the two consecutive three dimensional hypersurfaces is given by $d\tau=\lambda^N(t,x,y,z,)dt$. The base hypersurface is described by the 3 metric $h_{ij}$. The eventual position on the consecutive hypersurface $x^i_{next}$, of a point on the base hypersurface $x^i_{base}$ involves a shift $\lambda^i(t,x,y,z)$ with respect to its original position $dx^i_{next}=dx^i_{base}-\lambda^i(t,x,y,z)dt$. The contribution of the proper distance in the base hyper surface to the spacetime metric is thus $dl^2=h_{ij}d^i_{base}dx^j_{base}$ where $dx^i_{base}=dx^i_{next}+\lambda^idt$ \cite{Misner:1974qy}. Dropping the subscripts, we arrive at the 4 dimensional spacetime metric as
\begin{align}
ds^2&=-d\tau^2+dl^2\\
ds^2=-\left(\lambda^N\right)^2dt^2+&h_{ij}\left(dx^i+\lambda^idt\right)\left(dx^j+\lambda^jdt\right).
\end{align}

Here the spatial indices regarding the 3 hyper surface are raised and lowered by the 3 metric, for example $N^i=h^{ij}N_j$. The inverse metric elements are
\begin{align}
g^{00}=-\frac{1}{(\lambda^N)^2},~~ g^{0i}=g^{i0}=\frac{\lambda^i}{\left(\lambda^N\right)^2},~~g^{ij}=h^{ij}-\frac{\lambda^i\lambda^j}{\left(\lambda^N\right)^2},
\end{align} 
and the volume element is 
\be \sqrt{-g}dx^0dx^1dx^2dx^3=\lambda^N\sqrt{h}dtdx^1dx^2dx^3.\ee
For example in the case of flat spatial hypersurfaces $h_{ij}=a^2(t)\delta_{ij}$ and $\sqrt{-g}=\lambda^Na^3$.
This 4 dimensional spacetime posses the unit timelike normal vector 
\be {\bf n}=n_\mu dx^\mu=-\lambda^Ndt.\ee

Now that we have constructed the metric, it is time to talk about curvature. What we have done so far is that we considered the 4 dimensional spacetime as 3 dimensional hyper surfaces embedded in it. The hyper surfaces will have intrinsic curvature, as measured by the Ricci scalar of $h_{ij}$. We will denote this 3 dimensional intrinsic curvature by $R^{(3)}$. Looking at the hyper-surface within the four dimensional spacetime, it will also have an extrinsic curvature depending on how these surfaces are imbedded. This is measured by the extrinsic curvature tensor $K$. It measures the deformation of a figure lying on the hyper-surface when each point of the figure on the hyper-surface is carried forward by a unit of proper time into the four dimensional outer surface. This motion is along the direction normal to the hyper surface. The change in a vector ${\bf n}$ normal to the hyper-surface under this displacement will be perpendicular to ${\bf n}$, in other words ${\bf dn}$ lies along the hyper-surface. 
The extrinsic curvature K is defined as the linear operator that measures the change in ${\bf n}$ under being transported parallel to itself in the outer 4 dimensional geometry 
\be ({\bf dn})_i=\nabla_k n_i dx^k=-K_{ik}dx^k.\ee
It has the value
\be K_{ik}=-\frac{1}{2\lambda^N}\left[\nabla_k\lambda_i+\nabla_i\lambda_k-\frac{\partial h_{ik}}{\partial t}\right].\ee

Imagine a piece of paper. The paper is a 2 dimensional surface, and as the Ricci scalar always vanishes in 2 dimensions it is intrinsically flat. Imagine the paper lying on the surface of a table, if you draw a triangle on it, the sum of the internal angles of the triangle will add up to $\pi$, as they do in flat Euclidean geometry. However this piece of paper is lying in a 3 dimensional space. If we take it and fold it around without tearing it, the sum of the internal angles of the triangle we drew on it will still be $\pi$, confirming that the paper is still intrinsically flat. However from our point of view, it looks bend and folded, it has some extrinsic curvature within the 3 dimensional space.

The four dimensional Ricci scalar $R$ in the ADM formalism has the form
\be R=R^{(3)}+K_{ij}K^{ij}-K^2\ee
where $K=h^{ij}K_{ij}$.

If we consider pure General Relativity, with no matter fields the action is
\be S_{GR}=\frac{m_{pl}^2}{2}\int d^4x \lambda^N\sqrt{h}\left[R^{(3)}+K_{ij}K^{ij}-K^2\right].\ee
Notice that this action does not have any derivatives of the shift and lapse functions. The canonical conjugate momenta for them vanish
\be \pi_N\equiv\frac{\partial\mathcal{L}}{\partial\dot{\lambda}^N}=0,\ee
\be \pi_i\equiv\frac{\partial\mathcal{L}}{\partial\dot{\lambda}^i}=0.\ee
These are constraints that imply that the variables $\lambda^N$ and $\lambda^i$ are not physical variables, but that they are Lagrange multipliers. During the analysis of the system we must solve for them in terms of the other variables. Dirac has formulated how to quantize a constraint system based on an analysis of constraints. In his formalism, constraints such as these, that arise before the equations of motion have been employed are refereed to as \emph{primary constraints}, $b_a=\{\pi_i,\pi_N\}$. For the self consistency of the system these constraints must be preserved in time, $G_a\equiv \dot{b}_a=0$. This puts a further set of constraints on the system referred to as \emph{secondary constraints}.

As the primary constraints turned out to be canonical momenta, by the Hamilton equations of motion we know that the secondary constraints will be
\begin{align}
\dot{\pi}_N &=-\frac{\partial \mathcal{H}}{\partial \lambda^N}=0\\
\dot{\pi}_i &=-\frac{\partial \mathcal{H}}{\partial \lambda^i}=0
\end{align} 
In order to derive the secondary constraints we need to go to the Hamiltonian formalism. The conjugate momentum of $h_{ij}$ is
\be \pi^{ij}\equiv \frac{\partial \mathcal{L}}{\partial \dot{h}_{ij}}=\frac{m^2_{pl}}{2}\sqrt{h}\left(K^{ij}-Kh^{ij}\right).\ee
We find that the Hamiltonian density 
\be \mathcal{H} \equiv \int d^3x\left(\pi^{ij}\dot{h}_{ij}+\pi_N\dot{\lambda}^N+\pi_i\dot{\lambda}^i-\mathcal{L}\right)=\int d^3x\mathcal{H}_g\ee
becomes
\be \mathcal{H}_g=\frac{2}{m_{pl}}\frac{\lambda^N}{\sqrt{h}}\left(\pi_{ij}\pi^{ij}-\frac{1}{2}\pi^2\right)-\frac{m^2_{pl}}{2}\sqrt{h}\lambda^NR^{(3)}-2\sqrt{h}\lambda^j\nabla_i\left(\frac{{\pi^i}_i}{\sqrt{h}}\right).\ee
This means the Hamiltonian itself is a linear combination 
\be \mathcal{H}_g=\lambda^NG_N+\lambda^iG_i\ee
of the secondary constraints, who are
\be G_N\equiv-\frac{\partial \mathcal{H}_g}{\partial\lambda^N}=\frac{2}{m^2_{pl}}\frac{1}{\sqrt{h}}\left(\pi_{ij}\pi^{ij}-\frac{1}{2}\pi^2\right)-\frac{m^2_{pl}}{2}\sqrt{h}R^{(3)}=0,\ee
\be G_i\equiv-\frac{\partial \mathcal{H}_g}{\partial\lambda^i}=-2\sqrt{h}\nabla_j\left(\frac{{\pi_i}^j}{\sqrt{h}}\right)=0.\ee
This means the Hamiltonian of general relativity vanishes on shell and it does not generate time evolution. Rather then time evolution, the dynamical equations of this system, obtained by varying the action with respect to the metric, describe further coordinate changes. In a sense the ADM formalism captures evolution of the 3 dimensional hypersurface along a parameter called time, as opposed to time evolution of the four dimensional spacetime, by stacking together the versions of the hypersurface under spatial diffeomorphisms with a specific order. After all if one is interested in the physics on the spatial sections, time evolution does mean how the spatial section keeps deforming. Each version will be connected to another under spatial diffeomorphisms and hence time can be considered as a parameter that sets the order of these spatial diffeomorphisms. This issue is referred to as the problem of time in general relativity and we refer the reader to \cite{Isham:1992ms} for a detailed and pedagogical discussion. However this inherent property of the ADM formalism also makes it very convenient to study perturbations on a time dependent background, starting with a time diffeomorphism fixing, which we will look into in chapter \ref{chp:quantization methods}. The evolution equations of general relativity do not determine the variables $\{\lambda^N,\lambda^i\}$. They are left arbitrary. 

There is an important link between diffeomorphism invariant and constrained systems.  The presence of diffeomorphism invariance signifies that some of the variables in the original formulation of a theory are redundant. These variables are necessery so as to ensure that correct transformation properties are obeyed. Yet they are redundant in the sense that they can be set to zero via performing a diffeomorphism. Hence they are not physical. Their arbitrariness is removed by making a coordinate choice. The freedom to choose the coordinate system in general relativity is similar to the freedom to choose the potential in electromagnetism. Which is why we use the words diffeomorphism and gauge transformation interchangeably.

The constraints in a given theory are analyzed by being classified into two classes. This formalism for treating constraints was developed by Dirac with the purpose of quantizing gauge systems. A function or a constraint F, is said to be \emph{first class} if its commutator vanishes weakly with  each constraint that exists in the system. If there is at least one constraint that has a non vanishing commutator with F, then F is said to be \emph{second class}. All first class primary constraints generate gauge transformations. In addition, the Poisson bracket of any two first class primary constraints, and also the Poisson bracket of any first class primary constraint with the first class Hamiltonian generate gauge transformations. Thus in some cases first class secondary constraints can also generate gauge transformations. Because of this, the distinction between first and second class constraints is more important then the distinction between primary and secondary constraints. We will analyze the constraints with a scalar field coupled to gravity in section \ref{subsec:analysisconstraints} to see that they are first class on shell. And this will set the basis of quantization of cosmological perturbations with gauge fixing taken into account.

Having arbitrary variables, one needs a method of distinguishing the physical and the redundant ones in a constrained system. The variables which are not effected by the gauge transformations are the physical variables. The act of the gauge transformation on a variable $x$ is given by the Poisson bracket of the variable with the generator of the gauge transformation. As the generators are the constraints, this makes 
\be \Delta x_b= \{x_b,G_a\}\xi^a=\left(\frac{\partial x_b}{\partial q^i}\frac{\partial G_a}{\partial p_i}-\frac{\partial x_b}{\partial p_i}\frac{\partial G_a}{\partial q^i}\right)\xi^a,\ee
where $\xi^a$ is the parameter of the transformation. What is happening here is that the constraints of the system do two things for us, one they define a constraint surface in the space of variables, two they generate the gauge transformations. In other words, the Poisson bracket with the constraint defines a derivative on the constraint surface. Hence the gauge orbits defined by the gauge generators will lie on the constraint surface. Physical variables of the system, $O$, correspond to fixed points on a gauge orbit on the constraint surface
\be \Delta O=\{O,G_a\}\xi^a=0.\ee
We pictorially represent this in Figure \ref{fig:gaugeorbit}. 
\begin{figure}[h]
	\includegraphics[width=\textwidth]{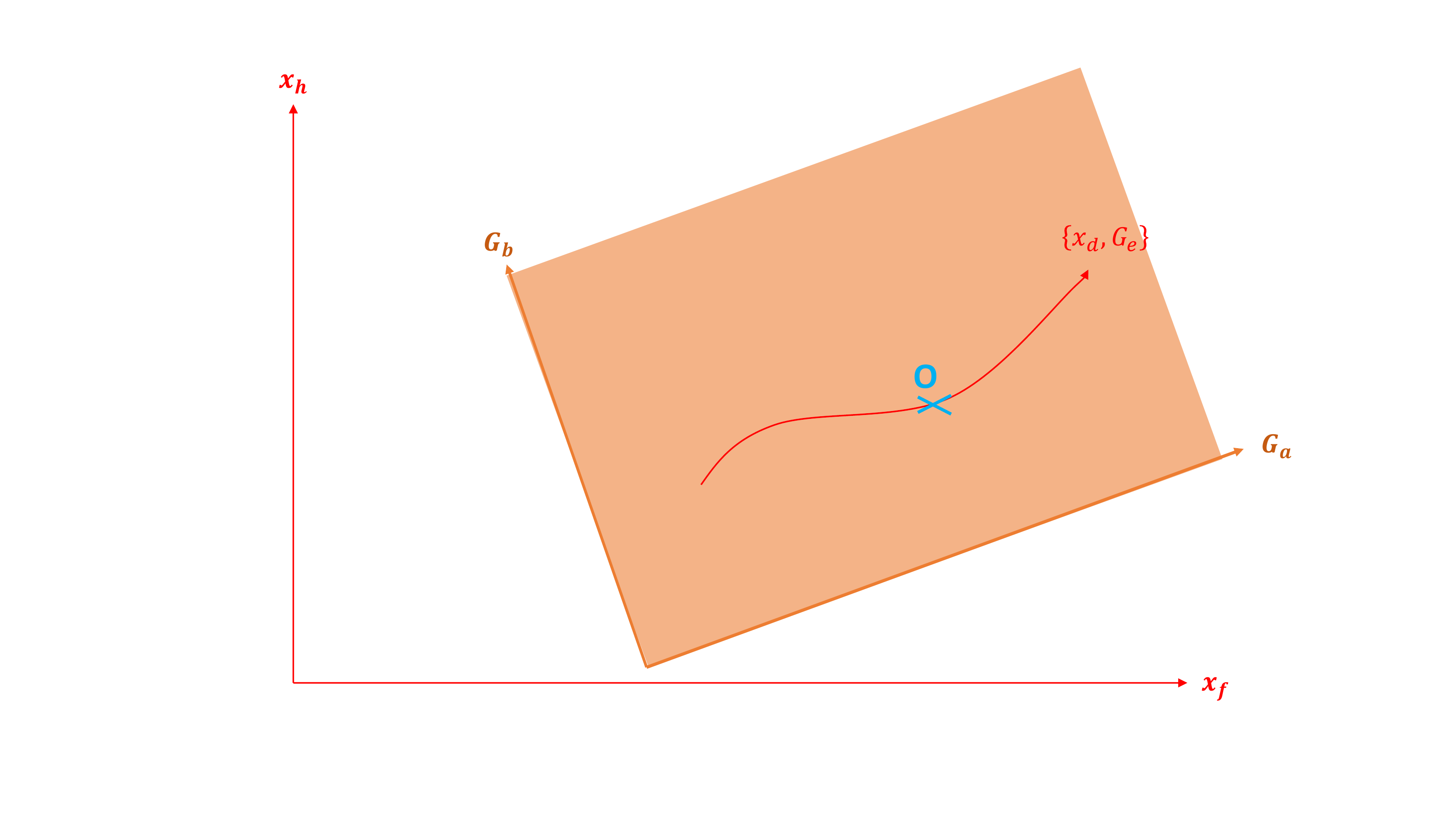}
	\caption{The orange rectangle represents the gauge surface with $\{G_e,x_f\}$ giving the gauge orbits. Physical observables like $O$ remain fixed along the gauge orbit.}
	\label{fig:gaugeorbit}
\end{figure}
In other words the observables are the set of gauge invariant functions on the constraint surface. If we carry on to quantize the theory in this form, we will have followed the Dirac quantization. 

In chapter 2 we will make use of this analysis of the constraints to find the physical observables among cosmological perturbations and their quantization. We refer to quantization with respect to gauge invariance as Dirac quantization. However we will also work out the quantization of observables in the case of gauge fixing. This is when a coordinate system is declared from the beginning as a gauge choice. This choice amounts to fixing the otherwise arbitrary variables of the theory. Once gauge fixed, the gauge invariance is broken by the removal of the arbitrariness. Then the system exhibits BRST invariance, a symmetry discovered by Becchi, Rouet, and Stora, and Tyutin and named after their initials. This is a fermionic symmetry generated by a nilpotent charge $\Omega_a$, where nilpotency means
\be \delta\Omega=\{\Omega_a,\Omega_b\}\xi^b=0,\ee of grading $\epsilon_\Omega=1$. This symmetry acts over an extended space of variables where one has Grassmann fields, $\eta,\mathcal{P}$, as well as complex commuting variables $x$. The main property of Grassmann variables is that they are anti commuting, if you consider two such variables $\eta$ and $\mathcal{P}$
\be \eta\mathcal{P}=-\mathcal{P}\eta.\ee
This is where the grading enters in. In a so called $Z_2$ algebra, where two different types of variables exist, the Grasmann variables have odd grading, ie $\epsilon_\eta=1$ , while the complex variables $x$ have even grading $\epsilon_x=0$. So the multiplication rule of the algebra is
\be AB=(-1)^{(\epsilon_A+\epsilon_B)}BA.\ee
The Poisson brackets of two functions $(A,B)$ in this extended space works as follows
\be \{A,B\}=\left[\frac{\partial A}{\partial q^i}\frac{\partial B}{\partial p_i}-\frac{\partial A}{\partial p_i}\frac{\partial B}{\partial q^i}\right]+(-1)^{\epsilon_A}\left[\frac{\partial_L A}{\partial \eta^a}\frac{\partial B}{\partial_L \mathcal{P}_a}-\frac{\partial_L A}{\partial \mathcal{P}_a}\frac{\partial_L B}{\partial \eta^a}\right],\ee
where $\partial_L$ denotes left derivative. 

The observables under in BRST invariant theory live in the cohomology of the BRST charge. The cohomology of a charge is the set of functions who are not obtained by the said transformation of another function, and who remain invariant under the said transformations. The BRST charge is constructed so that its cohomology consists of gauge invariant functions. Again the concept of having gauge invariant functions requieres two things, one needs to define the constraint surface and you need to define a derivative operator on this surface. In this extended space, the constraint surface is defined by purely imaginery Grassmann variables $\mathcal{P}_a$ with grading $\epsilon_a=-1$, also referred to as antighosts and the derivative on the constraint surface is defined by real Grassmann differential forms $\eta^a$ with grading $\epsilon_\eta=1$, referred to as ghosts.

\section{Time Evolution and Quantization of Perturbations}
\label{sec:time evolution}

In the previous section we introduced the variables we are interested in, the cosmological perturbations. Now we would like to consider their time evolution. This is inherently linked with quantization. From beginning onwards we have been considering fields, $F_a=\{\phi,g_{\mu\nu}\}$, as split into background and perturbation $\bar F_a+\delta F_a=\{\phi_0+\delta\phi,\bar{g}_{\mu\nu}+\delta g_{\mu\nu}\}$. We quantize perturbations around a given classical background solution. Following \cite{Weinberg:2005vy} this means, in a Hamiltonian formalism where each field $F$ and its conjugate momenta $\pi_a=\frac{\partial\mathcal{L}[F,\dot{F}]}{\partial F_a}$, are canonical variables, the classical equations of motion
\be \label{classical} \dot{\bar{F}}_a=\frac{\delta H[\bar{F}_a(t),\bar{\pi}_a(t)]}{\delta \bar{\pi}_a},~~~~ \dot{\bar{\pi}}_a=-\frac{\delta H[\bar{F}_a,\bar{\pi}_a]}{\delta \bar{F}_a}\ee
hold. Where as the time evolution is given by the commutator of the canonical variable with the full Hamiltonian $H[F_a(t),\pi_a(t)]$
\be \dot{F}_a=i[H[F(t),\pi(t)],F_a],~~~~ \dot{\pi}_a=i[H[F(t),\pi(t)],\pi_a],\ee
and the quantum perturbations obey the canonical commutation relations
\begin{subequations}
	\label{commutations}
	\begin{align} [\delta F_a(\vec{x},t),\delta\pi_b(\vec{x'},t)]&=i\delta_{ab}\delta^3(\vec{x}-\vec{x}'),\\
	[\delta F_a(\vec{x},t),\delta F_b(\vec{x'},t)]=[\delta&\pi_a(\vec{x},t),\delta\pi_b(\vec{x'},t)]=0.\end{align}
\end{subequations}
We can expand the Hamiltonian as
\be H[F_a(t),\pi_a(t)]=H[\bar{F}_a(t),\bar{\pi}_a(t)]+\sum \frac{\delta H}{\delta \bar{F}_a}\delta F_a+\sum\frac{\delta H}{\delta\bar{\pi}_a}\delta\pi_a+\delta H,\ee
where $\delta H$ is the perturbed Hamiltonian starting from second order in perturbations. Now if we consider
\be \dot{F}_a=i[H[F(t),\pi(t)],F_a] \ee 
and expand both sides
\be \dot{\bar{F}}_a+\delta\dot{F}_a=i[H[F(t),\pi(t)],\bar{F}_a]+i[H[F(t),\pi(t)]].\ee
The background solution is a complex valued function, hence it commutes with the Hamiltonian making the first term on the right hand side vanish. The same also holds for the Hamiltonian at the background level $H[\bar{F}(t),\bar{\pi}(t)]$, making the commutators $[H[\bar{F}(t),\bar{\pi}(t)],\delta F_a]$ and $[H[\bar{F}(t),\bar{\pi}(t)],\delta \pi_a]$ also drop out. This leaves us with
\be \dot{\bar{F}}_a(\vec{x},t)+\delta \dot F_a(\vec{x}.t)=i\frac{\delta H}{\delta \bar{\pi}_a}[\delta \pi_a(\vec{x}',t),\delta F_a(\vec{x},t)]+i[\delta H[F(t),\pi(t)],\delta F_a(\vec{x},t)].\ee
Via use of the background equations of motion and the canonical commutation relations we note that the first terms cancel each other, and we arrive at
\be\label{dotdeltaF} \delta\dot{F}_a(\vec{x},t)=i\left[\delta H[F(t),\pi(t)],\delta F_a(\vec{x},t)\right].\ee
The same relation also holds for momentum perturbation
\be \delta\dot{\pi}_a(\vec{x},t)=i\left[\delta H[F(t),\pi(t)], \delta \pi_a(\vec{x},t)\right].\ee
Thus the time evolution of the perturbations are determined by the perturbed Hamiltonian, while the time evolution of the full fields are determined by the full Hamiltonian.

Given an initial state $\delta F_a(t_0)$, we would like to be able to represent its time evolution by a unitary hermitian operator
\be \delta F_a(t)=U^{-1}(t,t_0)\delta F_a(t_0)U(t,t_0)\ee
and of course the same holds for the momentum conjugate
\be \delta\pi_a(t)=U^{-1}(t,t_0)\delta\pi_a(t_0)U(t,t_0).\ee
Demanding this form for $\delta F_a(t)$ to satisfy \eqref{dotdeltaF} leads to
\be [\delta H,\delta F_a]=i\left[U^{-1}(t,t_0)\frac{dU(t,t_0)}{dt},\delta F_a\right]\ee
implying that $U(t,t_0)$ is an operator that obeys the differential equation
\be\label{diffforU}\frac{U(t,t_0)}{dt}=-iU(t,t_0)\delta H=-i\delta HU(t,t_0).\ee 
One can define the initial conditions such that $U(t_0,t_0)=1$. The solution to \eqref{diffforU} is
\be U(t,t_0)=TE^{-i\int_{t_0}^{t}\delta H(t')dt'}\ee
where $T$ denotes time ordering.

The perturbed Hamiltonian $\delta H$ starts at quadratic order in fluctuations, let's refer to this part as $H_0$. This is the free Hamiltonian, and the higher order terms represent the interactions, let us denote these as $H_I$,
\be \delta H=H_0(\delta F_a^2,\delta \pi_a^2)+H_I.\ee
The time evolution of the free fields are governed only by the quadratic part of the Hamiltonian, $H_0$. Just as the Interaction picture in Quantum Mechanics, we can define the free cosmological perturbations as the interaction picture fields
\be \delta F^I_a(t)=U^{-1}_0(t,t_0)\delta F_a^I(t_0)U_0(t,t_0)\ee
and 
\be \delta\dot{F}^I_a=i\left[H_0,\delta F^I_a\right].\ee
Then the effect of further interactions are accounted for by an operator
\be K(t,t_0)=Te^{-i\int_{t_0}^{t}H_I(t't)dt'} \ee
such that
\be U(t,t_0)=U_0(t,t_0)K(t,t_0).\ee
The interaction picture fields are connected to the original fields as follows
\begin{align} \delta F_a(t)&= U^{-1}(t,t_0)\delta F_a(t_0)U(t,t_0)\\
&=K^{-1}\left[U_0^{-1}(t,t_0)\delta F_a(t_0)U_0(t,t_0)\right]K\\
&=K^{-1}(t,t_0)\delta F_a^I(t)K(t,t_0)\\
&=\bar{T}e^{i\int_{t_0}^{t}H_I(t't)dt'}\delta F_a^I(t) Te^{-i\int_{t_0}^{t}H_I(t't)dt'}
\end{align}
Taylor expanding the exponentials via the Bekker-Campbell-Hausdorff formula, the expectation value for a perturbation or any other operator $\delta\mathcal{O}$ build out of the perturbations becomes \cite{Weinberg:2005vy}
\be \langle\delta \mathcal{O}(t)\rangle= \sum^{\infty}_{N=0}i^N\int^t_{-\infty}dt_N \int^{t_N}_{\infty}dt_{N-1}\dots\int^{t_2}_{\infty}dt_1\langle\left[H_I(t_1),\left[H_I(t_2),\dots\left[H_I(t_N)\delta\mathcal{O}^I(t)\right]\dots\right]\right]\rangle\ee

\section{Gauge Invariant Variables and Quantization}
\label{sec:gaugeinvvar}
The previous section was a formal review of the in-in formalism on how to account for time evolution in cosmological perturbation theory. Here we will make use of this formalism to calculate expectation values in a practical application. We have seen in section \eqref{sec:diffeomorphisms} how important the diffeomorphism invariance is. While it is unavoidable to make a coordinate choice in order to carry on a calculation, at the end of our calculations we must make sure the results do not carry any redundancies coming from the coordinate system that was worked on. One way to make sure of this is to work in terms of gauge invariant variables. This will be the focus of this section and what follows in this chapter. Another way is to understand how the choice of coordinates can be correctly implemented. This is the subject of chapter 2.

In the next two sections we will  work in conformal time $\eta$ defined as
\be\label{etadef}
\eta=\int^{t}_{t_0=0}\frac{d\tilde t}{a(\tilde t)}.
\ee
We will denote time derivatives with respect to coordinate time with a dot $\dot{f}=\frac{df}{dt}$ and, derivatives with respect to conformal time with a prime $f'=\frac{df}{d\eta}$. The conformal time Hubble parameter is denoted as
\be \mathcal{H}(\eta)\equiv\frac{a'}{a}=aH(t).\ee

The initial conditions can be chosen so as to make $H(t)=\frac{\dot{a}(t)}{a(t)}$ constant for a finite time interval and give inflation. We will work on the quantization of perturbations both in exact and near de Sitter for comparison.  We pick power law inflation as an example for a near de Sitter situation. Our advantage is that the equations are solvable for both of these examples. 

In exact de Sitter the scale factor goes as $a_{dS}(t)\propto e^{Ht}$ where the Hubble rate is constant $H_{dS}=const$. This gives rise to an internal accelerated expansion. In terms of conformal time, by \eqref{etadef}
\be a_{dS}(\eta)=-\frac{1}{H\eta}.\ee
For power law inflation, the scale factor goes as $a_{pl}(t)\propto t^p$ where $p > 1$ and,  $a_{pl}(\eta)\propto \eta^{\frac{p}{1-p}}$. Conformal time can go down to negative infinity and has negative values during inflation, which raises the danger of imaginery values for the scale factor in this last form we have written. Since $p>1$, $\frac{p}{1-p}<0$ so we will use the combination $\frac{p}{1-p}\eta$ as the argument of $a_{pl}(\eta)$,
\be \label{aeta}
a_{pl}(\eta)=\left(\frac{p}{1-p}\eta\right)^{\frac{p}{1-p}}.\ee
This expression approaches that of de Sitter, in the limit $a\xrightarrow{p\to \infty} -\frac{1}{\eta}$.

Taking derivatives with respect to conformal time one obtains,
\be \label{H}
\mathcal{H}_{pl}=\frac{a'}{a}=\left(\frac{p}{1-p}\right)\frac{1}{\eta}, \ee

\be\mathcal{H}'_{pl}=\frac{p-1}{p}\mathcal{H}^2,\ee
and for higher derivatives
\be \mathcal{H}^{(n)}_{pl}=n!\left(\frac{p-1}{p}\right)^n\mathcal{H}^{(n+1)}\ee
for the power law inflation. For exact de Sitter
\be \Hmath_{dS}=\frac{a'_{dS}}{a_{dS}}=-\frac{1}{\eta},\ee
and 
\be \Hmath_{dS}^{(n)}=n!\frac{(-1)^{n+1}}{\eta^{n+1}}=n!(-1)^{n+1}\Hmath^{n+1}.\ee
\subsection{The Mukhanov Variable $v$}

Though the variables of actual interest are the field and metric perturbations, $\delta\phi$ and $\psi$, it is advantageous to work with the Mukhanov variable $v$ \cite{Mukhanov:1990me}. This simplifies the problem of quantizing metric and density perturbations of a system with gauge symmetries into quantizing a single gauge invariant variable, $v$. This way, calculations for multiple variables are combined into one variable whose use removes any gauge dependent ambiguities.  
The variable $v$ is a combination of $\delta\phi$ and $\psi$. For single scalar field matter content it has the form \cite{Mukhanov:1990me}, 
\be \label{vscalar} v=a\delta\phi+\frac{a\phi_0'}{\Hmath}\psi.\ee
It will be useful to define the variable
\be \label{eq:z} z=\frac{a\phi'_0}{\Hmath}.\ee

Originally we are interested in Einstein gravity with matter content
\begin{align} \label{EHmat} \mathcal{S}(g_{\mu\nu},\phi;t)&=\mathcal{S}_{EH}+\mathcal{S}_{m}\\
&\int d^4x\sqrt{-g}\left[\frac{1}{2}m^2_{pl}R+\mathcal{L}_m\right].
\end{align}
In coordinate time, at the background level, with $\phi=\phi_0(t)$ and the metric components $\bar{g}^{\mu\nu}=g^{\mu\nu}(t)$ that of LFRW metric, the Friedmann equations of motion obtained by varying this action with respect to the metric are\footnote{The first of these equations is the $00$th component, the second is a combination of the first equation and the trace of the field equations. A more convenient version is to write them as $-2m^2_{pl}\dot{H}=\dot{\phi}_0^2$ and $6m^2_{pl}H^2=\dot{\phi}_0^2+2V(\phi_0)$.}
\be \label{friedmann 1}3m^2_{pl}H^2=\frac{1}{2}\dot{\phi}_0^2(t)+V(\phi_0),\ee
\be \label{friedmann2}-3m^2_{pl}\left(\dot{H}+H^2\right)=\dot{\phi}_0^2-V(\phi_0).\ee  
Here the Hubble parameter $H(t)$ represent the metric since the only nontrivial component is $\bar{g}_{ij}=a^2(t)\delta_{ij}$.

We are interested in this action at second order in perturbations, with a single field matter content  whose Lagrangian in the convention of $(-,+,+,+)$ metric signatures is
\be \label{matter} \mathcal{L}_m(g_{\mu\nu},\phi;t)=-\frac{1}{2}g^{\mu\nu}\partial_\mu\phi\partial_\nu\phi-V(\phi).\ee
To study the evolution of perturbations, one substitutes
\begin{subequations}
	\label{linearperts}
	\begin{align}
	g_{\mu\nu}(t,\vec{x})=\bar{g}_{\mu\nu}(t)+h_{\mu\nu}(t,\vec{x})\\
	\phi(t,x)=\phi_0(t)+\delta\phi(t,\vec{x})
	\end{align}
\end{subequations}
into \eqref{EHmat} and expands the terms up to second order in $\delta\phi$, $h_{\mu\nu}$. Even though these perturbations are introduced linearly in \eqref{linearperts}, nonlinear terms arise due to the nonlinearities of \eqref{EHmat}. In other words the action at second order in perturbations is the free theory. In terms of the Mukhanov variable
\be \label{actionv}
\delta^2 S=\frac{1}{2}\int\left[v'^2-(\partial_iv)^2+\frac{z''}{z}v^2\right]d^4x,
\ee
where total derivative terms are dropped \cite{Mukhanov:1990me}. 

Before moving onto quantization, let us convince ourselves of the gauge invariance of $v$. 
As mentioned in the beginning we consider infinitesimal coordinate transformations, the diffeomorphisms,
\be x^a\to\tilde x^a=x^a+\xi^a\ee
where $\xi=(\xi^0,\vec{\xi})$ is the parameter of the transformation, as gauge transformations. 
The most general diffeomorphism that preserves the scalar nature of the metric fluctuations is
\begin{subequations}\begin{align}
	\eta \to \tilde{\eta}=\eta+\xi^0(\eta,\vec{x}),\\
	x^i \to \tilde{x^i}=x^i+\gamma^{ij}\nabla_j\xi(\eta, \vec{x})
	\end{align}
\end{subequations}
where the three vector has been split into $\xi^i=\xi^i_{tr}+\gamma^{ij}\nabla_j\xi$. Under this transformation, on a timedependent homogeneous background fluctuation in a scalar field $\delta q$, such as our $\delta\phi$, transform as
\be \delta q\to\delta\tilde{q}(\eta,x)=\delta q(\eta, x)-q'_0(\eta)\xi^0,\ee
as we found in \eqref{scalartr}. The metric perturbation transforms as\cite{Mukhanov:1990me}
\be \psi\to \tilde{\psi}=\psi+\mathcal{H}\xi^0.\ee
Plugging these in we see that 
\be \tilde{v}=a\delta\tilde{\phi}+\frac{a\phi'_0}{\mathcal{H}}\tilde{\psi}=v,\ee
the Mukhanov variable is invariant. Hence, the powerspectrum of the Mukhanov variable will also be invariant under gauge transformations.

The variable $v$ can be expanded in terms of eigenvectors of the Laplacian $\nabla \chi_k=-k^2\chi_k$, with $\chi_k=e^{-i\vec k\cdot \vec x}$ and promoted to an operator as
\be v=\int\frac{d\vec k^3}{(2\pi)^3}\left[v_k(\eta)\hat{a_k}e^{i\vec{k}\cdot \vec{x}}+{v_k}^*(\eta)\hat{a_k}^\dagger e^{-i\vec{k}\cdot\vec{x}}\right].\ee
The canonical commutation relations \eqref{commutations} with $\delta F_a=v$ implies
\be [\hat{a}_k,\hat{a}^\dagger_{k'}]=\delta(k-k'),~~[\hat{a}_k,\hat{a}_{k'}]=[\hat{a}^\dagger_k,\hat{a}^\dagger_{k'}]=0.\ee
In return we get the following equation of motion 
\be \label{v_k}
{v_k}''+\left[{k}^2-\frac{z''}{z}\right]v_k=0\ee
for the mode functions $v_k$ to obey. This is the equation of motion of a harmonic oscillator with a time dependent frequency $\omega_k(\eta)=k^2-\frac{z''}{z}$! So the scalar perturbations, the modes of interest are just harmonic oscillators. The fact that they have a time dependent frequency implies that they can transition from one state to another over time, simply because their frequency carries time dependence. This is the analogue of time dependent perturbation theory in quantum mechanics. Here time dependence is involved in the interactions of the perturbations because they are coupled to a time dependent background. This is a curved background because of its time dependence. So instead of perturbations to the potential being turned on or off in quantum mechanical examples, here there is time dependence due to considerations of a curved background. 



To solve of the mode functions of power law inflation, we need to calculate the ratio
\be \frac{z''}{z}=\frac{{\phi_0}'''}{\phi_0'}+\frac{2}{p}\frac{{\phi_o}''}{\phi_0'}\Hmath+\frac{1}{p}\Hmath^2.\ee
We can make use of the Friedmann equations \eqref{friedmann 1}-\eqref{friedmann2} to evaluate the higher derivatives on $\phi_0$. In terms of conformal time we have\footnote{where $l_p=\sqrt{\frac{\hbar G}{c^3}}=\left(m_{pl}\sqrt{8\pi}c\right)^{-1}$.} 
\be\frac{3}{2}l^2_{pl}{\phi_0'}^2=\Hmath^2-\Hmath'\ee 
which in our case give 
\be \label{backgr}
\frac{3}{2}l_{pl}^2{\phi_0'}^2=\Hmath^2-{\Hmath}'=\frac{1}{p}\Hmath^2.\ee 
By differentiating this we get
\be 
\frac{\phi_0''}{\phi_0'}=\frac{p-1}{p}\Hmath\ee and
\be \frac{\phi_0'''}{\phi_0'}=2\left(\frac{p-1}{p}\right)^2\Hmath^2.\ee Thus for power law inflation the time dependent frequency of the harmonic oscillator is 
\be \frac{z''}{z}|_{powerlaw}=\frac{2p-1}{p}~\Hmath^2=~\frac{p(2p-1)}{(1-p)^2}~\eta^{-2}.\ee

The mode function is the solution of
\be {v_k}''+\left[{k}^2+\frac{1-2p}{p}\Hmath^2\right]v_k=0 \ee or in terms of conformal time 
\be \label{conf} {v_k}''+\left[{k}^2+\frac{p(1-2p)}{(1-p)^2}\frac{1}{\eta^2}\right]v_k=0 \ee
as $p\to \infty$ \eqref{conf} does give the correct expression of ${v_k}''+(k^2-\frac{2}{\eta^2})v_k=0$. Equation \eqref{conf} looks like the Bessel equation of form
\be \label{bessel}w''+\left(\lambda^2-\frac{\nu^2-1/4}{\eta^2}\right)=0\ee
where $\lambda^2={k}^2$ and, $\nu^2=\frac{(1-3p)^2}{4(1-p)^2}$. This has a solution in terms of Bessel functions $\mathcal{C}_\nu(\lambda\eta)$, in the form of
\be\label{gensoln} v_k(\eta)=\sqrt{\eta}~~\mathcal{C}_\nu(\lambda\eta).\ee
As $\eta$ is negative during inflation, the $\sqrt{\eta}$ in \eqref{gensoln} looks problematic. However, because equation \eqref{bessel} is invariant under time reversal as $\eta \to -\eta$, we can just replace $\eta$
\be\label{gensol2}v_k(\eta)=\sqrt{-\eta}~~\mathcal{C}_\nu(-\lambda\eta).\ee

To solve a second order differential equation, one needs two boundary conditions. The first appropriate boundary condition is to choose the normalizaton so that\cite{Baumann:2009ds}
\be \label{bc1} \langle v_k(\eta),v_k(\eta)\rangle\equiv \frac{i}{\hbar}({v_k}^*{v_k}'-{v^*_k}'v_k)=1.\ee From here on we will set $\hbar=1$.
The second appropriate boundary condition is to assume that the vacuum was Minkowski vacuum in the past. In the far past all comoving scales are far inside the horizon, meaning as $\eta \to \-\infty$, $k\gg aH$. Since $\frac{z''}{z}$ is composed of $aH$'s it is negligable  and the mode equation has the form
\be {v_k}''=-{k}^2v_k.\ee
The general solution to this equation will be of the form $v_k(\eta)=N_-(k)e^{ik\eta}+N_{+}(k)e^{-ik\eta}$. We will consider the positive frequency solution, which means setting\footnote{Note that here we have make the replacement $\eta\to-\eta$ which makes us pick up a minus sign in comparison to the convention in literature. One can read equation \eqref{bc} as $\lim_{|\eta|\to\infty}v_k(\eta)\to\frac{1}{\sqrt{2k}}e^{-ik|\eta|} $.} $N_+=0$. The first boundary condition fixes the normalization to be $|N_-|^2=\frac{1}{2k}$. So the Minkowski vacuum satisfies $v_k=\frac{e^{+ik\eta}}{\sqrt{2k}}$.

Thus the solution of \eqref{conf} should satisfy
\be\label{bc} \lim_{\eta\to-\infty}v_k(\eta)\to\frac{1}{\sqrt{2k}}e^{+ik\eta}.\ee
This initial condition is the Bunch Davies vacuum.
Among the Bessel functions $H^{(2)}_\nu(-k\eta)~\to\sqrt{\frac{2}{-\pi k\eta}}e^{ik\eta}$ as $k\eta \to \infty$. Hence for power law inflation, the mode functions are
\be \label{modefuns} v_k(\eta)=\frac{1}{2}\sqrt{-\pi\eta}~H^{(2)}_\nu(-k\eta).\ee

Exact de Sitter means the background is fixed to be the de Sitter metric, and one only has the perturbations of the scalar field on this background. In otherwords $\psi=0$ and $v=a\delta\phi$. The action is
\be S=\frac{1}{2}\int d^4x \sqrt{-g}(m_{pl}^2R-g^{\mu\nu}\partial_\mu\phi\partial_\nu\phi-V).\ee
In terms of conformal time de Sitter metric is $ds_{dS}^2=\frac{1}{\eta^2H_{dS}^2}\left[-d\eta^2+d\vec{x}^2\right]$ with $a_{dS}(\eta)=-\frac{1}{\eta H_{dS}}$.
\be \mathcal{S}_{dS}=\frac{1}{2}\int d^4 x\frac{1}{\eta^4H^4_{dS}}\left[m^2_{pl}R+\eta^2H^2_{dS}\left[\phi'^2-\left(\vec{\nabla}\phi\right)^2\right]-2V(\phi)\right]\ee 
In terms of $v$, the second order action is
\be \delta^{(2)}\mathcal{S}_{dS}=\frac{1}{2}\int d^4x\left[v'^2-(\vec{\nabla}v)^2-m_\phi^2v^2+\frac{2}{\eta^2}v^2\right]\ee
where we made use of $H_{dS}=const$ and  $\Hmath_{dS}'=\Hmath_{dS}^2$. We can think of the mass as absorbed in wavenumber in equation \eqref{actionv}, which then implies that for exact de Sitter $\frac{z''}{z}|_{dS}=\frac{2}{\eta^2}$. Note that in the limit $p \to \infty$, power law inflation converges to exact de Sitter with $\frac{z''}{z}|_{pl}\to \frac{z''}{z}|_{dS}$. 
The solution 
\be v_k=\left(1-\frac{i}{k\eta}\right)\frac{e^{ik\eta}}{\sqrt{2k}},\ee
satisfies the two boundary conditions .
\section{Power spectra and the Long Wavelength Limit}
\label{sec:powerspectra}
The powerspectrum $P_v(k)$ is the quantity accessible more directly to the observations. It is defined in relation to the two point function $\langle v_k v_{k'}\rangle$ as 
\be\label{spectrum}
\langle v_k v_{k'}\rangle= (2\pi)^3\delta(\vec{k}+\vec{k'})P_v(k),
\ee
or
\be 
\Delta^2_{v_k}=\frac{k^3}{2\pi^2}P_{v}(k).
\ee
The two point function for the harmonic oscillator is
\be \label{harmosc}
\langle v_k v_{k'}\rangle= (2\pi)^3\delta(\vec{k}+\vec{k'})|v_k(\eta)|^2.\ee So in principle all that it takes to calculate the powerspectra of scalar perturbations is to solve \eqref{v_k} for the mode functions. Since the relation between $v_k$ and $\delta\phi$ or $\psi$ are known, for example \eqref{vscalar} for scalar field content, it is easy to obtain the powerspectrum of fluctuations from the power spectrum of the mode functions.

Now that the mode functions for power law inflation are known to be,
\be v_k(\eta)=\frac{1}{2}~\sqrt{-\pi\eta}~H^{(2)}_\nu(-k\eta),\ee
their power spectrum can be computed as
\be \Delta^2_{v_k}=\frac{k^3}{2\pi^2}P_{v_k}(k)=\frac{k^3}{8\pi}~(-\eta)~|H^{(2)}_\nu(-k\eta)|^2.\ee
And the power spectrum for de Sitter is
\be \label{psdesitter}\Delta^2_{v_k}=\frac{k^2}{4\pi^2}\left[1+\frac{1}{k^2\eta^2}\right]=\frac{1}{4\pi^2}[k^2+\mathcal{H}^2].\ee
We would like to refer the reader to \cite{Anninos:2012qw} for a good review on de Sitter and an alternative derivation of the two point function in the wavefunctional formalism. 

Let's look at the power spectra in comoving gauge, where $\delta\phi=0$ and all the perturbations are in the metric. In this case $v=z\psi$ so that
\be \Delta^2_{\psi_k}=\frac{1}{z^2}\Delta^2_{v_k}.\ee Using the background equation, for power law inflation
\be \label{zpowerlaw} \frac{1}{z^2}=\frac{\Hmath^2}{a^2\phi'^2_0}=\frac{3l^2_{Pl}}{2}\frac{(1-p)^2}{p}~ \eta^{\frac{2p}{p-1}}.\ee
With this the power spectrum for metric perturbations in comoving gauge becomes
\be \Delta^2_{\psi_k}=-\frac{3l^2_{Pl}}{16\pi}\frac{(1-p)^2}{p}k^3\eta^{\frac{3p-1}{p-1}}|H^{(2)}_\nu(-k\eta)|^2.\ee

Next interesting thing to calculate is the amplitude $A_\psi(\eta)$ and spectral index $n_\psi$, defined as
\be \label{a,ns} \Delta^2_{\psi_k}(\eta,k)=A_\psi(\eta)\left(\frac{k}{k_*}\right)^{n_\psi-1},\ee
where $k_*$ is some chosen scale with respect to which everything is normalized. One could choose $k_*$ to be the mode that exits the horizon N e-folds before the end of inflation. At a given time $t_*$, the mode that satisfies
\be k_*=a(t_*)H(t_*)=\mathcal{H}(\eta_*),\ee
will be the mode crossing the horizon. During inflation the Hubble parameter is almost constant so one can take $H(t_*)=H_{inf}$. And from the definition $dN=Hdt=d(ln a)$, for a given number of e-folds, $a=e^N$. Since $\mathcal{H}$ goes as $\frac{1}{\eta}$, as confirmed by equation \eqref{H}, one has $k_*\eta=1$, for a mode that exits the horizon sometime before the end of inflation. Keeping this in mind the amplitude is obtained to be
\be A_\psi(\eta)=\Delta^2_\psi(\eta,k_*),\ee
which for power law inflation gives
\be A_\psi(\eta)=\frac{3l^2_{Pl}}{16\pi^2}\frac{(1-p)^2}{p}{k^3}_*\eta^{\frac{3p-1}{p-1}}|H^{(2)}_\nu(k_*\eta)|^2.\ee

For the spectral index notice that, from \eqref{a,ns}
\be \ln{\Delta^2_\psi}=\ln{A_\psi}+(n_\psi-1)\ln{\frac{k}{k_*}}\ee
which suggests
\be \label{ns} n_\psi=1+\frac{d\ln{\Delta^2_\psi}}{d\ln{k}}. \ee
Thus for power law inflation one obtains
\be n_\psi=4+2\frac{d\ln{H^{(2)}_\nu(k\eta)}}{d\ln{k}}. \ee
\subsection{The Long Wavelength Limit}
At the background level, the scalar field we are interested in obeys the equation of motion
\be \ddot \phi_0+3H\dot\phi_0+V'(\phi_0)=0,\ee
where we have switched to cosmological time $t$ and a prime refers to derivative with respect to field $\phi$. This is obtained by varying the action \eqref{EHmat} with respect to $\phi$.
To have an intuition for the way the fluctuations that arise during inflation evolve, we need to look at the equation of motion for the scalar field perturbation $\delta\phi(x,t)$. Substituting $g^{\mu\nu}=\bar{g}^{\mu\nu}+\delta g^{\mu\nu}$ and $\phi(x,t)=\phi_0(t)+\delta\phi(x,t)$ and varying with respect to $\delta\phi$, and demanding background equations of motion to be satisfied, this is
\be \label{fluct} \delta \ddot{\phi}+3H\dot{\delta \phi}+V''(\phi_0)\delta\phi+\frac{k^2}{a^2}\delta\phi=-2\psi V'(\phi_0)+4\dot{\psi} \dot{\phi_0}.\ee
To make things simple, just for now let us also take the metric perturbation $\psi=0$, like in spatially flat gauge $\psi=E=0$,
\be\label{flatpert} \ddot{\delta \phi}+3H\dot{\delta \phi}+V''(\phi_0)\delta\phi+\frac{k^2}{a^2}\delta\phi=0.\ee

Modes for which the physical momentum is larger than the hubble parameter, $\frac{k}{a}\gg H$, or in other words whose wavelength is smaller than the horizon size, are called \emph{``subhorizon~modes"}. If for these modes $\left(\frac{k}{a}\right)^2\gg V''(\phi_0)$ is also satisfied they evolve according to
\be \label{subhor} \ddot{\delta\phi}+\frac{k^2}{a^2}\delta\phi=0.\ee

Subhorizon modes oscillate rapidly and, with small $a$ and negligable $H$ behave as if they are in Minkowski space. During inflation $a(t)$ grows fast while $H(t)$ evolves slowly. Hence $\frac{k}{a}$ will decrease comparably where as $H$ will stay almost the same. And modes that were subhorizon early on will eventually become \emph{``superhorizon"} with $\frac{k}{a}\ll H$. 

What happens when modes become superhorizon? We can estimate $\delta\ddot\phi\sim \left(\frac{k^2}{a^2}+m^2_\phi\right)\delta\phi$. During inflation $H\gg m_\phi$ and since we are considering modes for whom $\frac{k}{a}\ll H$, the $\delta\ddot{\phi}$ the equation of motion in this case is effectively
\be \label{superhor} 3H\dot{\delta\phi}=0.\ee
The amplitude of superhorizon modes doesn't change. Provided they don't reenter the horizon since the time they exit, or that we know how they evolve once they reenter, if we observe these modes we will know how to obtain information from the epoch at which they exited the horizon. This is why it is useful to calculate things for modes that exit the horizon during inflation to gain information about inflation. These are the adiabatic modes we mentioned earlier on. 

The fact that during inflation only subhorizon modes exit the horizon while no superhorizon modes enter is crucial. This mechanism is what lets inflation set the initial conditions. Earlier on we used the Minkowski spacetime solution as our initial condition. Minkowski spacetime is flat and non dynamic. But we know that later on spacetime becomes dynamic. Up untill inflation starts we can consider the spacetime to be Minkowski. Because superhorizon modes do not become subhorizon during inflation, all the subhorizon modes live inside the horizon at any time during inflation have been living there since the start of inflation. So we can safely set Minkowski vacuum to be our initial condition. If during inflation things happened in a way that made superhorizon modes become subhorizon we could not trust the subhorizon Minkowski vacuum, modes to set the initial conditions because there would be contributions to them that come from the later times.

Thus we will now consider the superhorizon, or long wavelength limit, $k\eta \to 0$ of some of the quantities we have calculated above. For exact de Sitter we found
\be \Delta^2_{v_k}=\frac{k^2}{4\pi^2}\left[1+\frac{1}{k^2\eta^2}\right]=\frac{1}{4\pi^2}[k^2+\mathcal{H}^2].\ee
In the long wavelength limit, $k\ll \mathcal{H}$ this becomes 
\be \label{dslong} \lim_{k\eta\to 0}\Delta^2_{v_k}=\frac{1}{4\pi^2\eta^2}=\frac{\mathcal{H}^2}{4\pi^2}.\ee

In comoving gauge the metric perturbations are $\psi=\frac{v}{z}$. In exact de Sitter $\mathcal{H}'=\mathcal{H}^2$ and by the background equations of motion $z=\frac{a\phi_0'}{\mathcal{H}}$ vanishes because ${\phi_o'}^2=\frac{2}{3{l_{pl}}^2}[\mathcal{H}^2-\mathcal{H}']$ vanishes.In general we can write $z^2=\frac{a^2{\phi_0'}^2}{\mathcal{H}^2}=\frac{2a^2}{3{l^2}_{pl}}\epsilon$ where $\epsilon\equiv-\frac{\dot H}{H^2}=1-\frac{\mathcal{H}'}{\mathcal{H}^2}$ and exact de Sitter corresponds to the limit $\epsilon\to 0$. Thus we can say that for quasi de Sitter spacetime the power spectrum of metric perturbations is
\be \Delta^2_{\psi_k}=\frac{1}{z^2}\Delta^2_{v_k}=\frac{1}{4\pi^2\epsilon}\frac{3l^2_{pl}}{2a^2}\mathcal{H}^2\left[k^2\eta^2+1\right],\ee
which in the long wavelength limit becomes
\be \label{dSlongps} \lim_{k\eta \to 0} \Delta^2_{\psi_k}=\frac{1}{4\pi^2\epsilon}\frac{3l^2_{pl}}{2a^2}\mathcal{H}^2=\frac{H^2}{4\pi^2\epsilon}\frac{3l^2_{pl}}{2}.\ee
Thus the long wavelength limit of power spectrum for metric perturbations in comoving gauge approach a constant in quasi de Sitter spacetime. And these comoving gauge results also apply for the powerspectrum of gauge invariant comoving curvature perturbation $\mathcal{R}=\frac{v}{z}$. 

A close look at \eqref{dSlongps} reveals that the powerspectrum $\Delta^2_{\psi_k}$ for exact de Sitter in the longwavelength  limit does not depend on comoving momenta $k$. This is a crucial fact, because it implies that each mode contributes the same amount of power. In other words the spectra is scale invariant. This is a reminant of the scale invariance of de Sitter, invariance under dilatations parametrized by \eqref{dilatation}. This property implies that among the observable quantities,
\be \label{nsdS} n^{dS}_\psi=1,\ee
by \eqref{ns}.

For power law inflation we found, in comoving gauge,\be \Delta^2_{\psi_k}=\frac{3l^2_{Pl}}{16\pi^2}\frac{(1-p)^2}{p}k^3\eta^{\frac{3p-1}{p-1}}|H^{(2)}_\nu(k\eta)|^2.\ee

In general $H^{(2)}_\nu(z)=J_\nu(z)-iY_\nu(z)$ and in the limit $z\to 0$ this expression becomes
\be \label{Hlong} H^{(2)}_\nu(-k\eta)\sim -\frac{i}{\pi}\Gamma(\nu)\left(\frac{-k\eta}{2}\right)^{-\nu}.\ee
Plugging this in gives the long wavelength limit of powerspectra for metric perturbations to be
\be \Delta^2_{\psi_k}=\frac{3l^2_{Pl}}{16\pi^2}\frac{(1-p)^2}{p} k^{2/(1-p)}\left[\frac{4^\nu\Gamma^2(\nu)}{\pi^2}(-k\eta)^{(\frac{3p-1}{p-1}-2\nu)}\right]\ee
in general.

Let's also consider the long wavelength limit of our result for the spectral index
\be n_\psi=4+2\frac{d\ln{H^{(2)}_\nu(-k\eta)}}{d\ln{k}}. \ee In the long wavelength limit 
\be \frac{dH^{(2)}_\nu(-k\eta)}{dk}=\frac{\nu}{k}\left[\frac{i\Gamma(\nu)}{\pi}\left(\frac{-k\eta}{2}\right)^{-\nu}\right]=\frac{\nu}{k}H^{(2)*}_\nu(-k\eta),\ee
and the spectral index is
\be n_\psi=4+2\nu\frac{H^{(2)*}_\nu(-k\eta)}{H^{(2)}_\nu}.\ee
Since $k\eta \to 0$, the second term in $H^{(2)}_\nu$ will dominate and the spectral index will be
\be n^{pl}_\psi\sim 4-2\nu, \ee
which is constant. Notice that depending on $\nu^2=\frac{(1-3p)^2}{4(1-p)^2}$, the spectral index for power law inflation need not be 1 and imply scale invariance which is different from exact de Sitter where $n^{dS}_\psi=1$.  

Lastly we will mention the power spectrum of the gauge invariant comoving curvature perturbation $\mathcal{R}$. It has a very simple definition in terms of $v$
\be \mathcal{R}=\psi+\frac{\mathcal{H}}{\phi'_0}\delta\phi=\frac{\Hmath}{a\phi'_0}v=\frac{v}{z}.\ee
Since the power spectrum of v is gauge invariant, it is clear that the power spectrum of $\mathcal{R}$ will also be gauge invariant
\be \Delta^2_{\mathcal{R}_k}=\frac{1}{z^2}\Delta^2_{v_k}. \ee
For power law inflation we have calculated that $\frac{1}{z^2}=\frac{3l^2_{Pl}}{2}\frac{(1-p)^2}{p}\eta^{\frac{2p}{p-1}}$, so that
\be \Delta^2_{\mathcal{R}_k}=\frac{3l^2_{Pl}k^3}{16\pi^2}\frac{(1-p)^2}{p}\eta^{\frac{3p-1}{p-1}}[H^{(2)}_\nu(k\eta)]^2.\ee

One does get the same answer in comoving gauge where $\mathcal{R}=\psi$ and
\be 
\Delta^2_{\mathcal{R}_k}=\Delta^2_{\psi_k}=\frac{3l^2_{Pl}k^3}{16\pi^2}\frac{(1-p)^2}{p}\eta^{\frac{3p-1}{p-1}}|H^{(2)}_\nu(k\eta)|^2. \ee
This expression in the long wavelength limit gives
\be
\Delta^2_{\mathcal{R}_k}=\frac{3l_{Pl}^2}{16\pi^4}\frac{(1-p)^2}{p}4^\nu\Gamma^2(\nu)k^{\frac{2}{1-p}}(-k\eta)^{(\frac{3p-1}{p-1}-2\nu)}.\ee
This again gives room for deviations from scalar invariance which are exact for exact de Sitter.

At the time of the writing of this thesis, the spectral index for superhorizon $\zeta$ modes as observed by Planck is
\be 1-n_s=0.0355\pm0.005 (68\% CL).\ee
This is an observational evidence in favor of the existance of primordial scalar perturbations of the type that arises in de Sitter. Yet it is an observation that leaves room for small deviations from de Sitter. We will not work out the tensor modes here but the ratio of the power spectrum of tensor modes, to the ratio of the power spectrum of scalars $r=\frac{\Delta_t^2}{\Delta_s^2}$ is another quantity accessible to observations.

\section{An Application: Primordial Black Hole Constraints and the Spectral Index}
\label{sec:BPH}

In this section we will talk about the growth of perturbations and differences between structure formation in radiation and matter domination. The two point correlation function, which we introduced as the power spectra, makes the connection between perturbation theory and observables. Previously, we saw how the power spectra can be characterized by its amplitude, and the scalar index which accounts for its scale dependence, in the sense that it parametrizes if different wavelengths $k$, hence different scales, carry more power then others.  In this section we will demonstrate how bounds on structure formation can be expressed in terms of bounds on the mass of the scalar field that sets the perturbations and the spectral index of their two point functions. Our main goal is to derive these bounds for primordial black hole formation in non-thermal eras driven by scalar fields besides the inflaton, which can dominate the energy density in the early universe. To be able to make use of these bounds, we need to understand what is the fraction of mass $M$ the mass of a black hole at the time it is formed, which makes $M$ the mass that went into the black hole to form it, to the over all mass present in the universe at that time. This fraction is denoted by $\beta(M)$. We will first talk about the motivation for such a primordial matter dominated era, then review structure formation and how formation of black holes are understood and end with the constraints. The main work related to this section has been published in \cite{Georg:2016yxa}.
\subsection{A Short Review of Non-thermal Histories}
In standard cosmology, inflation is followed by a radiation dominated era that lasts up to CMB. Non-thermal histories are motivated by split-SUSY scenarios with an attempt to explain the origins of CDM. We refer the reader to \cite{Kane:2015jia} for a more detailed review. In split-SUSY spectrum one has scalar superpartners at the 10-100 TeV range, whose mass is set by SUSY breaking. The fermion superpartners take place at lower scales on the order of 100-1000GeV, below the SUSY breaking scale. If the SUSY breaking scale  $\Lambda_{SUSY}$ is mediated by gravity, then there is the relation $m_{3/2}=\frac{\Lambda_{SUSY}^2}{m_{Pl}}$ between $\Lambda_{SUSY}$ and the graviton mass $m_{3/2}$. During inflation this becomes $m_{3/2}=\frac{\Lambda_{SUSY}^2}{m_{Pl}}\simeq H_{I}$, where $H_I$ is the Hubble rate during inflation.

In the range of scalar superpartners, one can also have the moduli, $\sigma$, which appear from requirements of UV completion. These are shift symmetric scalar fields, who gain mass by the breaking of the shift symmetry either during inflation or by quantum effects. In split-SUSY models the mass of the field can be very heavy. As such they can decay into radiation and dark matter, who are among fermion superpartners below the susy breaking scale. With this mass range of 10-100TeV the field will decay early enough without disturbing BBN. Thus it is important that the field be heavy and not disturb the well established physics that is to arrise later.

The key property of non-thermal histories is that they give rise to a new matter dominated phase when the oscillations of the field reach a frequency that is equal to the mass of the field. Breaking of the shift symmetry during inflation happens via low-energy SUSY breaking. Having had supersymmetry at one point implies corrections to the potential as
\begin{equation}
\label{eq:susypotential}
\Delta V=-c_1H^2_I\sigma^2-\frac{c_n}{M^{2n}}\sigma^{4+2n}+...
\end{equation} 
where $H_I$ is the Hubble rate during inflation and $M$ is the scale of new physics. The first term in \eqref{eq:susypotential} causes the symmetry to be broken during inflation in the early universe. It will be restored and broken again later. Earlier, at times of high inflation ($H_I>m_\sigma$), the minimum of the field is at
\begin{equation}
\label{eq:minimumearly} \langle \sigma \rangle \sim M \left(\frac{H_I}{M}\right)^{\frac{1}{n+1}}
\end{equation}
and being a function of the minimum of the potential the mass of the field is $m_\sigma \propto H_I$.

Later on at the time of low-energy SUSY breaking when $H_I \sim m_\sigma $ and the higher order corrections become negligable
\begin{equation}
\label{eq:laterpot}\Delta V =m^2_\sigma \sigma^2.
\end{equation}
The minimum in this case, is near $\langle \sigma \rangle \sim 0$ and the mass becomes $m\sigma = c_0 m_{3/2}\simeq 10-1000$ TeV.

Although the field always sits at the minimum of the potential, the minimum of the potential changes at earlier and later times during cosmic evolution. This displacement of the field between two energy scales leads to energy being stored in the form of coherent oscillations of the field and a scalar condensate being formed \cite{Iliesiu:2013rqa}. 

The equation of motion for a scalar field is 
\begin{equation}
\label{eq:scalar fieldeom} \ddot \sigma + 3H\dot \sigma =\frac{\partial V}{\partial \sigma}
\end{equation}
With the potential $V=\frac{1}{2}m_\sigma^2\sigma^2$ this gives
\begin{equation}
\label{eq:phisquare}\ddot \sigma +3H\dot \sigma +m^2_\sigma \sigma=0.
\end{equation}

If we make the following field redefinition of 
\begin{equation}
\sigma(t)=\frac{1}{a^{3/2}(t)}\chi (t)
\end{equation}
the equation simplifies to
\begin{equation}
\ddot \chi+\left(m_\sigma^2-\frac{3}{2}\frac{\ddot a}{a}-\frac{3}{4}\frac{\dot a^2}{a^2} \right)\chi=0
\end{equation}
Starting from $H \approx m_\sigma$, H will continue to decrease so that in time the regime of interest becomes $m_\sigma >> H$ and the Hubble friction term $\frac{\dot a}{a}$ is negligable. In this case the solution is of the form \cite{Gorbunov:2011zzc}
\begin{equation}
\sigma(t)=\sigma_*\frac{cos(m_\sigma t+\beta)}{a^{3/2}(t)}.
\end{equation}

Let's now calculate $T_{\mu \nu}=\partial_\mu \sigma \partial_\nu \sigma -g_{\mu \nu}\mathcal{L}$, where $\mathcal{L}=\frac{1}{2}\dot \sigma^2 +\frac{m_\sigma^2}{2}\sigma^2$. Since we are in the $m>>H$ regime and $\frac{\dot a}{a}$ terms are negligable
\begin{equation}
\dot \sigma(t)=-m_\sigma\sigma_*\frac{sin(m_\sigma t+\beta)}{a^{3/2}(t)}-\frac{3}{2}\sigma_*\frac{cos(m_\sigma t+\beta)}{a^{3/2}(t)}\frac{\dot a}{a}\sim -m_\sigma\sigma_*\frac{sin(m_\sigma t+\beta)}{a^{3/2}(t)}.
\end{equation}
And we obtain
\begin{equation}
\label{eq:rho}\rho_\sigma=T_{00}=\frac{1}{2}\dot \sigma^2+\frac{1}{2}m_\sigma^2\sigma^2=\frac{1}{2}m^2_\sigma \frac{\sigma_*^2}{a^3(t)}
\end{equation}
\begin{equation}
\label{eq:pressure}p_\sigma \delta_{ij}=T_{ij}=[\frac{1}{2}\dot \sigma^2-\frac{1}{2}m_\sigma^2 \sigma^2]=\left[\frac{1}{2}\sigma_*^2m_\sigma^2\frac{sin^2(m_\sigma t+\beta)}{a^3(t)}-\frac{1}{2}m_\sigma^2\sigma_*^2\frac{cos^2(m_\sigma t+\beta)}{a^3(t)}\right]
\end{equation}
Since the average over all angles of both $sin^2(m_\sigma t+\beta)$ and $cos^2(m_\sigma t+\beta)$ are $\frac{1}{2}$, the average pressure is $\langle p_\sigma \rangle=0$. This gives the equation of state $\frac{p_\sigma}{\rho_\sigma}=0$, which is the equation of state of matter. These averages are meaningful in the limit $m_\sigma >> H$ which is when the field oscillates with frequency $\omega=m_\sigma$, much faster than the universe expands. Thus the energy stored in the coherent oscillations of the scalar field leads to a matter dominated era. The field being heavy and the Hubble rate falling in the $m_\sigma >> H$ is what leads to this matter dominated era. The good thing about having this matter dominated era is that because of the way the fluctuations scale and the rate at which field decays, there is the possibility to form primordial structures in this era, as opposed to radiation domination. The heaviness of the field also ensures that it decays out before BBN. Thus for non-thermal cosmologies to be pausible models, it is important that the scalar field be heavy. 
\subsection{Jean's Scale and Horizon Size}
\label{sec:jeans scale}
For a non expanding fluid the evolution of small denstiy perturbations of nonrelativistic matter are governed by
\begin{equation}
\label{eq:euler}
\ddot{\delta\rho}-{c_s}^2\nabla^2\delta\rho=4\pi G\rho\delta\rho
\end{equation}
where $\rho$ is the background density, $\delta\rho$ denote the density perturbations, the speed of sound is ${c_s}^2=\frac{\partial p}{\partial\rho}$ and density and pressure are related by $p=w\rho$. The solution will be of the form $\delta \rho(r,t)=\rho_0e^{-i\vec{k}.\vec{r}+i\omega t}$, which when plugged in gives the following dispersion relation
\begin{equation}
\label{eq:dispersion}
\omega^2={c_s}^2k^2-4\pi G\rho.
\end{equation}
For coordinate momenta ${k_J}^2=\frac{4\pi G\rho}{{c_s}^2}$ the frequency is zero, $\omega^2=0$. This point at which the frequency squared changes sign is known as Jean's scale, and it has important consequences for structure formation. For $k^2>{k_J}^2$ the frequency is real, which means $\delta \rho(r,t)$ is proportional to $e^{i\omega t}$ and the perturbations oscillate as sound waves.

On the other hand for $k^2<{k_J}^2$ the frequency is purely imaginery and the perturbations will exponentially grow or decay. The exponentially growing modes of the density perturbation will eventually become comparable to the background density $\delta \rho \sim \rho$. In this regime the linear perturbation theory breaks down and this region that is now denser than the background collapses to form a compact object. 

In a nonexpanding universe the coordinate momentum, $k$, and physical momentum, $p$, are the same, yet for an expanding universe $q=\frac{k}{a(t)}$, where $a(t)$ is the scale factor. Thus in general
\begin{equation}
\label{eq:jeans momentum}
q_J=\sqrt{\frac{4\pi G\rho}{{c_s}^2}}.
\end{equation}
We will stick with the physical momentum from now on, where again it will be perturbations for which $q < q_J$, that collapse to form structures. Only this time the modes will grow as a power law and not as exponentials. All this analysis so far has been Newtonian. It holds for perturbations of nonrelativistic matter at length scales well below the horizon size. This momentum scale on the stability and instability of perturbation growth can be turned into a wavelength scale by the following equation
\begin{equation}
\label{eq:jeans wavelength}
\lambda_J \equiv \frac{2\pi}{q_J}.
\end{equation}

For a pressureless fluid, such as dust, the speed of sound is zero. In the limit ${c_s}^2 \rightarrow 0$, Jeans momentum goes to infinity and as can be seen from Eq. \eqref{eq:jeans wavelength}, the Jeans wavelength goes to zero. This means perturbations within the horizon will always be bigger than Jeans scale and everything can collapse to form structures.

We are interested in exploring structure formation, mainly primordial black holes (PBH), in a matter dominated era where density perturbations of a scalar field, $\sigma$ dominates over all other species present. At first glance it is very natural to assume that the average pressure of a scalar field will be zero. However there is the important subtlety that this only holds for scales $m_\sigma \gg\frac{q^2}{H}$, where $m_\sigma$ is the mass of the scalar field, and $H$ is Hubble's parameter. In this regime the field is nonrelativistic and is pressureless. Yet above this scale, $\frac{q^2}{H} \geq m$ the average pressure of the field obtains corresctions of order O($m_\sigma$) and can no longer be neglected. In this case the field behaves as relativistic matter with the sound speed \cite{Hu:1998kj}, \cite{Hu:2000ke}
\begin{equation}
\label{eq:sound speed}
{c_s}^2=\frac{k^2}{4a(t)^2m\sigma^2}=\frac{q^2}{4m_\sigma^2}
\end{equation}
and the Jeans momentum is 
\begin{equation}
\label{eq:jeans momentum pressurely}
q_J=(16\pi G\rho_\sigma m_\sigma^2)^{\frac{1}{4}}.
\end{equation}   
The scale at which the condition that the field is no longer is pressureless coincides with the Jeans scale \cite{Gorbunov:2011zzc}.
All the fluctuations in between $\lambda_J < \lambda < R_H$, where $R_H$ is the Hubble scale, will lead to compact structures. But for scales within $\lambda < \lambda_J$, it will be harder to form compact structures since, they will only be able to form from the modes whose gravitational attraction can overcome the repulsion coming from the average pressure of the scalar field. These scales and the difference between radiation, matter and oscillating scaler field are demonstrated in Figure \ref{jeans}.

\begin{figure}
	\begin{subfigure}[b]{0.5\textwidth}
		\includegraphics[width=\textwidth]{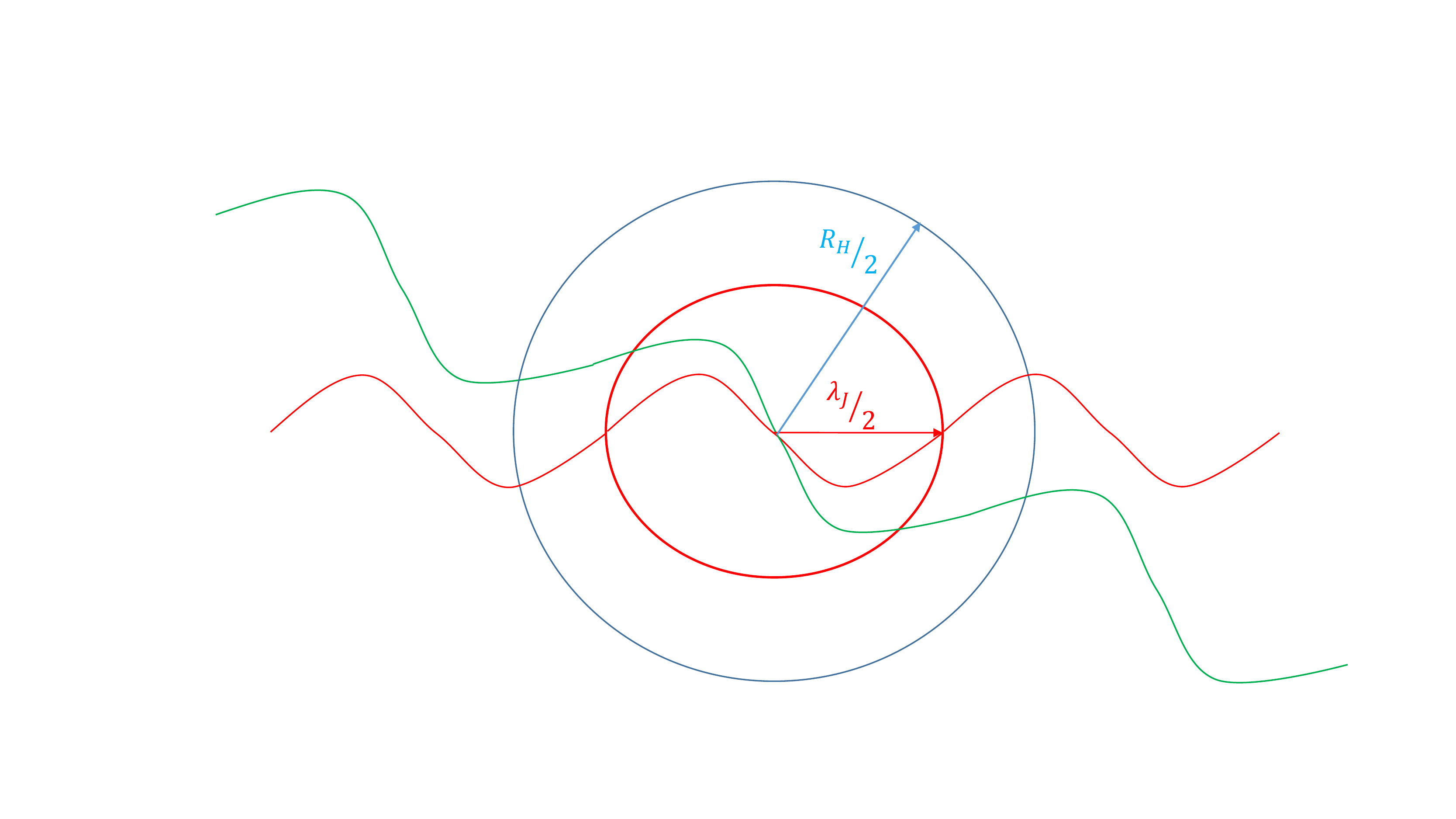}
		\caption{}
		\label{zeropresnum}
	\end{subfigure}
	~
	\begin{subfigure}[b]{0.5\textwidth}
		\includegraphics[width=\textwidth]{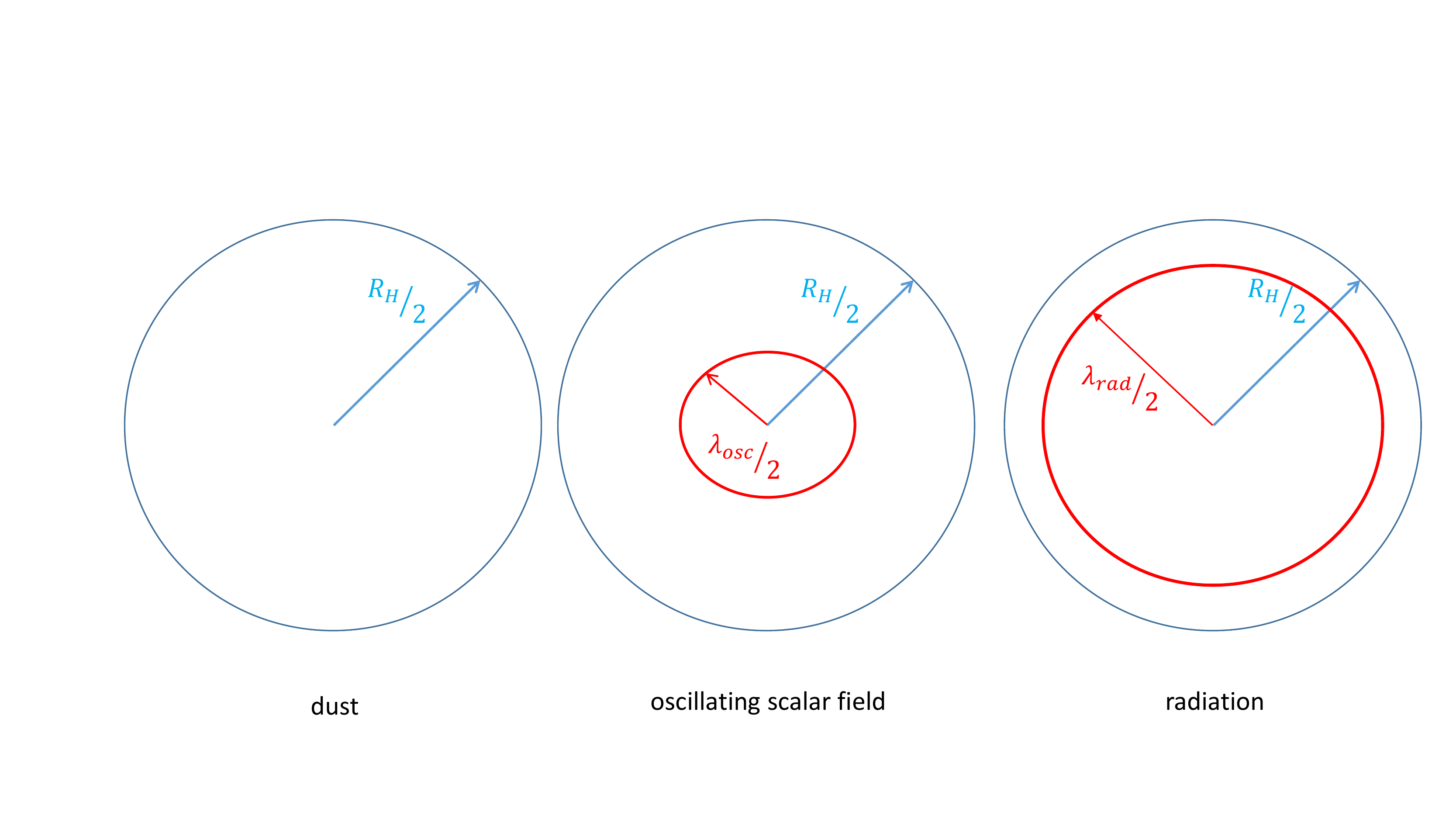}
		\caption{}
	\end{subfigure}
	\caption{Wavelengths for structure formation: (a) The blue and red circles represent the Hubble and Jeans spheres respectively. Modes with wavelengths of the size of Jeans sphere, shown here with as the red mode and below cannot over come the repulsive effect of pressure and hence cannot collapse form structures, where as the modes of bigger wavelengths such as the green one can. (b) The difference between the region of scales for collapse in a universe dominated by presureless dust, oscillating scalar field or radiation.}
	\label{jeans}
\end{figure}

As such the next point to consider is how inside the horizon, Jeans scale will be. For our case, the mass of the scalar field is $m_\sigma=10^5 GeV$. We are in the matter dominated era and working with natural units, $\hbar=c=1$ such that $m_P^2=G^{-1}$, where $m_p$ is Planck's mass and $G$ is the gravitational constant.The background density for the scalar field is \cite{Fan:2014zua} 
\be \rho \simeq 3H^2m_p^2=3\frac{m_\sigma^2m_P^2}{a(t)^3}\ee where $H=H_0 a(t)^{-\frac{3}{2}}$
and $H_0\simeq m_\sigma$. Using Eq. (\ref{eq:jeans momentum pressurely}) we get
\begin{equation}
\label{eq:lambda J}
\lambda_J=1.79\times 10^{-5}a(t)^{\frac{3}{4}} GeV^{-1}.
\end{equation}
The Hubble scale on the other hand is
\begin{equation}
\label{eq:hubble scale}
R_H=\frac{c}{H}=a(t)^{\frac{3}{2}}\times 10^{-5} GeV^{-1}
\end{equation}
Comparing equations (\ref{eq:lambda J}) and (\ref{eq:hubble scale}), with $a(t)=t^{\frac{2}{3}}$ for a matter dominated universe, one arrives at 
\begin{equation}
\label{eq:scale comparing}
R_H\simeq 0.6\sqrt{t}\lambda_J.
\end{equation}

The ratio $\frac{R_H}{\lambda_J}\simeq 0.6\sqrt{t}$ means that at some time $t=t*$ the Hubble scale and Jeans scale will be the same. For earlier times $t<t*$, this ratio will be smaller than unity and the Hubble scale will be smaller than Jeans scale. Since it is the modes within the horizon that form structures and in this interval they are all within the nonzero pressure regime, it is very unlikely that structures will form here. Yet as time passes it will be $t>t*$, and the horizon size will be greater than Jeans scale. We will have modes in between $\lambda_J$ and $R_H$, in the zero pressure regime. Thus in time it will become easier to form compact objects.

\subsection{Mass Fraction of Primordial Black Holes}
Up to this point our focus has been on formation of any compact object, be it a star or a black hole. We have talked about how the scalar field will have non zero average pressure below a certain scale and how this will affect structure formation. Now we would like to calculate the mass fraction of primordial black holes (PBH), $\beta$, that would have formed from the decay of a shift symmetric scalar field $\sigma$ (a moduli) at the end of inflation. During this epoch the field will decay into dark matter and radiation and there will be perturbations arising from them as well. At first glance, the dark matter particle perturbations could contribute to PBH formation and we would expect the overall mass fraction to be the sum coming from scalar field perturbations and dark matter perturbations. But as \cite{Fan:2014zua} shows, the dark matter perturbations are negligably small next to scalar field perturbations in this era. So we will only consider the contribution to PBH coming from scalar field perturbations. We mainly want to see how the fact that the pressure of the scalar field is nonzero up to $\lambda_J$ affects the mass density. For that we will be comparing the mass fraction obtained from the density perturbation of the scalar field in two different versions, with and without the pressure being taken into consideration.
\subsection{Density Perturbations}
\label{sec:Density Perturbations}
Before we can calculate $\beta$, we need to figure out how the density perturbations $\delta_\sigma$ evolve. Since our scalar field is relativistic, we need to solve the perturbed Einstein Equations, rather than the Euler equations. In the absence of anisotropic stress, with $\Gamma_\sigma$ as the decay rate of the scalar field, the equations to be solved are \cite{Fan:2014zua}
\begin{subequations}
	\begin{align}
	\dot{\delta_\sigma}+3H(c_\sigma^2-w_\sigma)\delta_\sigma+(1+w_\sigma)\left(\frac{\theta_\sigma}{a}-3\dot{\Phi}\right)=-\Gamma_\sigma\Phi \\
	\dot{\theta_\sigma}+H\theta_\sigma+\frac{c_\sigma^2}{1+w_\sigma}\frac{\nabla^2\delta_\sigma}{a}-3Hw_\sigma\theta_\sigma+\frac{\nabla^2\Phi}{a}=-\Gamma_\sigma\left[\frac{\theta_\sigma}{1+w_\sigma}-\theta_\sigma\right]
	\end{align}
	
\end{subequations}

where $\theta_\sigma$ is the velocity perturbation. With the matter dominated background, the equation of state is zero, we also require that $\Phi=\Phi_0$. In the momentum space these equations become
\begin{subequations}
	\begin{align}
	\label{ee1}\dot{\delta_\sigma}+3Hc_\sigma^2\delta_\sigma+\frac{\theta_\sigma}{a}=-\Gamma_\sigma\Phi_0\\
	\label{ee2}\dot{\theta_\sigma}-H\theta_\sigma-c_\sigma^2\frac{k^2}{a^2}\delta_\sigma-\frac{k^2}{a}\Phi_0=0
	\end{align}
\end{subequations}
Our startegy will be to solve Eq.(\ref{ee1}) for the velocity perturbation in terms of the density perturbation and obtain a second order differential equation for the density perturbations by inserting the expression for velocity perturbation into Eq. (\ref{ee2}). 

Let us first consider the $ m_\sigma \gg \frac{q^2}{H} $ regime, where $c_\sigma^2=0$, here the equation for the density perturbation becomes
\begin{equation}
\ddot {\delta_\sigma}+2H\dot \delta_\sigma +2\Gamma_\sigma H\Phi_0+\frac{k^2}{a^2}\Phi_0=0\end{equation}
We will switch to number of e-folds, $N$ as the time coordinate which is, $dN=dlna$. In a matter dominated universe $H=H_0e^{\frac{-3N}{2}}$ and $a=e^N$. During scalar domination $\Gamma_\sigma / H \ll 1$.  With derivatives with respect to $N$ denoted by prime we arrive at 
\begin{equation}
\label{eq:for zero pressure}\delta_\sigma''+\frac{1}{2}\delta_\sigma'=-\frac{k^2}{a^2H^2}\Phi_0
\end{equation}
whose solution, demanding that $\delta_\sigma$ goes to $-2\Phi_0$ as $k$ goes to zero, is
\begin{equation}
\label{eq:zeropressuresoln}\delta_\sigma(k,N)=-2\Phi_0-\frac{2}{3}\frac{k^2}{H_0^2}\Phi_0e^N,
\end{equation}
which is plotted in Figure \ref{zeropresnum}.


\begin{figure}
	\begin{subfigure}[b]{0.3\textwidth}
		\includegraphics[width=\textwidth]{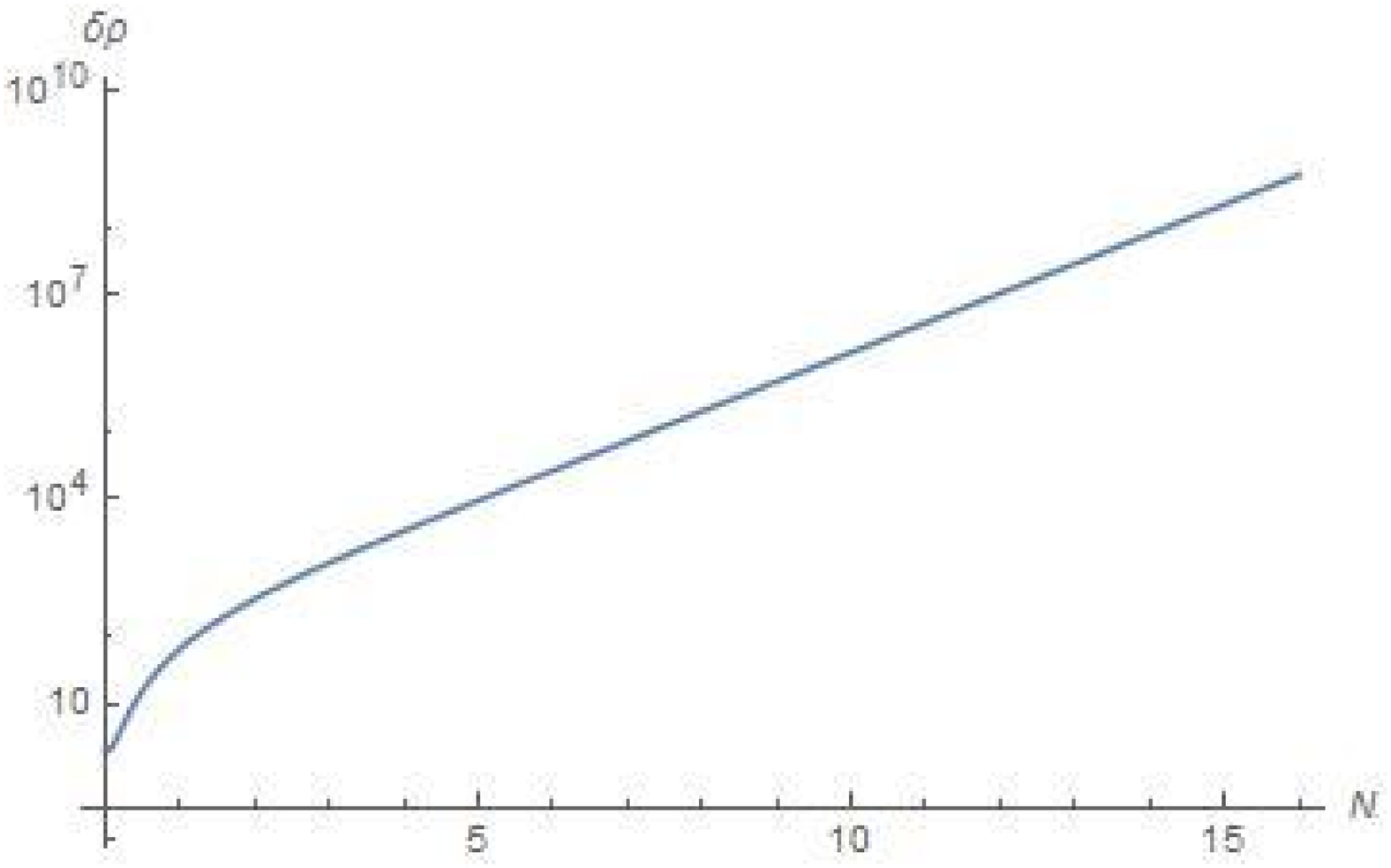}
		\caption{}
		\label{zeropresnum}
	\end{subfigure}
	~
	\begin{subfigure}[b]{0.3\textwidth}
		\includegraphics[width=\textwidth]{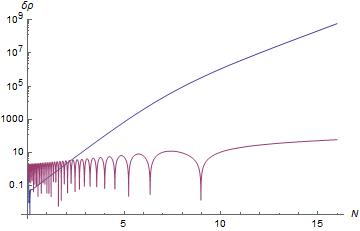}
		\caption{}
		\label{fig:nonzeropressure*}
	\end{subfigure}
	~
	\begin{subfigure}[b]{0.3\textwidth}
		\includegraphics[width=\textwidth]{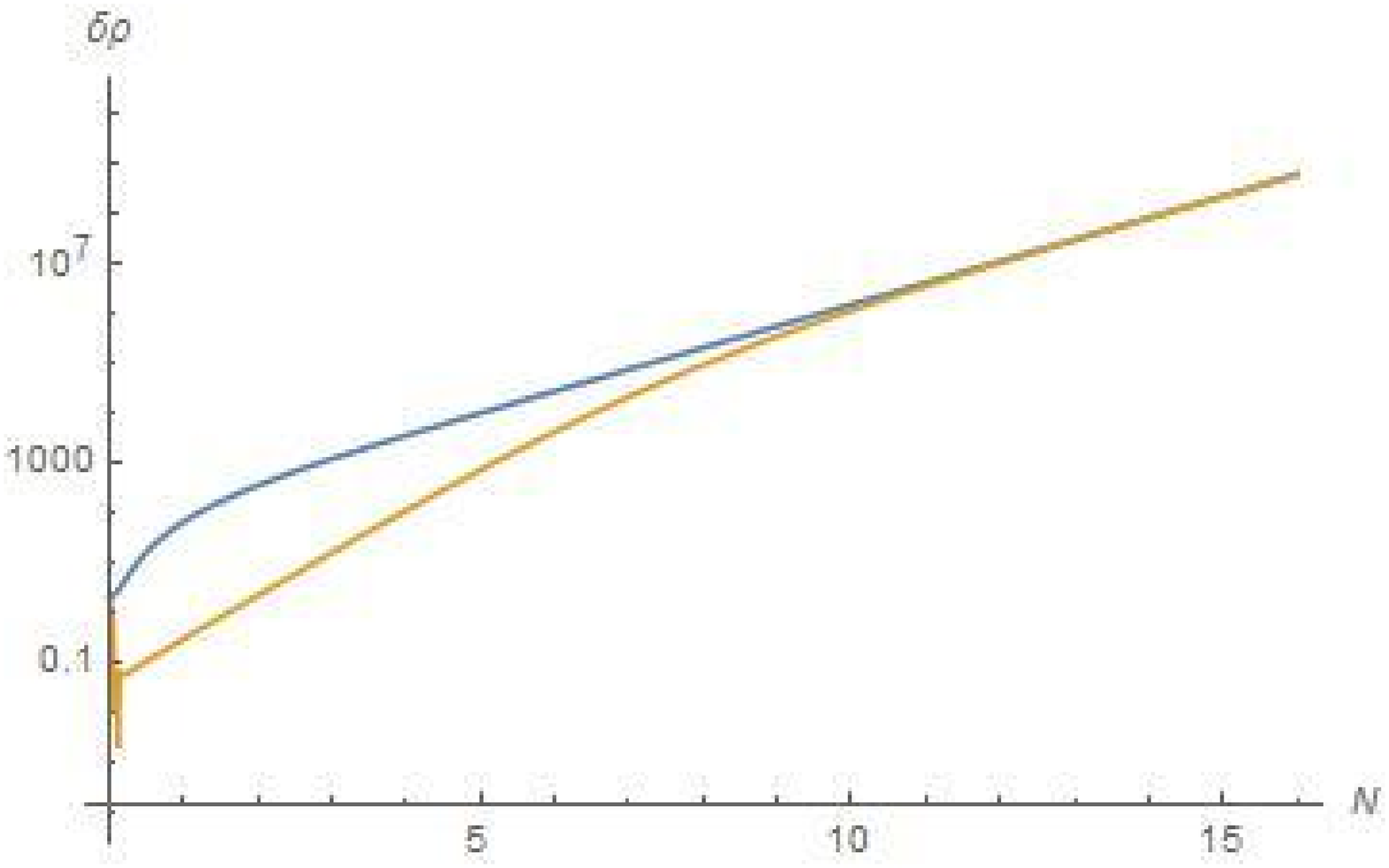}
		\caption{}
		\label{fig:comparing}
	\end{subfigure}
	\caption{Numerical evolution of the density perturbations:(a) The numerical solution for the evolution of the density perturbation in the case of zero pressure, which is governed by equation \eqref{eq:for zero pressure}. (b) The evolution of the density perturbation in the case of nonzero pressure, where the blue curve is the numerical solution to full equation \eqref{eq:nonzeropressureeqn} and the purple curve is the solution to the approximate equation \eqref{eq:appr} for large k, for values $k=10^6$ GeV, $m_\sigma=10^5$ GeV, $\Phi_0=10^{-5}$. (c) Comparison of the effect of pressure on growth via the numerical solutions, the blue curve is solution to zero pressure evolution equation \eqref{eq:for zero pressure} and the orange curve is the solution to \eqref{eq:nonzeropressureeqn} with the presence of pressure.}
\end{figure}

In the $\frac{q^2}{H} \geq m_\sigma$ regime where the average pressure of the scalar field is nonnegligable and the sound speed is $c_\sigma^2=\frac{k^2}{4m^2a^2}$ the equation that governs the evolution of density perturbations becomes,
\begin{equation}
\label{eq:nonzeropressureeqn}\delta_\sigma''+\left[\frac{1}{2}+\frac{3}{4}\frac{k^2}{a^2m_\sigma^2}\right]\delta_\sigma'-\frac{k^2}{4a^2m^2}\left[\frac{9}{2}+\frac{k^2}{a^2H^2}\right]\delta_\sigma=-\frac{k^2}{a^2H^2}\Phi_0
\end{equation}

It looks as though the k and N dependence of $\delta_\sigma(k,N)$ come as $\delta_\sigma(k,N)=f(k)e^{\alpha N}$. Naively we would expect only the term that goes as $k^4$ to contribute in the $k\rightarrow \infty$ limit, and the equation to be
\begin{equation}
\label{eq:appr}\delta_\sigma'' +\frac{k^4}{4a^4m^2H^2}\delta_\sigma=0.
\end{equation}
At first sight we can see that the solution to \eqref{eq:appr} will be oscillating sound waves, in which case the perturbations do not lead to much structure formation due to pressure. We indeed do see this oscillatory behavior if we plot the numerical solution in Figure \ref{fig:nonzeropressure*}. However from the same plot we see that the numerical solution of the full equation, \eqref{eq:nonzeropressureeqn} has actually a growing behaviour. 

If we plot the numerical solution to \eqref{eq:for zero pressure} and \eqref{eq:nonzeropressureeqn} we see that there is some difference in the growth of perturbations up to some time in number of e-folds. However because the terms that arrise from the effects of pressure are all inversely related to mass, and because the mass of the field is heavy, their contribution is not big enough to overcome the gravitational collapse. 

\subsection{Why the pressure effects are negligible}
It is the perturbations within the horizon, $q>H$, that can grow. In section \ref{sec:jeans scale} we also pointed out that for perturbations to lead to structure formation, they must be below Jean's momentum scale, $q_J>q$. Therefore the condition on the physical momenta of the modes that lead to structures, in our case PBHs, is
\begin{equation}
q_J>q>H.
\end{equation} 
From (\ref{eq:jeans momentum}) using the units $m_{pl}=\sqrt{\frac{1}{8\pi G}}$ and the Friedmann equation $\rho_\sigma=3m_{pl}^2H^2$ where we have assumed that the universe is dominated by the scalar field, we obtain
\begin{equation}
\label{eq:q_j^2}q_J^2=\sqrt{6}Hm_\sigma.
\end{equation}
On the other hand given an initial mode $q_0$, we know how it evolves in time. As the universe expands the physical momentum $q(t)$ decreases, yet the comoving momentum $k$ is constant, $k=a(t)q(t)=a_0q_0$. Thus we have
\begin{equation}
\label{eq:momentum evol}q(t)=q_0\frac{a_0}{a(t)}.
\end{equation}

We also know that the relation between scale factor $a=e^N$, and Hubble parameter $H=H_0e^{-3N/2}$ is 
\begin{equation}a(t)=\left( \frac{H(t)}{H_0}\right)^{2/3}.\end{equation}
Altogether we have
\begin{equation}
q(t)=q_0a_0\left( \frac{H(t)}{H_0}\right)^{2/3}.
\end{equation}
If we take the subscript zero to mean the beginning of oscillations, then $a_0=1$, $H_0\sim m_\sigma$,
\begin{equation}
\label{eq:q(t)}q(t)=q_0\left( \frac{H(t)}{m_\sigma}\right)^{2/3}.
\end{equation}
Now let us write our growth condiition as
\begin{equation}
\label{eq:growth}\frac{q_J^2}{H^2}>\frac{q^2}{H^2}>1,
\end{equation}
and plug in equations \ref{eq:q_j^2} and \ref{eq:q(t)} to get
\begin{equation}
\label{eq:for growth}\sqrt{6}>\frac{q_0^2}{m_\sigma^2}\left(\frac{H}{m_\sigma}\right)^{1/3}>\frac{H}{m_\sigma}.
\end{equation}
Let's consider a mode for whom $\frac{q_0}{m_\sigma}=1$. Such a mode will lead to structure formation if $\sqrt{6}>(\frac{H}{m_\sigma})^{1/3}>\frac{H}{m_\sigma}$. The rightside of the constraint requires that the ratio $\frac{H}{m_\sigma}$ be less than 1. This is always true, at the start of the oscillations the Hubble's parameter is as big as the mass of the field and in time it decreases where as the mass of the field stays constant. Therefore modes whose initial momenta are comparable to the mass of the field will always overcome pressure and lead to formation of black holes. For pressure to come into play, equation \ref{eq:for growth} must break down. This will only happen for modes whose initial momentum is so small compared to the mass of the field that even though $\frac{H}{m_\sigma}$ is less than one $\frac{q_0^2}{m_\sigma^2}\left(\frac{H}{m_\sigma}\right)^{1/3}<\frac{H}{m_\sigma}$ in which case the perturbations that evolve via \ref{eq:nonzeropressureeqn} should enter the calculations. Perturbations of momenta $k=q(t)a(t)$, correspond to mass $M\propto k^{-3}$, which suggests that such small momenta will correspond to large masses. As there is an upper bound on the mass of the black holes it is likely that these modes do not form black holes anyways and so pressure of the scalar field is negligable for PBH formation.

\subsection{Mass Fraction in the Pressureless Case for a Radiation Dominated Universe}

The mass fraction is the ratio of the mass that goes into PBH to the total mass within the horizon, $\beta=\frac{m_{PBH}}{m_{horizon}}$. To calculate $\beta$ we will use the Press-Schechter method, following the definitions of \cite{Young:2014ana}
\begin{equation}
\beta_{\text{PS}}(\nu_c)=2\int_{\nu_c}^{\infty}P(\nu)d\nu=\frac{2}{\sqrt{2\pi}}\int_{\nu_c}^{\infty}e^{-\frac{\nu^2}{2}}d\nu
\end{equation}
which considers the mass fraction to be the integral of the probability distribution function over peaks above a certain height $\nc$. This amplitude related variable is $\nu=\frac{\Delta}{\sigma^6}$. In \cite{Young:2014ana} the density contrast is taken to be $\Delta=\delta-1$, but we will consider $\delta_\sigma \equiv \frac{\delta\rho_\sigma}{\rho_\sigma}$ itself as the density contrast, which appears to be more common in literature. Thus out of all the density perturbations, it will be the ones whose ratio with respect to the background density, are bigger than some critical value $\delta_c$, that collapse into PBH's. Perturbations below this can still collapse to form other compact objects such as stars. As such the constraints on PBH formation enters into the calculation via this critical parameter with value $\delta_c=0.5$ \cite{Young:2014ana}. In which case
\begin{equation}
\nu=\frac{\delta}{\sigma^6}
\end{equation} 
where $\sigma$ is the variation
\begin{equation}
\sigma=\langle\Delta^2\rangle=\int_0^{\infty}\frac{dk}{k}\tilde{W}^2(R,k)\mathcal{P}_{\delta}(k)
\end{equation}

The physical aspects of the cosmological scenario under consideration enters into the mass fraction through the integral limit $\nc$. With $\nc=\frac{\delta_c}{\sigma^6}=\frac{0.5}{\sigma^6}$, the information related with the perturbations enters the calculation only via the power spectrum $\PS$, calculated as $\PS=k^3|\delta_\sigma|^2$. The window function,
\begin{equation}
\tilde{W}(R,k)=\exp\left(-\frac{k^2R^2}{2}\right)    
\end{equation}
picks the super-horizon modes, where 
the radius of the horizon is $R=\frac{1}{aH}$ 
\begin{equation}
\tilde{W}(R,k)=e^{\frac{-k^2}{2a^2H^2}}.
\end{equation}
In a radiation dominated universe the Jean's length scale is almost as big as the horizon size, therefore modes within will not lead to structures. Modes outside the horizon do not contribute since they are frozen. So it is only the modes that have just crossed over that can form PBHs in a radiation dominated universe. 




\subsection{The Matter Dominated Era}

The matter dominated era begins when the energy contribution of the matter particles that the scalar field decays into, $E_\sigma$, becomes comparable to the energy contribution of the relativistic particles, $E_r$, which is in our case photons. That is when the field oscillates as fast as the background expansion $H(t_0)\simeq m_\sigma$, where subscript zero denotes the beginning of matter domination. At temperature T the energy density of photons is $E_r\sim k_B T n_r$ where $n_r$ denotes the number of photons. If we denote $\nu =\frac{n_\sigma}{n_r}$ as the relative consentration of scalar particles, their energy density is $E_\sigma=m_\sigma c^2n_\sigma=m_\sigma c^2\nu n_r$. At the beginning of matter dominated era, if the temperature is $T_0$ we have from $E_\sigma=E_r$, $T_0=\frac{m_\sigma \nu c^2}{k_B}=m_\sigma \nu$, where the last part is in units $k_B=c=\hbar =1$.

We also know that $\rho=3m_{pl}^2H^2=\frac{\pi^2}{30}g_*T^4$, where $g_*$ is the effective number of degrees of freedom. Starting from Planck time, $t_{pl}$, up to $t_0$ when matter dominated era begins, the universe is radiation dominated with a scale factor of $a\sim\left(\frac{t}{t_{pl}}\right)^{1/2}$, so that $H(t_0)=\frac{1}{2t_0}$. As such from
\begin{equation}
\frac{3m_{pl}^2}{2t_0^2}=\frac{\pi^2}{30}g_*(m_\sigma \nu)^4\end{equation}
we get
\begin{equation}\label{eq:t0}
t_0\sim\left(\frac{m_{pl}}{m_\sigma \nu}\right)^2t_{pl}\end{equation}
with numerical factors ignored. 

Following \cite{Brandenberger:2003vk}, we can introduce the cosmological fluctuations in terms of mass fluctuations at the time of horizon crossing, $t_H$, as
\begin{equation}
\label{eq:mass fluc}
\left(\frac{\delta M}{M}\right)^2(k,t_H)\sim k^{n-1}.
\end{equation}
Here $n$ is the spectral index and $n=1$ corresponds to a scale invariant spectra. In a matter dominated era these modes will evolve as
\begin{equation} 
\left(\frac{\delta M}{M}\right)(k,t)=\frac{a(t)}{a(t_H)}\left(\frac{\delta M}{M}\right)(k,t_H).
\end{equation}
With $a(t)\sim t^{2/3}$ this becomes
\begin{equation}
\left(\frac{\delta M}{M}\right)^2(k,t)=\left(\frac{t}{t_H}\right)^{4/3}k^{n-1}.
\end{equation}
At the time of horizon crossing $a(t_H)H(t_H)=k$ in units where $c=1$, by this we can express $t_H$ in terms of comoving momenta $k$
$$t_H^{2/3}\frac{2}{3t_H}=k$$
\begin{equation}
t_H\sim k^{-3},
\end{equation}
\begin{equation}
\left(\frac{\delta M}{M}\right)^2(k,t)\sim k^{n+3}.
\end{equation}
For these modes the power spectrum can be written as $P_\delta(k)=\left(\frac{\delta M}{M}\right)^2(k,t)=k^3|\delta_k|^2$, and $|\delta_k|^2\propto k^n$ \cite{Brandenberger:2003vk}.

For a mass distribution of $M=\bar{M}+\delta M$, with fluctuations $\delta M$ around a mean mass $\bar{M}$, the density enclosed in a volume $v\sim t^3$ will be
$$\rho=\frac{M}{V}=\frac{\bar{M}}{V}+\frac{\delta M}{V}.$$
By $\rho=\bar{\rho}+\delta \rho$, one can recognize that $\delta\rho=\frac{\delta M}{t^3}$ and
\begin{equation}
\frac{\delta \rho}{\rho}=\frac{\delta M/t^3}{M/t^3}=\frac{\delta M}{M}.
\end{equation} 
So the rms amplitude of the fluctuations at the time the mode reenters the horizon, by \eqref{eq:mass fluc} will be
\begin{equation}
\label{eq:density at tH}
\frac{\delta \rho}{\rho}(k,t_H)=\frac{\delta M}{M}(k,t_h)\sim k^{(n-1)/2}.
\end{equation}

Now we can associate with each wavelength $\lambda (t)=\frac{2\pi a(t)}{k}$ of the modes, a mass M defined as the mass of non-relativistic particles contained in a sphere of radius $\frac{\lambda(t)}{2}$ \cite{Padmanabhan:2010zzb}
\begin{equation}M\equiv \frac{4\pi \rho(t)}{3}\left(\frac{\lambda(t)}{2}\right)^3=4\pi m^2_{pl}H^2(t)\left(\frac{a(t)}{k}\right)^3.\end{equation}
This suggests that $k\sim M^{-1/3}$. Thus we can write the rms amplitude of our fluctuations, at the time of horizon crossing, to be of the form
$$\frac{\delta \rho}{\rho}(M,t_H)=\delta(M)\delta(t_H)$$
where by \eqref{eq:density at tH}
$$\delta(M)=M^{(1-n)/6}$$
in agreement with \cite{Carr:1994ar}. If normalized on the COBE/DMR quadrupole scale
\begin{equation}
\delta(M)=\delta_C\left(\frac{M}{M_C}\right)^{(1-n)/6},
\end{equation}
where $\delta_C\sim 3.8\times 10^{-6}$.

For a matter dominated era, as noted in section \eqref{sec:jeans scale}, the Jeans scale falls deep inside the horizon and so more of the modes inside can grow big enough to form PBHs. Denoting the beginning of matter dominated era by $t_0$ and the end of it by $t_e$, the density perturbations that are initially of the order of metric perturbation $\frac{\delta \rho}{\rho}(t_0,M)\sim \Phi_0$, will evolve as
\begin{equation}
\frac{\delta \rho}{\rho}(t,M)=\frac{\delta \rho}{\rho}(t_0,M)\left(\frac{t}{t_0}\right)^{2/3}=\Phi_0\left(\frac{t}{t_0}\right)^{2/3}. 
\end{equation}
When density perturbations become order 1, their evolution is no longer linear. Such configurations will no longer follow the background evolution but begin to form structures. It is at these times that PBHs can also form. Meaning PBH formation starts at time $t_1$ when
$$\frac{\delta \rho}{\rho}(t_1,M)\sim 1$$
$$\Phi_0\left(\frac{t_1}{t_0}\right)^{2/3}\sim 1$$
\begin{equation}
\label{eq:t1}
t_1=\Phi_0^{-3/2}t_0.
\end{equation}
Out of these order one fluctuations that form structures, PBHs will be the ones that remain spherically symmetric and that have contracted to a size that corresponds to the Schwarzchild radius, $r_g$, of the configuration. If $r_1$ is the horizon size at time $t_1$ and $M$ denotes the mass within, the horizon density will be
\begin{equation}
\rho_1=\frac{3M}{4\pi r_1^3},
\end{equation}
and the maximal density of spherical configurations will be \cite{Polnarev:1986bi}
\begin{equation}
\rho_{max}\sim \rho_1s^{-3}
\end{equation}
where s determines the degree of deviation from spherical symmetry. Following \cite{Padmanabhan:2010zzb}, with the parameter $x$ defined as $x=\frac{r_g}{r_1}$, the density of mass within a sphere of radius $r_1$, that is the density of PBH's is
\begin{equation}
\rho_{PBH}=\frac{3M}{4\pi r_g^3}\sim\rho_1x^{-3}.
\end{equation}
And the condition of spherical symmetry translates into the requirement of $\rho_{PBH}$ to be within the density of spherical configurations, $\rho_{PBH}\leq \rho_{max}$, that is
\begin{equation}
s\leq x.
\end{equation}
Since the size of a black hole cannot be bigger than the horizon size, we also require that $x\leq 1$. Thus the configurations that satisfy
\begin{equation}
\label{eq:spherical}
s\leq x\leq 1
\end{equation}
will be the possible PBHs. Nonspherical, disk like structures will lie in the interval $s\geq x$, which can be bigger than the above interval. As such spherical structures and therefore black holes are less likely to form compared to everything else. The probability of a configuration to lie in the range of \eqref{eq:spherical} is \cite{Polnarev:1986bi}
\begin{equation}
\label{eq:prob. of spheres}
W_s\simeq 2.5\times10^{-2}x^5.
\end{equation}
One may be concerned that among the structures that initially collapse spherically, which are  within this factor, there might be ones with velocity perturbations and angular momenta big enough to distort them into minidisk halos. Such structures would not lead to black holes as they would be loosing their sphericity. However as noted in \cite{Doroshkevic:1970} it is very hard to have rotating structures. So we are safe to assume that spherical configurations will lead to blackholes and stick with the calculations of \cite{Polnarev:1986bi}. A more through consideration of formation of minidisk halos, which would reduce the factor of sphericity that goes into blackhole formation, is not likely to change the estimate much.

For black hole formation we need such spherical distributions to contract to their Schwarzschild radius. The condition on contraction of an almost spherical dust configuration with ultra compact mini halos and caustics excluded is a restriction on the inhomogeneity of the density distribution, 
$u \simeq \frac{\rho_{c1}-\rho_1}{\rho_1}$, with $\rho_{c1}$ being the density at the center of the distribution at time $t_1$, \cite{Polnarev:1986bi}
\begin{equation}
\label{eq:inhomogeneity}
u\leq x^{3/2}.
\end{equation}
And the probability of formation of such configurations with small inhomogeneity that obey \eqref{eq:inhomogeneity} is
\begin{equation}
W_u\sim\int\limits_{0}^{x^{3/2}}e^{-u^2}du \sim x^{3/2}.
\end{equation}

It will be more meaningful if we can write this expression in terms of density perturbations. To do that we rewrite $r_1$ and $r_g$ as
$$r_1=\left(\frac{3M}{4\pi\rho_1}\right)^{1/3},$$
$$r_g=\frac{2M}{m^2_{Pl}}$$
in units where $m_{Pl}=\frac{1}{\sqrt{G}}$, $c=1$. As such we have
\begin{equation}
x=\frac{2M}{^2_{Pl}}\left(\frac{4\pi \rho_1}{3M}\right)^{1/3}.
\end{equation}
In a matter dominated era the density evolves as
\begin{equation}
\rho(t)=\frac{3}{8\pi}m^2_{Pl}H^2(t)\sim \frac{m_{Pl}^2}{t^2}.
\end{equation}

So at $t_1$ we have $\rho(t_1)\sim\frac{m_{Pl}^2}{t_1^2}$ and by \eqref{eq:t1} we get
\begin{equation}
\rho(t_1)\sim\frac{m^2_{Pl}}{t_0^2}\left(\frac{\delta\rho(t_0, M)}{\rho}\right)^3\sim \rho(t_0)\Phi_0^3.
\end{equation}
As the perturbations reach order one they collapse into PBHs. This also means that the most massive PBHs will be formed from perturbations that reach order one by the time the matter era ends. We know the constraints on how big the maximum mass of these PBHs can be. Using these we can actually constrain the parameters of our model. 

To calculate $M_{max}$ consider $\frac{\delta\rho}{\rho}(M_{max},t_e)\sim 1$ that is 
$$\frac{\delta\rho}{\rho}(M_{max},t_H)\left(\frac{t_e}{t_H}\right)^{2/3}\sim 1$$
\begin{equation}
\label{eq:order1}
\delta_C\left(\frac{M_{max}}{M_C}\right)^{(1-n)/6}\left(\frac{t_e}{t_H}\right)^{2/3}\sim 1
\end{equation}
On the way to calculating $M_{max}$, we also have that 
\begin{equation}
\label{eq:rho}
\rho=\frac{3}{8\pi}m^2_{Pl}H^2\sim \frac{m^2_{Pl}}{t^2}\sim m_{Pl}\rho(t_{Pl})\frac{t^3_{Pl}}{t^2}
\end{equation}
where we have used $\rho(t_{Pl})= \frac{m_{Pl}}{t^3_{Pl}}$. On the other hand at $t=t_H$ we said that $M=M_{max}$ which means
\begin{equation}
\label{eq:rho2}
\rho(t_H)=\frac{M_{max}}{t^2_H}.
\end{equation}
From equations \ref{eq:rho} and \ref{eq:rho2} we can write $t_H$, the time of horizon crossing as
\begin{equation}
\label{eq:th}
t_H=\frac{M_{max}}{m_{Pl}}\frac{1}{t^3_{Pl}\rho(t_{Pl})}.
\end{equation}
Since $t_e$ denotes the time at which the scalar field has almost completly decayed into matter, it is related to the decay rate of the field $\Gamma^{-1}$, so from now on we assume $t_e\sim \Gamma^{-1}$. And with all of this \ref{eq:order1} becomes
\begin{equation}
\label{eq:mmax1}
\delta_C\left(\frac{M_{max}}{M_C}\right)^{(1-n)/6}\left(\frac{t^3_{Pl}}{\Gamma}\frac{m_{Pl}\rho(t_{Pl})}{M_{max}}\right)^{2/3}\sim 1.
\end{equation}
We can now turn equation \ref{eq:mmax1} into an equation for $M_{max}$, starting with $t^2_{Pl}\rho(t_{Pl})=m_{Pl}^2$,
$$\delta_C\left(\frac{M_{max}}{M_C}\right)^{(1-n)/6}\left(\frac{m_{Pl}t^2_{Pl}}{\Gamma t_{Pl}}\frac{m^2_{Pl}}{M_{max}}\right)^{2/3}\sim 1$$
and using $t_{Pl}\sim m_{Pl}^{-1}$,
$$\delta_C\left(\frac{M_{max}}{M_C}\right)^{(1-n)/6}\left(\frac{1}{\Gamma t_{Pl}}\frac{m_{Pl}}{M_{max}}\right)^{2/3}\sim 1.$$
From here it is straight forward that 
\begin{equation}
\label{eq:mmax}
M_{max}\sim \left(\frac{\delta_C^{3/2}}{\Gamma t_{Pl}}\right)^{4/(3+n)}\left(\frac{M_C}{m_{Pl}}\right)^{(n-1)/(n+3)}m_{Pl}.
\end{equation}
Via $\Gamma=\frac{m_\sigma^3}{m^2_{Pl}}$, the maximum mass of black holes that can form is related to the mass of the scalar field $\sigma$
\be\label{eq:maxmass} M_{max}=\alpha^{\frac{1}{3+n}}\left(\frac{M_C}{m_{Pl}}\right)^{\frac{n-1}{n+3}}\left(\frac{m_{Pl}}{m\sigma}\right)^{\frac{12}{n+3}}m_{Pl}  \ee
where $\alpha=3.6\times 10^{-22}$.

Equation \eqref{eq:mmax} is the same expression with \cite{Carr:1994ar} who also list that the relic constraint applies for cases of $M_{max}<10^{10}g$. This would imply
$$\left(\frac{\delta_C^{3/2}}{\Gamma t_{Pl}}\right)^{4/(3+n)}\left(\frac{M_C}{m_{Pl}}\right)^{(n-1)/(n+3)}m_{Pl}<10^{10}$$
and
\begin{equation}
\frac{4}{n+3}(log_{10}\delta_C^{3/2}-log_{10}\Gamma t_{Pl})+\frac{n-1}{n+3}log_{10}\left(\frac{M_C}{m_{Pl}}\right)+log_{10}m_{Pl}<10.
\end{equation}
With $M_C=10^{57}h^{-1}g$, $h^{-1}=0.7$, $\delta_C\simeq 3.8\times 10^{-6}$, $m_{Pl}=2.17\times 10^{-5}gr$ one obtains \cite{Carr:1994ar} 
\begin{equation}
-35+12n<log_{10}\Gamma t_{Pl}.
\end{equation}

The relic constraint on the maximum mass of allowed black holes is constrains the reheat temperature in relation to the spectral index. For density fluctuations that arise from the inflation field the spectral index has been measured to be around $n\simeq 0.96$ by Planck. This is the spectral index around 50 e-folds of inflation in a window of 7 e-folds. There are two important points to note here, firstly we have been considering the black holes that would arise from the perturbations of the scalar field $\sigma$ and not the inflaton. The Planck measurement would more directly apply if we consider black hole formation from inflationary perturbations during a non-thermal phase set by another field, which in itself would also be a valid constraint. Secondly, the epoch we are considering is well later then the window of Planck measurements, and leaves room for a change in the spectral index. All in all, in the very restricted case, if we assume the spectral index for both inflationary and secondary scalar perturbations to be the same and moreover assume that it remains the same even after the measured window of e-folds, the lifetime of the scalar field would be bounded by
\begin{equation}
\label{eq:lifetime}
10^{-24}<\Gamma t_{Pl}
\end{equation}
And the constraint for reheat temperature would be $T_{rh}\sim\sqrt{\Gamma t_{Pl}}m_{Pl}$ where $m_{Pl}=1.22\times10^{19}GeV/c^2$
\begin{equation}
\label{Treheat}
1.22\times 10^7 GeV/c^2<T_{rh}.
\end{equation}

The minimal PBH mass corresponds to the collapse of the Hubble volume at the onset of scalar oscillations (matter domination) $H_{osc}\simeq m_\sigma$. Using the Hubble
equation and the energy density in the volume at that
moment is $\rho = M_{min}H^3$, we have
\be \label{eq:mmin} M_{min}=\frac{3H^2_{osc}m^2_{Pl}}{H^3_{osc}}=3\frac{m^2_{Pl}}{H_{osc}}=3\frac{m^2_{Pl}}{m_\sigma}.\ee
Hence the allowed range of primordial black hole masses, $M_{min}\leq M \leq M_{max}$ in these scenarios depend on $m_\sigma$. We refer the reader to \cite{Georg:2016yxa} for these ranges on two specific models and where they stand with respect to observations \cite{Carr:2009jm}, based on distortions in gamma ray and CMB measurements due to evaporation or requirements for primordial black holes not to exceed the critical density.  

\subsection{Pressure considerations}
In order to grow, a mode must have crossed the horizon. The minimal value of the modes physical momentum is determined by the value of the Hubble parameter at the time of horizon crossing. Since we are only interested in growth during the matter dominated phase, which begins when the scalar field starts to oscillate when $H\approx m_{\sigma} \equiv H_{\text{osc}}$, the minimal value for $q$ will be $q_{\text{min}}=H_{\text{osc}}\equiv q_{\text{osc}}$.\\
The upper bound on the momentum is given by the Jeans momentum. At values above this scale gravitational collapse is not possible due to pressure. Therefore, $q$ must lie in the following range:
\begin{equation}
q_{\text{J}} >q>q_{\text{osc}}
\end{equation}
We rewrite this slightly to get
\begin{equation}
\left(\frac{q_{\text{J}}}{H}\right)^2>\left(\frac{q}{H}\right)^2>1
\end{equation}
Assuming the universe is dominated by the scalar field $\sigma$, the Jeans momentum is given by 
\begin{equation}
\label{eq:jeans}
q_\text{J}=\left(\frac{2}{m_{\text{pl}}^2}\rho_{\sigma}m_{\sigma}^2\right)^{\frac{1}{4}}
\end{equation}
with the energy density given by the Friedmann equation
\begin{equation}
\label{eq:fried}
\rho_{\sigma}=3H^2m_{\text{pl}}^2
\end{equation}
With Eqs.~(\ref{eq:jeans}) and (\ref{eq:fried}) the upper bound on the physical momentum becomes
\begin{equation}
\left(\frac{q_{\text{J}}}{H}\right)^2=\sqrt6\frac{m_{\sigma}}H
\end{equation}
Noting that $q$ scales with $a$ as $q=q_0\left(\frac{a_0}a\right)$ the condition for growth becomes after some algebra
\begin{equation}
\sqrt6>\left(\frac{q_0}{H_0}\right)^2\left(\frac{H}{m_{\sigma}}\right)^{\frac13}>\frac{H}{m_{\sigma}}
\end{equation}
$\frac{q_0}{H_0}$ is constant. The ratio $\frac{H}{m_{\sigma}}$ is always smaller than $1$ since the maximum value for the the Hubble parameter is $m_{\sigma}$ and after that $H$ decreases. Therefore, $\frac{H}{m_{\sigma}}$ will always be smaller than $\left(\frac{H}{m_{\sigma}}\right)^{\frac13}$ and the condition is fulfilled. In other words, pressure does not play a role for mode growth, when the universe is dominated by a oscillating scalar field.

\section{Spontaneously broken symmetries}
\label{sec:spontaneously broken}
Spontaneous symmetry breaking will be a key concept for handling possible interactions of perturbations in section \ref{sec:EFTfor perts}. So let us review here the familiar textbook example of spontaneous breaking of the global $U(1)$ symmetry in Linear Sigma model. The Lagrangian
\be\label{linear sigma} \mathcal{L}=\frac{1}{2}(\partial_\mu\phi)(\partial^\mu\phi*)+\frac{1}{2}m^2(\phi\phi^*)-\frac{\lambda}{4}(\phi\phi^*)^2\ee
is invariant under the $U(1)$ transformation $\phi \to e^{i\alpha}\phi$, where $\alpha$ parametrizes the transformation. For constant $\alpha$'s this is a global transformation. So far nothing has been quantized. The values $\phi_+=\sqrt{\frac{6m^2}{\lambda}}$ and $\phi_-=-\sqrt{\frac{6m^2}{\lambda}}$ that minimize the classical potential $V(\phi)=-\frac{1}{2}m^2|\phi|^2+\frac{\lambda}{4!}|\phi|^4$, also extremize the action for \eqref{linear sigma}, and they are the solutions of the classical equations of motion. But being a constant number, this classical solution doesn't respect the $U(1)$ symmetry, $\phi_\pm\neq e^{i\alpha}\phi_\pm$.

The physical observables that can be probed by experiments are related to perturbations, as we have already seen in an example. As much as we would like to study the physics that gives rise to a background, at times we would like to be able to focus on the perturbations, $\delta\phi$, themselves  around a background solution, $\phi_0$, we take for granted. The important thing is we cannot randomly choose the background. Given a quantum theory, the questions one asks boil down to expectation values. And at the end of the day what is obtained from the quantum theory should give the result of the classical theory in the classical limit $\hbar\to 0$. 

In the quantum theory, the expectation values of operators are averages over field configurations weighted by the action
\be \label{expect}\langle\Omega|\phi|\Omega\rangle=\int \mathcal{D}\phi~\phi~e^{\frac{iS}{\hbar}}. \ee
In the limit $\hbar \to 0$, the biggest weight comes from the solution that minimizes $S_{cl}$, thus the expectation value for the field goes to the value of the classical solution. The classical solution sets the leading order behavior. Hence choosing one of the $\phi_\pm$ as the ground state $\phi_0$ is appropriate. Yet, this also happens to be the choice that doesn't respect the symmetry of the Lagrangian.  Hence for the energy scale at which we are interested in the physics of fluctuations and experimental results, the background is said to spontaneously brake the symmetry that is respected by the Lagrangian. In other words, the background field $\phi_0$ does not transform as a scalar under the transformation that leaves the whole field $\phi=\phi_0+\delta\phi$ and hence the Lagrangian invariant, $\phi_0$ changes.

If we had chosen $\phi_0=0$ as the ground state then the $U(1)$ symmetry would have been respected by the ground state as well as the Lagrangian. The symmetry wouldn't have been broken but we wouldn't be guaranteed the classical expectation values in the $\hbar\to 0$ limit. So in a sense breaking of the symmetry picks the background to be the one that ensures the classical expectation values are obtained at lower energy scales in the classical limit of the higher energy theory. Now we can write down a theory for the fluctuations by plugging
\be \phi=\phi_0+\delta\phi, \ee
into the Lagrangian $\mathcal{L}(\phi)$ with the form we expect to be valid at higher energies, with not any random thing but the classical level solution for $\phi_0$, and obtain the Lagrangian for the fluctuations $\mathcal{L}(\delta\phi)$. This careful selection of what the background should be is referred to as \emph{``integrating out the background''} and the obtained $\mathcal{L}(\delta\phi)$ gives the \emph{``effective field theory of fluctuations''}. For the linear sigma model, the field $\phi$ is a spinor, its perturbations are denoted by $\delta\phi=\{\sigma, \phi\}$, and the whole field can be expressed as
\be \phi=\left(\sqrt{\frac{2m^2}{\lambda}}+\frac{\sigma(x)}{2}\right)e^{\frac{i\pi(x)}{\Fbar_\pi}}.\ee
Inserting this in \eqref{linear sigma} gives,
\be\label{U1broken} \mathcal{L}=\frac{1}{2}\frac{\phi_0^2}{F^2_\pi}(\partial\pi)^2+\frac{1}{2}(\partial\sigma)^2+\frac{\sqrt{2}\phi_0\sigma}{F^2_\pi}(\partial\pi)^2+\frac{\sigma^2}{2\Fbar^2_\pi}(\partial\pi)^2-\frac{\lambda}{4}\left(\phi_0^4+2\sqrt{2}\phi_0^3\sigma+3\phi_0^2\sigma^2+\sqrt{2}\phi_0\sigma^3+\frac{\sigma^4}{4}\right).\ee 

One can write down a bunch of models that respects the symmetry at high energies and gives the specific value for $\phi_0$ determined by the classical theory. But once we have $\mathcal{L}(\delta\phi)$, we don't need to worry about the specific mechanisms of the models and from our results on $\delta\phi$ we can figure out what interactions will give what results for a general group of such models. The effective field theory developed in this way will hold for energy scales below the scale at which spontaneous symmetry breaking happens, for the linear sigma model the EFT holds below $U(1)$ symmetry breaking scale. Notice that by expanding the field around the background that spontaneously breaks the global symmetry, we got two different types of perturbations; the field $\sigma$ obtained mass $m^2_\sigma=\frac{3\lambda}{4}\phi_0^3$ coming from the background, where as $\pi$ turned out to be massless. A massless field, the $\pi$ of our example, always appears in the case of spontaneous breaking of global symmetries, and is referred to as the Goldstone boson. 

The original symmetry of $U(1)$ for our example, is no longer respected by the background, yet it is still a symmetry of the Lagrangian. To ensure this, the whole field should still be invariant under $U(1)$,
\be \label{wholephi}
\left(\phi_0+\frac{\sigma}{\sqrt{2}}\right)e^{\frac{i\pi}{F_\pi}}~\to~\left(\tilde{\phi_0}+\frac{\tilde{\sigma}}{\sqrt{2}}\right)e^{\frac{i\tilde{\pi}}{F_\pi}}=\left(\phi_0+\frac{\sigma}{\sqrt{2}}\right)e^{\frac{i\pi}{F_\pi}}. \ee
The Goldstone boson is the important character who transforms in a way so as to keep $\phi$ invariant under the original transformation. That is, while $\tilde{\phi_0}= e^{i\alpha}\phi_0$ and $\tilde{\sigma}= e^{i\alpha}\sigma$ pick up a factor of $e^{i\alpha}$, to make the whole field invariant
\be \left(\phi_0+\frac{\sigma}{\sqrt{2}}\right)e^{\frac{i\pi}{F_\pi}}~=~\left(\phi_0+\frac{\sigma}{\sqrt{2}}\right)e^{\frac{i\tilde{\pi}}{F_\pi}-i\alpha}
\ee 
the Goldstone boson $\pi$ should transform as 
\be\pi\to\tilde{\pi}=\pi+F_\pi\alpha.\ee
The phase rotation of $\phi$ implies a shift of $\pi$, and to realize the symmetry, the Goldstone boson should be shift symmetric. And the whole transformation under which the Lagrangian is invariant is
\be \phi_0\to e^{i\alpha}\phi_0,~~\sigma\to e^{i\alpha}\sigma,~~\pi\to\pi+F_\pi\alpha\ee
with $F_\pi=2\phi_0=2\sqrt{\frac{2m^2}{\lambda}}$. Hence the lagrangian (5) gives a description of the linear sigma model via the Goldstone boson, $\pi$. Because Goldstone bosons arise by the spontaneous breaking of global symmetries, this description is valid when working at energies below the symmetry breaking scale $\Lambda_{\not{sym}}$. The symmetry breaking scale can be obtained by calculating the scale at which the charge associated with the symmetry becomes ill defined.

\subsection{For cosmological backgrounds:}
In the case of cosmology, at the background level we work with FRW metrics 
\be ds^2=-dt^2+a^2(t)d\vec{x}^2,\ee
and the linear action of a scalar field coupled to gravity
\be S=\int d^4x\sqrt{-g}\left[\frac{m^2_{Pl}}{2}R+\frac{1}{2}g^{\mu\nu}\partial_\mu\phi\partial_\nu\phi-V(\phi)\right].\ee
So at the background level the equations of motion are the Friedmann equations
\begin{subequations}\begin{align}
	-3m^2_{Pl}(\dot H+H^2)=\dot{\phi_0}^2-V(\phi_0),\\
	3m^2_{Pl}H^2=\frac{\dot\phi_0^2}{2}+V(\phi_0).\end{align}\end{subequations}
During inflation the scalar field is said to roll slowly, which means $\epsilon\equiv\frac{\dot H}{H^2}\sim 0$ and $H\sim const$ hence $\dot \phi_0\sim 0$. Inflation ends as $\epsilon\to 1$, the time dependence of the fields, $\phi_0(t)$ and $H(t)$ become important! In a sense these are the high energy physics integrated out classical solutions, because they are just the solutions coming from classical GR.
As we have seen, under time diffeomorphisms
\be t \to t+\xi^0(t,\vec{x}),\ee
scalar perturbations, such as $\delta\phi$ transform as \cite{Mukhanov:1990me}
\be
\delta\phi \to\delta\phi-\dot{\phi_0}(t)\xi^0(t,\vec{x}).\ee
Since the time dependence of the background solution, $\phi_0=\phi_0(t)$ becomes important at the end of inflation, the $\dot\phi_o\xi^0$ term in the transformation doesn't vanish and hence the perturbation is no longer a scalar under time diffeomorphisms. The background field $\phi_0$ will transform as
\be\phi_0(t)\to\phi_0(t+\xi^0)=\phi_0(t)+\dot{\phi_0}(t)\xi^0,\ee
which is just a Taylor expansion. Notice that the perturbation $\delta\phi$ and the background field $\phi_0$ transforms very conveniently so as to keep the whole field $\phi=\phi_0+\delta\phi$ invariant. As such the Lagrangian is also invariant under all diffeomorphisms. Time diffeomorphism is still a symmetry of the Lagrangian, but it is no longer a symmetry of the ground state $\phi_0$. In other words, time diffeomorphisms are spontaneously broken. This observation has been intorduced to study generalizations of interactions for single field eras at the level of perturbations in \cite{Creminelli:2006xe}.  We will make use of this formalism to study interactions during preheating in a general manner. This way of developing theories at the energy scale of interest, without knowing the high energy theory that gives rise to them, but knowing only the symmetries to be obeyed at the level of interest, is referred to as a bottom up construction.

\chapter{BRST Quantization of Cosmological Perturbations}
\label{chp:quantization methods}
\section{Introduction}
\label{sec:introfor BRST}

In order to explain the properties of the primordial fluctuations, in many models of the origin of structure one needs to quantize general relativity  coupled to a scalar field. Because the action is invariant under diffeomorphisms, the challenge one faces is the quantization of a theory with  local symmetries, very much like  those in non-Abelian gauge theories. Given the formal similarity between diffeomorphisms and the latter we shall refer to all of them as ``gauge theories."

The quantization of gauge theories is somewhat subtle, but becomes relatively straight-forward if one is interested in tree-level calculations alone: One can either fix the gauge and break the local symmetry, thus clearing the way, say,  to  canonical quantization, or one can work  with the essentially equivalent method of quantizing an appropriate set of gauge-invariant variables, a procedure also known as reduced phase space quantization.  This is the way  primordial spectra were originally calculated in the free theory \cite{Mukhanov:1990me}.  Although BRST quantization provides an elegant and powerful method  of quantization, it is not strictly necessary in those cases. 

Yet problems arise when one attempts to go beyond the free (linearized) theory. Beyond the linear order, it becomes increasingly difficult to work  with gauge-invariant variables alone (see for instance \cite{Malik:2008im,Malik:2008yp} for a Lagrangian treatment of second order perturbation theory), and beyond tree-level one needs to take into account the interactions of the ghosts associated with the gauge-fixing procedure.  From a phenomenological point of view, loop calculations in cosmological perturbation theory may not be important at this point, because observations are not sensitive enough  to these corrections (yet), but from a theoretical perspective they arguably are the distinct feature of quantum gravity, in the same way as the anomalous magnetic moment of the electron  is regarded as one of the distinct quantum signatures of QED. Among other reasons, this is why researchers  have relatively recently begun to explore loop corrections to primordial spectrum calculations  \cite{Weinberg:2005vy,Sloth:2006az,Chaicherdsakul:2006ui,Seery:2007we,Adshead:2008gk,Gao:2009fx,Senatore:2009cf,Campo:2009fx,Bartolo:2010bu,Kahya:2010xh,Onemli:2013gya}. But due to the inherent complexity of these calculations, the contribution from the ghosts has been mostly ignored so far. 

The  most common method of quantizing a gauge theory is that of Faddeev and Popov (and also DeWitt.) This method is inherently linked to the choice of gauge-fixing ``conditions." In the case of cosmological perturbations, these can be taken to be four functions $f^\mu$ of the inflaton and metric perturbations $\delta\varphi,\delta h_{\mu\nu}$ that are \emph{not} invariant under the four independent diffeomorphisms with infinitesimal parameters $\xi^\nu$. Once these conditions have been chosen, the action of the theory needs to be supplemented with appropriate ghost terms,
\begin{equation}\label{eq:FP0}
S_\mathrm{ghost} =\int d^4 x \, d^4 y \, \,
\bar \eta_\mu  (x)
\frac{\delta f^\mu (x)}{\delta \xi^\nu(y)}
\eta^\nu(y),
\end{equation}
which typically couple the fermionic  Faddeev-Popov ghosts  $\bar\eta_\mu$ and $\eta^\nu$ to the metric and inflaton perturbations. The Faddeev-Popov method works well in renormalizable gauge theories, provided that the gauge-fixing functionals are linear in the fields. But if the gauge-fixing functionals are non-linear, or the theory is non-renormalizable,  the Lagrangian needs to be supplemented with terms that are not just quadratic in the ghosts. Since these factors are absent in the Faddeev-Popov prescription (\ref{eq:FP0}), the method fails.  

A very general and powerful  quantization method that avoids these problems, and reduces to the one  of Faddeev and Popov in appropriate cases, is that of Becchi, Rouet, and Stora, and Tyutin (BRST) \cite{Becchi:1975nq,Tyutin:1975qk}.  The BRST method not only justifies Faddeev and Popov, but  also endows it with a geometric interpretation. In its Hamiltonian formulation, BRST quantization manifestly results in a unitary theory, and it also manifestly preserves a global supersymmetry known as BRST symmetry, even after gauge-fixing.  This quantization method has been successfully applied in many different contexts, ranging from electrodynamics to string theory, and  is arguably  the best way to quantize a theory with local symmetries.

There are many important reasons for pursuing BRST quantization in the context of cosmological perturbations. As we already mentioned,   the method of Faddeev and Popov fails when applied to gravity. In addition,   BRST quantization  provides us with an enormous freedom to choose gauge-fixing terms. While in field theories in Minkowski spacetime the demand of Lorentz invariance and renormalizability severely restricts the possible gauge-fixing choices, in cosmological perturbation theory the smaller degree of symmetry allows for a much wider set of gauge conditions that have remained essentially unexplored so far.  Calculations in cosmological perturbation theory are notoriously involved, and an eventual simplification of the propagators and the structure of the ghost interactions facilitated by appropriate generalized gauge choices may render loop calculations in the BRST method much  more manageable. The BRST global symmetry preserved even after  gauge fixing may  also place interesting constraints on the structure of the theory that describes the cosmological perturbations and its implications too (see \cite{Binosi:2015obq} for a  discussion in the context of the  antifield formalism.)  On a related topic, we should also note that whereas typical gauge choices in cosmological perturbation theory break locality,   BRST quantization allows for gauge-fixing conditions that manifestly preserve the latter. Locality (analyticity in the spatial momenta) was for instance an important ingredient in the derivation of consistency relations between cosmological correlators derived in \cite{Berezhiani:2013ewa,Armendariz-Picon:2014xda}.

In this article we study the BRST quantization of cosmological perturbations. We mostly concentrate on the general formalism and illustrate many of the results in the free theory. We approach the quantization from the Hamiltonian perspective, which is manifestly unitary, and makes the role of boundary conditions more explicit.  In a cosmological background, the split into space and time required by the Hamiltonian formulation does not conflict with any spacetime isometry, and thus does not pose any immediate significant drawback.  If it exists, the Lagrangian formulation can be recovered from the Hamiltonian one by integration over the canonical momenta as usual.  For lack of space, however, we do not discuss  the  powerful Lagrangian  antifield formalism of Batalin and Vilkovisky, which we hope to explore in future work. 

We have tried to make the article relatively self-contained, which is why we quote  main results in BRST quantization, at the expense of making the manuscript  longer than strictly necessary. Our presentation mostly follows the excellent monograph  by Henneaux and Teiltelboim \cite{Henneaux:1992ig}, which the reader may want to consult for further background and details. To our knowledge, our work is  the first to focus on the Hamiltonian BRST quantization of cosmological perturbations, although Barvinsky has discussed the BRST formalism in the context of a cosmological density matrix \cite{Barvinsky:2013nca}, and Binosi and Quadri have used the  antifield formalism to reproduce some of the consistency relations satisfied by cosmological correlators \cite{Binosi:2015obq}. Other authors have analyzed somewhat related issues, mostly  within the path integral quantization of   cosmological perturbations \cite{Anderegg:1994xq,Prokopec:2010be}, or within the loop quantum cosmology program \cite{Ashtekar:2011ni}. The Hamiltonian of cosmological perturbations    was  calculated  to quadratic order in reduced phase space in \cite{Langlois:1994ec}, and  to cubic order  only in spatially flat gauge  \cite{Nandi:2015ogk,Nandi:2016pfr}, due to the above-mentioned complexity of the gauge-invariant formalism at higher orders. An intriguing approach that aims at formulating the Hamiltonian of the theory directly in terms of gauge-invariant variables (to all orders) is discussed in \cite{Giesel:2007wi,Giesel:2007wk}. The latter introduces a pressureless fluid parameterized by four spacetime scalars, which are used to define gauge-invariant observables and a gauge-invariant Hamiltonian by deparameterization. Such approach is in many ways complementary to the one of the BRST formalism, which is built around gauge-variant fields, and ultimately relies on a choice of gauge. Both share the property that a pair of additional  fields is introduced for each of the four constraints of the theory,  but whereas BRST-invariance guarantees that the ghosts do not change the gauge-invariant content of the theory, the dust fields of \cite{Giesel:2007wi,Giesel:2007wk}  do seem to ultimately survive as additional degrees of freedom in the system.

\section{Action}
\label{sec:action}

Our main goal is the quantization of cosmological perturbations in a spatially flat universe dominated by a canonical scalar field. This is for instance what is needed to calculate primordial perturbation spectra in conventional inflationary models, although our results do not really depend on any particular scalar field background, as long as the latter is homogeneous and time-dependent. 

We begin with the action of general relativity minimally coupled to a scalar field $\varphi$ in Hamiltonian form.
As  is well known, the Hamiltonian action takes its most natural form in the ADM formulation \cite{Arnowitt:1962hi}, in which the metric components are written as 
\begin{equation}\label{eq:ADM}
ds^2=-(\lambda^N)^2 dt^2+h_{ij}(dx^i+\lambda^i dt)(dx^j+\lambda^j dt),
\end{equation}
where $\lambda^N$ is the lapse function and $\lambda^i$ the shift vector.  With this  choice of variables the action of the theory becomes
\begin{equation}\label{eq:Hamiltonian S}
S=\int dt\, \int d^3x \, \left(\pi^{ij}\, \dot h_{ij} +\pi_\varphi \, \dot \varphi\right) -H,        
\end{equation}
where the Hamiltonian  $H$  is linear in  $\lambda^N$ and $\lambda^i$,
\begin{equation}\label{eq:H}
H=\int d^3x\,\left[\lambda^N \, G_N +\lambda^i\, G_i\right],
\end{equation} 
with coefficients given by
\begin{subequations}\label{eq:secondary}
	\begin{align}
	G_N &\equiv 
	\frac{2}{M^2}\frac{1}{\sqrt{h}} \left(\pi_{ij}\pi^{ij} -\frac{\pi^2}{2}\right)-\frac{M^2}{2}\sqrt{h}R^{(3)}
	+\frac{\pi_\varphi^2}{2\sqrt{h}}
	+\frac{\sqrt{h}}{2}h^{ij}\partial_i\varphi \partial_j\varphi
	+\sqrt{h}\,V(\varphi), \label{eq:Hamiltonian C}
	\\
	G_i&\equiv 
	-2\sqrt{h}\nabla_j\left(\frac{\pi_i{}^j}{\sqrt{h}} \right)+\pi_\varphi \partial_i \varphi
	. \label{eq:momentum C}
	\end{align}
\end{subequations}
To arrive at these expressions  we have discarded a surface term at the spatial boundary. Indices are raised and lowered with the spatial metric $h_{ij}$, and $M$ is the reduced Planck mass.  

Because the action does not contain derivatives of the lapse function $\lambda^N$ and shift vector $\lambda^i$,  their conjugate momenta vanish, $b_N\equiv 0, b_i\equiv 0$. 
In Dirac's analysis, these would be interpreted as primary constraints.  But in the end, this interpretation mostly leads to unnecessary complications. In the Hamiltonian formulation it is much simpler  to think of $\lambda^N$ and $\lambda^i$ as Lagrange multipliers, and restrict phase space to the appropriately constrained  canonical pairs
$\{(h_{ij},\pi^{ij}), (\varphi,\pi_\varphi)\}$. 

Variation of the action  (\ref{eq:Hamiltonian S}) with respect to  the $\lambda^N$ and $\lambda^i$  yields the secondary constraints $G_N=G_i=0$, which is why we loosely refer to $G_N$ and $G_i$ as ``the constraints." These constraints define gauge transformations on phase space functions $F$ through  their Poisson brackets,
$
\Delta_a F\equiv\{F,G_a\}. 
$
In this way, the  $G_i$ generate spatial diffeomorphisms.  The constraint $G_N$ generates diffeomorphisms along the normal to the equal-time hypersurfaces only after the equations of motion are imposed on the Poisson bracket.  For our purposes, what matters most is that the constraints are first class, and that they define an  open algebra  with field-dependent structure constants (the Dirac algebra.) In practice this means that their Poisson brackets are just proportional to the constraints themselves \cite{Isham:1992ms}, 
\begin{subequations}\label{eq:algebra}
	\begin{align}
	\left\{G_N(\vec{x}),G_N(\vec{y})\right\}&=
	\frac{\partial\delta(\vec{y}-\vec{x})}{\partial x^i}h^{ij}(\vec{x}) G_j(\vec{x})-\frac{\partial\delta(\vec{x}-\vec{y})}{\partial y^i}h^{ij}(\vec{y})G_j(\vec{y}), \label{eq:GN GN}
	\\
	\left\{G_N(\vec{x}),G_i(\vec{y})\right\}&=\frac{\partial\delta(\vec{x}-\vec{y})}{\partial x^i}G_N(\vec{y}),
	\\
	\left\{G_i(\vec{x}),G_j(\vec{y})\right\}&=\frac{\partial \delta(\vec{x}-\vec{y})}{\partial x^i}G_j(\vec{x})-\frac{\partial \delta(\vec{y}-\vec{x})}{\partial y^j} G_i(\vec{y}),
	\end{align}
\end{subequations}
with coefficients that,  in the case of equation (\ref{eq:GN GN}), depend on the inverse spatial metric $h^{ij}$. Because of this dependence, the commutator of two gauge transformations is another gauge transformation only after the constraints are imposed on the result. In that sense, the algebra of the constraints only closes on-shell, which is why one refers to it as  an open algebra.  For simplicity, we shall nevertheless refer to the constraints as the generators of diffeomorphisms.  The Hamiltonian (\ref{eq:H})  itself  is thus a linear combination of the  four secondary constraints, and  vanishes identically on shell.  Because its  Poisson brackets with the secondary constraints vanish weakly,\footnote{A function of the canonical variables is said to vanish weakly, when it vanishes after the application of the constraints.} no further constraints appear in the theory. Whereas  in the Hamiltonian formalism the algebra of constraints is open, in the Lagrangian formalism the generators of diffeomorphisms along the four spacetime coordinates define a closed algebra: The commutator of two diffeomorphisms with  infinitesimal parameters $\xi_1^\mu$ and $\xi_2^\mu$ is  a diffeomorphism with infinitesimal parameters  $\xi_1^\nu\partial_\nu\xi_2^\mu-\xi_2^\nu\partial_\nu \xi_1^\mu$.

\subsection{Perturbations}

We are actually interested in quantizing the perturbations around a cosmological background. Therefore, we split the ADM  variables into  background plus perturbations,
\begin{subequations}\label{eq:background+perturbations}
	\begin{align}
	\varphi&=\bar{\varphi}(t)+\delta\varphi,
	&\pi_\varphi&=\bar\pi_\varphi+\delta \pi_\varphi,\\
	h_{ij}&=\bar h_{ij}+\delta h_{ij},
	&\pi^{ij}&=\bar\pi^{ij}+\delta\pi^{ij},\\
	\lambda^N&=\bar{\lambda}^N+\delta\lambda^N,\\
	\lambda^i&=\bar{\lambda}^i+\delta\lambda^i,
	\end{align}
\end{subequations}
where the background quantities are
\begin{equation}\label{eq:background}
\bar{\lambda}^N=a,
\quad
\bar \lambda^i=0,
\quad
\bar{h}_{ij}=a^2\delta_{ij},
\quad
\bar{\pi}^{ij}=-M^2 \mathcal{H} \, \delta^{ij},
\quad
\bar{\pi}_\varphi=a^2 \dot{\bar{\varphi}},
\end{equation}
and $a$ is the scale factor. Note that our choice of background lapse function, $\bar{\lambda}^N\equiv a$, implies that the time coordinate $t$ is actually conformal time, rather than the conventional cosmic time. In addition, our definition of the metric perturbations $\delta h_{ij}$ does not separate a factor of $a^2$ from the perturbations. A dot denotes a derivative with respect to coordinate time (conformal time), and
$\mathcal{H}=\dot{a}/a$ stands for the comoving Hubble scale. The background quantities obey the equations of motion
\begin{subequations}\label{eq:background motion}
	\begin{align}
	\ddot{\bar\varphi}+2\mathcal{H}\dot{\bar\varphi}+a^2 \bar{V}_{,\varphi}&=0,\\
	2\dot{\mathcal{H}}+\mathcal{H}^2+\frac{1}{M^2}\left(\frac{\dot{\bar\varphi}^2}{2}-a^2\bar{V}\right)&=0,
	\end{align}
	as well as the single constraint
	\begin{equation}
	\mathcal{H}^2-\frac{1}{3M^2}\left(\frac{\dot{\bar\varphi}^2}{2}+a^2\bar{V}\right)=0.
	\end{equation}
\end{subequations}
We shall often use these relations to  simplify some of the resulting expressions throughout this work. 

To simplify the notation, a transition to DeWitt notation proves to be advantageous.  In this notation equations (\ref{eq:background+perturbations}) become
\begin{equation}\label{eq:canonical T}
q^i\equiv \bar{q}^i+\delta q^i,
\quad 
p_i\equiv \bar{p}_i+ \delta p_i,
\quad
\lambda^a\equiv \bar\lambda^a+\delta\lambda^a,
\end{equation}
where the $q^i$ stand for the canonical variables $ h_{ij}(\vec{x})$ and $\varphi(\vec{x})$, the $p_i$ for their conjugate momenta, and the  $\lambda^a$ for the Lagrange multipliers $\lambda^N(\vec{x})$ and $\lambda^i(\vec{x})$. In particular, in DeWitt notation indices like $i$ and $a$ stand both for field indices and spatial coordinates.

To obtain the action for the perturbations we simply substitute the expansion (\ref{eq:background+perturbations}) into  the action (\ref{eq:Hamiltonian S}). The Hamiltonian action for the perturbations then  becomes
\begin{equation}\label{eq:S pert I}
\delta S=\int dt \left[\delta p_i \delta\dot q^i-\delta H\right],
\end{equation}
in which the  Hamiltonian for the perturbations is obtained from the original  one by removing linear terms in the canonical perturbation variables. In the case at hand, the Hamiltonian of the full theory  is just a linear combination of the constraints, so the  Hamiltonian of the perturbations becomes 
\begin{equation}\label{eq:delta H}
\delta H=-\dot{\bar q}^i \delta p_i+\dot{\bar p}_i\delta q^i +(\bar\lambda^a+\delta\lambda^a)G_a(\bar{p}_i+\delta p_i,\bar{q}^i+\delta q^i).
\end{equation}
The first two terms in the Hamiltonian (\ref{eq:delta H}) are just the result of a time-dependent canonical transformation applied to a vanishing Hamiltonian.  Of course, although we have changed variables, the original constraints are still satisfied. In particular,  variation with respect to $\delta\lambda^a$ yields again 
\begin{equation}\label{eq:constraints pert}
\delta G_a(\delta p_i, \delta q^i,t)\equiv G_a(\bar{p}_i+\delta p_i,\bar{q}^i+\delta q^i)=0.
\end{equation}
Our notation aims to emphasize that the $\delta G_a$ are functions of the perturbations and time.  They have the same value as the constraints $G_a$, but  their functional forms differ. 

Enforcing the constraints in equation (\ref{eq:delta H}) returns a Hamiltonian linear in the perturbation variables that is not particularly useful for cosmological perturbation theory calculations. Instead, it shall  prove to be more useful to separate only the contribution of the Lagrange multipliers $\delta\lambda^a$ to the Hamiltonian. We thus write
\begin{equation}\label{eq:H HD}
\delta H=\delta H_D+\delta\lambda^a \delta G_a,
\end{equation}
where the ``first-class Hamiltonian" is 
\begin{equation}\label{eq:H Dirac}
\delta H_D=-\dot{\bar q}^i\delta p_i+\dot{\bar p}_i\delta q^i +\bar\lambda^a \delta G_a.
\end{equation}
Note that because the background satisfies the equations of motion, the linear terms in this expression cancel. As we show below,  under the time evolution defined by $\delta H_D$  the constraints are also preserved weakly, or, alternatively, the Hamiltonian $\delta H_D$ is weakly gauge-invariant (hence its name.) The first-class Hamiltonian plays an important role in the BRST theory.

The time derivative of any function of the perturbations $\delta F$ follows from its Poisson brackets with the Hamiltonian,
\begin{equation}\label{eq:delta t pert}
\frac{d\delta F}{dt}=\{\delta F,\delta H\}+\frac{\partial \delta F}{\partial t}. 
\end{equation} 
We include a partial derivative with respect to time to capture a possible explicit time dependence of $\delta F$, say, due to its dependence on  background quantities. Nevertheless, if $\delta F$ is of the form $\delta F(\delta q^i,\delta p_i)=F(\bar{q}^i+\delta q^i,\bar{p}_i+\delta\bar{p}_i),$ its time evolution obeys
\begin{equation}\label{eq:dt delta F}
\frac{d\delta F}{dt}=\{F,H\},
\end{equation}
with the understanding that after calculation of the Poisson bracket on the right hand side in the original unperturbed variables, the substitution (\ref{eq:canonical T}) is supposed to be made.  The perturbation Hamiltonian (\ref{eq:delta H}) weakly equals $-\dot{\bar{q}}^i \delta p_i+\dot{\bar{p}}_i\delta q^i$, which is generically   not weakly conserved.  On the other hand, the original Hamiltonian $H$, understood as function of the perturbations does remain constant. In fact, it identically vanishes.

\subsection*{Analysis of the Constraints}
\label{subsec:analysisconstraints}
Variation of the action with respect to the perturbed Lagrange multipliers yields equations (\ref{eq:constraints pert}). These are simply the original constraints expressed in terms of the perturbations.  Because the transition to the perturbations is a canonical transformation, the Poisson brackets of any phase space function is preserved under such a change of variables. In particular, it follows that 
\begin{equation}
\{\delta q^i, \delta G_a\}=\{q^i, G_a\}, \quad
\{\delta p_i, \delta G_a\}=\{p_i, G_a\},
\end{equation}
where, again, on the expressions on the right hand sides $q^i$ and $p_i$ are to be expanded as in equations (\ref{eq:canonical T}) after calculation of the bracket. Therefore, the constraints act on the perturbations as they did on the original variables. Since, as we mentioned, the constraints $G_a$ generate diffeomorphisms when they act on $p_i$ and $q^i$, so do the $\delta G_a$ when they act on $\delta p_i$ and $\delta q^i$.  It also    follows  immediately that
$
\{\delta G_a,\delta G_b\}=\{G_a,G_b\}(\bar{p}_i+\delta p_i,\bar q^i+\delta q^i).
$
Because the Poisson bracket $\{G_a,G_b\}$ is linear in the constraints, and because of  equation (\ref{eq:constraints pert}), the algebra of perturbed constraints remains the same. In particular, the set of constraints $\delta G_a=0$ is first class. In DeWitt notation we write the algebra of constraints as
\begin{equation}\label{eq:structure}
\{\delta G_a,\delta G_b\}=C_{ab}\,{}^c\delta G_c,
\end{equation} 
and call the field-dependent coefficients $C_{ab}{}^c$ the ``structure constants" of the theory. But a bit of care must be taken because the constraints $\delta G_a$ now depend explicitly on time, so the Poisson brackets alone do not  fully determine their time  evolution. Nevertheless, using equation (\ref{eq:dt delta F}) we find
\begin{equation}\label{eq:dt delta C}
\frac{d \delta G_a}{dt}=(\bar\lambda^b+\delta \lambda^b)\{\delta G_a,\delta G_b\},
\end{equation}
which indeed vanishes weakly because of equation (\ref{eq:structure}). As a result, no equation in the theory fixes the value of the Lagrange multipliers $\delta \lambda^a$, which remain undetermined by the equations of motion, as  in the background-independent theory. 

\subsection*{Gauge Symmetry}

By definition, in the Hamiltonian formulation the gauge transformations generated by the constraints  act only on the canonical variables $\delta q^i$ and $\delta p_i$.  They capture the redundancy in the description of the state of the system at any particular time.  It is also interesting to elucidate to what extent the first-order action (\ref{eq:S pert I}) is symmetric under diffeomorphisms, and how the latter act on the Lagrange multipliers. By direct substitution, the reader can check that, indeed, the transformations
\begin{subequations}\label{eq:Delta delta}
	\begin{align}
	\Delta \delta q^i&=\xi^a \{\delta q^i,\delta G_a\}, \\
	\Delta \delta p_i&=\xi^a\{\delta p_i, \delta G_a\},\\
	\Delta \delta \lambda^a&=\dot{\xi}^a
	+\xi^b(\bar{\lambda}^c+\delta\lambda^c) C_{bc}{}^a,
	\label{eq:Delta delta lambda}
	\end{align}
\end{subequations}
leave the Hamiltonian action for the perturbations invariant for arbitrary $\xi^a(t)$, provided that the latter vanish at the time boundary.  The structure of the transformation laws  (\ref{eq:Delta delta}) follows from the properties of the constraints and the first-class Hamiltonian $\delta H_D$. Indeed, using  equations (\ref{eq:H HD}) and  (\ref{eq:dt delta C}) we find
\begin{equation}\label{eq:first class}
\{\delta G_a,\delta H_D\}+\frac{\partial \delta G_a}{\partial t}=\bar\lambda^b C_{ab}{}^c \delta G_c.
\end{equation}
The term $\xi^b\bar\lambda^c C_{bc}{}^a$ in equation (\ref{eq:Delta delta lambda}) is what one expects in a theory in which the first-class Hamiltonian satisfies equation (\ref{eq:first class}), whereas the term $\xi^b\delta\lambda^c C_{bc}{}^a$  is what one should have in a gauge theory with structure constants $C_{ab}{}^c$. 

\section{BRST Symmetry}
\label{sec:BRSTsym}

In the classical theory, the existence of local symmetries leads to equations of motion that do not admit unique solutions with the prescribed boundary conditions, because the Lagrange multipliers remain arbitrary. There are mainly two ways of dealing with such ambiguities: One can either work with gauge-invariant variables alone, as in the reduced phase space method, or one can simply fix the gauge. Working with gauge-invariant variables becomes increasingly complex and cumbersome at higher orders in perturbation theory, so we choose here the much simpler alternative of fixing the gauge.  One of the key aspects of gauge theories is that, even after gauge fixing, they retain a global symmetry known as BRST symmetry, so named after Becchi, Rouet, Stora, and Tyutin \cite{Becchi:1975nq,Tyutin:1975qk}. We introduce this symmetry next, mostly by following the presentation of \cite{Henneaux:1992ig}.  An alternative  way to reach some of the results here consists of identifying  the BRST symmetry directly in the background-independent theory, and then making a transition to the theory of the perturbations by canonical transformation. 

\subsection{Ghosts and Antighosts}

To bring the BRST symmetry to light, we introduce for each  constraint $\delta G_a=0$ a  real Grassmann  variable  $\delta\eta^a$ known as the ``ghost", and its purely imaginary  conjugate momentum $\delta \mathcal{P}_a$. The ghosts $\delta \eta^a$ carry ghost number one, and their conjugate momenta carry ghost number minus one.   These variables commute with all bosonic fields and anticommute with all fermionic fields.\footnote{The previous statements  imply for instance that a term $\delta\dot\eta^a \delta\mathcal{P}_a $ in the first-order ghost action is real and has ghost number zero.} In particular, ghosts and their momenta obey the  Poisson bracket relations
\begin{equation}
\{\delta \mathcal{P}_a,\delta \eta^b\}\equiv
-\delta_a{}^b. 
\end{equation} 
From now on, the brackets $\{\cdot,\cdot\}$ refer to the graded bracket of any two functions defined on the extended phase space spanned by the canonical pairs $(\delta p_i,\delta q^i)$ and $(\delta\mathcal{P}_a, \delta\eta^a)$. We summarize the properties of the graded bracket in Appendix \ref{sec:Graded Commutator}.   

We could also think of these ghosts as  perturbations around a background with vanishing ghosts. The transformation properties of the former under the isometries of the background are those expected from their indices. Namely, $\delta\eta^N$ and $\delta\eta^i$ respectively transform as scalars under translations, and as a scalar and a three-vector under rotations.

For some purposes, it shall prove to be convenient to introduce yet another set of  ghosts associated with the primary constraints $\delta b_N\equiv\delta b_i\equiv 0$ that we opted to bypass for simplicity earlier on. We shall denote these (Grassmann) ghosts by  $\delta C_a$, and their conjugate momenta by $\delta\rho^a$. The  $\delta C_a$ are real and carry ghost number minus one; they  are thus known as ``antighosts." Their conjugate momenta are purely imaginary and carry ghost number one. These variables accordingly obey the Poisson bracket relations
\begin{equation}
\{\delta \mathcal{\rho}^a,\delta C_b\}=-\delta^a{}_b.
\end{equation}

In the work of Faddeev, the ghosts were simply introduced to represent the Faddeev-Popov determinant $\det \{f^a,\delta G_b\}$ as a quadratic functional integral in phase space \cite{Faddeev:1969su}. Such a determinant renders the  expectation value of an observable independent of the gauge-fixing conditions $f^a=0$, so in this approach the ghosts were simply regarded as a necessary by-product  of any gauge-fixing procedure. But the   ghosts $\delta\eta^a$ and their conjugates $\delta\mathcal{P}_a$ also have a geometrical interpretation that only  emerged after the discovery of the BRST symmetry. In the BRST formalism one identifies the observables of the gauge theory with the cohomology of an appropriately defined BRST differential. In a gauge theory, the observables consist of those functions defined on the constraint surface $\delta G_a=0$ that  remain invariant under the transformations generated by the  constraints. On one hand, in order to identify functions on the constraint surface with the cohomology of a differential operator, one needs to introduce as many Grassmann variables $\delta\mathcal{P}_a$ as there are conditions  defining the constraint surface. On the other hand,  in order to identify the cohomology of a differential operator with those functions invariant under the constraints, one needs to introduce as many differential forms $\delta\eta^a$ as there are gauge generators. These forms can be thought of as the  duals to the tangent vector fields  defined by the generators, and, along with their exterior product,  they define a Grassmann algebra.  In a theory with first-class constraints the functions that define the constraint surface are  the same as the gauge generators, which allows one to identify the $\delta\mathcal{P}_a$ as the canonical conjugates of the $\delta\eta^a$.

\subsection{BRST Transformations}

The main idea behind BRST quantization is the replacement of the local symmetry under gauge transformations by a nilpotent  global symmetry generated by what is known as the BRST charge.   This  BRST charge is defined to be a real, Grassmann-odd, ghost number one and   nilpotent function in the  extended phase space introduced above. For many purposes, it is convenient to expand this BRST charge in the antighost number $p$, that is, in powers of the fields $\delta\mathcal{P}_a$,
\begin{equation}\label{eq:omega}
\delta \Omega=\delta \eta^a \delta G_a-\frac{1}{2}\delta \eta^b \delta \eta^c C_{cb}{}^a \delta \mathcal{P}_a+\delta \Omega_{p\geq 2},
\end{equation}
where the structure constants $C_{ab}{}^c$ are those of equation (\ref{eq:algebra}), and $\delta \Omega_{p\geq 2}$ is of antighost number $p\geq 2$.  The latter vanishes for closed algebras, and can be otherwise determined recursively by demanding that the BRST charge be of ghost number one and obey the nilpotence condition
\begin{equation}\label{eq:nilpotence}
\{\delta\Omega,\delta\Omega\}=0.
\end{equation}
The BRST charge $\delta\Omega$ captures the gauge symmetries of a theory, and not the variables we choose to describe it. In particular, we could have arrived at equation (\ref{eq:omega}) by identifying the gauge symmetries and their algebra in the background-independent theory, and then performing a canonical transformation to the perturbations around our cosmological background.  Such a procedure manifestly underscores that $\delta\Omega$ has the structure
\begin{equation}\label{eq:omega structure}
\delta\Omega(\delta p_i,\delta q^i; \delta\mathcal{P}_a,\delta \eta^a)=\Omega (\bar{p}_i+\delta p_i,\bar{q}^i+\delta q^i ;\delta\mathcal{P}_a,\delta \eta^a),
\end{equation}
where $\Omega$ is the BRST charge of the background-independent theory. As we show below, this structure guarantees that the BRST action for the perturbations remains BRST-symmetric.

The BRST charge defines a transformation $s$ on any function of the extended phase space through its graded bracket,
\begin{equation}\label{eq:BRST T}
s \,  \delta F=\{\delta F,\delta\Omega\}.
\end{equation}
Its action on the canonical variables contains a gauge transformation with the ghosts as gauge parameters, plus terms of antighost number $p\geq 1$ (denoted by dots), 
\begin{subequations}
	\begin{align}
	s \, \delta q^i & =\delta \eta^a \{ \delta q^i,\delta G_a\}+\cdots
	, 
	&s \,   \delta p_i &=\delta \eta^a \{\delta  p_i,\delta G_a\}+\cdots
	,\\ 
	s \, \eta^a&=\frac{1}{2}\delta \eta^b \delta\eta^c C_{cb}{}^a+\cdots
	,
	&s\, \delta\mathcal{P}_a&=-\delta G_a+\cdots.
	\end{align}
\end{subequations}
These  BRST transformations are  nilpotent, $s^2=0$, because of equations (\ref{eq:nilpotence}) and (\ref{eq:Jacobi}). If under  translations and rotations the ghosts transform as indicated by their indices, we expect the BRST charge to be a scalar.

In a gauge theory, the observables consist of the set of all gauge-invariant functions on the constraint surface. One of the fundamental results in BRST theory is that this set is in one-to-one correspondence with  the cohomology of the BRST operator $s$ at  ghost number zero.
The cohomology  consists of all those functions in the extended phase space  of ghost number zero that are BRST-closed ($s \delta\mathcal{O}=0$) modulo those functions that are BRST-exact ($\delta \mathcal{O}=s\delta \mathcal{Q}$.)  If a member of the zero ghost cohomology $\delta\mathcal{O}_\mathrm{BRS}$  satisfies
\begin{equation}
(\delta\mathcal{O}_\mathrm{BRS})|_{p=0}=\delta \mathcal{O}, 
\end{equation}
where $\delta\mathcal{O}=\delta\mathcal{O}(p_i, q^i)$ is an observable, and $|_{p=0}$ is the restriction  to those terms of antighost number zero, one speaks of $\delta\mathcal{O}_\mathrm{BRS}$ as a ``BRST-invariant extension" of the observable $\delta \mathcal{O}$. In theories with Abelian constraints, such as the linearized theory of cosmological perturbations, gauge-invariant operators are automatically BRST-invariant, and one can construct further BRST-invariant extensions by multiplication with  a BRST-invariant extension of the identity operator (see below.) In theories with a non-minimal sector, the BRST charge is
\begin{equation}\label{eq:omega nm}
\delta\Omega^\mathrm{nm}=\delta \Omega-i \delta\rho^a \delta b_a,
\end{equation}
where $\delta\Omega$ is the same as in equation (\ref{eq:omega}). The extra contribution can be thought to originate from the additional constraints $\delta b_a=0$ that appear when one regards the Lagrange multipliers as phase space variables. The only difference is that  we think of the $\delta\rho^a$ as canonical momenta, rather than configuration variables.

\subsection{Time Evolution}

Because of the explicit time dependence, neither $\delta H$  nor $\delta H_D$ in equation (\ref{eq:H HD}) are gauge-invariant, so we cannot define their BRST-invariant extensions in the previous sense.  Instead, we define a ``BRST extension" of the Hamiltonian to be any  function $\delta H_\mathrm{BRS}$ of ghost number zero that generates  a time evolution under which the BRST charge is conserved, and whose zero antighost component agrees with the first-class Hamiltonian, 
\begin{subequations}\label{eq:BRST extension}
	\begin{align}
	\{\delta\Omega,\delta H_\mathrm{BRS}\}+\frac{\partial\delta \Omega}{\partial t}&=0,
	\label{eq:H extension}\\
	(\delta H_\mathrm{BRS})_{p=0}&=\delta H_D. 
	\end{align}
\end{subequations}
Because the BRST transformation is nilpotent, the BRST extension of the Hamiltonian is not unique. Indeed, if $\delta H_\mathrm{BRS}$ satisfies equation (\ref{eq:H extension}), so does 
\begin{equation}\label{eq:H K}
\delta H_K\equiv \delta H_\mathrm{BRS}+\{\delta K,\delta\Omega\}
\end{equation}
for any  Grassmann-odd extended phase space function $\delta K$ of ghost number  minus one.  In addition, up to terms proportional to the constraints, the $p=0$ terms of both $\delta H_K$ and $\delta H_\mathrm{BRS}$ agree. 

The function $\delta K$ is known as the gauge-fixing fermion, and the Hamiltonian (\ref{eq:H K}) as the gauge-fixed Hamiltonian. This gauge-fixed Hamiltonian determines the time evolution of any function $\delta F$ in the extended phase space though its Poisson bracket as usual,
\begin{equation}\label{eq:dt BRST}
\frac{d\delta F}{dt}=\{\delta F,\delta H_K\}+\frac{\partial\delta F}{\partial t},
\end{equation}
By definition, under such  evolution the BRST charge is conserved.   The conservation of  $\delta\Omega$ in the theory of the perturbations essentially follows from the conservation of the background-independent  $\Omega$  under a time evolution generated by an identically vanishing Hamiltonian. 

\subsection{BRST symmetry}

Because of the explicit time dependence, the gauge-fixed Hamiltonian $\delta H_K$  is not BRST-invariant. Nevertheless, by definition, the transformation induced by $\delta\Omega$ is a global symmetry, in the sense of the conservation law (\ref{eq:H extension}).  In fact, the BRST equations of motion in the minimal sector can be derived from the variational principle
\begin{equation}\label{eq:delta S BRST}
\delta S_K=\int dt\left[\delta\dot{q}^i \delta p_i+\delta\dot{\eta}^a\delta\mathcal{P}_a-\delta H_K\right].
\end{equation}
By direct substitution, one can check that, up to boundary terms, this action is invariant under the infinitesimal BRST transformation  $\Delta_\theta F\equiv \{F,\theta \delta\,\Omega\}$, where $\theta$ is a constant  Grassmann-odd parameter. Note that the variables $\delta\lambda^a$  do not appear in the action (\ref{eq:delta S BRST}). They show up in the non-minimal sector of the theory, whose dynamics is generated by the action
\begin{equation}\label{eq:delta S BRST non-min}
\delta S^{\mathrm{nm}}_K=\int dt\left[\delta\dot{q}^i \delta p_i+\delta\dot{\eta}^a\delta\mathcal{P}_a
+\delta\dot\lambda^a \delta b_a+\delta\dot{C}_a\delta\rho^a
-\delta H_K\right].
\end{equation}
In this case, the variables $\delta\lambda^a$ and their conjugate momenta $\delta b_a$ belong to the phase space of the theory.  The BRST Hamiltonian still is that of equation (\ref{eq:H K}), although the BRST charge that determines the contribution of the gauge-fixing fermion is that of the non-minimal sector (\ref{eq:omega nm}). 

\subsection{Gauge Fixing}

The time evolution equations (\ref{eq:dt BRST}) contain no arbitrary functions, and thus define a unique trajectory in phase space for appropriately specified initial conditions. In that sense, we can think of the action (\ref{eq:delta S BRST}) as a gauge-fixed action, with a gauge-fixing condition that is associated with the choice of $\delta K$. 

It is important to realize that, the choice of $\delta K$ is completely arbitrary,  up to Grassmann parity and ghost number. There is hence a huge amount of freedom to gauge fix the theory. In 
field theories in Minkowski spacetime, manifest relativistic invariance and renormalizability severely restrict the choice of the gauge fixing fermion, but in a gravitational theory in a time-dependent cosmological background the set of possible $\delta K$ is much larger, even if one insists on manifest invariance under translations and rotations. 

At this point we shall mostly focus on the simplest  gauge fixing fermions, namely, those linear in the fields of ghost number minus one,
\begin{equation}\label{eq:delta K}
\delta K=\chi^a\, \delta\mathcal{P}_a
+i \sigma^a \, \delta C_a.
\end{equation}
The coefficients $\chi^a$ and $\sigma^a$ are assumed to be arbitrary functions of the bosonic fields of the theory, that is, the original canonical pairs, the Lagrange multipliers and their conjugate momenta in the non-minimal sector. With this choice, by repeated use of the identities (\ref{eq:P symmetry}) and (\ref{eq:P Leibnitz}), we arrive at
\begin{multline}\label{eq:BRST correction}
\Delta\delta H\equiv\{\delta K, \delta \Omega\}=
-\chi^a \delta G_a-\sigma^a \delta b_a
\\
+\delta_\eta \chi^a \delta\mathcal{P}_a
+\chi^a \delta\eta^b C_{ba}{}^c \delta\mathcal{P}_c
-i\delta_\rho \chi^b\delta\mathcal{P}_b
+i\delta_\eta \sigma^a \delta C_a
+\delta_\rho \sigma^a\delta C_a
\\
-\frac{1}{2}\{C_{cb}{}^a,\chi^d\}\delta\eta^b \delta \eta^c \delta\mathcal{P}_a\delta\mathcal{P}_d
-\frac{i}{2}\delta C_a \{\sigma^a,C_{cb}{}^d\} \delta\eta^b\delta\eta^c\delta\mathcal{P}_d
\\
+\{\chi^a\delta\mathcal{P}_a,\delta\Omega_{p\geq 2}\}
-i\{\sigma^a,\delta\Omega_{p\geq 2}\}\delta C_a.
\end{multline}
where $\delta_\eta \chi$ and $\delta_\rho \sigma$ respectively are the changes of $\chi$ and $\sigma$ under gauge transformations with parameters $\delta\eta^a$ and $\delta\rho^a$,  ${\delta_\eta \chi \equiv\delta\eta^a \{\delta G_a,\chi\}}$ and ${\delta_\rho \sigma \equiv\delta\rho^a \{\delta b_a,\sigma\}}$.  

It is illustrative to check how this formalism  reproduces the well-known Faddeev-Popov action in the case of canonical gauge conditions.  Setting $\chi^a=-\delta\lambda^a$ and $\sigma^a=-f^a(\delta p_i, \delta q^i)/\varepsilon$ in equation (\ref{eq:BRST correction}), and changing variables  $ \delta b_a\to\varepsilon \delta b_a$, $\delta C_a\to\varepsilon \delta C_a$ in the  gauge-fixed action (\ref{eq:delta S BRST non-min}) returns, in the limit $\varepsilon\to 0$,
\begin{multline}\label{eq:Faddeev-Popov-action}
\delta S_\mathrm{FP}=\int dt\Big[\delta\dot{q}^i \delta p_i+\delta\dot{\eta}^a\delta\mathcal{P}_a
-\delta H_D-\delta\lambda^a\delta G_a
-\delta b_a f^a
+i\delta_\eta f^a \delta C_a\\
+i\delta\rho^a \delta\mathcal{P}_a
+\delta\lambda^a \delta\eta^b C_{ba}{}^c \delta\mathcal{P}_c
-\frac{i}{2}\delta C_a\{f^a,C_{cb}{}^d\}\delta\eta^b\delta\eta^c\delta\mathcal{P}_d
+\delta\lambda^a\{\delta\mathcal{P}_a,\delta\Omega_{p\geq 2}\}+i\{f^a,\delta\Omega_{p\geq 2}\}\delta C_a
\Big].
\end{multline}
Because the conjugate momenta $\delta b_a$  appear linearly in this action, integration over these variables produces a delta function $\delta(f^a)$ that enforces the canonical gauge-fixing conditions $f^a=0$. Hence, the first line of equation (\ref{eq:Faddeev-Popov-action}) is nothing but the gauge-fixed Faddeev-Popov  action, which includes the original Hamiltonian, the gauge-fixing term  and the corresponding ghost contribution $i\delta\eta^b\{\delta G_b,f^a\}\delta C_a$. Similarly, integration over the (imaginary) variable $\delta\rho^a$ enforces the conditions $\delta\mathcal{P}_a=0$, which eliminates all the remaining terms on the second line of the equation, since they are all of antighost number $p\geq 1$.  In that sense, the imposition of canonical gauge conditions is just a special case of gauge fixing in the BRST formalism.  Note, however, that because we are dealing with an open algebra, equation  (\ref{eq:BRST  correction}) implies that  the gauge-fixed Hamiltonian  will generically contain  cubic and higher order terms in the ghosts that do not appear in the Faddeev-Popov formalism in generalized gauges. 

Equation (\ref{eq:BRST correction}) also enables us to finally determine a BRST-invariant extension of the first-class Hamiltonian $\delta H_D$. Under the assumption that the BRST charge has the structure (\ref{eq:omega structure}), it is easy to check that $\delta H_\mathrm{BRS}\equiv-\dot{\bar q}^i \delta p_i+\dot{\bar p}_i \delta q^i$ satisfies equation (\ref{eq:BRST extension}). In the background-independent formulation of the theory, this is just the statement that an identically vanishing Hamiltonian is BRST-invariant.  Adding to the latter the BRST-exact expression obtained by setting $\chi^a=-\bar{\lambda}^a$ and $\sigma^a\equiv 0$ in equation (\ref{eq:BRST correction}) we arrive at
\begin{equation}\label{eq:BRST Hamiltonian}
\delta H_\mathrm{BRS}\equiv 
-\dot{\bar q}^i \delta p_i+\dot{\bar p}_i \delta q^i -\bar\lambda^a\{\delta\mathcal{P}_a,\delta\Omega\}=
\delta H_D
-\bar{\lambda}^b \delta\eta^c C_{cb}{}^a \delta\mathcal{P}_a
-\bar\lambda^a\{\delta\mathcal{P}_a,\delta\Omega_{p\geq 2}\},
\end{equation}
which carries ghost number zero and satisfies the two properties (\ref{eq:BRST extension}).
The Hamiltonian (\ref{eq:BRST Hamiltonian}) shall be the starting point of our perturbative calculations. Similarly, we can construct a BRST-invariant extension of the identity operator from any gauge-fixing fermion $\delta J$ by the prescription
\begin{equation}
\mathbbm{1}_{J}=
\exp\left(i\{\delta J,\delta\Omega\}\right).
\end{equation}
This extension shall prove to be useful in defining BRST-invariant extensions of gauge-invariant operators.

\section{Quantization}
\label{sec:BRST Quantization}

In Dirac's  approach to the quantization of gauge theories in the Schr\"odinger representation, the Hilbert space consists of wave functionals of the configuration variables $\delta q^i$. The constraints are imposed on the physical states of the theory, $\delta\hat{G}_a |\psi_\mathrm{phys}\rangle=0$, and the time evolution operator is taken to be
\begin{equation}\label{eq:U Dirac}
\hat{\mathcal{U}}_D(t;t_0)=\mathcal{T}\exp\left[-i\int_{t_0}^{t} \delta H_D(\delta \hat q^i(t_0),\delta \hat p_i(t_0),\tilde{t}\,) \, d\tilde{t}\right],
\end{equation}
where $\mathcal{T}$ is the time-ordering operator, and we have assumed that the Hamiltonian is of the form (\ref{eq:H HD}). We shall not pursue  Dirac quantization in the Fock representation here, which proceeds by enforcing only half of the constraints on the physical states. 

BRST quantization in the Schr\"odinger representation follows basically the same steps as  Dirac quantization, basically by replacing gauge-invariance by BRST-invariance:

\begin{enumerate}
	\item The real even variables in  phase space are replaced by bosonic hermitian field operators, and the real odd variables are replaced by real fermionic operators. The field operators $\delta\hat{q}^i$, $\delta\hat{p}_i$,  $\delta\hat{\eta}^a$  and $\delta \hat{C}_a$ are hermitian, while the fermionic fields $\delta\hat{\mathcal{P}}_a$ and $\delta\hat\rho^a$ in the non-minimal sector are anti-hermitian.
	
	\item The graded Poisson brackets of the canonical variables are replaced by the graded commutators of the corresponding operators,
	\begin{equation}\label{eq:graded commutations}
	[\hat{q}^i,\hat{p}_j]\equiv i \delta^i{}_j,
	\quad
	[\delta\hat{\eta}^a,\delta\hat{\mathcal{P}}_b]\equiv -i \delta^a{}_b.
	\end{equation} 
	\item  Physical states $|\Psi\rangle$ belong to the cohomology of $\delta \hat{\Omega}$ at  ghost number zero,
	\begin{equation}
	|\Psi\rangle\in \left(\frac{\mathrm{Ker}\, \delta\hat\Omega}{\mathrm{Im}\, \delta\hat\Omega}\right)_0.
	\end{equation}
	
	\item The gauge-fixed Hamiltonian $\delta H_K$ generates time evolution,
	\begin{equation}
	i\frac{d\delta \hat F}{dt}=[\delta\hat F, \delta\hat{H}_K]+i\frac{\partial \delta \hat F}{\partial t}.
	\end{equation}
\end{enumerate}
Hence, the Heisenberg representation fields at time $t$ can be recovered from the time-evolution operator as usual,
\begin{align}
\delta \hat F(t)&=\hat{\mathcal{U}}_K^\dag(t,t_0) \delta \hat F(t_0) \hat {\mathcal{U}}_K(t,t_0),
\\ 
\hat{\mathcal{U}}_K(t,t_0)&\equiv \mathcal{T}\exp\left(-i\int_{t_0}^{t} \, \delta\hat{H}_K\, d\tilde{t} \right). \label{eq:BRST U}
\end{align}
In this formulation time evolution is  manifestly unitary. In the Schr\"odinger picture, it is also manifestly insensitive to the gauge-fixing fermion $\delta \hat{K}$, because the latter only changes the  physical state $|\Psi\rangle$ by vectors in the image of the BRST charge, which do not affect the cohomology of $\delta\hat{\Omega}$. In the following we remove the hats on top of the operators for simplicity.

\subsection{Expectation Values}

\subsubsection{BRST Quantization} In cosmological perturbation theory, we are  interested in the expectation values of observables in appropriately chosen quantum states. A convenient basis  is furnished by the BRST-invariant states of ghost number zero
\begin{equation}\label{eq:BRST inv}
|\delta q^i_\mathrm{BRS}\rangle=| \delta q^i, \delta \eta^a=0, \delta b_a=0, \delta C_a=0\rangle,
\end{equation}
where the labels indicate the eigenvalues of the corresponding operators.  An important theorem in BRST theory\footnote{See Section 14.5.5 in \cite{Henneaux:1992ig}.} states that if $\delta \mathcal{O}_\mathrm{BRS}$ is a BRST-invariant extension of a given observable $\delta\mathcal{O}$, its matrix elements between BRST-invariant states of the form (\ref{eq:BRST inv}) return the projected kernel of $\delta \mathcal{O}$, up to an irrelevant overall sign, 
\begin{equation}\label{eq:projected kernel}
\langle \delta q^i_\mathrm{BRS}| \delta \mathcal{O}_\mathrm{BRS}| \delta \tilde q_\mathrm{BRS}^j\rangle = \delta \mathcal{O}_P(\delta q^i, \delta\tilde q^j). 
\end{equation}
The projected kernel $\delta \mathcal{O}_P$ on the right hand side is a regularized version of the kernel of $\delta\mathcal{O}$  projected onto the space of gauge-invariant states.  As a consequence of this projection,  $\delta\mathcal{O}_P$ satisfies the constraints, in the sense that 
\begin{equation}
\delta G_a(\delta q^i, -i\partial/\partial \delta q^i)\delta\mathcal{O}_P(\delta q^i, \delta\tilde q^j)=
\delta G_a(\delta\tilde  q^i, -i\partial/\partial \delta\tilde q^i)
\delta\mathcal{O}^*_P(\delta q^i, \delta\tilde q^j)=0.
\end{equation}
In general, in order for the matrix element (\ref{eq:projected kernel}) to be well-defined, it is necessary that the BRST-invariant extension $\delta\mathcal{O}_\mathrm{BRS}$ mix the minimal and non-minimal sectors.   With these definitions, the expectation value of a gauge-invariant operator $\delta\mathcal{O}(t)$ becomes
\begin{subequations}\label{eq:diagonal vev all}
	\begin{equation}\label{eq:diagonal vev}
	\langle \Psi_\mathrm{in} | \delta \mathcal{O}(t) | \Psi_\mathrm{in}\rangle
	=
	\int d\delta q \, d\delta \tilde q\,
	\psi^*_\mathrm{in}(\delta q) \mu(\delta q)\,
	\delta\mathcal{O}_P(t)(\delta q,\delta \tilde{q})
	\mu(\delta\tilde q)  \,
	\psi_\mathrm{in}(\delta \tilde q),
	\end{equation}
	where the projected kernel of the operator equals, using equation (\ref{eq:projected kernel}),
	\begin{equation}\label{eq:vev extension}
	\delta \mathcal{O}_P(t)(\delta q, \delta \tilde{q})=
	\langle \delta q_\mathrm{BRS}| \, 
	\big[\mathcal{U}_D^\dag(t,t_0) 
	\delta \mathcal{O}(t_0) 
	\mathcal{U}_D(t,t_0)\big]_\mathrm{BRS}\, 	
	| \delta \tilde q_\mathrm{BRS} \rangle.
	\end{equation}
	In equation (\ref{eq:diagonal vev}), $\mu$ is a  regularization functional needed to render  the integrals finite\footnote{\textit{ibid.}, Section 13.3.4.}. In simple cases in which the constraints are linear in the conjugate momenta  it is determined by  a set of conditions $f^a(\delta q,\delta \pi)=0$ through the relation 
	\begin{equation}\label{eq:mu}
	\mu=\det\{f^a,\delta G_b\}\delta(f^c).
	\end{equation}
	In the absence of this regularization factor, the integral over the gauge-invariant directions would diverge, since  the projected kernel satisfies the constraints. We illustrate this divergence with a concrete example in Section \ref{sec:Quantization: Vectors}.  Although the constraints in general relativity are not linear in the momenta, we shall only need  the form of $\mu$ in the limit in which interactions are turned off, and the form (\ref{eq:mu}) does apply.  
	
	The BRST extension of $\mathcal{U}_D^\dag(t,t_0) \delta \mathcal{O}(t_0) \mathcal{U}_D(t,t_0)$ in equation (\ref{eq:vev extension}) is pretty much arbitrary, up to the requirement that it mix the minimal and non-minimal sectors. A relatively natural choice would be 
	$
	\mathcal{U}_K^\dag(t,t_0) 
	\delta\mathcal{O}(t_0)_\mathrm{BRS} \,
	\mathcal{U}_K(t,t_0),
	$ 
	for a $\delta K$ that mixes the minimal and non-minimal sectors. This is the choice typically made in approaches in which one is interested in calculating in-out matrix elements of the time evolution operator. 
	But since we are interested in in-in matrix elements,   in some instances this prescription fails    because the terms that mix  sectors in $\mathcal{U}^\dag_K$ and $\mathcal{U}_K$ cancel each other. It  is thus convenient to work instead with the slightly more general expression
	\begin{equation}\label{eq:observable extension}
	[\mathcal{U}_D^\dag(t,t_0) 
	\delta \mathcal{O}(t_0) 
	\mathcal{U}_D(t,t_0)]_\mathrm{BRS}=
	\mathbbm{1}_J(t_0)\,
	\mathcal{U}_K^\dag(t,t_0) 
	\delta\mathcal{O}_\mathrm{BRS}(t_0) \,
	\mathcal{U}_K(t,t_0)\,
	\mathbbm{1}_J(t_0),
	\end{equation}
\end{subequations}
for appropriately chosen and unrelated gauge-fixing fermions $\delta J$ and $\delta K$. The formalism guarantees that the projected kernel is insensitive to the particular choice of BRST extension. Although  it is not necessary to include two copies of $\mathbbm{1}_J$ in the extension (\ref{eq:observable extension}), their inclusion renders it slightly more symmetric.  

\subsubsection{Dirac Quantization}

Within the BRST formalism it is also possible to recover the  matrix elements of the Dirac time evolution operator (\ref{eq:U Dirac}) from those of the BRST-invariant operator in equation (\ref{eq:BRST U}). In those cases, it is not necessary to choose a gauge-fixing fermion that mixes the minimal and non-minimal sectors, and one can indeed set $\delta K=0$. With a vanishing gauge-fixing fermion, the Hamiltonian  of the theory reduces to the first-class Hamiltonian $\delta H_D$, plus the ghost terms needed to construct its BRST extension. Since the  first-class Hamiltonian $\delta H_D$  is that of the original theory with Lagrange multipliers $\delta\lambda^a$ set to zero, the bosonic sector of the action  is the one in equation (\ref{eq:Hamiltonian S}) for a metric of the form 
\begin{equation}\label{eq:synchronous coords}
ds^2=-\bar{N}^2 dt^2+(\bar{h}_{ij}+\delta h_{ij})dx^i dx^j.
\end{equation}
Up to the irrelevant factor $\bar{N}^2=a^2$, this is just the metric in synchronous coordinates, or 
synchronous gauge. 

But if one is interested in the  unprojected kernel of the time evolution operator in the Dirac method anyway, one can dispense of ghosts and BRST-invariance altogether, and simply rely on the original representation of the operator in equation (\ref{eq:U Dirac}).  Quantization of cosmological perturbations following Dirac's method thus amounts to quantization of the theory with a metric of the form (\ref{eq:synchronous coords}), with no ghosts in the action.  For that reason, we shall use ``Dirac quantization" and ``synchronous gauge quantization" as synonyms, in spite of our previous remarks concerning synchronous gauge in the BRST formalism.  In particular, in synchronous  gauge the expectation value of a gauge-invariant operator reads
\begin{subequations}\label{eq:non-diagonal vev all}
	\begin{equation}\label{eq:non-diagonal vev}
	\langle \Psi_\mathrm{in} | \delta \mathcal{O}(t) |  \Psi_\mathrm{in}\rangle
	=
	\int d\delta q \, d\delta \tilde q\,
	\psi^*_\mathrm{in}(\delta q) \mu(\delta q) 
	\delta \mathcal{O}(t)(\delta q,\delta \tilde{q})
	\psi_\mathrm{in}(\delta \tilde q),
	\end{equation}
	where the kernel of the operator $\delta \mathcal{O}(t)$ equals
	\begin{equation}\label{eq:O Heisenberg}
	\delta \mathcal{O}(t)(\delta q,\delta \tilde{q})=
	\langle \delta q | 
	\mathcal{U}^\dag_D(t,t_0) 
	\delta \mathcal{O}(t_0)\,
	\mathcal{U}_D(t,t_0)	
	| \delta \tilde q \rangle,
	\end{equation}
\end{subequations}
and the time evolution operator is that of equation (\ref{eq:U Dirac}). Note that it is only necessary to insert a single regularization factor $\mu$ in the integrand of (\ref{eq:non-diagonal vev}), since it contains the non-projected kernel of the operator $\delta\mathcal{O}$.  The factor of $\mu$ then regularizes the scalar product between the gauge-invariant states $|\Psi_\mathrm{in}\rangle$ and $\delta\mathcal{O}(t)|\Psi_\mathrm{in}\rangle$. Clearly, one of the main advantages of Dirac (or synchronous gauge) quantization is the absence of ghosts. 

\section{Free Theory}
\label{sec:freetheory}

Our discussion of gauge-invariance and BRST symmetry so far has been mostly exact. Even though we have expanded  the fields around a non-trivial cosmological background, the expressions that we have derived apply to all orders in the perturbations. In practice, however, such calculations do not occur. Instead, one typically expands the action to a certain order in the perturbations, and carries out calculations only to that order. In that case, all the expressions we have previously derived hold only to the appropriate order in the perturbation expansion. As we shall see, such perturbative calculations  obscure the  symmetries and simplicity of the underlying theory even further. 

We shall first  illustrate such a  calculation in the simplest case, that of linear perturbation theory. This case is  relevant because it provides the foundation upon which perturbative calculations are built, and because it suffices  to  calculate the primary observables of inflationary theory, namely, primordial power spectra. The linear analysis can be easily generalized to  arbitrary higher orders, at least formally, although specific calculations become increasingly cumbersome as the perturbation order increases.

\subsection{Einstein-Hilbert Action}

We begin by expanding the Einstein-Hilbert action in Hamiltonian form (\ref{eq:S pert I})  to quadratic order in the perturbations. The action therefore becomes 
\begin{equation}\label{eq:delta S Hamiltonian}
\delta S=\int dt\left[\delta p_i \delta\dot{q}^i-\delta^{(2)}\!H\right],
\end{equation}
where $\delta^{(2)}H$ is the Hamiltonian expanded to second order,
\begin{equation}
\delta^{(2)}\!H=-\dot{\bar{q}}^i \delta p_i+ \dot{\bar{p}}_i \delta q^i+\bar{\lambda}^a\delta^{(2)} G_a+\delta\lambda^a\delta^{(1)}G_a,
\end{equation}
and the $\delta^{(n)}G_a$ are the constraints expanded to $n$-th order in the perturbations. Note that the Hamiltonian somewhat simplifies because the background satisfies the classical equations of motion, so the linear terms in the perturbations cancel.  Hence, if $\delta G^{(k)}$ is the term of $k$-th order in $\delta G$,
we can also write
\begin{equation}
\delta^{(2)}\!H=\bar\lambda^a \delta G_a^{(2)}+\delta\lambda^a \delta G_a^{(1)}\equiv \delta^{(2)}\!H_D+\delta\lambda^a \delta G_a^{(1)},
\end{equation}

In order to find $\delta G_a^{(2)}$ and $\delta G_a^{(1)}$, we simply expand  expressions (\ref{eq:secondary}) to the desired order.  We begin with the terms linear in the perturbations, 
\begin{subequations}\label{eq:linear constraints}
	\begin{align}
	\delta G^{(1)}_N &=
	2a \mathcal{H}\delta\pi
	+\frac{\dot{\bar\varphi}}{a}\delta\pi_\varphi
	+\frac{M^2}{2a}\left(\nabla^2 \delta h-\partial_i\partial_j \delta h^{ij}\right)
	+\frac{M^2}{a}\dot{\mathcal{H}}\delta h+a^3 \bar{V}_{,\varphi}\delta\varphi,\\
	\delta G^{(1)}_i&=-\left[2a^2\partial_j\delta\pi_i{}^j-M^2 \mathcal{H}(2\partial_j \delta h_i{}^j-\partial_i \delta h)\right]
	+a^2\dot{\bar\varphi}\, \partial_i\delta\varphi,
	\end{align}
\end{subequations}
in which indices are raised and lowered  with a Kronecker delta. To derive these expressions, we have used the background equations of motion (\ref{eq:background motion}), and that the relevant background quantities  satisfy (\ref{eq:background}).

Because in our background the shift vector vanishes ($\bar{\lambda}^i\equiv 0$), in order to obtain the quadratic Hamiltonian we only need to calculate $\delta G_N$ to second order. After a somewhat long but straight-forward calculation we arrive at 
\begin{multline}\label{eq:H0 pert}
\delta H_D^{(2)}=\int d^3 x\bigg[\frac{2a^2}{M^2}\delta\pi^{ij}\delta\pi_{ij}-\frac{a^2}{M^2}\delta\pi^2+\frac{1}{2a^2}\delta\pi_\varphi^2
-2\mathcal{H}\delta\pi^{ij}\delta h_{ij}+\mathcal{H}\delta\pi\delta h-\frac{\dot{\bar\varphi}}{2a^2}\delta\pi_\varphi \delta h\\
-\frac{M^2}{8a^2}\delta h^{ij}\nabla^2 \delta h_{ij}+\frac{M^2}{8a^2}\delta h\nabla^2 \delta h+\frac{M^2}{4a^2}\delta h^{ij} \partial_j\partial_k \delta h_i^k-\frac{M^2}{4a^2}\delta h \partial_j \partial_k \delta h^{jk}
\\
+\frac{a^2}{2}\partial_i\delta\varphi \partial^i\delta\varphi
+\frac{M^2\left(\mathcal{H}^2-\dot{\mathcal{H}}/2\right)}{a^2}\delta h_{ij}\delta h^{ij} +\frac{a^4}{2}\bar{V}_{,\varphi\varphi}\delta\varphi^2+\frac{a^2 \bar{V}_{,\varphi}}{2}\delta\varphi\delta h
\bigg].
\end{multline}

\subsection{Constraints}

Variation of the action (\ref{eq:delta S Hamiltonian}) with respect to $\delta\lambda^a$  results in the linear constraints $\delta G_a^{(1)}=0$, or, in standard notation,
\begin{equation}
\delta G_N^{(1)}=\delta G_i^{(1)}=0.
\end{equation}
The reader can readily check that the linear constraints (\ref{eq:linear constraints}) generate linear diffeomorphism when acting on the configuration variables. In particular, $\delta G_N^{(1)}$ generates linear diffeomorphisms along the unit normal to the constant time hypersurfaces $n^\mu=(a^{-1},\vec{0})$. But  as opposed to that of the full theory,  the algebra of the linearized constraints (\ref{eq:linear constraints}) is Abelian. This  follows for instance by expanding equation (\ref{eq:structure}) in powers of the perturbations, and noting that the background satisfies the zeroth order constraints. 

Because the full action for the perturbations (\ref{eq:S pert I})  is invariant under the transformations (\ref{eq:Delta delta}), the quadratic action (\ref{eq:delta S Hamiltonian}) is invariant under a truncation of those transformations to linear order,
\begin{subequations}\label{eq:Delta delta linear}
	\begin{align}
	\Delta \delta q^i&=\{\delta q^i,\xi^a \delta G^{(1)}_a\}, \\
	\Delta \delta p_i&=\{\delta p_i,\xi^a \delta G^{(1)}_a\},\\
	\Delta \delta \lambda^a&=\dot{\xi}^a
	+\xi^b\bar{\lambda}^c \bar{C}_{bc}{}^a.
	\label{eq:Delta delta lambda linear}
	\end{align}
\end{subequations}
Readers familiar with  gauge transformations in the Hamiltonian formalism will recognize in equation (\ref{eq:Delta delta lambda linear})  the transformation properties of the Lagrange multipliers under a  set of Abelian first class constraints with a weakly gauge-invariant Hamiltonian.  

It is in fact illustrative to see how  the linear constraints 
(\ref{eq:linear constraints}) are preserved by the time evolution. On the one hand, from the equations of motion in the linearized theory, their time derivatives are
\begin{equation}
\frac{d\delta G_a^{(1)}}{dt}\equiv \{\delta G_a^{(1)},\delta^{(2)}\!H\}+\frac{\partial\delta G_a^{(1)}}{\partial t}.
\end{equation}
On the other hand, in the theory to all order in the perturbations, the time derivatives of the full constraints obey, from equations (\ref{eq:structure}) and (\ref{eq:dt delta C}),
\begin{equation}\label{eq:full evolution}
\frac{dG_a}{dt}\equiv \{\delta G_a,\delta H\}+\frac{\partial\delta G_a}{\partial t}=(\bar\lambda^b+\delta\lambda^b) C_{ab}{}^c \delta G_c.
\end{equation}
Expanding the equation on the right hand side of (\ref{eq:full evolution}) to first order in the perturbations, and bearing in mind that $\delta H^{(1)}$ vanishes, we immediately get that 
\begin{equation}\label{eq:dGa 1 a}
\{\delta G_a^{(1)},\delta^{(2)}\!H\}+\frac{\partial\delta G_a^{(1)}}{\partial t} =\bar\lambda^b \bar{C}_{ab}{}^c  \delta G_c^{(1)}.
\end{equation}
Therefore, the time derivatives of the linear constraints  vanish weakly, and the Hamiltonian itself is not  invariant under diffeomorphism, even on-shell. In the full theory, the linearity of the Hamiltonian in the first class constraints is essentially responsible for  the preservation of the constraints under the time evolution. In the linear theory, this simple structure is hidden  and distorted by the expansion  to second order in the perturbations around a time-dependent background. Note that the coefficient $\bar\lambda^b \bar{C}_{ab}{}^c$ multiplying the constraints on the right hand side of (\ref{eq:dGa 1 a}) is precisely the coefficient that appears in the transformation law (\ref{eq:Delta delta lambda linear}). This is what one expects in a theory in which the Hamiltonian is weakly gauge-invariant.

\subsection{Classical BRST Symmetry}

The BRST charge of the linearized theory (in the non-minimal sector) is that of the full theory, equation (\ref{eq:omega nm}), expanded to second order in the  perturbations,
\begin{equation}\label{eq:delta omega linear}
\delta\Omega ^{(2)}\equiv \delta\eta^a \delta G_a^{(1)}-i\delta\rho^a\delta b_a.
\end{equation}
Again, this is consistent with our observation that the constraints in the linear theory are Abelian. In order to write down the BRST-invariant action in the free theory, we need to find the BRST Hamiltonian. Expanding equation (\ref{eq:BRST Hamiltonian}) to second order, we readily arrive at 
\begin{equation}\label{eq:delta 2 H BRST}
\delta^{(2)}\!H_\mathrm{BRS}\equiv \delta^{(2)}\!H_D-\bar{\lambda}^a \delta\eta^b \bar{C}_{ba}{}^c\delta\mathcal{P}_c,
\end{equation}
which  indeed is BRST-invariant, in the sense that it obeys
\begin{equation}\label{eq:Omega 1 conserved}
\{\delta\Omega^{(2)},\delta^{(2)}H_\mathrm{BRS}\}+\frac{\partial\delta\Omega^{(2)}}{\partial t}=0,
\end{equation}
as the reader can verify using equation (\ref{eq:dGa 1 a}).  The BRST-invariant action of the linearized theory in the non-minimal sector still has the form (\ref{eq:delta S BRST non-min}), but the gauge-fixed Hamiltonian is that obtained from equation (\ref{eq:delta 2 H BRST}) and the usual relation
\begin{equation}
\delta^{(2)}\!H_K=\delta^{(2)}\!H_\mathrm{BRS}+\{\delta K,\delta\Omega^{(2)}\},
\end{equation}
where the gauge-fixing fermion $\delta K$ is an arbitrary Grassmann-odd function of the canonical variables of ghost number minus one. The gauge-fixed Hamiltonian still satisfies  equation (\ref{eq:Omega 1 conserved}), which  guarantees that, up to total derivatives, the quadratic action  is invariant under the linear BRST transformations generated by $\delta\Omega^{(2)}$.   To preserve the quadratic structure of the Hamiltonian, $\delta K$ needs to be quadratic in the perturbations too. 

\subsubsection*{Comohology}

In a theory with a gauge symmetry the algebra of observables consists of gauge invariant functions of the canonical variables. As we mentioned earlier, this algebra agrees with the set of BRST-invariant functions of ghost number zero. To see how this works in practice, let us determine the cohomology of the linearized BRST charge (\ref{eq:delta omega linear}).  

We begin by calculating the action of $\delta\Omega^{(2)}$ on the different fields in the non-minimal sector of the theory,
\begin{align}
s\, \delta\lambda^a&=-i\delta\rho^a, 
&s\,\delta\rho^a&=0,
\\
s\,\delta C_a&=i\delta b_a,  &s \, \delta b_a&=0.
\end{align}
Since all the fields that are closed ($s\delta F=0$) are exact ($\delta F=s\delta G$), the cohomology  in the non-minimal sector is trivial. Hence, observables can be assumed not to depend on the variables of the non-minimal sector. In particular, they  do not depend on the Lagrange multipliers  $\delta\lambda^a\equiv\{\delta \lambda^N,\delta\lambda^i\}$. 

In the minimal sector, the action of the BRST charge amounts to a gauge transformation with gauge parameters $\delta\eta^a$.  To study the cohomology, it proves to be useful to move to a different field basis. First, we carry out  a canonical transformation to a set of fields that transform irreducibly, as described in Appendix \ref{sec:Irreducible Representations}. The latter are classified according to  their transformation properties under rotations as scalars, vectors and tensors.  In linear perturbation theory the three sectors decouple from each other. In each sector, then, we perform another canonical transformation to a basis of fields with particularly simple properties under gauge transformations.  

\paragraph{Tensor Sector} In the tensor sector this basis consists of the conjugate pairs
\begin{equation}
\zeta_{\pm2} \equiv \frac{ \delta h_{\pm2}}{a^2}, \quad \Pi^{\pm2}\equiv a^2\delta\pi^{\pm2}.
\end{equation}
Because they are invariant under the linearized diffeomorphisms, and there is no helicity two field of ghost number minus one, the helicity two fields are exact  and cannot be written as a BRST transformation of any other fields. They span the observables of the tensor sector of the linear theory, and the BRST charge in this sector identically vanishes. The Hamiltonian (\ref{eq:H0 pert})  in the tensor sector reads
\begin{equation}
\delta H^{(2)}_{D,t}=
\frac{1}{2}\sum_{\sigma=\pm 2}\int d^3 p 
\left[
\frac{\Pi^\sigma (\vec{p})\Pi^\sigma (-\vec{p})}{ a^2 M^2}
-8\mathcal{H}\,\Pi^\sigma(\vec{p})\zeta_\sigma(\vec{p})
+a^2 M^2\left(\vec{p}\,^2
+8\mathcal{H}^2-4\dot{\mathcal{H}}\right)\zeta_\sigma(\vec{p})\zeta_\sigma(-\vec{p})\right].
\end{equation}
There is no contribution from the tensor sector to the BRST-charge,
$
\delta\Omega^{(2)}_t=0.
$

\paragraph{Vector Sector} In the vector sector the new  field basis is
\begin{equation}
x^{\pm 1}\equiv -\frac{\delta h_{\pm1}}{a^2 p}(\vec{p}), 
\quad
G_{\pm1}\equiv 
-p\left[a^2\delta\pi^{\pm1}(\vec{p})-2M^2 \mathcal{H} \delta h_{\pm1}(-\vec{p})\right].
\end{equation}
Among other properties, this basis is useful because the constraints in the vector sector simply read $G_{\pm1}=0$.  The action of a BRST transformation on each of these helicity one fields is
\begin{subequations} 
	\begin{align}
	s \, G_{\pm1} &=0,
	\quad
	&s\,\delta\mathcal{P}_{\pm1}&= -G_{\pm1},\\
	s \, \delta\eta^{\pm1}&=0, &s\, x^{\pm1}&=\delta\eta^{\pm1}. 
	\end{align} 
\end{subequations}
Hence, all BRST-closed fields are BRST-exact. Clearly, the BRST cohomology in the vector sector is trivial and  there are no observables in this sector.  The Hamiltonian is
\begin{equation}\label{eq:Hv}
\delta H^{(2)}_{D,v}=\sum_{\sigma=\pm 1} \int d^3 p\,
\bar{A}_\sigma G_\sigma (\vec{p})G_\sigma (-\vec{p})
\quad \bar{A}_{\pm1}=\frac{1}{a^2 M^2 p^2},
\end{equation}
and  the BRST charge in the vector sector reads
\begin{equation}\label{eq:Omegav}
\delta\Omega^{(2)}_v=\sum_{\sigma=\pm1} \int d^3 p \,\big[\delta \eta^\sigma(\vec{p}) G_\sigma(\vec{p})-i\delta\rho^\sigma (\vec{p}) \delta b_\sigma(\vec{p})\big].
\end{equation}
Since the Poisson bracket of the first-class Hamiltonian with the BRST charge  vanishes, $\delta H^{(2)}_{D,v}$  is BRST-invariant. 

\paragraph{Scalar Sector} The bosonic scalar sector is spanned by the three canonical pairs  $(\delta h_L,  \delta\pi^L), (\delta h_T, \delta\pi^T)$ and $(\delta\varphi, \delta\pi_\varphi)$.  A new choice of conjugate pairs  that happens to be particularly convenient is
$(\zeta,\Pi), (x^L,G_L)$, $(x^N, G_N)$, where the latter are defined by
\begin{subequations}\label{eq:new scalar variables} 
	\begin{align}
	\zeta &\equiv\frac{\delta h_L}{a^2}+\frac{\delta h_T}{3a^2}-\frac{\mathcal{H}}{\dot{\bar{\varphi}}}\delta\varphi,
	\\
	\Pi &\equiv
	\frac{6 M^2 \mathcal{H}^2-2M^2p^2}{\mathcal{H}}\delta h_L
	-\frac{2 M^2p^2}{3\mathcal{H}}\delta h_T
	-3a^2\dot{\bar\varphi}\,\delta\varphi
	+a^2\delta\pi^L
	\\
	x^L&\equiv-\frac{\delta h_T}{p a^2}, 
	\\
	G_L &\equiv 2\mathcal{H}M^2 p\, \delta h_L+\frac{8}{3} \mathcal{H} M^2 p\, \delta h_T
	-a^2 \dot{\bar\varphi}\, p \, \delta\varphi
	+\frac{a^2}{3} p\, \delta\pi^L-a^2 p\,  \delta \pi^T,
	\\
	x^N&\equiv \frac{a \,\delta\varphi}{\dot{\bar\varphi}}
	,\\
	G_N &\equiv -\frac{2M^2}{a}\left(p^2-3\dot{\mathcal{H}}\right)\delta h_L-\frac{2M^2 p^2}{3a}\delta h_T+a^3 \bar{V}_{,\varphi} \delta\varphi+a\mathcal{H}\,\delta \pi^L+\frac{\dot{\bar\varphi}}{a}\delta\pi_\varphi.
	\end{align}
\end{subequations}
We quote the inverse relations that express the old variables in terms of the new variables in equations (\ref{eq:inverse relations}).

The new variables $G_L$ and $G_N$ are nothing but the scalar components of the original linearized constraints (\ref{eq:linear constraints}), which now simply read $G_L=G_N=0$. The variable $\zeta$ is the standard curvature perturbation in comoving slices, and  setting $x^L=x^N=0$ amounts to working in comoving gauge. The variables $\zeta$, $\Pi$, as well as the constraints $G_L$ and $G_N$ are gauge-invariant. The configuration fields canonically conjugate to the latter, $x^L$ and $x^N$,  then span the two independent gauge variant directions in field space.  The action of a BRST transformation on these fields is
\begin{subequations}\label{eq:BRST scalar delta}
	\begin{align}
	s\,\zeta &=0,\\
	s\,\Pi&=0, \\
	s\, G_N&=0, &s\, \mathcal{P}_N&=- G_N\\
	s\, G_L&=0, &s\, \mathcal{P}_L&=- G_L,\\
	s\,  \eta^L&=0, &s\,  x^L&=\delta\eta^L,\\
	s\,  \eta^N&=0, & s\, x^N&= \delta\eta^N.
	\end{align}
\end{subequations}
Hence, the only BRST-closed fields in our set that are not BRST-exact are $\zeta$ and $\Pi$. The latter are thus the only observables in the scalar sector of the linear theory. Note that we did not have to invoke the zero ghost number condition to identify the space of observables. To conclude, let us write down the contribution of the scalar sector to the BRST charge in the new field basis,
\begin{equation}\label{eq:Omega s}
\delta\Omega^{(2)}_s=\sum_{\sigma=L,N}\int d^3 p  \big[\delta \eta^\sigma(\vec{p})  G_\sigma(\vec{p})
-i \delta\rho^\sigma(\vec{p})\delta b_\sigma (\vec{p})\big], 
\end{equation}
from which we could also have easily derived equations (\ref{eq:BRST scalar delta}). In terms of the new canonical fields, the  Hamiltonian (\ref{eq:H0 pert}) in the scalar sector   becomes
\begin{multline}\label{eq:Hs}
\delta H_{D,s}^{(2)}= \int d^3p \, \bigg[
\frac{\mathcal{H}^2}{2a^2  \dot{\bar{\varphi}}^2} \Pi^2
+ \frac{p^2 a^2 \dot{\bar{\varphi}}^2}{2\mathcal{H}^2} \zeta^2
+\frac{3}{4M^2p^2 a^2}G_L^2
+\frac{1}{2 \dot{\bar{\varphi}}^2}G_N^2
\\
-\frac{p}{\mathcal{H}}\zeta\,G_L
-\Pi\left(\frac{G_L}{2M^2 a^2 p}+\frac{\mathcal{H}G_N}{a \dot{\bar{\varphi}}^2}\right)-\frac{p}{a}x^N G_L
\bigg].
\end{multline}
In this form, the Poisson brackets of the Hamiltonian with the constraints can be read off  immediately.  Because the former do not vanish, the Dirac Hamiltonian is not BRST-invariant. Instead, from equations (\ref{eq:Omega s}) and (\ref{eq:Hs}), or directly from (\ref{eq:dGa 1 a}) and (\ref{eq:delta 2 H BRST}),  a BRST-invariant Hamiltonian is
\begin{equation}
\delta H^{(2)}_{\mathrm{BRS},s}=\delta H_{D,s}^{(2)}-\int d^3 p \, \frac{p}{a}\delta\eta^N \delta\mathcal{P}_L.
\end{equation}
For simplicity, we have dropped the momentum labels of the different fields in the previous integrals. The latter can be reintroduced by demanding that the corresponding expression be a scalar under translations.  Enforcing the constraints on the scalar sector action by setting $G_L= G_N=0$  yields the Hamiltonian  action for the gauge-invariant conjugate variables $\zeta$ and $\Pi$,
\begin{equation}\label{eq:S zeta R}
S_\mathrm{GI}[\Pi,\zeta]=\int d^3p \, \left[
\Pi(\vec{p})\dot{\zeta}(\vec{p})-
\frac{\mathcal{H}^2}{2a^2  \dot{\bar{\varphi}}^2} \Pi(\vec{p})\Pi(-\vec{p})
- \frac{a^2 p^2 \dot{\bar{\varphi}}^2}{2\mathcal{H}^2} \zeta(\vec{p})\zeta(-\vec{p})\right].
\end{equation}
This is the action one would use in the standard gauge-invariant (or reduced phase space) approach, in which only a complete set of gauge-invariant variables is kept in the  theory. Replacing $\Pi$ by the solution of its own equation of motion yields the well-known Lagrangian action for the gauge-invariant variable $\zeta$ in the gauge-invariant formalism \cite{Garriga:1999vw}. For an alternative discussion of the reduced phase space  in linearized cosmological perturbation theory, see \cite{Langlois:1994ec}.

The transition to the new variables (\ref{eq:new scalar variables}) considerably simplifies the structure of the scalar Hamiltonian, which now is  almost diagonal. The Hamiltonian (\ref{eq:Hs})  does not depend on  $x^L$, and only depends on $x^N$ through the combination $x^NG_L$, which manifestly shows that the constraints $G_L$ and $G_N$ are preserved by the time evolution. It is in fact possible to fully diagonalize the Dirac Hamiltonian by an appropriate canonical transformation.  We discuss the diagonalization of the scalar Hamiltonian in Appendix \ref{sec:Diagonalization of the Scalar Hamiltonian}, where we show that it can be cast in the form 
\begin{equation}\label{eq:Hs diagonal}
\delta H_{D,s}^{(2)}=\int d^3p 
\left[
\frac{\mathcal{H}^2}{2a^2 \dot{\bar\varphi}^2} \underline{\Pi}^2
+\frac{p^2 a^2 \dot{\bar{\varphi}}^2}{2\mathcal{H}^2} \underline{\zeta}^2
+ \bar{A}_L\, \underline{G}_L^2
+ \bar{A}_N\, \underline{G}_N^2
\right],
\end{equation}
where $\bar{A}_L$ and $\bar{A}_N$ are the time-dependent coefficients quoted in equations (\ref{eq:As}).  The variable $\underline{\zeta}$ is still the comoving curvature, but only when the constraints $\underline{G}_L=\underline{G}_N=0$ are satisfied.

\subsection{Quantization: Vectors}
\label{sec:Quantization: Vectors}

The vector sector in the free theory offers a simple setting to illustrate the methods behind Dirac and BRST quantization, and how they differ from other common approaches. It will also serve as a useful warm-up for  the quantization of the scalar sector. The Hamiltonian in the vector sector is given by equation (\ref{eq:Hv}), and the BRST charge by equation (\ref{eq:Omegav}). In this sector the constraints read $G_\pm=0$.

\subsubsection{Dirac Quantization}

As we mentioned above, Dirac quantization amounts to the quantization of the theory  in synchronous gauge.  In the Dirac approach to the quantization of first class constraints, the generator of the time evolution is  the first-class Hamiltonian $\delta H_D$ in equation (\ref{eq:Hv}). The constraints are then imposed on the physical states of the theory by demanding ${G_{\pm 1}(t_0) |\Psi\rangle=0}$. If we represent these states by wave functionals in configuration space $\psi[x^{+}(\vec{p}),x^{-}(\vec{p})]$, this implies that the latter do not depend on the fields $x^{\pm1}\equiv x$. 

Observables  are functions of   gauge-invariant operators alone. The only such operators in the vector sector are the  $G_\pm$.  Since  the Hamiltonian of the theory commutes with $G_\pm$, the action of these operators on any physical state will hence vanish. But one needs to be careful when evaluating expectation values of gauge-invariant operators that do not involve powers of $G_\pm(t)$, because the latter are naively ill-defined. Consider for example the expectation of the (gauge-invariant) identity operator in the vector sector for a physical state $|\Psi\rangle$ with wave function $\psi$, a.k.a. the norm, $\int dx \,  \psi^*[x] \psi[x].$ If the state is physical, its wave function  does not depend on $x$, and the integral diverges. In order to avoid this problem, one needs to regularize the inner product  by inserting an appropriate regularization factor $\mu$, as in equation (\ref{eq:non-diagonal vev}). The latter typically involves choosing a slice through the space of gauge-variant configurations. In the present context $\mu$  can be  chosen to simply be $\delta[x]\equiv \delta[x_+]\delta[x_-]$. In that case the expectation value becomes
\begin{equation}
\langle \Psi | \mathbbm{1}|\Psi\rangle \equiv\int dx \,  \psi^*[x]\delta[x]\psi[x]=|\psi[0]|^2, 
\end{equation}
which equals one, as it should, if $\psi$ is properly normalized. 

\subsubsection{BRST Method:}

Things look quite different in BRST quantization. We already know that in this sector the  cohomology of fields is trivial, so there are no non-trivial observables in the vector sector. Nevertheless, for illustration, let us proceed with the calculation of  expectation values outlined by equations (\ref{eq:diagonal vev all}).  One begins by choosing a convenient BRST-invariant extension of the observable at hand. For illustration we choose  the vector power spectrum, 
\begin{equation}\label{eq:vector power}
\delta\mathcal{O}_\mathrm{BRS}(t)=
G_+(t,\vec{p}_2)G_+(t,\vec{p}_1),
\end{equation}
which already is BRST-invariant in the free theory. We choose next a BRST extension of the time-evolution operator. Since the Hamiltonian (\ref{eq:Hv}) is already BRST-invariant, we simply set $\delta K=0$ in equation (\ref{eq:observable extension}). If we had chosen for instance $\delta K=-\delta\lambda^a \delta G_a$, the contributions from the gauge-fixing fermion  in $\mathcal{U}^\dag_K$ and $\mathcal{U}_K$ would have cancelled each other. Because the ensuing time-evolution operator commutes with $\delta\mathcal{O}_\mathrm{BRS}(t_0)$, the problem reduces to the calculation of the expectation value of the operator (\ref{eq:vector power}) at $t_0$. 

It is now obvious that for BRST-invariant states of the form (\ref{eq:BRST inv}), such an expectation value is ill-defined, because $\delta\mathcal{O}_\mathrm{BRS}(t)=\delta\mathcal{O}_\mathrm{BRS}(t_0)$ does not mix minimal and non-minimal sectors, and the BRST-invariant states
$
|x^\pm_\mathrm{BRS}\rangle\equiv  |x^\pm, \delta b_\pm=0,
\delta \eta^\pm=0, \delta C_\pm=0
\rangle
$
have an ill-defined norm.  We thus select a non-zero $\delta J$ that mixes sectors, say,
\begin{subequations}\label{eq:J vector}
	\begin{align}
	\delta J&=-T\sum_{\sigma=\pm 1} \int d^3p\, \delta \lambda^\sigma(\vec{p}) \delta \mathcal{P}_\sigma(\vec{p}),
	\\ 
	\mathbbm{1}_J&\equiv \exp\left(i \{\delta J, \delta^{(2)} \Omega\}\right)=
	\exp\left[T \sum_{\sigma=\pm 1}\int d^3 p\, 
	\left(i\delta \lambda^\sigma G_\sigma
	+\delta\rho^\sigma \delta\mathcal{P} _\sigma
	\right)
	\right],
	\end{align}
\end{subequations}
where $T$ is an arbitrary constant with dimensions of time.  With this choice, it is a straight-forward exercise in Fourier transforms to show that
\begin{equation}\label{eq:vector expectation}
\langle\tilde  x{}^\pm_\mathrm{BRS}|
\mathbbm{1}_J(t_0)
G_+(t_0,\vec{p}_2)G_+(t_0,\vec{p}_1)
\mathbbm{1}_J(t_0)
| x^\pm_\mathrm{BRS}\rangle
=0,
\end{equation}
regardless of the values of $\tilde x^\pm$ and $x{}^\pm.$  Along the same lines, the matrix element of the BRST-invariant extension of the identity operator can be  seen to equal 
\begin{equation}\label{eq:vector identity}
\langle\tilde x{}^\pm_\mathrm{BRS}|\mathbbm{1}_J|x^\pm_\mathrm{BRS}\rangle=
\int d(TG_\sigma) \,
\delta[T\, G_\sigma]\exp\left(i\sum_{\sigma=\pm} \int d^3 p\,  G_\sigma(\vec{p})[\tilde x{}^\sigma(\vec{p})-x^\sigma(\vec{p})]\right),
\end{equation}
which is ill-defined for $T=0$ (it equals $0\times \infty$), but simplifies to  one for $T\neq 0$. The factor of $T$ inside the integral measure is the contribution of the fermionic sector, and the factor of $T$ inside the delta function is that of the  bosonic sector.  Note that in both equations (\ref{eq:vector expectation}) and (\ref{eq:vector identity})  the matrix element does not depend on the values of $\tilde x^\pm$ and $x^\pm$, as expected from the projected kernel.

\subsection{Quantization: Scalars}

We proceed now to illustrate Dirac and BRST quantization in the scalar sector, in which both methods are non-trivial. The fields in this sector consist of the gauge-invariant $\zeta$ and its conjugate $\Pi$ and the two gauge-variant variables $(x^L,x^N)\equiv x$  with the constraints $(G_L,G_N)\equiv G$ as their conjugates.  

\subsubsection{Dirac Quantization:}
\label{sec:Free Synchronous Gauge}

In the Dirac method, one quantizes the metric in synchronous coordinates.  Although synchronous gauge was the gauge initially chosen by Lifshitz in his seminal article on cosmological perturbation theory \cite{Lifshitz:1945du}, synchronous gauge has been  widely criticized most notably  because the conditions $\delta\lambda^N=\delta\lambda^i=0$ do not fix the gauge uniquely, since  it is still possible to perform non-trivial gauge transformations that preserve the  synchronous gauge conditions \cite{Mukhanov:1990me}. To see  this, consider the transformation properties of the  linear perturbations under the linear  gauge transformations (\ref{eq:Delta delta linear}),
\begin{subequations}\label{eq:gauge transformations}
	\begin{align}
	\Delta \delta \lambda^N&=a\left(\dot\xi^0+\mathcal{H}\xi^0\right),\\
	\Delta \delta \lambda_i&=\dot{\xi}_i-\partial_i \xi^0,\\
	\label{eq:Delta h}
	\Delta \delta h_{ij}&=a^2\left(2\mathcal{H}\xi^0 \delta_{ij}+\partial_i\xi_j+\partial_j \xi_i\right),\\
	\Delta \delta\varphi&=\xi^0 \, \dot{\bar\varphi},
	\end{align} 
\end{subequations}
where we have set $\xi^N=a\, \xi^0$ (as mentioned above, $\delta G_N$ generates diffeomorphisms along the unit normal to the constant time hypersurfaces.)  The most general gauge transformation that preserves synchronous gauge therefore is
\begin{equation}\label{eq:gauge parameters}
\xi^0=\frac{A(\vec{x})}{a}, \quad \xi_i=B(\vec{x})+\partial_i A \int^t \frac{d\tilde{t}}{a},
\end{equation}
which in fact is non-trivial for any non-zero choices of the free functions  $A$ and $B$. This residual gauge freedom implies that solutions to the equations of motion for the perturbations in synchronous gauge are not unique, since $\Delta \delta h_{ij}$  in (\ref{eq:Delta h}) with gauge parameters (\ref{eq:gauge parameters}) is always a solution of the synchronous linear perturbation equations. That is mostly why synchronous gauge has been essentially abandoned for analytical studies of cosmological perturbations, although it still plays a prominent role in numerical calculations.

Yet when we solve the equations of motion for the perturbations, we need to impose appropriate boundary conditions to single out a unique solution. When we extremize the action, for instance, we are typically interested in boundary conditions in which the perturbations at the endpoints are specified. Similarly,  in an initial value problem we prescribe the  values of the fields and their time derivatives at some initial time. Clearly, from equations (\ref{eq:gauge transformations}), the only choice that does not alter such  boundary conditions is $\xi^0=\xi_i=0$. In this context, the apparent gauge freedom of synchronous gauge disappears, and does not pose any particular problem. 

With this understanding in mind, let us hence consider the time evolution operator $\mathcal{U}_D$ for the perturbations (\ref{eq:U Dirac}).   The main advantage of this approach is that there are no ghosts to deal with, at any order. In addition, in synchronous gauge the action is a local functional of the perturbations.

Let us calculate the expectation of a gauge-invariant operator $\delta \mathcal{O}(t)$ such as the power spectrum of $\zeta$ in synchronous gauge. In this case, from equations (\ref{eq:non-diagonal vev all}) the expectation value is
\begin{equation}\label{eq:synchronous power}
\langle \delta \mathcal{O}(t)\rangle=
\int d\zeta\, dx \,
\mathcal \psi^*_\mathrm{in}[\zeta, x]  \, 
\delta[x]
\langle \zeta, x | \mathcal{U}^\dag_D(t,t_0) \delta \mathcal{O}(t_0)\, \mathcal{U}_D(t,t_0) | \Psi_\mathrm{in}\rangle,
\end{equation}
where the in-state satisfies the constraints 
$
G_L(t_0) |\Psi_\mathrm{in}\rangle=G_N(t_0) |\Psi_\mathrm{in}\rangle=0,
$
and we have chosen $f=x$ in the regularization factor $\mu=\det\{f,G\} \delta[f]$.
Because $\delta \mathcal{O}(t_0)$ commutes with  $G_{L,N}(t_0)$ by gauge-invariance,  both time evolution operators in equation (\ref{eq:synchronous power}) act on a zero eigenstate of the constraints, so we can set $G_L=G_N=0$ in the scalar Hamiltonian (\ref{eq:Hs}).  The time evolution operator reduces then to that of the gauge-invariant approach
\begin{equation}\label{eq:scalar GI}
\mathcal{U}^\dag_D(t,t_0) \delta  \mathcal{O}(t_0) \, \mathcal{U}_D(t,t_0) |\Psi_\mathrm{in}\rangle=\mathcal{U}_\mathrm{GI}^\dag(t,t_0)\delta  \mathcal{O}(t_0) \, \mathcal{U}_\mathrm{GI}(t,t_0) |\Psi_\mathrm{in}\rangle,
\end{equation}
in which $\mathcal{U}_\mathrm{GI}$ is determined by the action of equation (\ref{eq:S zeta R}). Inserting equation (\ref{eq:scalar GI})  into equation (\ref{eq:synchronous power})  finally yields
\begin{equation}\label{eq:vev GI}
\langle \delta \mathcal{O}(t) \rangle=\int
d\zeta \, d\tilde\zeta \,
\mathcal \psi^*_\mathrm{in}[\zeta]  \, 
\langle \zeta |
\mathcal{U}_\mathrm{GI}^*(t, t_0)
\delta \mathcal{O}(t_0)\, 
\mathcal{U}_\mathrm{GI}(t, t_0)
|\tilde\zeta\rangle \,
\psi_\mathrm{in}[\tilde\zeta],
\end{equation}
where we have set $\psi_\mathrm{in}[\zeta]\equiv \psi_\mathrm{in}[\zeta,x]$ (because of the constraints, the in-state wave function is $x$-independent).  Equation (\ref{eq:vev GI})  is what one would write down in the gauge-invariant formalism. Hence, from now on  the calculation follows the standard route, and expectation values of gauge-invariant operators in synchronous gauge automatically agree with those obtained in the gauge-invariant formalism. 

One of the main disadvantages of  synchronous gauge is that our scalar variables  are still coupled to each other, which renders the calculation of some of the  propagators in the scalar sector difficult.   As we describe in Appendix \ref{sec:Diagonalization of the Scalar Hamiltonian}, to decouple the scalar variables we need to carry out a canonical transformation whose coefficients contain time integrals, as opposed to local expressions in time. In the new variables, the scalar sector Hamiltonian takes the form (\ref{eq:Hs diagonal}). In the new variables, the calculation of the propagators in the scalar sector is straight-forward. Recall that  in the in-in formalism, there are four different types of propagators \cite{Sloth:2006az,Chaicherdsakul:2006ui,Seery:2007we,Adshead:2008gk,Gao:2009fx,Campo:2009fx,Senatore:2009cf,Bartolo:2010bu,Kahya:2010xh}. The four types can be constructed from appropriately ordered expectation values of  field bilinears, which, from the structure of equations (\ref{eq:non-diagonal vev all}) with $\mu=\delta[\underline x]$ require the calculation of matrix elements of the form 
\begin{equation}
\int d\underline\zeta\, d\tilde{\underline\zeta} \, 
\psi^*_\mathrm{in}[\underline\zeta]\,
\langle \underline{\zeta},\underline{x}=0 |
\underline{z}^i(t_2) \underline{z}^j(t_1)
|\tilde{
	\underline\zeta},\tilde{\underline G}=0\rangle\,
\psi_\mathrm{in}[\tilde{\underline\zeta}],
\end{equation}
where the $z^i$ stand for any of the fields or conjugate momenta in the theory.  Therefore, quadratic matrix elements in the gauge-invariant sector spanned by $\underline{\zeta}$ and $\underline{\Pi}$ are the same as in the gauge-invariant method. Among the remaining bilinears, the only  non-vanishing matrix elements are
\begin{subequations}\label{eq:synchronous propagators}
	\begin{align}
	\langle \Psi_\mathrm{in}| \underline{G}_{\sigma_2} (t_2,\vec{p}_2) \, \underline{x}^{\sigma_1} (t_1,\vec{p}_1)|\Psi_\mathrm{in}\rangle&=-i\,\delta_{\sigma_2}^{\sigma_1}\,\delta(\vec{p}_2-\vec{p}_1),\\
	\langle \Psi_\mathrm{in}| \underline{x}^{\sigma_1} (t_2,\vec{p}_2) \, \underline{x}^{\sigma_2} (t_1,\vec{p}_1)|\Psi_\mathrm{in}\rangle&=2i\,\delta^{\sigma_1 \sigma_2} \delta(\vec{p}_2+\vec{p}_1) \int_{t_0}^{t_1} dt \, \bar{A}_\sigma(t) .
	\end{align}
\end{subequations}
These results  not only apply in the scalar sector, but also in the vector sector, where  the Hamiltonian already has the required diagonal structure (\ref{eq:Hv}). Although we shall not do so here,  for calculations beyond the free theory it may  be more convenient to use the inverse relations (\ref{eq:inverse relations}) to cast the propagators in terms of the original variables  $\delta h_{ij}$, $\delta\varphi$, and their canonical conjugates. Once the latter are known, one can study interactions between cosmological perturbations at any order without the need of any additional field transformations.

\subsubsection{Derivative Gauges}
\label{sec:Free Derivative Gauges}

As we have  emphasized earlier, BRST quantization allows for a much wider set of gauge choices. To illustrate the flexibility of the BRST formalism,  let us show how to dramatically simplify the structure of the Hamiltonian (\ref{eq:Hs}) by an appropriate choice of gauge-fixing fermion. First, since the gauge-fixed Hamiltonian should be quadratic and of ghost number minus one, it has to be linear in $\delta\mathcal{P}$ and $\delta C$, as in equation (\ref{eq:delta K}). Second,
because we do not intend to calculate matrix elements of the time-evolution operator between BRST-invariant states, it is not necessary for the gauge-fixing fermion $\delta K$ to  couple minimal and non-minimal sectors. One of the simplest choices that satisfies these conditions  has  $\sigma^a=0$. Choices for which $\sigma^a\neq 0$ can be used to enforce  conventional canonical gauge conditions, as we described earlier. 

By choosing  $\chi$ in equation (\ref{eq:delta K}) appropriately it is possible to remove terms in the Hamiltonian proportional to the gauge-invariant fields $G$  that are constrained to vanish in the original formulation of the theory. The freedom to alter  the evolution of the system in such a way  just captures our  expectation that only the dynamics of the gauge-invariant pairs $\zeta$ and $\Pi$ is physically relevant. In particular, picking
\begin{equation}
\delta K=\int d^3 p\left[\left(\frac{3G_L}{4M^2 p^2 a^2}-\frac{p \, \zeta }{\mathcal{H}} -\frac{\Pi}{2M^2 a^2 p}-\frac{p \, x^N}{a } \right)\delta\mathcal{P}_L
+\left(\frac{G_N}{2\dot{\bar\varphi}^2}-\frac{\mathcal{H}\,  \Pi}{a\dot{\bar\varphi}^2}\right)\delta \mathcal{P}_N
\right]
\end{equation} 
we find that the gauge-fixed Hamiltonian in the scalar sector $\delta^{(2)}\! H_{0,s}+\{\delta K,\delta^{(2)}\Omega_s\}$ becomes
\begin{equation}\label{eq:gauge fixed Hs}
\delta H^{(2)}_K=\frac{1}{2}\int d^3 p \left[
\frac{\mathcal{H}^2}{a^2 \dot{\bar\varphi}^2} \Pi^2
+\frac{a^2  p^2 \dot{\bar\varphi}^2}{\mathcal{H}^2}\zeta^2
\right].
\end{equation}
By a suitable choice of variables and gauge-fixing fermion we have thus  diagonalized the Hamiltonian.  At first sight it may appear that we took a long detour to arrive at the gauge-invariant formalism, but, in fact, the  situation here is quite different, because $\delta\lambda^{(N,L)}, x^{(L,N)}, \delta\eta^{(L,N)}, \delta C^{(L,N)}$ and their canonical conjugates remain part of the phase space.    

The  Hamiltonian  (\ref{eq:gauge fixed Hs}) does not depend quadratically on the momenta $G$ or $b$, so the theory  does not admit a Lagrangian formulation: Integrating over these momenta yields delta functionals $\delta[\dot{x}]$ and $\delta[\delta\dot{\lambda}]$, rather than quadratic terms in the velocities. In that sense, our choice of gauge-fixing fermion corresponds to a derivative gauge in which we implicitly impose the on-shell conditions $\delta\dot{\lambda}^a=0$ on the Lagrange multipliers. This property is shared by many other gauge-fixing fermions. By appropriate choices of $\chi$ and $\sigma$ we could have obtained a theory with a Lagrangian formulation  in which the $\delta\dot{\lambda}^a$ still vanish on-shell, at the expense of making the Hamiltonian slightly less simple.

We would like to calculate the expectation value of a gauge-invariant operator $\delta\mathcal{O}(t)$ along the lines of equations (\ref{eq:diagonal vev all}).  
Because the operator  $\delta\mathcal{O}(t)$ is gauge-invariant, in the free theory its BRST-invariant extension in equation (\ref{eq:observable extension}) can be taken to be the operator itself, ${\delta\mathcal{O}_\mathrm{BRS}(t_0)=\delta\mathcal{O}(t_0)}$. Then, since  by gauge choice  the Hamiltonian (\ref{eq:gauge fixed Hs})  equals that in the gauge-invariant formulation, the matrix elements of the projected operator  in equation (\ref{eq:vev extension}) factorize, thus becoming
\begin{multline}
\langle \delta q|
\delta\mathcal{O}_P(t)
|\delta\tilde q\rangle
=\langle \zeta | 
\mathcal{U}_{GI}^\dag(t,t_0) \delta\mathcal{O}(t_0)\mathcal{U}_\mathrm{GI}(t,t_0)
|\delta\tilde\zeta\rangle
\\
\times
\langle x, \delta\eta=0, \delta b=0, \delta C=0| \mathbbm{1}_J \mathbbm{1}_J 
|\tilde x,\delta\tilde\eta=0, \delta\tilde b=0, \delta\tilde C=0\rangle.
\end{multline}
In order for the last matrix-element to be well-defined, $\mathbbm{1}_J$ needs to couple the minimal and non-minimal sectors. We choose the analog of the gauge-fixing fermion (\ref{eq:J vector}) in the vector sector,
\begin{subequations}
	\begin{align}
	J&=-T\sum_{\sigma=N, L} \int d^3p\, \delta\lambda^\sigma(\vec{p}) \delta \mathcal{P}_\sigma(\vec{p}),
	\\ 
	\mathbbm{1}_J&\equiv \exp\left(i [\delta J, \delta \Omega]\right)=
	\exp\left[T \sum_{\sigma=N,L}\int d^3 p\, 
	\left(i\delta \lambda^\sigma G_\sigma
	+\delta\rho^\sigma \delta\mathcal{P} _\sigma
	\right)
	\right],
	\end{align}
\end{subequations}
which, as in equation (\ref{eq:vector identity}), implies that 
\begin{equation}
\langle x, \delta\eta=0, \delta b=0, \delta C=0| \mathbbm{1}_J \mathbbm{1}_J 
|\delta\tilde x,\delta\tilde\eta=0, \delta\tilde b=0, \delta\tilde C=0\rangle=1.
\end{equation}
As expected,  then, the projected kernel of the operator $\delta \mathcal{O}(t)$ does not depend on the variables $ x$ and $\tilde x$, and agrees with that of the gauge-invariant method. To complete the calculation, we just need to fold the kernel of  $\delta\mathcal{O}_P(t)$ with the appropriate  wave functional of the in state, $\psi_\mathrm{in} [\zeta_i, x_i]$,  following equation (\ref{eq:diagonal vev}).  Because of gauge invariance,  the wave function is $x$-independent, so the convolution needs to be regularized by inserting factors of $\mu[x]$, say $\mu[x]=\delta[x]$. With $\psi_\mathrm{in}[\zeta]\equiv \psi_\mathrm{in}[\zeta,x=0]$ the expectation value of the operator becomes
\begin{equation}\label{eq:in in}
\langle \Psi_\mathrm{in}| \delta \mathcal{O}(t)|\Psi_\mathrm{in} \rangle
=\int  d\tilde\zeta\,  d\zeta\,
\psi^*_\mathrm{in}[\tilde\zeta] \delta \mathcal{O}_P(t)[\tilde\zeta,0;\zeta, 0]\psi_\mathrm{in}[\zeta].
\end{equation}
Therefore, the BRST  returns the same expectation value as the gauge-invariant formalism, as it should. 

\subsubsection{Propagators in Derivative Gauges}
\label{sec:Propagators in Derivative Gauges}

Our  next goal  is to determine the propagators of the in-in formalism in the scalar sector, with a gauge-fixed Hamiltonian determined by equation (\ref{eq:gauge fixed Hs}).  As  in synchronous gauge, the propagators can be constructed from  
various expectations of field bilinears. But in this case the states are BRST-invariant, and one needs to regularize the scalar products by inserting BRST-invariant extensions of the identity operator that mix minimal and non-minimal sectors, as we did above.  We shall thus calculate 
\begin{subequations}
	\begin{equation}
	\langle \Psi_\mathrm{in}| \mathbbm{1}_J(t_0)\,   
	z^i(t_2) z^j(t_1) \mathbbm{1}_J(t_0)\, |\Psi_\mathrm{in} \rangle,
	\end{equation}
\end{subequations}
in which,  because we are working in the Hamiltonian formulation,  the fields $z^i$ and $z^j$  run over the configuration variables and their conjugate momenta. 

Because the variables $\zeta$ and $\Pi$ decouple from the rest, all the field bilinears involving the latter agree with those of the gauge-invariant free theory, and, by symmetry, any mixed bilinear with a single factor of $\zeta$ or $\Pi$ vanishes. Because the conjugates $\delta b_a$ do not appear anywhere in the gauge-fixed Hamiltonian, there is no need to calculate bilinears that contain these fields.  Then, the only  remaining non-vanishing expectation values  are 
\begin{subequations}
	\begin{align}
	\langle \Psi_\mathrm{in}| \mathbbm{1}_J(t_0)\,   
	\delta \lambda^{\sigma_2} (t_2) G_{\sigma_1}(t_1) \mathbbm{1}_J(t_0)\, |\Psi_\mathrm{in} \rangle
	&=\frac{i}{2T}\delta^{\sigma_2}_{\sigma_1},
	\\
	\langle \Psi_\mathrm{in}| \mathbbm{1}_J(t_0)\,   
	x^{\sigma_2} (t_2) G_{\sigma_1}(t_1) \mathbbm{1}_J(t_0)\, |\Psi_\mathrm{in} \rangle
	&=\frac{i}{2}\delta^{\sigma_2}_{\sigma_1}.
	\end{align}
\end{subequations}
As in the case of synchronous gauge,  it may prove more convenient to calculate the propagators of the original fields $\delta h_{ij}$ and $\delta\varphi$ using the last expressions.  Note that because physical states carry ghost number zero, and both $\mathcal{U}_K$ and $\mathbbm{1}_J$ conserve ghost number, ghosts only appear in loop diagrams: No connected diagram  can be disconnected by cutting a single ghost line.

\subsubsection{Comparison with Other Gauges}

Leaving issues of renormalizability aside,  the reader may be wondering at this point  whether the BRST formalism has any advantages  with respect to the traditional approaches. One way to answer this question is to compare the gauge-fixed  Hamiltonian (\ref{eq:gauge fixed Hs}) with the one in the  often-employed comoving gauge, that of equation (\ref{eq:Hs}) with  $x^L\equiv x^N\equiv 0$,
\begin{equation}\label{eq:comoving Hs}
\delta H_\mathrm{com}^{(2)}= \int d^3p \, \bigg[
\frac{\mathcal{H}^2\,  \Pi^2}{2a^2  \dot{\bar{\varphi}}^2} 
+ \frac{p^2 a^2 \dot{\bar{\varphi}}^2\,  \zeta^2}{2\mathcal{H}^2}
+\frac{3G_L^2}{4M^2p^2 a^2}
+\frac{G_N^2}{2 \dot{\bar{\varphi}}^2}
\\
-\frac{p \,\zeta\,G_L}{\mathcal{H}}
-\Pi\left(\frac{G_L}{2M^2 a^2 p}+\frac{\mathcal{H}G_N}{a \dot{\bar{\varphi}}^2}\right)\bigg].
\end{equation} 
By integrating over the canonical momenta in the Hamiltonian action we could   eliminate all the conjugate momenta and arrive at the Lagrangian formulation of the theory, but since our gauge-fixed Hamiltonian (\ref{eq:gauge fixed Hs}) does not admit such a formulation, we restrict ourselves to the Hamiltonians for comparison purposes. Enforcing the constraints $G_L=G_N=0$   in equation (\ref{eq:comoving Hs}) by integrating over the Lagrange multipliers  takes us back to the action (\ref{eq:S zeta R}). But the elimination of the multipliers  is not useful beyond the free theory,  because in comoving gauge the terms proportional to the multipliers (the constraints) contain quadratic and higher order  terms in the canonical variables, and it is more convenient to deal with them perturbatively. In the Lagrangian formulation one often integrates out the multipliers by replacing them by the solutions of their own equations of motion \cite{Maldacena:2002vr,Weinberg:2005vy}, but this procedure does not apply beyond  tree level,   because in the interacting theory the Lagrangian action is not quadratic in the multipliers.  Whatever the case, at this stage it is  obvious that the Hamiltonian (\ref{eq:gauge fixed Hs}) has a simpler structure than the Hamiltonian (\ref{eq:comoving Hs}). In particular, although the latter contains fewer variables, the former is diagonal.  As in many other instances, we have traded a larger number of variables for a simpler description of the theory. Although the question of simplicity  in the free theory is moot, it is important when one considers interactions, because a simpler structure of the propagators significantly eases perturbative  calculations.

\section{Interacting Theory}
\label{sec:interactingtheory}

We proceed now to study  interactions among cosmological perturbations.  Rather than studying these in all generality, we restrict ourselves for illustration to cubic order. The generalization to higher orders is then (conceptually) straight-forward. Even in the cubic theory there are subtleties such as operator ordering issues that we shall  gloss over. 

Our starting point is the original Hamiltonian (\ref{eq:H HD}) expanded to cubic order. Its Hamiltonian  is
\begin{equation}\label{eq:delta H 3}
\delta^{(3)}\!H=\delta^{(3)}\!H_D+\delta\lambda^a \, \delta^{(2)}G_a,
\end{equation}
where the first-class Hamiltonian is that of the original theory (\ref{eq:H Dirac}) expanded to the same order
\begin{equation}
\delta^{(3)}\!H_D=\bar\lambda^a(\delta G^{(2)}_a+\delta G^{(3)}_a).
\end{equation}
Note the absence of linear terms in $\delta^{(3)}\!H_D$, because we assume that the background satisfies the classical equations of motion. The constraints in the cubic theory read $\delta^{(2)}G_a=0$.  We shall  regard the expansion  to cubic order just as an approximation to the original theory, rather than as a theory on its own. Although we were able to interpret the quadratic theory literally as a gauge-invariant theory under a set of Abelian constraints,  in the cubic theory exact invariance under the appropriately truncated transformations is lost, and is replaced by invariance modulo terms of cubic order. Consider for instance the time derivative of the constraints in the cubic theory, under the time evolution generated by the cubic Hamiltonian (\ref{eq:delta H 3}),
\begin{equation}
\frac{d\delta^{(2)}G_a}{dt}=
\{\delta^{(2)}G_a,\delta^{(3)}H_D\}+\delta\lambda^b\{\delta^{(2)}G_a,\delta^{(2)}G_b\}+\frac{\partial\delta^{(2)}G_a}{\partial t}.
\end{equation}
As in the quadratic theory, by expanding the time derivative of the full constraints in the full theory  we obtain
\begin{equation}
\{\delta^{(2)}G_a,\delta^{(3)}H_D\}+\delta\lambda^b\{\delta^{(2)}G_a,\delta^{(2)}G_b\}+\frac{\partial\delta^{(2)}G_a}{\partial t}=
(\bar\lambda^b+\delta\lambda^b)C_{ab}{}^c\delta^{(2)}G_c+\mathcal{O}^{(3)}(\delta p_i,\delta q_i,\delta\lambda^a),
\end{equation}
which shows that in the cubic theory, the quadratic  constraints are preserved only up to terms of cubic order. 

\subsection{Dirac Quantization}

In Dirac quantization, the Hamiltonian of the theory is taken to be $\delta^{(3)}\!H_D$, and the constraints are imposed on the physical states,
$
\delta^{(2)}G_a |\Psi_\mathrm{in}\rangle=0
$
(it is at this stage where factor-ordering ambiguities begin.) From equations (\ref{eq:non-diagonal vev}) and (\ref{eq:O Heisenberg}), expectation  values of gauge-invariant operators $\delta\mathcal{O}(t)$ are then determined by
\begin{equation}\label{eq:vev synchronous}
\langle \delta \mathcal{O}(t)\rangle=
\int  dq_0  \,  d\tilde{q}_0 \, dq  \, d\tilde q\,
\psi^*_\mathrm{in}(q_0)\,
\mu(q_0)\,
\mathcal{U}_D^*(q, t; q_0,t_0)
\delta\mathcal{O}(t_0)(q,\tilde{q})\,
\mathcal{U}_D(\tilde q, t;\tilde q_0,t_0)\,
\psi_\mathrm{in}(\tilde q_0),
\end{equation}
where $\mathcal{U}_D$ is the time evolution operator determined by the first-class Hamiltonian  (\ref{eq:H BRST 3}).

The nature of the in-state in the presence of interactions deserves special attention. Ideally, we would like to choose the in-state to be the vacuum of the interacting theory. A well-known theorem of Gell-Mann and Low \cite{GellMann:1951rw} states that by adiabatically switching interactions off as one moves into the asymptotic past, one can recover an eigenstate of the full Hamiltonian from that of the free theory.  In particular, if the free eigenstate is chosen to be the free vacuum, Gell-Mann and Low's prescription is expected to return the vacuum of the interacting theory. Adiabatically switching off interactions in the asymptotic past amounts to time evolution  on a time contour with an infinitesimal positive imaginary component, $t\to t(1-i\epsilon)$. For our purposes, what matters is that with interactions turned off at the infinite past, we can assume the theory to be free as  $t_0\to -\infty$. In that case, our results of Section \ref{sec:Free Synchronous Gauge} apply, and we can take the wave-function $\psi_\mathrm{in}[\delta q]$ to be that of the free vacuum.  In addition, because the theory  is free in the asymptotic past, we can choose the regularization factor $\mu(\delta q)$ to be that of the free theory.  

With these choices, the calculation of the expectation value (\ref{eq:vev synchronous}) proceed as usual, either by switching to the interaction picture, or by relying on the path integral. In particular, all the propagators of the theory can be determined from the matrix elements quoted around equations (\ref{eq:synchronous propagators}). 

\subsection{Derivative Gauges}

To explore derivative gauges, we begin by expanding  the full BRST-invariant action (\ref{eq:delta S BRST non-min}) to cubic order  in the perturbations,
\begin{equation} 
\delta^{(3)} S_K=\int dt\left[\delta p_i \delta\dot{q}^i+\delta\dot{\eta}^a\delta\mathcal{P}_a
+\delta\dot{\lambda}^a\delta b _a+\delta\dot{C}_a\delta\rho^a-\delta^{(3)}\!H_K\right],
\end{equation}
where we have also included the contribution of the variables in the non-minimal sector.  The BRST charge here is that of equation (\ref{eq:omega nm}) expanded to the same order, 
\begin{equation}\label{eq:Omega 3}
\delta^{(3)}\Omega=\delta\eta^a \,\delta^{(2)}G_a-\frac{1}{2}\delta\eta^b \delta\eta^c \bar C_{cb}{}^a \delta\mathcal{P}_a-i\delta\rho^a \delta b_a,
\end{equation}
which corresponds to a theory in which the structure constants $\bar C_{cb}{}^a$  are non-vanishing but perturbation-independent. As  can be also seen from the BRST charge (\ref{eq:Omega 3}), the constraints are the original ones expanded to second order.
The BRST extension of the  Hamiltonian is that of equation (\ref{eq:BRST Hamiltonian})  to cubic order
\begin{equation}\label{eq:H BRST 3}
\delta^{(3)}\!H_\mathrm{BRS}=
\delta^{(3)}\!H_D
-\bar{\lambda}^b \delta\eta^c (\bar{C}_{cb}{}^a+\delta C_{cb}^{(1)}{}^a)\delta\mathcal{P}_a,
\end{equation}
and the gauge-fixed Hamiltonian is constructed  from the former by adding a BRST-exact term as usual,
$
\delta^{(3)}\!H_K=\delta^{(3)}\!H_\mathrm{BRS}+\{\delta K,\delta^{(3)}\Omega\}.
$
To preserve the cubic structure of the action $\delta K$ should be quadratic at most. In particular, we can choose the same gauge-fixing fermions as in the free theory. Note that at this stage, the Hamiltonian already contains interactions between the ghosts and the cosmological perturbations. 

We shall concentrate on the class of gauge-fixing fermions (\ref{eq:delta K}) that resulted in the free scalar Hamiltonian (\ref{eq:gauge fixed Hs}). These were characterized  by functions $\chi^a$ linear in the canonical variables, and  vanishing $\sigma^a$. With this choice, the gauge-fixed Hamiltonian becomes
\begin{equation}
\delta^{(3)}\! H_K=\delta H^{(2)}_K+\delta H^{(3)}_\mathrm{BRS}-\chi^a \delta G_a^{(2)}+\delta\eta^a \{\delta G_a^{(2)},\chi^b\} \delta\mathcal{P}_b-\chi^a\delta\eta^b \bar C_{ab}{}^c\delta\mathcal{P}_c.
\end{equation}
Therefore, the original gauge-fixing fermion preserves the gauge-fixed Hamiltonian we derived in the free theory, $\delta H^{(2)}_K$, and introduces new interactions between the canonical variables and the ghosts.  

Our ultimate goal is to calculate expectation values of gauge-invariant operators in the interacting theory. In derivative gauges, the latter is determined by  equations (\ref{eq:diagonal vev all}).  As in synchronous gauge, the theory becomes free in  the limit $t_0\to -\infty\times(1-i\epsilon)$ so the factors of $\mu$ can be taken to be those in the free theory, $\mu=\delta[x]$. The same comments apply to the extensions $\mathbbm{1}_J(t_0)$, which can be taken to be those of the free theory. Therefore, from now on calculations  proceed perturbatively as usual: The expectation value  is split into the exponential of a quadratic piece, which includes  quadratic contributions from the $\mu$ and the $\mathbbm{1}_J$,  plus interactions, which include the cubic pieces from the Hamiltonian. By expanding the integrand in powers of these interactions one gets different moments of Gaussian integrals, which can be evaluated using the field bilinears described in  Section \ref{sec:Propagators in Derivative Gauges}. Whereas at cubic order a gauge-invariant operator is automatically BRST-invariant, at higher orders in perturbation theory one would need to replace the gauge-invariant operator $\delta\mathcal{O}$ by an appropriate BRST  extension. 

To conclude let us note that in the cubic theory neither the gauge-fixed action nor the BRST-extension of $\delta\mathcal{O}$ depend on the Lagrange multipliers $\delta\lambda^a$. In particular, quite remarkably, nothing in the action indicates that the perturbations need to satisfy the quadratic constraints $\delta^{(2)}G_a=0$, even though the in-state only obeys the linear constraints $\delta^{(1)}G_a=0$ in the asymptotic past.

\section{Summary and Conclusions}
\label{summaryandconcl}
There are basically two approaches to quantize a gauge theory: One can quantize a complete set of gauge-invariant variables in phase space, a method known as reduced phase space quantization, or one can simply fix the gauge by appropriately modifying the action of the theory.  The non-linear nature of general relativity   makes the first approach impractical, so in this article we have pursued the second approach. 

In this article we have studied the BRST quantization of cosmological perturbations in a theory with a scalar field  minimally coupled to gravity.  BRST quantization is not just yet another way of quantizing cosmological perturbations, but  is in fact necessary in any canonical gauge calculation beyond one loop. BRST quantization also offers a very general and flexible framework to quantize cosmological perturbations. For special choices of the gauge-fixing fermion in the time-evolution operator it produces the standard canonical gauge conditions of the standard quantization methods. When the gauge-fixing fermion is taken to vanish,  for appropriate state choices, it reproduces the kernel of the evolution operator in Dirac quantization, which is  the same one would write down in synchronous gauge. Finally,  when the BRST-extension of the observable mixes minimal and non-minimal sectors it allows for a wide variety of derivative gauges that cannot be reached by  other methods. We have mostly explored synchronous and derivative gauges here. 

The main advantage of Dirac quantization is the absence of  ghosts, and the relative simplification of the action implied by the vanishing of the Lagrange multipliers. Although it is often argued that the synchronous conditions implicit in Dirac quantization do not fix the gauge uniquely, this residual gauge freedom disappears when boundary conditions need to be preserved.   Even though it appears that the Dirac action does not contain information about the constraints of the theory, the latter are actually imposed on the states of the system, and the structure of the action guarantees that they are preserved by the time evolution. Actually, when interactions are adiabatically switched off in the asymptotic past, the in-state only needs to satisfy the free constraints.  The main disadvantage of Dirac quantization is that in order to fully diagonalize the Hamiltonian one needs to perform canonical transformations with coefficients that depend on the expansion history, which makes the calculation of the propagators more cumbersome. 

We have also explored derivative gauges here, in which new terms  are added or subtracted from  the first-class (or Dirac) Hamiltonian of the theory. Rather then restricting the values of the canonical variables or the Lagrange multipliers, these gauges impose \emph{on-shell} restrictions on the time derivatives of the latter.  One advantage of these derivative gauges is that by appropriate choice of the gauge-fixing fermion one can diagonalize the free Hamiltonian immediately.  The resulting simplification of the propagators then allows one  to easily carry out perturbative calculations in the interacting theory. But as in the standard approaches, the drawback of this method is that the ghosts do not decouple from the variables in the bosonic sector, and one has to include their contributions in loop calculations. 

We have shown that the structure of the Hamiltonian in derivative gauges can be much simpler than in the popular comoving gauge, even if the former contains more variables. Although the free Hamiltonian of comoving gauge dramatically simplifies when one imposes the constraints, this simplification is not particularly useful in the interacting theory,  because the constraints change with the inclusion of interactions. As in Dirac quantization, the simplified structure of the Hamiltonian in derivative gauges, combined with the theorem of Gell-Mann and Low, allows one to bypass the solution of  the constraints beyond  the linearized  theory.  

An almost unavoidable technical disadvantage of both Dirac and BRST quantization in derivative gauges is that observables are required to be gauge-invariant. Therefore, only expectation values of gauge-invariant operators have an immediate physical interpretation, and only these are guaranteed not to depend on the choice of gauge-fixing fermion.  By contrast, in canonical gauges one can always assume that any operator is the restriction of a gauge-invariant function to that gauge, so any operator can be identified with an observable. Another advantage of canonical gauges is that the antighosts $\delta\mathcal{P}_a$ are constrained to vanish in the gauge-fixed Hamiltonian, so one can ignore  that the algebra of diffeomorphisms is open. 

Our analysis has mostly focused on the operator quantization of the perturbations. The transition to the Lagrangian path integral formulation is straight-forward, as it only involves integration over the conjugate momenta of the variables in the Hamiltonian. In some of the gauges we have discussed the action is linear in the momenta, and such a Lagrangian formulation does not exist.  

Finally, we should also point out that many of  our results may be useful beyond the context of BRST quantization. We have written down for instance  the free classical Hamiltonian in a form that immediately allows one to identify the gauge-invariant variables, the constraints and their algebra. This form could be useful to shed further insights into the dynamics of cosmological perturbations in the Hamiltonian formulation and beyond.
\section{Appendices}
\label{sec:appendices}
\subsubsection{Generalized Bracket and Graded Commutator}
\label{sec:Graded Commutator}

Given two functions $F$ and $G$ of the extended phase space variables $( q^i,p_i)$ and $(\eta^a, \mathcal{P}_a)$ of definite Grassmann parity $\epsilon_F$ and $\epsilon_G$, we define their generalized Poisson bracket by
\begin{equation}
\{F,G\}=\left[\frac{\partial F}{\partial q^i}\frac{\partial G}{\partial p_i}-\frac{\partial F}{\partial p_i}\frac{\partial G}{\partial q^i}\right]
+(-)^{\epsilon_F}\left[\frac{\partial_L F}{\partial \eta^a}\frac{\partial_L G}{\partial \mathcal{P}_a}+\frac{\partial_L F}{\partial \mathcal{P}_a}\frac{\partial_L G}{\partial \eta^a}\right],
\end{equation}
where  $\partial_L/\partial$ denotes a left derivative. 
The generalized bracket obeys the algebraic properties
\begin{align}
\{F,G\}&=-(-)^{\epsilon_F\epsilon_G} \{G,F\}
\label{eq:P symmetry} \\
\{F,G H\}&=\{F,G\}H+(-)^{\epsilon_F\epsilon_G}G\{F,H\} \label{eq:P Leibnitz}
\end{align}
and the Jacobi identity
\begin{equation}\label{eq:Jacobi}
\{\{F_1,F_2\},F_3\}
+(-)^{\epsilon_1(\epsilon_2+\epsilon_3)}
\{\{F_2,F_3\},F_1\}
+(-)^{\epsilon_3(\epsilon_1+\epsilon_2)}
\{\{F_3,F_1\},F_2\}=0.
\end{equation}
When multiplied by $i$, the algebraic properties of the generalized bracket match those of the graded commutator of two operators $\hat{F}$ and $\hat{G}$,
\begin{equation}
[\hat{F},\hat{G}]=\hat{F}\hat{G}-(-1)^{\epsilon_F \epsilon_G} \hat{G}\hat{F}.
\end{equation}
The formal analogy between the generalized bracket and the graded commutation is exploited in  canonical quantization.

\subsubsection{Irreducible Representations}
\label{sec:Irreducible Representations}

In cosmological perturbation theory it is sometimes convenient to work with perturbations that transform irreducibly under the isometries of the cosmological background: spatial rotations and translations. We thus introduce a set of seven irreducible tensors $Q_{ij}{}^\sigma(\vec{x};\vec{p})$ and $Q(\vec{x};\vec{p})$ that we use as basis elements in an expansion of arbitrary cosmological perturbations,
\begin{equation}
\delta h_{ij}(\eta,\vec{x})=\sum_{\sigma} \int d^3 p \, Q_{ij}{}^\sigma (\vec{x};\vec{p})
\,  \delta h_\sigma(\eta,\vec{p}), \quad
\delta\varphi(\eta,\vec{x})=\int d^3 p\,  Q(\vec{x};\vec{p})\delta\varphi(\eta,\vec{p}).
\end{equation}
These tensors are plane waves, 
\begin{equation}
Q_{ij}{}^\sigma(\vec{x};\vec{p})\equiv  \frac{e^{i\vec{p}\cdot \vec{x}}}{(2\pi)^{3/2}} Q_{ij}{}^\sigma(\vec{p}), \quad
Q(\vec{x};\vec{p})\equiv  \frac{e^{i\vec{p}\cdot \vec{x}}}{(2\pi)^{3/2}},
\end{equation}
with  momentum-dependent components 
\begin{subequations}\label{eq:Q cov}
	\begin{align}	 
	Q_{ij}{}^{L} &=2 \delta_{ij},\\
	Q_{ij}{}^{T} &=2 \left(\frac{1}{3}\delta_{ij}-\frac{p_i p_j}{p^2}\right),
	\\	
	Q_{ij}{}^{\pm1} &=-i\left(\frac{p _i}{p}\hat{\epsilon}^\pm_j+\frac{p_j}{p}\hat{\epsilon}^\pm_i\right),\\
	Q_{ij}{}^{\pm 2} &= 2\hat{\epsilon}^\pm_i\hat{\epsilon}^\pm_j.
	\end{align}
\end{subequations}
Here, $\hat{\epsilon}^\pm (\vec{p})$ are two orthonormal transverse vectors with\footnote{These vectors can be taken to be $\hat{\epsilon}^\pm=R(\hat{p})\frac{1}{\sqrt{2}}(\hat{e}_x\pm i \hat{e}_y)$, where $R(\hat{p})$ is a standard rotation mapping the $z$ axis to the $\hat{p}$ direction.}
\begin{subequations}
	\begin{align}
	&\vec{p}\cdot \hat{\epsilon}^\pm=0,\\
	&\vec{p} \times \hat\epsilon^\pm =\mp \, i \,  p \,  \hat{\epsilon}^\pm.
	\end{align}
\end{subequations}
Note that the polarization vectors are complex, and that $(\hat{\epsilon}^\pm)^*=\hat{\epsilon}^\mp$. Hence, it follows that $(\hat{\epsilon}^\pm)^*\cdot  \hat{\epsilon}^\pm=\hat{\epsilon}^\mp\cdot  \hat{\epsilon}^\pm=1$, but $\hat{\epsilon}^\pm\cdot  \hat{\epsilon}^\pm=(\hat{\epsilon}^\mp)^* \cdot \hat{\epsilon}^\pm=0$. The fields $\delta h_\sigma(\vec{p})$ and $\delta\varphi(\vec{p})$ are eigenvectors of spatial translations by $\vec{a}$ with eigenvalues $\exp(-i \vec{p}\cdot \vec{a})$, and spatial rotations by an angle $\theta$ around the $\vec{p}$ axis with eigenvalues $\exp(-i m \theta)$, where $m=0$ for $\delta\varphi, \delta h_L,\delta h_T$ (scalars), $m=\pm1$ for $\delta h_\pm$ (vectors) and $m=\pm 2$ for $\delta h_{\pm\pm}$ (tensors).

We similarly decompose the canonical momenta in irreducible representations, 
\begin{equation}
\delta \pi^{ij}(\eta,\vec{x})=\sum_{\sigma} \int d^3 p \, \tilde{Q}^{ij}{}_\sigma (\vec{x};\vec{p})
\,  \delta\pi^\sigma(\eta,\vec{p}), \quad
\delta\pi^\varphi(\eta,\vec{x})=\int d^3 p\,  \tilde{Q}(\vec{x};\vec{p})\delta\pi^\varphi(\eta,\vec{p}),
\end{equation}
where this time the projection tensors are plane waves of opposite momentum,
\begin{equation}
\tilde{Q}^{ij}{}_\sigma(\vec{x};\vec{p})\equiv \frac{e^{-i\vec{p}\cdot \vec{x}}}{(2\pi)^{3/2}} \tilde{Q}^{ij}{}_\sigma(\vec{p}),\quad
\tilde{Q}(\vec{x};\vec{p})\equiv \frac{e^{-i\vec{p}\cdot \vec{x}}}{(2\pi)^{3/2}} ,
\end{equation}
with components given by
\begin{subequations}\label{eq:Q con}
	\begin{align}
	\tilde{Q}^{ij}{}_L &=\frac{1}{6} 
	\delta^{ij},\\
	\tilde{Q}^{ij}{}_T &=\frac{3}{4}
	\left(\frac{1}{3}\delta^{ij}-\frac{p^i p^j}{p^2}\right),\\
	\tilde{Q}^{ij}{}_{\pm1} &=\frac{i}{2} \left(\frac{p^i}{p}\hat{\epsilon}_\mp^j+\frac{p^j}{p}\hat{\epsilon}_\mp^i\right),\\
	\tilde{Q}^{ij}{}_{\pm2} &=\frac{1}{2}\hat{\epsilon}_\mp^i\hat{\epsilon}_\mp^j.
	\end{align}
\end{subequations}
In these expressions vector and tensor indices are raised with the Euclidean metric $\delta^{ij}$.  

The projection operators (\ref{eq:Q cov}) and (\ref{eq:Q con})  satisfy the completeness relation
\begin{equation}\label{eq:completeness}
\sum_{ij}\int d^3x\,  \tilde{Q}^{ij}{}_{\sigma_1}(\vec{x};\vec{p}_1) Q_{ij}{}^{\sigma_2}(\vec{x};\vec{p}_2)=\delta_{\sigma_1}{}^{\sigma_2} \,\delta^{(3)}(\vec{p}_1-\vec{p}_2), 
\end{equation}
which guarantee that the transition to the variables in the helicity representation is a canonical transformation. Because by definition  the field $\delta\pi^\sigma (\vec{p})$ is canonically conjugate to $\delta h_\sigma (\vec{p})$, momentum conservation demands that the field $\delta\pi^\sigma (\vec{p})$ carry the opposite momentum and helicity as $\delta h_\sigma (\vec{p})$.

Given arbitrary metric and scalar perturbations $\delta h_{ij}(\vec{x})$ and $\delta\varphi(\vec{x})$ it is straight-forward to find their components in the basis of tensors above. Because of the completeness relations (\ref{eq:completeness}), we have that
\begin{equation}
\delta h_\sigma(\eta,\vec{p})=\int d^3 x \, 
\delta h_{ij}(\eta,\vec{x})\tilde{Q}^{ij}{}_\sigma(\vec{x};\vec{p}) ,
\quad
\varphi(\eta,\vec{p})=\int d^3 x\,  \delta\varphi(\eta,\vec{x})\tilde{Q}(\vec{x};\vec{p}),
\end{equation}
and, similarly,
\begin{equation}
\delta \pi^\sigma(\eta,\vec{p})=\int d^3 x \, \delta \pi^{ij}(\eta,\vec{x}) Q_{ij}{}^\sigma(\vec{x};\vec{p}) ,
\quad
\delta\pi^\varphi(\eta,\vec{p})=\int d^3 x\,  \delta\pi^\varphi(\eta,\vec{x})Q(\vec{x};\vec{p}).
\end{equation}
It is also convenient to work with the irreducible components of the spatial vectors $\delta\eta^i$ and $\delta\lambda^i$, and those of their conjugate momenta $\delta\mathcal{P}_i$  and $\delta b_i$  We thus write for instance
\begin{align}
\delta\eta^i(\eta,\vec{x})&=\int  d^3p \,
Q^i{}_\sigma(\vec{x};\vec{p})\, \delta\eta^{\sigma}(\eta,\vec{p}),
&\delta\eta^{\sigma}(\eta,\vec{p})&=\int  d^3x \,
\delta\eta^i(\eta,\vec{x})
\tilde{Q}_i{}^\sigma(\vec{x};\vec{p}),
\\
\delta\mathcal{P}_i(\eta,\vec{x})&=\int  d^3p \,
\tilde{Q}_i{}^\sigma(\vec{x};\vec{p})\, \delta\mathcal{P}_{\sigma}(\eta,\vec{p}),
&\delta\mathcal{P}_{\sigma}(\eta,\vec{p})&=\int  d^3x \,
\delta\mathcal{P}_i(\eta,\vec{x})
Q^i{}_\sigma(\vec{x};\vec{p}),	
\end{align}
where the components of these tensors are
\begin{subequations}
	\begin{align}
	Q^i{}_L(\vec{x};\vec{p})&=-\frac{i p^i}{p}\frac{e^{i\vec{p}\cdot \vec{x}}}{(2\pi)^{3/2}}, \quad
	\tilde{Q}_i{}^L(\vec{p};\vec{x})=\frac{i p_i}{p}\frac{e^{-i\vec{p}\cdot \vec{x}}}{(2\pi)^{3/2}},\\
	Q^i{}_\pm(\vec{x};\vec{p})&=\epsilon^i_{\pm}(\vec{p})\frac{e^{i\vec{p}\cdot \vec{x}}}{(2\pi)^{3/2}}, \quad
	\tilde{Q}_i{}^\pm{}(\vec{p};\vec{x})=\epsilon_i^{\mp}(\vec{p})\frac{e^{-i\vec{p}\cdot \vec{x}}}{(2\pi)^{3/2}}.
	\end{align}
\end{subequations}
We define the Fourier components of the scalars $\delta\eta^N$ and $\delta\lambda^N$, and their conjugate momenta  $\delta\mathcal{P}_N$  and $\delta b_N$ like those of $\delta\varphi$ and $\delta\pi^\varphi$. The projected linearized constraints $G_\sigma$  are defined like their ghost counterparts $\delta\mathcal{P}_\sigma$. The BRST charge $\delta\Omega^{(2)}$ is a spatial scalar, so it does not change the transformation properties of the perturbations. 

\subsubsection{Scalar Inverse Relations}

In equations (\ref{eq:new scalar variables}) we introduced a set of variables in which the Hamiltonian simplifies considerably. In  some cases it may be convenient to return to the original variables with the help of the inverse transformations
\begin{subequations}\label{eq:inverse relations}
	\begin{align}
	\delta h_L &=a^2\zeta+\frac{a^2 p}{3}x^L+a \mathcal{H} x^N,\\
	\delta h_T &=-a^2 p\, x^L,\\
	\delta\varphi &=\frac{\dot{\bar\varphi}}{a}x^N,\\
	\delta\pi_L&=\frac{2M^2 (p^2-3\mathcal{H}^2)}{\mathcal{H}}\zeta-2M^2 p \,\mathcal{H}\, x^L+\frac{3\dot{\bar\varphi}^2+2M^2(p^2-3\mathcal{H}^2)}{a}x^N+\frac{\Pi}{a^2},\\
	\delta\pi_T &=\frac{2M^2 p^2}{3\mathcal{H}}\zeta-\frac{8 M^2 p\, \mathcal{H}}{3}x^L+\frac{2M^2 p^2}{3a}x^N+\frac{\Pi}{3a^2}-\frac{G_L}{a^2 p},\\
	\delta\pi_\varphi&=3a^2 \dot{\bar\varphi}\,\zeta+a^2 p \,\dot{\bar\varphi}\,x^L-a^3 \bar{V}_{,\varphi}\, x^N-\frac{\mathcal{H}}{\dot{\bar\varphi}}\Pi+\frac{a}{\bar{\dot\varphi}}G_N.
	\end{align}
\end{subequations}
We have suppressed here the momentum arguments for simplicity. The latter can be restored by noting that conjugate momenta have momenta opposite to those of their conjugates.

\subsubsection{Diagonalization of the Scalar Hamiltonian}
\label{sec:Diagonalization of the Scalar Hamiltonian}

By an appropriate canonical transformation it is possible to further simplify the structure of the Hamiltonian (\ref{eq:Hs}). Rather than performing such a canonical transformation at once, it is more convenient to carry out a chain of transformations, each one chosen to eliminate a targeted set of non-diagonal terms. The first canonical transformation  eliminates the mixed term $x^NG_L$ in the Hamiltonian. 

\begin{subequations}\label{eq:canonical 1}
	\begin{align}
	x^L&\to x^L-\alpha_G\,x^N+\beta_G\, G_N,
	\\
	x^N&\to x^N+\beta_G\,G_L,
	\\
	G_N&\to G_N+\alpha_G\, G_L,
	\end{align} 
\end{subequations}
where 
\begin{align}
\alpha_G&=\int ^t \frac{p}{a}\,
d\tilde{t},
\\
\beta_G&=\int^t \frac{\alpha_G}{\dot{\bar{\varphi}}^2}  \, d\tilde{t}.
\end{align}
Note that in order to find the transformed Hamiltonian it is not necessary to calculate the generating function. Instead, one simply substitutes the transformation (\ref{eq:canonical 1}) into the action and reads off the transformed Hamiltonian. 

In order to remove the coupling between  $\Pi$ and $G_N$, while keeping the term we just eliminated absent, we redefine
\begin{subequations}\label{eq:canonical 2}
	\begin{align}
	\zeta&\to\zeta+\alpha_N\, G_N, \\
	x^L & \to x^L+\beta_N \, G_N, \\
	x^N &\to x^N-\gamma_N\,  \zeta+ \alpha_N \, \Pi+\beta_N \,G_L, \\
	\Pi&\to\Pi +\gamma_N \,G_N,
	\end{align}
	where $\alpha_N$, $\beta_N$  and $\gamma_N$ satisfy
	\begin{align}
	\ddot{\alpha}_N &=
	\left(4H+\frac{2a^2 \bar{V}_{,\varphi}}{\dot{\bar{\varphi}}}-\frac{\dot{\bar{\varphi}}^2}{M^2 \mathcal{H}}\right)
	\dot\alpha_N
	-p^2 \alpha_N
	-\frac{1}{2M^2a}, \label{eq:Dalpha2}
	\\
	\beta_N&=-\int^t \left(
	\frac{1}{2M^2 p\,  a \mathcal{H}}+\frac{\alpha_G}{\dot{\bar{\varphi}}^2}
	+\frac{p\,\alpha_N}{\mathcal{H}}
	+\frac{\dot{\bar{\varphi}}^2\,\dot{\alpha}_N}{2M^2 p \,\mathcal{H}^2}+\frac{a\, \alpha_G\,\dot{\alpha}_N}{\mathcal{H}}\right) d\tilde t,
	\\
	\gamma_N&=\frac{a}{\mathcal{H}^2}\left(\mathcal{H}+a\, \dot{\bar{\varphi}}^2 \dot{\alpha}_N\right).
	\end{align}
\end{subequations}

Finally, to get rid of the mixing between the $\zeta$ and $x^L$ sectors we introduce
\begin{subequations}\label{eq:canonical 3}
	\begin{align}
	\zeta&\to\zeta+\beta_L \, G_L, \\
	x^L &\to x^L-\alpha_L \, \zeta+ \beta_L \, \Pi+\gamma_L\,G_L, \\
	\Pi&\to\Pi +\alpha_L \,G_L,
	\end{align}
\end{subequations}
where the functions $\alpha_L, \beta_L$ and $\gamma_L$ obey
\begin{align}
\ddot{\alpha}_L&=\left(\frac{\dot{\bar{\varphi}}^2}{M^2 \mathcal{H}}-4\mathcal{H}-\frac{2a^2 \bar{V}_{,\varphi}}{\dot{\bar{\varphi}}}\right)
\dot\alpha_L
-p^2\alpha_L
+\frac{p^2 a\,\alpha_G}{\mathcal{H}}
+3p+\frac{2p\, a^2\, \bar{V}_{,\varphi}}{\dot{\bar\varphi}\mathcal{H}},
\\
\beta_L&=\frac{\mathcal{H} \left(p-\mathcal{H}\dot\alpha_L\right)}{p^2\, a^2\, \dot{\bar\varphi}^2},
\\
\gamma_L&=-p \int^t \frac{\beta_G}{a} d\tilde{t}.
\end{align}
The end-result of the successive canonical  transformations (\ref{eq:canonical 1}), (\ref{eq:canonical 2}) and (\ref{eq:canonical 3})  is a Hamiltonian that in the final variables, denoted by a tilde, has the form of equation (\ref{eq:Hs diagonal}), where the time-dependent coefficients $\bar{A}_L$ and $\bar{A}_N$ are
\begin{equation}\label{eq:As}
\bar{A}_L= \frac{(3-p\, \alpha_L)\dot{\bar\varphi}^2-2M^2 p\, (p+p\, a\, \mathcal{H}\, \alpha_G\,  \alpha_L-\mathcal{H}\,\dot\alpha_L)}{4M^2 p^2 a^2 \dot{\bar\varphi}^2},
\quad
\bar{A}_N=-\frac{a\, \dot\alpha_N}{2\mathcal{H}}.
\end{equation}
To bring the Hamiltonian to this form, we need to solve essentially two decoupled linear inhomogeneous second order differential equations, and a set of first integrals. During power-law inflation, the differential equations admit closed solutions in terms of Bessel functions.
\chapter{Effective Field Theory Methods for Preheating}
\label{chp:EFTpreheating}
\section{Introduction: A general look at cosmological perturbations}
\label{sec:intropertsforeft}
Our conventional way to study cosmological perturbations so far have been to start with an action for the full fields
\be S=\int d^4x \sqrt{-g}\left[\frac{1}{2}m^2_{pl}R-\frac{1}{2}G^{ab}(\psi)g^{\mu\nu}\partial_\mu\psi_a\partial_\nu\psi_b+V(\psi)\right]\ee
after which the perturbations can be introduced at linear order in the fields
\be g_{\mu\nu}=\bar{g}_{\mu\nu}+\delta g_{\mu\nu}~~\psi=\psi_0+\delta\psi.\ee
A complementary way is to develop an action for the perturbations from the start. In constructing this action, understanding which symmetries the background possess and which ones it doesn't, is a useful guide. From inflation to dark energy, the cosmological eras all have a dynamical background that posses some time dependence. If a given background has temporal or spatial dependence, and does not remain invariant under some of the diffeomorphisms, it means there is a preferred temporal or spatial slicing in which the interactions for the perturbations will have a simple form. The guide then, is to construct the action in the gauge that is aligned with the time or spatial dependence of the background, and include interactions in accordance with the remaining diffeomorphisms that are to be respected. This is a noncovariant formulation where either a temporal or spatial diffeomorphism has been fixed from the start. Once the action in the preferred gauge has been constructed further diffeomorphisms along the direction where the background doesn't remain invariant will involve a scalar degree of freedom that transforms nonlinearly. For example, if the background has time dependence, in a gauge where time is fixed, further time diffeomorphisms will involve a scalar degree of freedom that nonlinearly transforms so as to guarantee that the action on the whole respects all diffeomorphisms. We will refer to this way of constructing an action for the perturbations as the effective field theory formalism for cosmological perturbations.

This effective field theory formalism of perturbations, which was first developed for single field inflation is applicable to all eras individually. In the general convention, we are  working with time dependent backgrounds. What sets the specifics of the era of interest is the strength of the time dependence and therefore any further scales and hierarchies it may involve, and the number and types of perturbations of interest. There will always be a scalar perturbation that nonlinearly transforms under time diffeomorphisms. This is the perturbation to the field who dominates the energy density and whose time dependence sets a preferred time direction. Notice that if the dominant source of energy density at a given era had a spatially dependent background, then we would be talking about a preferred slicing for space. Then the perturbation associated to this dominant field would transform nonlinearly under spatial diffeomorphisms, and of course the interactions would be built on linearly realized time diffeomorphisms. An example to this later case is the Solid Inflation \cite{Endlich:2012pz}.

In this chapter our focus will be the end of inflation. We would like to understand how the mechanisms with which the inflaton can transfer its energy to another scalar field $\chi$ can be generalized. The first thing that comes to mind would be for the inflaton to perturbatively decay into fields lighter than itself \cite{Abbott:1982hn}. Being light, these decay products would act like radiation, and lead to a radiation dominated era. 
In \cite{Kofman:1997yn} it has been pointed out that although these types of decays are important to fully reheat the universe they alone cannot be sufficient and infact, the end of inflation has convenient properties for the inflaton to excite exponential growth in perturbations of another scalar through resonance. This resonant particle production would take place in the first stages of reheating, right at the end of inflation and has thus been named \emph{``preheating''}. The canonical example of preheating is the $g^2\phi^2\chi^2$ interaction. At the end of inflation, $\phi_0(t)$ exhibits oscillatory behavior. Therefore $\chi$'s coupling to the background $\phi_0$, gives it a time dependent effective mass which exhibits resonant instabilities for certain modes. Here we wish to explore how interactions which lead to resonance can be generalized and captured in an EFT point of view.    

\section{Background evolution from Inflation to Preheating}
\label{sec:background evolution}
Let us begin this section by considering a canonical scalar field, minimally coupled to gravity with the action
\be S=\int d^4 x \sqrt{-g}\left[m^2_{Pl}R-\frac{1}{2}g^{\mu\nu}\partial_\mu\phi\partial_\nu\phi-V(\phi)\right].\ee
To be in accordance with spatial homogeneity and isotropy, at the background level we demand that the scalar field can depend only on time $\phi_0(t)$, and the metric to be that of FRWL universe $ds^2=-dt^2+a^2(t)dx^2$. With these considerations the evolution of the metric and the scalar field are governed by
\be 3m^2_{Pl}H^2=\frac{1}{2}\dot{\phi}_0+V(\phi_0),\ee
\be \ddot{\phi}_0+3H\dot{\phi}_0+V'(\phi_0)=0\ee
respectively. From the energy momentum tensor $T_{\mu\nu}=-\frac{2}{\sqrt{-g}}\frac{\delta S}{\delta g^{\mu\nu}}$ the energy density and pressure for the scalar field can be found as
\be \label{eqngen:rho} \bar{\rho}\equiv T_{00}=\frac{1}{2}\dot{\phi}_0+V(\phi_0)=3m^2_{Pl}H^2,\ee
\be \label{eqngen:p} \bar{p}\equiv T_{ii}=\frac{1}{2}\dot{\phi}_0-V(\phi_0)=-m^2_{Pl}\left(3H^2+2\dot{H}\right)\ee
where in the last terms we made use of the Friedmann equations. Now we are ready to describe different eras as driven by a scalar field with a different equation of state $\omega_\phi\equiv \frac{\bar{p}}{\bar{\rho}}$.

One can achieve an era of accelerated expansion from the equation of state $\bar{p}\sim -\bar{\rho}$. The ratio of equations \eqref{eqngen:p} and \eqref{eqngen:rho} satisfies this for $ \dot{\phi}_0^2 \ll V(\phi_0)$. Notice that this condition is equivalent to 
\be \label{epsilon} \epsilon\equiv -\frac{\dot{H}}{H^2}\ll 1.\ee
The fact that the potential energy dominates over the kinetic energy of the scalar field means that the time dependence of the scalar field is weak. This in return implies that the time dependence of the Hubble parameter is also weak and the potential energy can be approximated as
\be V\simeq 3m^2_{Pl}H^2\simeq\text{constant}.\ee
The variable $\epsilon$ is defined to parametrize the strength of the time dependence of the background in terms of the Hubble parameter. This system has an attractor solution if there is one more hierarchy between scales such that, the Hubble friction dominates over the kinetic energy of the scalar field. This is the condition that
\be \label{eta} \eta\equiv \frac{\dot{\epsilon}}{\epsilon H}\ll 1.\ee
In terms of Hubble parameter it means $\ddot{H}\ll H\dot{H}$, and in terms of the scalar field $\ddot{\phi}_0\ll -H\dot{\phi}_0$. Either way it leads to the attractor solution \footnote{Note that with $d\phi_0=\dot{\phi}_0dt$, $V'=\frac{m_{Pl}}{\sqrt{-2\dot{H}}}\left(6H\dot{H}+\ddot{H}\right)$.} 
\be V'=-3H\sqrt{-2m^2_{Pl}\dot{H}}=-3H\dot{\phi}_0.\ee
Finally, from $Hdt=d(lna)$ with $H\simeq constant$, the solution for the scale factor is 
\be a(t)\simeq e^{Ht}.\ee
This is the behavior of the scale factor for de Sitter universe. What we demonstrated here, a weakly time dependent scalar field with equation of state $\omega_\phi=-1$, is the simplest case to realize inflation. One scalar field slowly rolling along an almost constant potential is enough to give a positive cosmological constant term. The key point we want to emphasize from this example is that inflation corresponds to an era of weak time dependence for the background.

In the strict case of $\phi_0=const$ and hence $H=const$, the spacetime at hand corresponds to de Sitter. Exact de Sitter poses time translation invariance symmetry and gives rise to scale invariant perturbations. The weak time dependence of the background $\phi_0(t)$ and hence $H(t)$, gives rise to quasi scale invariant perturbations. It is the quasiscale invariant perturbations that match the observations. A spacetime with weak time dependence can be approximated by de Sitter universe. This is the behavior during inflation and dark energy dominated eras \footnote{One main difference between the two is that during inflation the scalar field that drives inflation is the only matter source where as during dark energy the universe contains other matters fields which have important consequences in observations. This is also why inflationary observations are different from dark energy related observations \cite{Linder:2015rcz}.}. However one must be aware that exact de Sitter background posses no time dependence, where as the weak time dependence in these eras give rise to scalar perturbations, as we mentioned in section \ref{sec:cosmological perts}. 
In a sense the background sets a preferred choice for the time direction because of its time dependent amplitude. The scalar perturbation arises to keep track of the difference in this choice. It transforms nonlinearly under time translations so as to insure the invariance of the overall action. Hence the existence of a  specific time direction imposed by the background can be understood as a spontaneous breaking of time diffeomorphism and the scalar perturbation is referred to as the Goldstone boson associated with this spontanously broken time symmetry \cite{Creminelli:2006xe}. 

With this point of view, it should be natural to expect scalar perturbations in different eras as long as one starts with a, time dependent background field to describe the era. The overall behavior of a given era, such as the accelerated expansion we considered for inflation and dark energy, will determine the applicable symmetries that effect the physics of the scalar perturbation. This idea sets the essence of effective field theory for cosmological perturbations and in this chapter we will examine how the behavior of scalar perturbations change from era to era.

As $\epsilon,\eta \to 1$ inflation ends leading to strong time dependence in the era that follows, namely preheating. The end of inflation means, the friction term looses its dominance over the kinetic term. We will consider the potential to be $V=\frac{1}{2}m_\phi^2\phi_0^2$. This approximation is justified by being the leading term in the Taylor expansion of any potential for a small enough $\phi_0$ that stabilizes its potential. And we can always redefine the overall amplitude of the field so that at the minimum of the potential $V(\phi_0)=0$. To gain some intuition, if we could completely neglect the friction term, the equation of motion for the background field would be $ \ddot{\phi_0}+m_\phi^2\phi_0=0$
which has oscillatory solutions $\phi_0(t)\sim\Phi sin(m_\phi t)$. Below we will see that the effect of the friction term is to give a decaying time dependence to the amplitude.  Being homogeneous we can approximate the time derivative for the canonical inflaton field as $\dot{\phi}_0\sim m_\phi\phi_0$\footnote{The more precise expression is $\dot{\phi}=\omega\phi$, but since the homogenous background has spatial dependence only the mass enters the frequency in the case of $\phi_0$.}, with which the friction term dropping below the kinetic term $H\dot{\phi}_0\ll\ddot{\phi}_0$ implies $H\ll m_\phi$.

To summarize while inflation is an era of weak time dependence and where $H_{I}\gg m_\phi$, preheating is an era where the inflaton oscillates at the minimum of its potential with $H_{p}\ll m_\phi$. In this era $H_p\sim 1/t$, which makes the hierarchy $H_p\ll m_\phi$ imply that we are focusing on times $1\ll m_\phi t$.

During preheating, the inflaton is expected to transfer its energy to other sectors through its oscillations via couplings which were negligible during inflation but are important now. For our purposes we will consider the inflaton to be coupled to a single scalar field, and denote it by $\chi$, as the reheating sector. Through this energy transfer the universe reheats. Our main interest is in the initial stages of this process during which the inflaton still dominates the overall energy density of the universe and hence the time dependent amplitude of the inflaton background still sets the time direction. To study the details of this background we consider 
\begin{subequations}
	\begin{align} \label{h2}6m^2_{Pl}H^2=\dot{\phi}_0^2+m_\phi^2\phi_0^2,\\
	\label{hdot}2m^2_{Pl}\dot{H}=-\dot{\phi}^2_0.
	\end{align}
\end{subequations}
In writing these equations we have neglected the background of the second field, $\chi_0$ completely. This is justified only during the initial stages and implies that we will introduce the second field at the level of perturbations. 

In solving this system we will follow \cite{Mukhanov:2005sc} and switch from the variable $\phi_0$ to $\theta$ defined as
\begin{subequations}
	\begin{align}
	\label{theta1}\phi_0=\sqrt{6}m_{Pl}\frac{H}{m_\phi}sin\theta,\\
	\label{theta2}\dot{\phi}_0=\sqrt{6}m_{Pl}Hcos\theta.
	\end{align}
\end{subequations}
This definition automatically satisfies \eqref{h2} and gives
\be \dot{H}=-3H^2cos^2\theta.\ee 
As an internal consistency the derivative of \eqref{theta1} should give \eqref{theta2}. This condition leads to
\be \dot{\theta}=m_\phi+\frac{3}{2}Hsin(2\theta).\ee
For the era we are considering $H\ll m_\phi$ and hence $\theta\simeq m_\phi t+\Delta$. We use this approximation in \eqref{hdot} 
\be -\int^{H(t)}_{H_{end}}\frac{dH}{H^2}=3\int^{t}_{t_{end}}cos^2(m_\phi t'+\Delta)dt',\ee
to solve for H. The end of inflation occurs when 
\be \epsilon(t_{end})=\frac{-\dot{H}_{end}}{H^2_{end}}=1.\ee 
From equations \eqref{h2} and \eqref{hdot} with $\phi_0(t_{end})\sim m_{Pl}$, this means $H_{end}\simeq \frac{m_\phi}{2}$. The solution for H(t) reads as
\begin{align} \nn H_p&=\frac{2}{3t}\left[1+\frac{sin(2mt+2\Delta)}{2mt}\right]^{-1}\\
&\simeq H_m\left[1-\frac{3H_m}{4m}sin(2mt+2\Delta)+\frac{9}{16}\left(\frac{H_m}{m}\right)^2sin^2(2mt)+....\right]\end{align}
where $\alpha \equiv\frac{sin(2m_\phi t+2\Delta)}{2m_\phi t}$ is small at times $1\ll m_\phi t$ and hence we can consider a series expansion around $\alpha=0$. So the end of inflation represents a matter dominated era with oscillatory corrections.\footnote{The presence of these oscillatory corrections is what makes an era dominated by oscillating scalar field different then an era of dust which exactly behaves as $H_{dust}=H_m$ with zero pressure.} From \eqref{theta1} we see that this gives the following behaviour for the inflaton
\be\label{phibackg}
\phi_0(t)\simeq \sqrt{6}M_{pl}\frac{H_m}{m_\phi}\left[sin(m_\phi t)+\frac{3}{8}\frac{H_m}{m_\phi}(cos3m_\phi t-cosm_\phi t)+...\right].\ee
The behavior of the inflaton perturbations on this background and the duration of this oscillatory era have been studied to understand the end of single field inflation with canonical kinetic term and minimal coupling to gravity \cite{Easther:2010mr,Jedamzik:2010dq}. The canonical example of preheating \cite{Kofman:1997yn} considers only the zeroth order terms in this background 
\begin{subequations}
	\begin{align}
	H_{pc}&=H_m,\\
	\phi^{pc}_0(t)&=\sqrt{6}m_{Pl}\frac{H_m}{m_\phi}sin(m_\phi t+\Delta)
	\end{align} 
\end{subequations}
and involves the coupling $g^2\phi^2\chi^2$.  

Here we will address mechanisms with which the inflaton transfers its energy to a secondary field, and the general behavior of perturbations given this background H(t) with the oscillatory corrections to first order in $\frac{H_m}{m_\phi}$. Hence we want to study the dynamics of two scalar fields on this background and we want to be able to consider interactions more general then the simplest couplings. To achieve these goals, we question how the effective field theory of quasi single field inflation for perturbations can be reinterpreted to address preheating and use this systematic approach to point out the hierarchies in the scales important for various processes during preheating. We will conclude by discussing the short comings in developing an effective field theory for the background fields for this era.    

\section{Symmetries and Background Scales from Inflation to Preheating} 
\label{sec:symmetries}
Now that we have established inflation as an era of weak time dependence and preheating as one of strong time dependence, we can work towards understanding them more generally beyond the simplest examples we considered.

\subsection{Inflation and discrete symmetry breaking}
\label{sec:inflationsym}
In the case we considered for inflation above, we obtained de Sitter universe with the scale factor $a(t)\sim e^{Ht}$. On one hand, the de Sitter universe with exact time translation invariance and a constant background realizes accelerated expansion. On the other hand, as mentioned previously, we observe quasi scale invariant inflationary perturbations which are expected to arise in a quasi de Sitter universe with a time dependent background. A time dependent background is necessary for inflation to end and hence is important. 

The hints we have are that inflation ended at some point and universe moved on to other eras, and we observe the effects of quasi de Sitter universe associated with scales of order $H_I$. We can claim that for time scales of the order $H^{-1}_I$, the spacetime that realizes inflation posses an approximate time translation invariance, behaves as a quasi de Sitter universe and gives rise to quasi scale invariant perturbations in accordance with the observations. Rather then the expectation that deep into inflation corresponds to an exact de Sitter universe, it could be that on shorter time scales it looks completely different then de Sitter and has broken time translations. In other words, the UV completion for inflation need not correspond to exact de Sitter as long as it looks quasi de Sitter when averaged on time scales of order $H_I$. With this point of view, the following form for the Hubble parameter has been been proposed \cite{Behbahani:2011it}  
\be \label{Hinf} H_I(t)=H_{sr}(t)+H_{osc}(t)sin(\omega t),\ee
to generalize slow roll to towards involving oscillations. As this resembles the background of preheating, it is good to review the properties of this proposal here.

This form for $H(t)$ needs to be restricted in terms of the scales it involves in order to lead to a slow roll evolution with sensible degrees of freedom. To have inflation the slow roll part must dominate over the oscillatory part and the time dependence of these parts should be small. These requirements bring out the conditions
\be \label{HIcond1} H_{osc}\ll H_{sr}~~\&~~ \epsilon=-\frac{\dot{H}}{H^2}\sim\frac{\dot{H}_{sr}}{H^2_{sr}}\sim\frac{\dot{H}_{osc}}{H^2_{osc}}\ll 1.\ee

Being periodic, $sin(\omega t)$ in the oscillatory part of $H(t)$, respects a discrete time translation invariance as $t\to t+\frac{2\pi}{\omega}$. This symmetry will be prominent for scales $\omega < E$, if the time dependence in $H_{sr}$ and $H_{osc}$ is also weak. Since the strength of time dependence in these terms is governed by $\epsilon$, this is the parameter that controls the level of breaking the discrete time symmetry.  At very low energies compared to the scale of oscillations $E\ll \omega$, because $sin(\omega t)$ will be oscillating so rapidly it will look constant, and this will imply an approximate continuous time translation invariance with a very small $\epsilon$.   

By requiring the cutoff related to the gravity induced interactions ($\sim m_{Pl}$) be grater then the cutoff related to the symmetry breaking induced by this background ($F\simeq\frac{\sqrt{-2\dot{H}m^2_{Pl}}}{\omega}$) it is noted that \cite{Behbahani:2011it} 
\be \alpha\equiv \frac{\omega}{H}\gg \sqrt{\epsilon}.\ee
Lastly one must make sure that the oscillatory part does not dominate $\dot{H}$ to avoid ghost degrees of freedom\footnote{As we will see in the next section the Goldstone boson associated with the time translations gets canonically normalized by a factor of $\sqrt{-2m^2_{Pl}\dot{H}}$.}. If one considers 
\be \frac{\dot{H}}{H^2}=\frac{\dot{H}_{sr}}{H^2}+\frac{\dot H_{osc}}{H^2}sin(\omega t)+\frac{\omega H_{osc}}{H^2}cos(\omega t)\ee 
by condition \eqref{HIcond1}  we have $H\sim H_{sr}$ making first term approximately $\epsilon$ and the coefficient of the second term here can be rewritten as $\frac{{H^2}_{osc}}{H^2}\frac{\dot{H}}{H^2}$, which is negligible. Hence the requirement that the slow roll part in $\dot{H}_I$ dominate over the oscillatory can be expressed as 
\be \label{nonghostcond} \frac{\omega H_{osc}}{H^2}=\frac{\alpha H_{osc}}{H}\le \epsilon.\ee
All these restrictions imply hierarchies on the scales involved,
\begin{align}
\nn \text{hierarchies on inflationary scales:}&\\ \omega\gg \sqrt{\epsilon} H_I,&~~H_{sr}\gg H_{osc},~~\sqrt{\epsilon}H^2_I\gg\omega H_{osc} \end{align}

\subsection{Preheating and its symmetries}
\label{sec:preheatingsym}             
We found the evolution of the background during preheating to be
\be \label{Hp} H_p=H_m(t)-\frac{3}{4}\frac{H^2_m(t)}{m_\phi}sin(2m_\phi t+\Delta).\ee
As this is also an era where the background has time dependence, the first property we can note is that time translation invariance is also broken during preheating. Equation \eqref{Hp} suggests that the evolution of the background right after inflation continues to follow a pattern similar to \eqref{Hinf}
\be H(t)=H_{FRW}(t)+H_{osc}P(\omega t)\ee
where $P(\omega t)$ can be any periodic function. Hence the breaking of time translation invariance during preheating follows a pattern similar to one that can also apply to inflation with $H_{FRW}=H_m$, $H_{osc}=-\frac{3}{4}\frac{H^2_m}{m_\phi}$ and $\omega=2m_\phi$.

Here we need the restriction 
\be \label{FRWvsosc} H_{osc}\ll H_{FRW}\ee
to achieve an FRW evolution on the whole instead of the oscillations dominating on the overall, which have been shown to lead to pathologies \cite{Easson:2016klq}. Our previous definition of preheating as an era where $H_p\sim H_m\ll m_\phi$ automatically fits this requirement and hence is justified.  Different then the case of inflation here the time dependence of the coefficients $H_{FRW}$ and $H_{osc}$ is not weak, but we do require their time dependence to be small compared to the scale of oscillations $\omega$
\be \label{omegavscoeff} \frac{\dot{H}}{\omega H}\sim\frac{\dot{H}_{FRW}}{\omega H_{FRW}}\sim \frac{\dot{H}_{osc}}{\omega H_{osc}}\ll 1.\ee

During preheating there will be no approximate time translation invariance at low energies $E\ll \omega$, because the $H_{FRW}$ part will always have strong time dependence. And at scales $\omega < E$, we will not see much the effects of the discrete symmetry because of the time dependence of the other parts. So during preheating time translation invariance is completely broken but there is the hierarchy $\frac{\omega_p}{H_p}\gg 1$ similar to the hierarchy $\frac{\omega_I}{H_I}\gg \sqrt{\epsilon}$ during inflation. To summarize
\begin{align}
\nn\text{hierarchies on preheating scales:}&\\
\omega\gg H_p\sim H_m,&~~H_{FRW}\gg H_{osc},~~\omega H_m\gg\dot{H}_m.
\end{align}

With $m_\phi\gg H_m$, the time derivatives of $H_p$ go as\footnote{The dominant terms in these derivatives are\\ $\dot{H}_p=-\frac{3}{2}\left(1+cos(2m_\phi t+\Delta)\right)H^2_m,$\\
	$\ddot{H}_p=3m_\phi H^2_msin(2m_\phi t+\Delta),$\\
	$\dddot{H}_p=6m_\phi^2H^2_mcos(2m_\phi t+\Delta).$\\
	$\ddddot{H}_p=-12m_\phi^3H^2_{m}sin(2m_\phi t).$
	Since we are focusing on times $m_\phi t\gg1$, the oscillations are frequent enough for us to approximate $\ddot{H}_p$ and $\dddot{H}_p$ by their amplitudes. In the main text we also neglect the overall numerical factors in these amplitudes. Our main objective in the next sections will be to emphasize how the frequency of the oscillations $\omega=2m_\phi$ becomes explicit in the scales of the problem.}
\be \dot{H}_p\sim -H^2_p,~~\ddot{H}_p\sim\omega H^2_p,~~\dddot{H}_p\sim \omega^2H^2_p\ee
In this section we focused on broken time translation invariance and mechanisms that give the details of this breaking. In the next section we will write down the interactions for perturbations in both of these eras based on the observation of broken time translation invariance. We will then focus on preheating alone and explore the implications of the hierarchies we discovered for various phenomena.

\section{Effective Field Theory Interactions for the Perturbations}
\label{sec:EFTfor perts}

The important observation from the above discussion is that the background during inflation and preheating posses time dependence. A time dependent background implies a preferred definition for the time coordinate. This definition is the gauge where the coordinates are set so that at each time slice lies a surface of constant $H$, which corresponds to a surface of  $\phi_0$ constant if we consider a scalar field to realize this background. In other words, the amplitude of $H(t)$ or $\phi_0(t)$ can set the definition for the time coordinate $t$. This means the background field $\phi_0$ itself will not be invariant if the time coordinate is redefined by a time diffeomorphism. Time translations are still a symmetry of the action for the field $\phi$ but they are not respected by the background $\phi_0$. In considering the physics of the perturbations around this time dependent background, the time diffeomorphisms are spontaneously broken by the background. Among the degrees of freedom involved, there exists a scalar perturbation which transforms nonlinearly under time diffeomorphisms. This transformation property for the scalar perturbation guarantees to keep the Lagrangian for the field $\phi$ invariant on the whole. The scalar degree of freedom is the Goldstone boson associated with the spontaneously broken time diffeomorphisms. and it is the theory of the scalar perturbation we are interested in. This idea has been brought to cosmology firstly by \cite{Creminelli:2006xe} in the context of violating the null energy condition. It has been extended with the purpose of setting a general framework for studying inflationary perturbations at the scale of horizon crossing in \cite{Cheung:2007st}.   

We can consider the theory for the scalar perturbation after the time diffeomorphisms have been fixed. While the field $\phi$ is a scalar under all diffeomorphisms, the scalar perturbation remains invariant only under spatial diffeomorphisms. Hence the Lagrangian for the scalar perturbation is restricted by spatial diffeomorphism. Under spatial diffeomorphisms, $g^{00}$ and any polynomials of it are scalars, the extrinsic curvature $K_{\mu\nu}$ of surfaces of constant time transforms as a tensor. With time diffeomorphisms fixed, functions of time are invariant under spatial diffeomorphisms. Hence the Lagrangian for the scalar perturbation is allowed to contain polynomials of $g^{00}$, scalars obtained by contracting the indices of $K_{\mu\nu}$, and functions of time $f(t)$. Moreover, if a spatial diffeomorphism is performed on the action for the scalar perturbation, the gradient $\partial_\mu\tilde{t}$ that will arise is simply $\delta^0_\mu$ since the time coordinate is preset. This means any tensor is allowed to enter the Lagrangian with an upper index $0$ \cite{Cheung:2007st}. 

\subsection{Effective Field Theory of Single Field Inflation and the Symmetry Breaking Scale}

Here we give a short review of \cite{Cheung:2007st}. Single field inflation contains a single scalar degree of freedom which is the Goldstone boson that nonlinearly transforms under time diffeomorphisms. In the gauge with the time coordinate $t$, where surfaces of constant $t$ are along the surfaces of constant $\phi_0(t)$, the scalar degree of freedom is contained within the metric. In the literature this gauge is referred to as the unitary gauge
\be \text{unitary gauge:}~~\phi(x)=\phi_0(t),~g_{ij}=a^2(t)\left[(1+2\zeta)\delta_{ij}+h_{ij}\right],\ee
in the following sections we will also refer to it as the $\zeta$-gauge. Since the scalar degree of freedom takes place in the metric, here it is the variable $\zeta$, the interactions have the least complicated form in this gauge. With the terms mentioned in the previous section the most general Lagrangian in unitary gauge is \cite{Cheung:2007st}
\begin{align} \label{unitaryinf} 
\nn  S=\int d^4x\sqrt{-g}&\Bigg[\frac{1}{2}m^2_{Pl}R+m^2_{Pl}\dot{H}g^{00}-m^2_{Pl}\left(3H^2+\dot{H}\right)+\frac{M_2(t)^4}{2!}\left(g^{00}+1\right)^2\\
\nn &+\frac{M_3(t)^4}{3!}\left(g^{00}+1\right)^3-\frac{\bar{M}_1(t)^3}{2}\left(g^{00}+1\right)\delta {K^\mu}_\mu\\
&-\frac{\bar{M}_2(t)^2}{2}(\delta{K^\mu}_\mu)^2-\frac{\bar{M}_3(t)^2}{2}\delta{K^\mu}_\nu\delta{K^\nu}{_\mu}+...\Bigg] \end{align}
where the coefficients of the linear terms are fixed by requiring that the FRW evolution is respected at zeroth order in perturbations. 

The scalar degree of freedom is made explicit by performing the time diffeomorphism 
\be \label{topigauge} t \to \tilde{t}=t+\xi^0(x),~~\vec{x}\to \vec{\tilde{x}}=\vec{x}.\ee
The Goldstone boson $\tilde{\pi}$, is introduced by promoting the diffeomorphism parameter $\xi^0$ to a field as
\be \xi^0(x(\tilde{x}))\to -\tilde{\pi}(\tilde{x}).\ee
In these new coordinates $\tilde{x}$ the action reads
\begin{align}
\nn  S=&\int d^4\tilde{x}\sqrt{\tilde{g}(\tilde{x})}\Bigg[\frac{1}{2}m^2_{Pl}\tilde{R}-m^2_{Pl}\left(3H^2(\tilde{t}+\tilde{\pi})+\dot{H}(\tilde{t}+\tilde{\pi})\right)\\
&+m^2_{Pl}\dot{H}(\tilde{t}+\tilde{\pi})\frac{\partial(\tilde{t}+\tilde{\pi})}{\partial\tilde{x}^\mu}\frac{\partial(\tilde{t}+\tilde{\pi})}{\partial\tilde{x}^\nu}{\tilde{g}}^{\mu\nu}(\tilde{x})+...\Bigg]. \end{align}
This action is invariant under diffeomorphisms provided that the Goldstone boson transforms as
\be \pi(x)\to \tilde{\pi}(\tilde{x})=\pi(x)-\xi^0(x).\ee
To summarize, the scalar degree of freedom can be made explicit by performing the diffeomorphism $t\to t+\pi$ to the unitary gauge, where $\pi$ is a field that transforms as a scalar field plus an additional shift, $\pi\to \tilde{\pi}=\pi-\xi^0$ under time diffeomorphisms $t\to \tilde{t}=t+\xi^0$. We will refer to this gauge as the $\pi$-gauge
\be \label{pigaugeI}\text{$\pi$-gauge:}~~ \phi(x)=\phi_0(t)+\delta\phi(x),~~g_{ij}=a^2(t)\left[\delta_{ij}+\tilde{h}_{ij}\right]\ee
and in this gauge the effective field theory action for single field inflation is

\begin{align} \label{pigaugeinf}
\nn S=&\int d^4 x \sqrt{-g}\Bigg[\frac{1}{2}m^2_{Pl}R-m^2_{Pl}\left(3H^2(t+\pi)+\dot{H}(t+\pi)\right)+\\
\nn &+m^2_{Pl}\dot{H}(t+\pi)\left((1+\dot{\pi})^2g^{00}+2(1+\dot{\pi})\partial_i\pi g^{0i}+g^{ij}\partial_i\pi\partial_j\pi\right)\\
&+\frac{m_2(t+\pi)^4}{2!}\left((1+\dot{\pi})^2g^{00}+2(1+\dot{\pi})\partial_i\pi g^{0i}+g^{ij}\partial_i\pi\partial_j\pi+1\right)^2+...\Bigg].\end{align}

Notice that here the EFT parameter $m_2$ introduces temporal and spatial derivaties of $\pi$ at different orders, therefore it gives rise to a sound speed for $\pi$. At second order in $\pi$ the leading derivative terms in the action are
\begin{align}
S^{(2)}_\pi\supset\int d^4x a^3\left(-m^2_{pl}\dot{H}(t)+2m^4_2(t)\right)\dot{\pi}^2+m^2_{pl}\dot{H}(t)g^{ij}\partial_i\pi\partial_j\pi
\end{align}
With the definition
\be\label{pic} \pi_c\equiv\sqrt{-2m^2_{pl}\dot{H}} \sqrt{1-\frac{m^4_2}{m^2_{pl}\dot{H}}}\pi=\sqrt{2m^2_{pl}|\dot{H}|}\frac{\pi}{c_\pi}\ee
the action takes its canonical form
\be\label{Spic2} S^{(2)}_\pi\supset-\frac{1}{2}\int d^4xa^3g^{\mu\nu}\partial_\mu\pi_c\partial_\nu\pi_c+\dots=\frac{1}{2}\int d^4xa^3\left[\dot{\pi}_c^2-\frac{c^2_\pi}{a^2}\left(\vec{\nabla}\pi_c\right)^2+\dots\right]\ee
where the sound speed for $\pi$ is defined to be 
\be c^{-2}_\pi=1-\frac{m^4_2}{m^2_{pl}\dot{H}}, \ee
to account for the coefficient that now accompanies the spatial derivatives. This sound speed arises because of the Lorentz invariance is broken via spontaneously broken time translation invariance. However the coordinates and Lagrangian density can be rescaled, $\vec{x}\to\tilde{\vec{x}}\equiv c^{-1}_\pi\vec{x}$, $\mL\to\tilde{\mL}=c^3_\pi\mL$, and the canonically normalized field redefined as $\tilde{\pi}_c^2\equiv-2m^2_{pl}\dot{H}c_\pi\pi^2$ to restore it.

The time diffeomorphisms parametrized by the coordinate dependent parameter $\xi^0(\vec{x},t)$ are local symmetries. However if these are broken, then so is the special case of time translations where the parameter $\xi^0=constant$. With a constant parameter the time translations are global symmetries. Global symmetries with parameter $\xi$ have associated Noether currents
\be J_\mu=\sum_{n}\frac{\partial\mL}{\partial(\partial_\mu F_n)}\frac{\delta F_n}{\delta \xi}\ee
which involve a sum over all fields $F_n$ \cite{Schwartz:2013pla}. This current is conserved even if the symmetry is spontaneously broken. In the case of spacetime diffeomorphisms the associated currents are components of the energy momentum tensor.\footnote{Notice that $T_{\mu\nu}$ is the variation of the action with respect to the metric. The field content we are looking at includes scalar modes which would intuitively be associated with scalar matter fields. But as we saw even these can be written as components of the metric, which becomes explicit in a convenient gauge choice. Hence all the field content is indeed the metric and the energy momentum tensor makes up the whole of the Noether current.} For time translations this is $J_\mu=T_{\mu0}$. 

There also exists an associated charge
\be \label{charge} Q\equiv\int d^3J^0\ee 
that is well defined at the scales where the symmetry holds. The integral \eqref{charge} over all space is well defined so long as the zeroth component of the current falls off as $x^{-3}$. Otherwise the integrand will blow up at the boundaries as $x\to\pm\infty$. So by examining $T_{00}$ in $\pi$-gauge we can determine the scale at which the charge will become ill defined. This sets the scale of spontaneous breaking of time translations and hence the scale below which $\pi$-description is applicable.
As we explicitly compute in section \ref{sec:adiabaticmodes} the leading order $\pi$ contribution comes from the terms
\be \label{T00current}  T_{00}\supset\left(-2m^2_{pl}\dot{H}(t+\pi)+2m^4_2(t+\pi)\right)\dot{\pi}.\ee
Reintroducing the Lorentz invariance also involves rescaling the energy momentum tensor $\tilde{T}_{00}=c^2_\pi T_{00}$. At first order in $\tilde{\pi}_c$ 
\be \tilde T_{00}\supset\sqrt{-2m^2_{pl}\dot{H}c_\pi}\dot{\tilde{\pi}}_c-\frac{1}{2}\frac{\ddot{H}}{\dot{H}}\sqrt{-2m^2_{pl}\dot{H}c_\pi}\tilde\pi_c+\frac{\dot{c}_\pi}{c_\pi}\sqrt{-2m^2_{pl}\dot{H}c_\pi}\tilde{\pi}_c.\ee
The second term here that involves $\ddot{H}$, takes into account the oscillatory behavior of the background with frequency $\omega=2m_\phi$. This term introduces a discrete shift symmetry because of its periodic nature. It is negligible during slow roll inflation where the symmetry breaking scale can be read off of the first term in the current 
\be \tilde{J}_\mu=-\Lambda_{sb}^2\tilde{\partial}_\mu\tilde{\pi}_c,\ee
as 
\be \Lambda^2_{sb}=\sqrt{-2m^2_{pl}\dot{H}c_\pi}.\ee
By definition $\pi$ has the same dimension as time, which gives the canonically normalized filed $\tilde{\pi}_c$ the dimension of mass. The $\dot{\tilde{\pi}}_c$ term goes as $x^{-2}$. It does not fall off fast enough at the boundary and signals the scale at which spontaneous breaking of continuous time translations take place \cite{Baumann:2011su}. Also note that with the definition of \eqref{pic}, all higher derivative terms in the Lagrangian come suppressed with the scale $\Lambda_{sb}$.

In section \ref{sec:inflationsym} we mentioned a discrete symmetry breaking pattern in the spirit of \cite{Behbahani:2011it} which also consider a background of the form
\be H(t)=H_{FRW}+H_{osc}(t)sin(\omega t)\ee
with $H_{FRW}<H_{osc}$. Lets go back to discuss the scale associated to this discrete breaking. If the time dependence of $H_{FRW}$ contribution is negligable, such backgrounds respect a discrete time translation invariance up to some scale, because of the presence of $sin(\omega t)$. In \cite{Behbahani:2011it} this is interpreted as a discrete shift symmetry on the Goldstone boson. On one hand, we know that with such a background $\ddot{H}\sim\omega H^2$ and hence further powers of $\pi$ can be invoked by Taylor expansion of the coefficients $\dot{H}(t+\pi)=\dot{H}+\ddot{H}\pi+\dots$. In the Lagrangian this gives rise to terms
\be \mL\supset -m^2_{pl}\dot{H}(t+\pi)\dot{\pi}^2=-m^2_{pl}\dot{H}(t)\dot{\pi}^2-m^2_{pl}\ddot{H}(t)\pi\dot{\pi}^2+\dots \ee
On the other hand, these terms that come from Taylor expansion, ie $\ddot{H}\pi$ do not respect the shift symmetry of $\pi$ and hence will not respect the discrete time translation invariance. Technically, these discrete symmetry breaking terms would be expected to arise suppressed by the scale associated to the breaking of the symmetry. That is in terms of canonically normalized fields, one expects that
\be -m^2_{pl}\ddot{H}(t)\pi\dot{\pi}^2~~~\text{corresponds to}~~~\frac{1}{\Lambda_{db}}\tilde{\pi}_c\dot{\tilde{\pi}}_c^2.\ee 
Ignoring time dependence of the sound speed, and noting that $\frac{\ddot{H}}{\dot{H}}\sim \omega$ the scale below which discrete time translation invariance is broken can be read off as
\be \Lambda_{db}=\frac{c_\pi}{\omega}\sqrt{-2m^2_{pl}\dot{H}c_\pi},\ee
which lies a bit below the breaking scale for  the continuous time translation symmetry. For inflation there is such a range of scales over which the discrete time translation symmetry is respected, because the $H_{FRW}=H_{sr}$ part has negligable time dependence and hence the characteristics of the oscillatory part in the background sets the characteristics of the symmetries involved. In the case of preheating, $H_{FRW}=H_m$ which already has time dependence strong enough that it will not let the discrete time translation invariance be respected at any scale. 

Just like $m_2(t)$, the effect of the extrinsic curvature terms
\be K_{ij}=-\frac{1}{\lambda^N}\left[\nabla_k\lambda_i+\nabla\lambda_k-\frac{\partial h_{ik}}{\partial t}\right],\ee
is also to modify the dispersion relation. As $h_{ik}$ is the traceless part of the metric tensor it does not contribute to ${K^\mu}_\mu$ term. Via involving $(\dot{h_{ik}})^2$ in a different amount then $\nabla{h_{ik}}$, 
\begin{align} \mL&\supset \frac{1}{2}\left(m^2_{pl}(R^{(3)}+K_{ij}K^{ij})-\bar{M}_3^2\delta{K^i}_j\delta{K^j}_i\right)\\
&\sim\frac{a^{-6}}{2}\left[\left(m^2_{pl}-\bar{M}_3^2\right)\left(\dot{h}_{ij}\right)^2-m^2_{pl}\left(\vec{\nabla}h_{ij}\right)^2\right] \end{align}
the presence of $\bar{M}_3$ introduces a sound speed for the tensor degrees of freedom associated with the dominant source of energy density, the inflationary gravitational waves in this case
\be c^2_g\equiv\left[ 1-\frac{\bar{M}_3^2}{m^2_{pl}}\right]^{-1}.\ee
This tensor sound speed for example has observable consequences in for dark energy physics with in this effective field theory frame work \cite{Linder:2015rcz}. In general these extrinsic curvature terms also modify the $k$ dependence in the dispersion relation for $\pi$. The effects of such terms have been well noted in again dark energy physics \cite{ArkaniHamed:2003uy}.

\subsection{Effective Field Theory of Preheating and the Particle Production Scale}
As we explored in section \ref{sec:symmetries} preheating, just like inflation, is an era with a time dependent background. Therefore we expect the theory for perturbations during preheating to be also formulated via spontaneously broken time diffeomorphisms. During this era the presence of secondary sources is important. The inflaton still dominates the energy density of the universe and sets the clock at the initial stages of this era, which is where we will mainly focus on. But it transfers its energy to the secondary sources, exciting their perturbations through resonance. Hence we expect these secondary fields to be introduced at the level of perturbations. 
We will later on briefly discuss how the preheating sector can start to dominate the overall energy density towards the end of preheating. For simplicity we consider it to be composed of a single scalar field. This means in the effective field theory of preheating we have two scalar degrees of freedom where one is the Goldstone boson, $\pi$ that nonlinearly realizes time diffeomorphisms and is associated with the inflationary sector, and the other is the perturbations in the preheating sector, which we will denote by $\chi$\footnote{When needed we denote the background for the reheating field as $\chi_0$.}.    

The effective field theory for the Goldstone boson and a second scalar field have been developed in \cite{Noumi:2012vr} for the purpose of studying effects of heavy fields during inflation. We expect the same action to capture preheating, with a different behavior of EFT parameters; since the symmetries satisfied, and the pattern of breaking time diffeomorphisms as studied in sections \ref{sec:inflationsym} and \ref{sec:preheatingsym} are the similar to those during inflation. This action in the unitary gauge where the inflaton perturbations are encoded in $g^{00}$ is
\begin{align}
\nn S_g&=\int d^4x\sqrt{-g}\Bigg[\frac{m^2_{Pl}}{2}R-m^2_{Pl}\left(3H^2(t)+\dot{H}(t)\right)+m^2_{Pl}\dot{H}(t)g^{00}+\frac{m^4_2(t)}{2}(\delta g^{00})^2\\
~~~&-\frac{\bar{M}^2_2(t)}{2}(\delta {K^\mu}_\mu)^2+...\Bigg]\\
S_\chi&=\int d^4 x \sqrt{-g}\left[-\frac{\alpha_1(t)}{2}\partial^\mu\chi\partial_\mu\chi+\frac{\alpha_2(t)}{2}\left(\partial^0\chi\right)^2-\frac{\alpha_3(t)}{2}\chi^2+\alpha_4(t)\chi\partial^0\chi\right]\\
S_{g\chi}&=\int d^4 x \sqrt{-g}\left[\beta_1(t)\delta g^{00}\chi+\beta_2(t)\delta g^{00}\partial^0\chi+\beta_3(t)\delta^0\chi-(\dot{\beta}_3(t)+3H(t)\beta_3(t))\chi\right]
\end{align}

Here $\alpha_2(t)$ has a role similar to $m_2(t)$ and gives sound speed to the preheating sector. The parameters $\alpha_3(t)$ and $\alpha_4$ encode how $\chi$ couples to the background. Therefore it captures all the canonical examples. The interactions between the two sectors become more apparent in the $\pi$-gauge and their strengths are encoded via the parameters $\alpha_i(t)$. The $\beta_i(t)$ parameters imply additional mixings between the two sectors. Here we will consider them as voluntary complications and mostly set them to zero by choice. Let us move on to consider the $\chi$ sector on its own and discuss the scales below which $\chi_k$ modes can be excited. 

The preheating field is canonically normalized as
\be \tilde{\chi}_c\equiv \sqrt{\alpha_1+\alpha_2}\chi a^{3/2}\ee
At second order in $\chi$, without the metric fluctuations, the action for the canonically normalized field is
\begin{align}
\nn &S_c=\frac{1}{2}\int d^4x\Bigg[\dot{\chi}_c^2-\frac{\alpha_1}{\alpha_1+\alpha_2}\frac{(\vec{\nabla}\chi_c)^2}{a^2}-\frac{\alpha_3-\dot{\alpha}_4(t)}{\alpha_1+\alpha_2}\chi^2_c+\\
&+\left[\frac{3}{2}\dot{H}+\frac{9}{4}H^2+\frac{3H}{2}\frac{\dot{\alpha}_1+\dot{\alpha}_2}{\alpha_1+\alpha_2}-\frac{3H\alpha_4}{\alpha_1+\alpha_2}+\frac{1}{2}\frac{\ddot{\alpha}_1+\ddot{\alpha}_2}{\alpha_1+\alpha_2}-\frac{1}{4}\left(\frac{\dot{\alpha}_1+\dot{\alpha}_2}{\alpha_1+\alpha_2}\right)^2\right]\chi^2_c\Bigg].\end{align}
We can read off the effective mass of the reheat field as
\be \label{mchi} m^2_\chi(t)=\frac{\alpha_3(t)-\dot{\alpha}_4(t)}{\alpha_1(t)+\alpha_2(t)}-\frac{1}{2}\frac{\ddot{\alpha}_1+\ddot{\alpha}_2}{\alpha_1+\alpha_2}+\frac{1}{4}\left(\frac{\dot{\alpha_1}+\dot{\alpha}_2}{\alpha_1+\alpha_2}\right)^2.\ee
Moreover the generality of the EFT couplings allow for the reheat field to have a sound speed as 
\be\label{cchi} c_\chi^2(t)=\frac{\alpha_1(t)}{\alpha_1(t)+\alpha_2(t)},\ee
which lets us discover a new class of reheating models \cite{Giblin:2017qjp}. The role of sound speed for particle production in sectors other then the inflaton does seem to be a feature that does arise in super symmetric settings \cite{Hasegawa:2017hgd, Ema:2015oaa}.

As such the Lagrangian is;
\begin{align} \nn \mathcal{L}_\chi&=\frac{1}{2}\Bigg[\dot{\tilde\chi}_c^2-c^2_\chi(t)\left(\frac{\vec{\nabla}\tilde{\chi_c}}{a}\right)^2-m^2_\chi\tilde{\chi}_c^2\\
&~~~~+\left(\frac{3\dot{H}}{2}+\frac{9H^2}{4}+\frac{3H}{2}\frac{\dot{\alpha}_1+\dot{\alpha_2}}{\alpha_1+\alpha_2}-\frac{3H\alpha_4}{\alpha_1+\alpha_2}\right)\tilde{\chi}_c^2\Bigg].\end{align}
The terms in the last line appear due to the Hubble expansion, from the factor of $a^{3/2}$ we have in the definition of $\tilde{\chi}_c$. This field normalization is convenient to bring the equation of motion for $\chi$ in the form of a harmonic oscillator. Later on we will define the canonically normalized field as $\chi_c=a^{-3/2}\tilde{\chi}_c$, the difference between the two normalizations is having terms proportional to $H\tilde{\chi}_c^2$ or $a^{3/2}\chi_c\dot{\chi}_c$. 

Carrying on a mode decomposition, the equation of motion for each mode is
\be \label{chieom}\ddot{\tilde\chi}_k+\left[c^2_\chi(t)\frac{k^2}{a^2}+m^2_\chi(t)-\left(\frac{3\dot{H}}{2}+\frac{9H^2}{4}\right)\right]\tilde\chi_k^2=0.\ee
The modes $\chi_k$ are harmonic oscillators with the time dependent frequency $\omega^2_k=c^2_\chi(t)\frac{k^2}{a^2}+m^2_\chi-\left(\frac{3\dot{H}}{2}+\frac{9H^2}{4}\right)$. Part of the time dependence comes from the time dependence of the background and the interactions of the $\chi$ with this back ground, as is expected. In addition, we have an extra source for time dependence through the sound speed $c_\chi$, simply  because the symmetries allow us to introduce terms that give rise to it. As long as the time dependence of the $\chi$ oscillator is adiabatic, it will have a smooth behavior, which can be approximated by the WKB approximation
\be \label{chik} \chi_{c,k}(t)=\frac{\alpha_k(t)}{\sqrt{2\omega_k}}e^{-i\int^t dt' \omega_k(t')}+\frac{\beta_k(t)}{\sqrt{2\omega_k}}e^{i\int^t dt' \omega (t')}.\ee
This approximation breaks down at times when $\omega_k(t)$ develops nonadiabaticities, characterized by
\be \label{nonadiabatic} \dot{\omega}_k\geq \omega_k^2(t).\ee
The specific k-modes that satisfy the nonadiabaticity violation at specific times, and exhibit instabilities can be understood as $\chi_k$ particles being produced. These are the modes who undergo resonant instabilities and hence are referred to as produced particles during preheating. Note that the nonadiabaticity in $\chi$ behavior does not mean a nonadiabaticity on the background set by $H(t)$! 

Particle production should be a local transfer of energy from the background energy density to the $\chi_k$'s in which the total energy is fixed. This energy transfer should not care about the overall expansion, in other words it happens at a smaller length scale then $H^{-1}$. Thus we can set $a\to 1$, $H\to 0$ while keeping the combinations $m^2_{pl}\dot{H}\sim m^2_{pl}H^2$, which have dimensions $E^4$ and reflect the total energy, constant. Making use of these limits to neglect the overall expansion, the condition \eqref{nonadiabatic} can be turned into a range $K^2>k^2$ for producible k-modes. We will define the left hand side of this condition, $K$, as the scale below which particle production occurs. As we will see in specific examples, $\alpha_i(t)$'s can be written proportional to $m^2_{pl}H^2$. The second part of the limit guarantees that these interactions are of importance while the terms coming purely from the background expansion such as the ones in rightmost parenthesis of equation \eqref{chieom} are negligible.

A general expression for the production scale $K$ is complicated. With the expansion neglected this would involve calculating the term
\begin{align} \label{omegadomega}
2\omega_\chi\dot{\omega}_\chi=&\left[\frac{\dot{\alpha}_1}{\alpha_1+\alpha_2}-\frac{\alpha_1}{\alpha_1+\alpha_2}\frac{\dot{\alpha}_1+\dot{\alpha}_2}{\alpha_1+\alpha_2}\right]k^2+\frac{\dot{\alpha}_3-\ddot{\alpha}_4}{\alpha_1+\alpha_2}-\frac{\alpha_3-\dot{\alpha}_4}{\alpha_1+\alpha_2}\frac{\dot{\alpha}_1+\dot{\alpha}_2}{\alpha_1+\alpha_2}\\
&-\frac{1}{2}\frac{\dddot{\alpha}_1+\dddot{\alpha}_2}{\alpha_1+\alpha_2}+\frac{1}{2}\frac{\dot{\alpha}_1+\dot{\alpha}_2}{\alpha_1+\alpha_2}\frac{\ddot{\alpha}_1+\ddot{\alpha}_2}{\alpha_1+\alpha_2}-\frac{1}{2}\left(\frac{\dot{\alpha}_1+\dot{\alpha}_2}{\alpha_1+\alpha_2}\right)^3.
\end{align}
At this point what assumptions can we make for the $\alpha_i$'s? From dimensional analysis, where $\chi$ has the dimensions of mass, $\alpha_1$ and $\alpha_2$ are dimensionless, $[\alpha_1]=[\alpha_2]=1$,$\alpha_4$ has mass dimension one, $[\alpha_4]=M$ and $\alpha_3$ has mass dimension two, $[\alpha_3]=M^2$. In the spirit of the EFT, one would expect similar time dependence in all the EFT coefficients.  At linear order we have the terms $m^2_{pl}H^2$ and $m^2_{pl}\dot{H}$. These are the factors we expect to give rise to the time dependence in $\alpha_i(t)$. Remember that these combinations stay finite even when the background expansion is neglected. When we studied the background behavior at the end of single field inflation, we found two important scales, one that gave decelerated expansion $H_m$ and one that exhibited oscillations $m_\phi$. We saw that the scale associated with the oscillations is higher then the scale associated with the expansion, and this guaranteed that on the whole the background evolves monotonically. In other words the highest scale in the background physics is $m_\phi$. It is this $m_\phi$ that we can use sort of like a cutoff scale to adjust for the dimensions of $\alpha_i$. 

For convenience we can express $\alpha_i$ in the following form that meet our concerns above
\begin{align}
\label{convention1_a1}\alpha_1(t)&=c_0+\frac{c_1}{m^4_\phi}m^2_{pl}H^2+\frac{c_2}{m^4_\phi}m^2_{pl}\dot{H}\\
\label{convention1_a2}\alpha_2(t)&=d_0+\frac{d_1}{m^4_\phi}m^2_{pl}H^2+\frac{d_2}{m^4_\phi}m^2_{pl}\dot{H}\\
\label{convention1_a3}\alpha_3(t)&=a_0m^2_\chi+\frac{a_1}{m^2_\phi}m^2_{pl}H^2+\frac{a_2}{m^2_\phi}m^2_{pl}\dot{H}\\
\label{convention1_a4}\alpha_4(t)&=b_0m_\chi+\frac{b_1}{m^3_\phi}m^2_{pl}H^2+\frac{b_2}{m^3_\phi}m^2_{pl}\dot{H}
\end{align}
where $a_i$, $c_i$, $d_i$ are just constants of same order.Of course the last two terms for each EFT parameter above is of the same order, when considering estimations that involve time derivatives of the EFT coefficients the last term always gives the leading contribution. For now we can think of them as names of the terms involved in $\alpha_i$, they can also be thought of as parameterization of random models which will involve different factors. Note that with this assumption we have made the sound speed for $\chi$ constant $c^2_\chi=\frac{\alpha_1}{\alpha_1+\alpha_2}\sim\mathcal{O}\left(\frac{c_2}{c_2+d_2}\right)$, at leading order. However because of the way $\alpha_1(t)$ works into the normalization of $\chi$, the time dependence of these coefficients still contribute to particle production. 

Since $\alpha_1$ and $\alpha_2$ have mass dimension 1, we can always define a normalization such that $\alpha_1(t)+\alpha_2(t)=1$. In terms of our assumptions \eqref{convention1_a1} and \eqref{convention1_a2} this implies $c_0+d_0=1$, and $c_i=-d_i$ for $i\neq0$. With this the time dependence that was carried by the canonical normalization is traded to a time dependence of the sound speed, which is now simplified to be $c^2_\chi(t)=\alpha_1(t)$. Let us estimate the range of resonant modes this normalization. The term \eqref{omegadomega} that we need to calculate is simplified tremendously to
\begin{align} 2\omega_k\dot{\omega}_k=\dot{\alpha}_1k^2+\dot{\alpha}_3-\ddot{\alpha}_4&\propto \frac{m^2_{pl}\ddot{H}}{m^4_\phi}k^2+a_2\frac{m^2_{pl}\ddot{H}}{m^2_{\phi}}-b_2\frac{m^2_{pl}\dddot{H}}{m^3_\phi}\\
&\propto\frac{m^2_{pl}H^2}{m^3_\phi}k^2+\frac{m^2_{pl}H^2}{m_{\phi}}\end{align}
Note that t this point the presence of $\alpha_4$ works on equal footing as $\alpha_3$, so we can just think of them as one parameter $\tilde{\alpha}_3(t)=\alpha_3(t)-\dot{\alpha}_4(t)\sim\mathcal{O}\left(\frac{m^2_{pl}\dot{H}}{m^2_\phi}\right)$. The frequency of the oscillations themselves is of the order
\begin{align} \omega^2_k(t)&=\alpha_1(t)k^2+\alpha_3(t)-\dot{\alpha}_4(t)\\
&\propto \frac{m^2_{pl}H^2}{m^4_\phi}k^2+\frac{m^2_{pl}H^2}{m^2_\phi}
\end{align} 
The range of modes that grow non adiabatically is
\begin{align}\label{genscale} \dot{\omega}_k > \omega_k^2 &:\dot{\alpha}_1k^2+\dot{\alpha}_3-\ddot{\alpha}_4>2\left[\alpha_1 k^2+\alpha_3-\dot{\alpha}_4 \right]^{3/2}.\end{align}
where the largest mode that satisfies this inequality would set the scale of particle production.

At this point let us note that from the generality of EFT couplings, we discovered the possibility of having a sound speed for the reheating field, and moreover the possibility of production reheat particles from the time dependence of this sound speed. The general solution to \eqref{genscale} where the highest mode that satisfies the inequality would set the scale of $\chi$-production is not very easy. So let us consider two separate production scales, one coming purely from the sound speed and the other from background.   

Production solely due to the sound speed corresponds to the scale
\be k<\frac{1}{2\sqrt{\alpha_1}}\frac{\dot{\alpha}_1}{\alpha_1}\equiv K_{c_\chi}.\ee
And we can estimate this scale to be $K_{c\chi}\propto \mathcal{O}\left(\frac{m^3_\phi}{m_{pl}H}\right)$ which is highly suppressed by $\frac{m_\phi}{m_{pl}}$. Particle production with constant sound speed happens up to the scale
\be k^2<\frac{1}{c^2_\chi}\left[\left(\frac{\dot{\alpha}_3-\ddot{\alpha}_4}{2}\right)^{2/3}-\frac{\alpha_3-\dot{\alpha}_4}{\alpha_1}\right]\equiv K^2_{bck}.\ee
With our assumptions on the EFT coefficients this boils down to
\be K^2_{bck}\sim \frac{1}{\alpha_1}\left[\left(\frac{m^2_{pl}H^2}{m_\phi}\right)^{2/3}+\frac{m^2_{pl}H^2}{m^2_\phi}\right].\ee
We started with $\frac{m_\phi}{m_{pl}}\ll1$ and $\frac{H_p}{m_\phi}\leq 1$ this would make the ratio of $\frac{m_{pl}}{m_\phi}$ determine the strength above. However we are also considering the limit where $m^2_{pl}H^2$ is finite, with finite implying order 1. As such the first term seems to determine the order of magnitude and so we will take the particle production scale with constant sound speed to be of the order
\be K^2_{bck}\sim\mathcal{O}\left(\left[\frac{m^2_{pl}H^2}{m_\phi}\right]^{2/3}\right)\ee
acknowledging that there may be certain models in the parameter space for whom the ratios are adjusted to give $K^2_{bck}\sim\mathcal{O}\left(\frac{1}{c^2_\chi}\frac{m^2_{pl}H^2}{m^2_\phi}\right)$ with $m^2_\phi\ll m_{pl}H$. And we note that $K^2_{c\chi}<K^2_{bck}$.

Before we close this section, let us remind ourselves that in section \ref{sec:background evolution} we saw that the Hubble parameter at the end of single field inflation behaves according to
\be \dot{H}=-3H^2cos^2(m_\phi t)\ee
from where we got the solution in the following form
\be H(t)=H_{FRW}+H_{osc}sin(2m_\phi t).\ee
As this is also the behavior expected from the $\alpha_i(t)$, there will be oscillatory terms in $\omega^2_k$ which puts the equation of motion for $\chi_c$ in the form of a Mathieu equation
\be\label{Mathieu} X''_k+\left(A_k-2qcos(2z)\right)X_k=0.\ee
In other words complimentary to \eqref{convention1_a3}, we can also parametrize $\alpha_i(t)$ as
\be\label{convention2} \alpha_i(t)=M^p_iP(2m_\phi t)\ee
where $p$ denotes the mass dimension of each EFT coefficient\footnote{Of course the same logic applies to the other EFT coefficients $m_i(t)$ and $\beta_i$.}. For example, if for convenience we neglect the time dependence in $\alpha_1$, $\alpha_2$, and say
\be \label{convention2_a3} \alpha_3(t)=a_0m^2_\chi+M^2cos(2m_\phi t)\ee
neglecting the background expansion, from \eqref{chieom} we have for $\chi_c$
\be \ddot{\chi}_c+\left(\frac{\alpha_1}{\alpha_1+\alpha_2}k^2+\frac{a_0m^2_\chi}{\alpha_1+\alpha_2}+\frac{M^2}{\alpha_1+\alpha_2}cos(2m_\phi t)\right)\chi^2_c=0.\ee
Thus we see that the Mathieu parameters are connected to the EFT parameters, under the assumption of constant sound speed, as follows
\begin{align}
z&=m_\phi t,\\
A_k&=\frac{\alpha_1}{\alpha_1+\alpha_2}\frac{k^2}{m_\phi^2}+\frac{a_0}{\alpha_1+\alpha_2}\frac{m^2_\chi}{m^2_\phi},\\
q&=-\frac{1}{2m^2_\phi}\frac{M^2}{\alpha_1+\alpha_2}.
\end{align}

The solution \eqref{chik}, with the general form $\chi_c\sim e^{i\mu_k t}$, can exhibit exponential growth for modes which develop an imaginary component to $\mu_k$. The values of $\mu_k$ for which this happens have long been charted out for the Mathieu equation. So in principle, by knowing how the EFT parameters are connected to the Mathieu equation parameters, one can consult these charts for the precise range of the resonant modes from these charts, or an updated code version of these charts  \cite{Amin:floqex}.  While the nonadiabaticity condition \eqref{nonadiabatic} is the general condition for particle production, this exponential growth can occur and be efficient even without it in the case of narrow resonance \cite{Kofman:1997yn} which is the case with $q\ll1$. Including the time dependence of $\alpha_1$ and $\alpha_2$ will bring further contributions to $A_k$ and $q$. 

Back reaction effects, among the two fields can be captured from terms such as $\alpha_3(t+\pi)\chi$.

\subsection{Self Resonance of the Inflaton during Preheating}
The self interactions of the inflaton, which in the $\pi$ gauge is captured by $\pi_c$, arise from two sources. One contribution to self interactions come via the metric perturbations $\delta g^{00}$ and $\delta g^{0i}$. These are the ADM variables $\lambda^N$ and $\lambda^i$, which we occasionally also note as $\delta N$ and $\delta N^i$. Hence they are not independent variables but are related to $\pi$ via the constraint equations. Interactions that look like mixing between $\pi_c$ and scalar and vector metric perturbations lead to self resonance of $\pi_c$, the perturbation to the inflaton, during preheating \cite{Easther:2010mr}. The timedependence of the background gives rise to further self interactions of the inflaton. These effects captured by Taylor expanding the time dependent EFT coefficients, such as $H(t)$, $\dot{H}(t)$, $m_2(t)$
\be H^2(t+\pi)=H^2(t)+2H(t)\dot{H}(t)\pi+(\dot{H}^2+H\ddot{H})\pi^2+...\ee
Keeping track of these terms, the second order Lagrangian for the canonically normalized field $\pi_c$ is
\begin{align} \mL_{\pi_c}&=\frac{1}{2}\left(\dot{\pi}_c^2-\frac{c^2_\pi}{a^2}(\partial_c\pi_c)^2\right)-\frac{1}{2}m^2_\pi(t)\pi_c^2\\
&-\frac{\sqrt{-2\dot{H}}}{2c_\pi}\left(\dot{\pi}_c\delta g^{00}_c-\frac{1}{2}\left(\frac{\ddot{H}}{\dot{H}}-2\frac{\dot{c}_\pi}{c_\pi}\right)\pi_c\delta g^{00}_c\right)+\frac{3H}{2}c_\pi\sqrt{-2\dot{H}}\delta g^{00}_c\pi_c\dots\end{align}
where the metric perturbations are canonically normalized as $\delta g^{\mu\nu}_c=m_{pl}\delta g^{\mu\nu}$ and
\be m^2_\pi(t)=-3\dot{H}c^2_\pi-\frac{1}{4}\left(\frac{\ddot{H}}{\dot{H}}-2\frac{\dot{c}_\pi}{c_\pi}\right)^2-\frac{3H}{2}\left(\frac{\ddot{H}}{\dot{H}}-2\frac{\dot{c}_\pi}{c_\pi}\right)-\frac{1}{2}\partial_t\left(\frac{\ddot{H}}{\dot{H}}-2\frac{\dot{c}_\pi}{c_\pi}\right).\ee
During inflation the higher derivatives on $H$ would be negligable because of the de Sitter nature of the background. However during preheating, both $\pi$ and the canonically normalized field $\pi_c$ gain mass\footnote{After a straight forward calculation that involves the substitution of solutions for the ADM variables and a coupe of integration by parts one arrives at the term $-\frac{1}{2}m^2_{pl}a^2\left[2\dddot{H}-2\ddot{H}-3H\ddot{H}-5\frac{\dot{H}^3}{H^2}+\frac{\dot{H}\ddot{H}}{H}-5\dot{H}^2\right]\pi^2$ in the action at second order in $\pi$. All of these terms would be negligable during inflation but they are relevant for preheating}. We adress this issue in section \ref{sec:adiabaticmodes}, where our conclusion will be that $\zeta$ remains massless, and hence is the Goldstone boson at all times.

From our investigation of the background we know that $\ddot{H}\sim m_\phi H^2$ and $m_\phi > H$. This means that the leading contribution of metric mixing comes from the term
\be \frac{1}{c_\pi}\frac{\ddot{H}}{\sqrt{-\dot{H}}}\pi_c\delta g^{00}_c\sim m_\phi H \pi_c\delta g^{00}_c.\ee
Hence the mixing with metric and the self resonance it leads to become important starting from the scale
\be E^2_{\delta\phi}\sim\frac{1}{c_\pi}m_\phi H\ee
and below.

\subsection{Hidden Preheating}
\label{subsec:hidden}
So far we have seen that among the EFT parameters, $\alpha_i(t)$ control resonant $\chi_c$ production, and we studied how $H(t)$ and $m_i(t)$ work into $\pi_c$ behaviour. The coefficients $\beta_i(t)$ denote additional couplings between the inflaton fluctuations and $\chi$. These kinds of interactions are unavoidable  if the inflaton and the reheat field are derivatively coupled to each other. In the unitary gauge, the interactions under consideration are
\be\label{hiddenS} S^c_{g\chi}=\int d^4x\sqrt{-g}\left[\beta_1(t)\delta g^{00}\chi+\beta_2(t)\delta g^{00}\partial^0\chi+\beta_3(t)\partial^0\chi-(\dot{\beta}_3(t)+3H(t)\beta_3(t))\chi\right]\ee
where the mass dimensions of the parameters are $[\beta_1]=M^3$ and $[\beta_2]=[\beta_3]=M^2$. 

The inflaton fluctuations appear explicitly in the $\pi$-gauge, which is achieved via the time diffeomorphism $t\to t+\pi$. Under this nonlinear time diffeomorphism these terms transform according to
\begin{align}
g^{00}&\to g^{00}+2g^{0\mu}\partial_\mu\pi+g^{\mu\nu}\partial_\mu\pi\partial_\nu\pi\\
\beta_i(t)&\to\beta_i(t+\pi)\\
\partial^0\chi&\to\partial^0\chi+g^{\mu\nu}\partial_\mu\chi\partial_\nu\pi\\
\int d^4x\sqrt{-g}&\to\int d^4x\sqrt{-g},
\end{align}
and the second order action becomes \cite{Noumi:2012vr}
\begin{align}
\label{hiddenS1}& S^{(2)}_{g\chi}=\int d^4 x\sqrt{-g}\Bigg[ \beta_1(t+\pi)\left(\delta g^{00}+2\partial^0\pi+\partial_\mu\pi\partial^\mu\pi\right)\chi\\
\label{hiddenS2}&+\beta_2(t+\pi)\left(\delta g^{00}+2\partial^0\pi+\partial_\mu\pi\partial^\mu\pi\right)\left(\partial^0\chi+\partial_\nu\pi\partial^\nu\chi\right)\\
\label{hiddenS3}&+\beta_3(t+\pi)\left(\partial^0\chi+\partial_\mu\pi\partial^\mu\chi\right)-\left(\dot{\beta}_3(t+\pi)+3H(t+\pi)\beta_3(t+\pi)\right)\chi\Bigg].\end{align}
Let's begin by focusing on the re-normalizable quadratic couplings
\begin{align} 
S^{(2)}_{g\chi}\supset\int& d^4xa^3\Bigg[\beta_1(t)\left(\delta g^{00}\chi-2\dot{\pi}\chi\right)+\beta_2(t)\left(-\delta g^{00}\dot{\chi}+2\dot{\pi}\dot{\chi}\right)\\
&-\delta N\beta_3(t)\dot{\chi}-\dot{\beta}_3(t)\pi\dot{\chi}+\beta_3(t)\delta g^{0\mu}\partial_\mu\chi-\beta_3(t)\dot{\chi}\dot{\pi}\\
&+\beta_3(t)\partial_i\chi\partial_i\pi-\ddot{\beta}_3(t)\pi\chi-(3\dot{\beta}_3H+3\beta_3\dot{H})\pi\chi\Bigg].
\end{align}
Remember that the connection to inflaton particles is
\be \delta\phi=\pi_c=\frac{\pi}{c_\pi}\sqrt{-2m^2_{pl}\dot{H}}=\frac{\pi}{c_\pi}\dot{\phi}_0.\ee

Let's focus on the terms with temporal derivatives. After canonical normalization with $\alpha_1+\alpha_2=1$, and $\delta g^{00}=\frac{\delta g^{00}_c}{m_{pl}}=\frac{\delta N_c}{m_{pl}}$ \footnote{For reference, in the case where $m_i=\beta_i=0$, which would mean no strong coupling, the constraints give $\delta N=-\frac{\dot{H}}{H}\pi$. This makes $\delta N_c=m_{pl}\delta N_c=\frac{\sqrt{-\dot{H}}}{\sqrt{2}H}\pi_c\sim\frac{\pi_c}{\sqrt{2}}$.} we have
 
\begin{align}
S^{(2)}_{g\chi}&\supset\int d^4 xa^3\Bigg[\beta_1\delta N_c\frac{\chi_c}{m_{pl}}-2\beta_1\frac{c_\pi}{\sqrt{-2m^2_{pl}\dot{H}}}\left(\frac{\dot{c}_\pi}{c_\pi}\pi_c+\dot{\pi}_c-\frac{\ddot{H}}{2\dot{H}}\pi_c\right)\chi_c\\
&-\beta_2\frac{\delta N_c}{m_{pl}}\dot{\chi}_c+2\beta_2\frac{c_\pi}{\sqrt{-2m^2_{pl}\dot{H}}}\left(\frac{\dot{c}_\pi}{c_\pi}\pi_c+\dot{\pi}_c-\frac{\ddot{H}}{2\dot{H}}\pi_c\right)\dot{\chi}_c-\beta_3\frac{\delta N_c}{m_{pl}}\dot{\chi}_c\\
&-\frac{c_\pi}{\sqrt{-2m^2_{pl}\dot{H}}}\dot{\beta}_3\dot{\chi}_c\pi_c-+\beta_3\frac{\delta N_c}{m_{pl}}\dot{\chi}_c-\beta_3\frac{c_\pi}{\sqrt{-2m^2_{pl}\dot{H}}}\left[\frac{\ddot{H}}{\dot{H}}\pi_c+\dot{\pi}_c+\frac{\dot{c}_\pi}{c_\pi}\pi_c\right]\dot{\chi}_c\\
&-\frac{c_\pi}{\sqrt{-2m^2_{pl}\dot{H}}}\ddot{\beta}_3\pi_c\chi_c-\frac{c_\pi}{\sqrt{-2m^2_{pl}\dot{H}}}(3\dot{\beta}_3H+3\beta_3\dot{H})\pi_c\chi_c\Bigg]\end{align}
Our study of the background taught us that higher derivatives on $H$ are stronger because they involve more powers of $m_\phi$. Derivatives of $H$ appear in $S^{(2)}$ after canonical normalization. Looking at all the terms in $S^{(2)}$, the terms that involve the most number of derivatives will be stronger among the terms of same order in perturbations.\footnote{This does not imply that the next order action will be stronger then the previous. For example at third order one has the term \cite{Noumi:2012vr}.  $S^{(3)}_{g\chi}\supset \gamma \ddot{\pi}\chi\dot{\chi}$, which will involve $\gamma\left(-\frac{\dddot{H}}{\dot{H}}+\frac{\ddot{H}^2}{\dot{H}^2}\right)\frac{\dot{\pi}_c}{\sqrt{-2m^2_{pl}\dot{H}}}\chi_c\dot{\chi}_c\sim\gamma \frac{m^2_\phi\omega_k}{m_{pl}H}\pi_c\chi_c^2$where $\gamma$ is dimension zero and this term is highly suppressed via the symmetry breaking scale, compared to terms in $S^{(2)}$.} Among the terms parametrized by $\beta_1$, these are $\beta_1\frac{c_\pi}{\sqrt{-2m^2_{pl}\dot{H}}}\frac{\ddot{H}}{\dot{H}}\pi_c\dot{\chi}_c$ and $-2\beta_1\frac{c_\pi}{\sqrt{-2m^2_{pl}\dot{H}}}\frac{\ddot{H}}{\dot{H}}\pi_c\dot{\chi}_c$. Similarly we have $-\beta_2\frac{c_\pi}{\sqrt{-2m^2_{pl}\dot{H}}}\frac{\ddot{H}}{\dot{H}}\pi_c\dot{\chi}_c$ and $2\beta_2\frac{c_\pi}{\sqrt{-2m^2_{pl}\dot{H}}}\dot\pi_c\dot{\chi}_c$ for $\beta_2$ parametrization and $-\frac{c_\pi}{\sqrt{-2m^2_{pl}\dot{H}}}\dot{\beta}_3\dot{\chi}_c\pi_c$, $-\beta_3\frac{c_\pi}{\sqrt{-2m_{pl}\dot{H}}}\frac{\ddot{H}}{\dot{H}}\pi_c\dot{\chi}_c$, $-\frac{c_\pi}{\sqrt{-2m^2_{pl}\dot{H}}}\ddot{\beta}_3\pi_c\chi_c$ for $\beta_3$ parametrization.

Some of these terms involve derivative couplings,\footnote{We can ignore the terms coming from metric perturbations via $\delta N_c$ because they are Planck Suppressed at this order.} 
\be \frac{\beta_1}{\sqrt{-2m^2_{pl}\dot{H}}}\dot{\pi}_c\chi_c,~~\frac{\beta_2}{\sqrt{-2m^2_{pl}\dot{H}}}\frac{\ddot{H}}{\dot{H}}\pi_c\dot{\chi}_c,~~\frac{\beta_3}{\sqrt{-2m^2_{pl}\dot{H}}}\left(1+\frac{\ddot{H}}{\dot{H}}\right)\pi_c\dot{\chi}_c.\ee
These are terms of the form $R_1\dot{\pi}_c\chi_c$ and $\rho_\pi\pi_c\dot{\chi}_c$.
They can compete with the kinetic terms. Notice that $\rho_{\pi,\chi}$ has dimensions of energy, so it sets the energy scale for these derivative interactions. Below the scale set by $\rho_{\pi,\chi}$, if these derivative couplings dominate over the kinetic terms $\dot{\pi}_c^2$, $\dot{\chi}_c^2$, the system will effectively have a single degree of freedom. This is the same situation that is addressed in \cite{Baumann:2011su}, where they consider its implications during inflation. Following the same line of thought, we will now consider the consequences during preheating. We want to focus on the consequences of these interactions. For now we are not interested in how the sound speed can amplify or reduce their strength, so we have set $c_\pi=c_\chi=1$, which amounts to setting $m_2=0$, $\alpha_1=1$.

To see how the number of effective degrees of freedom goes down to being single, let us consider the quadratic Lagrangian for the canonically normalized perturbations in the presence of these interactions one by one. Keep in mind that we still consider scales within the limit $a\to 1$ that was meaningful for particle production purposes.So  whenever we need to compare terms with each other, we treat the combination $m_{pl}H$ to be an order one number, as was implied by the flat spacetime limit.
\subsubsection{Hidden Preheating by $\beta_1$:}

In the presence of $\beta_1$ we have,
\be \mathcal{L}^{(2)}=\frac{1}{2}\dot{\chi}_c^2+\frac{1}{2}\dot{\pi}_c^2-\frac{1}{2a^2}(\partial_i\chi_c)^2-\frac{1}{2a^2}(\partial_i\pi_c)^2-\frac{1}{2}m^2_\chi\chi_c^2-\frac{1}{2}m^2_\pi\pi_c^2-2R_1\dot{\pi}_c\chi_c-R_1\frac{\ddot{H}}{\dot{H}}\pi_c\chi_c\ee
where $R_1\equiv\frac{\beta_1(t)}{\sqrt{-2m^2_{pl}\dot{H}}}$. The derivatives of the fields go as $\dot\pi_c=\omega_\pi\pi_c$ and $\dot{\chi}_c=\omega_\chi\chi_c$. Note that, for inflationary observations the scale of interest is the Horizon crossing, when $\omega_\pi=H_I$ and $H_I$ is the highest scale involved in terms of the scales of the background, hence one can focus on modes with $\dot{\pi}_c=H_I\pi_c$ and $\dot{\chi}_c=H_I\chi_c$, which is different then the case for preheating.

In comparison, the kinetic terms go by $\omega_\pi^2$, $\omega_\chi^2$, while the kinetic coupling goes as $R_1\omega_\pi$ . For energies in the range of 
\be \label{hiddenbeta1scale} R_1>\omega_\pi~~ \text{and}~~ R_1>\frac{\omega^2_\chi}{\omega_\pi}\ee
the theory is 
\be \label{hiddenbeta1lag} \mathcal{L}\simeq -2R_1\dot{\pi}_c\chi_c-\frac{1}{2a^2}(\partial_i\chi_c)^2-\frac{1}{2a^2}(\partial_i\pi_c)^2-\frac{1}{2}m^2_\chi\chi_c^2-\frac{1}{2}m^2_\pi\pi_c^2-\frac{\ddot{H}}{\dot{H}}R_1\pi_c\chi_c.\ee
At first sight it looks like the kinetic term here has a wrong sign, but this depends on the sign of $R_1$, which is not necessarily positive! In fact we will consider a specific example that will demonstrate this point. 

In this range $\chi_c$ is no longer a dynamical field, since it doesn't have any kinetic terms. In fact it plays the role of canonical momenta for $\pi_c$,
\be p_\pi\equiv\frac{\partial\mathcal{L}}{\partial\dot{\pi_c}}=-2R_1\chi_c.\ee
So in this regime the perturbations of the reheating field are effectively heavier than inflaton perturbations and hence they are integrated out. This is an interesting point. By intuition from the canonical examples, we would have liked to see more and more of $\chi_c$ modes being the effective degrees of freedom during preheating, yet we are seeing that in the presence of $\beta_1$ interactions, the inflaton perturbations are the effective, in otherwords, lightest degrees that propagate! However it is important to note that the presence of the reheat perturbations is important, as they determine the canonical momenta of $\pi_c$. So in a sense this is a type of reheating where there is a range of energies in which the reheating field determines the dynamics of the inflaton perturbations, while it itself stays hidden. This is why we have called this range as the ``Hidden Preheating''. It is the preheating version of the situation discussed in \cite{Baumann:2011su}.

Let us try to make an estimate on the likeliness of such a range occuring, by making assumptions on the form of the EFT coefficients. As we have already noted, the mass dimensions of $\beta_i$ are $[\beta_1]=M^3$, $[\beta_2]=[\beta_3]=M^2$. Since the quadratic terms that were determined by the background are at order $m^2_{pl}\dot{H}$, we expect a similar form for $\beta_i$. Unless we know more about the background, we have $m_\phi$, which is the highest scale in the background evolution, that we can use to make the dimensions fit
\be\label{convbeta1}\beta_1=b_1\frac{m^2_{pl}\dot{H}}{m_\phi},\ee \be\label{convbeta2}\beta_2=b_2\frac{m^2_{pl}\dot{H}}{m^2_\phi},\ee
\be\label{convbeta3}\beta_3(t)=b_3\frac{m^2_{pl}\dot{H}}{m^2_\phi}.\ee
As such
\be\label{rho1est} R_1\equiv\frac{\beta_1}{\sqrt{-2m^2_{pl}\dot{H}}}=\mathcal{O}\left(\frac{m_{pl}H}{m_\phi}\right)=\mathcal{O}\left(\frac{\Lambda^2_{sb}}{m_\phi}\right).\ee
So while in the range $\Lambda_{sb}>\omega_{\pi,\chi}>R_1$ there are 2 effective degrees of freedom $\pi_c$ and $\chi_c$, in the range $R_1>\omega_{\pi,\chi}$, the inflaton perturbations $\pi_c$ are the only effective degree. We sumarize this distribution of effective modes with respect to scale in Figure \ref{fig:beta1modes}.

\begin{figure}[h]
	\includegraphics[width=\textwidth]{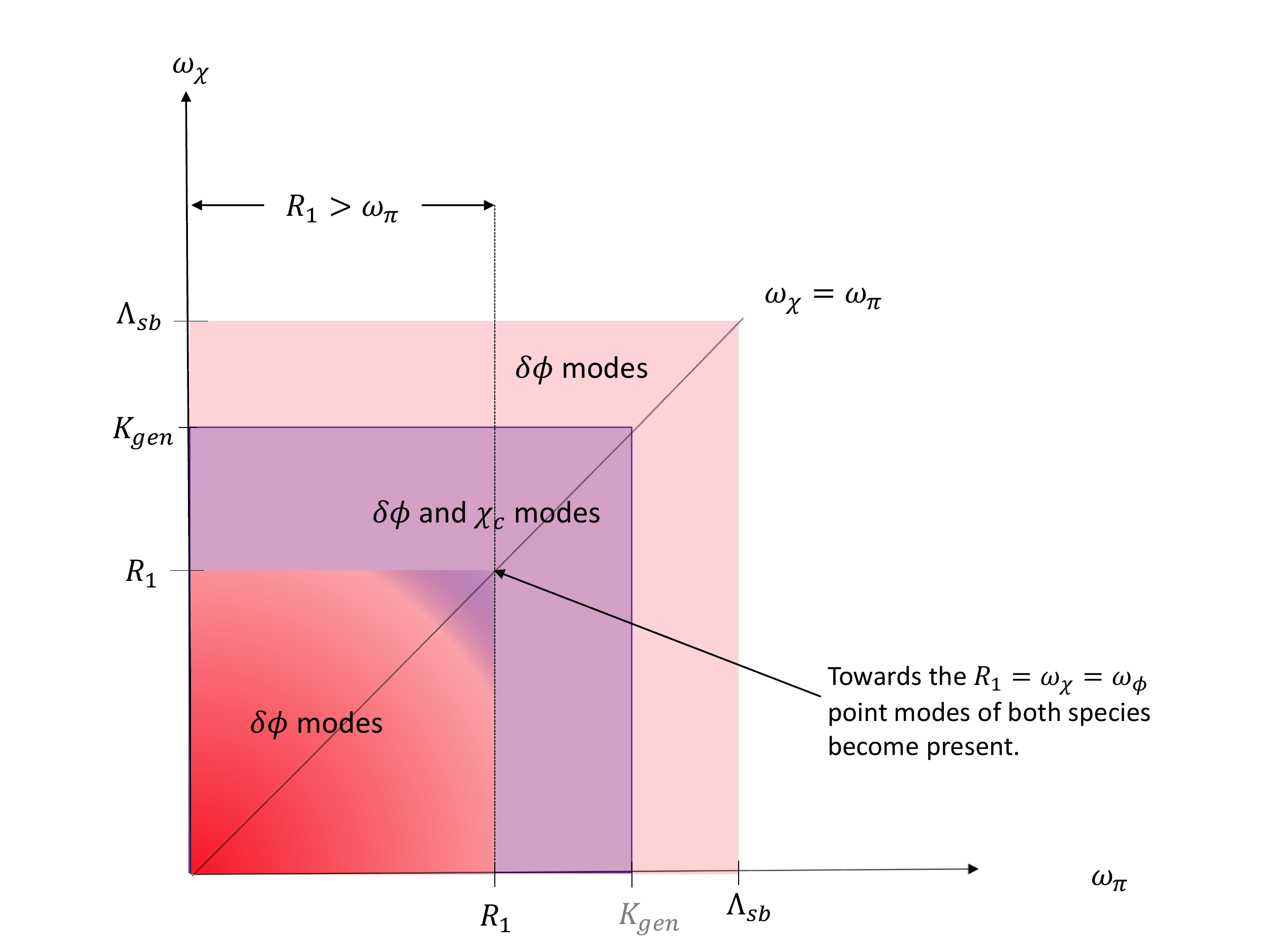}
	\caption{Here we show up to which scales the inflaton modes $\delta\phi=\pi_c$ and reheating modes $\chi_c$ appear as effective degrees of freedom.}
	\label{fig:beta1modes}
\end{figure}

Now we need to be a bit careful, does this leave any range for $\chi_c$-production? We found that the $\chi_c$ particles can be produced up to the scale
\be K^2_{bck}=\frac{1}{c_\chi^2}\left[\left(\frac{\dot{\alpha}_3-\ddot{\alpha}_4}{2}\right)^{2/3}-\alpha_3+\dot{\alpha}_4\right]\sim\mathcal{O}\left(\left[\frac{m^2_{pl}H^2}{m_\phi}\right]^{2/3}\right)=m_\phi^{2/3}R_1^{4/3}\ee
Considering \eqref{rho1est}, this suggests that the scale of particle production lies above the scale $R_1$, and so in the range $R_1<E<K_{bck}$ both $\pi_c$ and $\chi_c$ modes are effective degrees of freedom. 

These $\beta_1$ interactions are present in models where the inflaton is derivatively coupled to the reheating sector. As a solid example, in context of preheating derivative couplings have first been studied in \cite{ArmendarizPicon:2007iv}, with
\be\mathcal{L}=-\frac{1}{2}\partial^\mu\phi\partial_\mu\phi-\frac{1}{2}\partial^\mu X\partial_\mu X-\frac{1}{2}m^2_\phi\phi^2-\frac{1}{2}m^2_\chi X^2-\frac{1}{F^2}(\partial_\mu\phi\partial^\mu\phi)X^2\ee
where $F$ is the cutoff for this effective field theory. In the original work, the authors consider Chaotic Inflation in particular, to govern the inflaton sector in this low energy theory, in which case $F\simeq m_{pl}$. We will focus on this Lagrangian now to demonstrate our previous development of scales. This Lagrangian can describe preheating in general backgrounds where the reheating field can also contribute significantly to the background energy density, and be time dependent. This can raise some issues on the stability of the background, which we will come back to later. For the purposes of this section we can set $\chi_0=const$. This raises no stability questions and for example resembles geometric destabilization of inflation \cite{Renaux-Petel:2015mga}. For our purposes this effective field theory could have been obtained from a different UV completion, and $F$ is undetermined.  The background equations of motion for this system are
\be -2m^2_{pl}\dot{H}=\left(1+2\frac{\chi_0^2}{F^2}\right)\dot{\phi}_0^2\equiv R^2\dot{\phi}_0^2\ee
\be 3m^2_{pl}H^2=\frac{1}{2}\left(1+2\frac{\chi_0^2}{F^2}\right)\dot{\phi}_0^2+V(\phi_0)+U(\chi_0)\ee
and for $\chi_0$
\be U'(\chi_0)-2\frac{\dot{\phi}_0^2}{F^2}\chi_0=0.\ee

In unitary gauge, the fields are expanded in terms of linear perturbations as $\phi=\phi_0(t)$, $X=\chi_0+\chi(\vec{x},t)$ and, $g^{\mu\nu}=\bar{g}^{\mu\nu}(t)+\delta g^{\mu\nu}(\vec{x},t)$. With $V''(\phi_0)=m^2_\phi$ and $U''(\chi_0)=m^2_\chi$ the Lagrangian up to second order in perturbations is
\begin{align} \mathcal{L}^{(2)}&=-m^2_{pl}(3H^2(t)+\dot{H}(t))+m^2_{pl}\dot{H}g^{00}\\
&-\frac{1}{2}\partial_\mu\chi\partial^\mu\chi-\frac{1}{2}\left(m^2_\chi-2\frac{\dot{\phi}_0^2}{F^2}\right)\chi^2-2\frac{\chi_0}{F^2}\dot{\phi}^2\delta g^{00}\chi\end{align}
where the background equations of motion have been taken into account. This matches the EFT Larangian as a model with 
\be \alpha_1=1,~~\alpha_2=0,~~\alpha_3=m^2_\chi-2\frac{\dot{\phi}_0}{F^2}=m_\chi^2+\frac{4m^2_{pl}\dot{H}}{F^2},~~\alpha_4=0,~~m_i=0.\ee
Comparing the last line with \eqref{hiddenS} we can also read off that
\be \beta_1(t)=-2\frac{\chi_0}{F^2}\dot{\phi}_0^2=4\frac{\chi_0}{F^2R^2}m^2_{pl}\dot{H}.\ee
This sets the scale for Hidden preheating to be
\be R_1=\frac{\beta_1}{\sqrt{-2m^2_{pl}\dot{H}}}=-2\frac{\chi_0}{F^2R^2}\sqrt{-2m^2_{pl}\dot{H}}\sim\frac{\chi_0}{F^2R^2}\Lambda_{sb}^2.\ee
Also note that in the range $R_1>\omega_\pi$ and $R_1>\frac{\omega_\chi^2}{\omega_\pi}$, the canonical momentum of the effective degree of freedom $\pi_c$ is
\be p_\pi=-R_1\chi_c=4\frac{\chi_0}{F^2R^2}\sqrt{-2m^2_{pl}\dot{H}}~\chi_c.\ee

The $\chi_c$ production scale here is
\be K^2_{bck}=\left(\frac{2m^2_{pl}\ddot{H}}{F^2}\right)^{2/3}-m^2_\chi-4\frac{m^2_{pl}}{F^2}\dot{H}\ee
\be K^2_{bck}\sim\left(m_\phi\frac{\Lambda_{sb}^4}{F^2}\right)^{2/3}-m_\chi^2-4\frac{\Lambda_{sb}^4}{F^2}\ee
Corrections to $\phi$ dynamics here come with coefficients of $\frac{\chi_0}{F}$, which makes them perturbative corrections as long as $\chi_0\ll F$. This in return implies that $\frac{\chi_0}{R^2}=\frac{\chi_0}{\left(1+2\frac{\chi_0^2}{F^2}\right)^2}\sim\chi_0$, and the particle production scale will lie above the coupling $R_1^2\sim\frac{\chi_0^2}{F^2}\frac{\Lambda_{sb}^4}{F^2}$.

The derivative couplings preserve the shift symmetry of the inflaton. Hence they provide a very likely candidate for couplings of the inflaton with other fields. This also makes them more likely to be present in the later stages then nonderivative couplings, such as the original $g^2\phi^2X^2$ interaction considered for preheating. Here we have studied what type of perturbative degrees of freedom they are capable of and at what levels, which is complementary to the question of their efficiency in resonance.
\subsubsection{Hidden Preheating with $\beta_2(t)$:}

The quadratic Lagrangian in the presence of $\beta_2$ is  
\be \mathcal{L}^{(2)}=\frac{1}{2}\dot{\chi}_c^2+\frac{1}{2}\dot{\pi}_c^2-\frac{1}{2a^2}(\partial_i\chi_c)^2-\frac{1}{2a^2}(\partial_i\pi_c)^2-\frac{1}{2}m^2_\chi\chi_c^2-\frac{1}{2}m^2_\pi\pi_c^2+2\rho_2\dot{\pi}_c\dot\chi_c-\rho_2\frac{\ddot{H}}{\dot{H}}\pi_c\dot\chi_c\ee
where $\rho_2\equiv\frac{\beta_2}{\sqrt{-2m^2_{pl}\dot{H}}}$. In the previous case $R_1$ had mass dimension one and hence it defined a scale, but $\rho_2$ is dimensionless. Moreover different then the case with $\beta_1$, here $\chi_c$ appears with time derivatives and there are two derivative couplings. During preheating we can approximate $\omega_\pi\sim m_\phi$, moreover we know that $\ddot{H}\sim m_\phi H^2$. This puts the two terms $\rho_2\frac{\ddot{H}}{\dot{H}}\pi_c\dot{\chi}_c$ and $\rho_2\dot{\pi}_c\chi_c$ on equal footing. Comparing the term $\rho_2\dot{\pi}_c\dot{\chi}_c$ with the kinetic terms we notice that this derivative coupling dominates the kinetic energy of $\chi_c$ for  $\rho_2>\frac{\omega_\chi}{\omega_\pi}$ and dominates the kinetic energy of $\pi_c$ for $\rho_2>\frac{\omega_\pi}{\omega_\chi}$. The coefficient $\rho_2$ is a dimensionless number, which can at most be order 1. In terms of our assumptions for EFT coefficients, its value would depend on the ratio between the inflaton mass to the planck scale and the Hubble rate $\rho_2\sim\mathcal{O}\left(\frac{m_{pl}}{m_\phi}\frac{H}{m_\phi}\right)$. Rather then defining a scale, effects of $\rho_2$ become important in terms of the ratio between frequencies of the two species present. Unless $\omega_\chi=\omega_\pi$, there exist two opposite regimes
\be \text{{\bf regime 1:}}~~ 1\geq\rho_2>\frac{\omega_\chi}{\omega_\pi}~~\text{implying}~~\omega_\pi>\omega_\chi\ee
\be\text{{\bf regime 2}}~~1\geq\rho_2>\frac{\omega_\pi}{\omega_\chi}~~\text{implying}~~\omega_\chi>\omega_\pi.\ee

In regime 1, the derivative couplings are more dominant to the kinetic energy of $\chi_c$, but subdominant to kinetic energy of $\pi_c$. Which makes
\be \label{lagreg1} \mathcal{L}_{reg1}^{(2)}\simeq  \frac{1}{2}\dot{\pi}_c^2-\frac{1}{2a^2}(\partial_i\chi_c)^2-\frac{1}{2a^2}(\partial_i\pi_c)^2-\frac{1}{2}m_\chi^2(t)\chi^2-\frac{1}{2}m^2_\pi(t)\pi^2_c.\ee
Here $\chi_c$ is not a dynamical degree nor does it play into the canonical momenta for the inflaton perturbations. Since it won't go into resonance, we can also neglect the $(\partial_i\chi_c)^2$ terms and conclude that in this regime $\chi_c$ only contributes to the overall cosmological constant term.

In regime 2, with $\rho_2>\frac{\omega_\pi}{\omega_\chi}$, where $\omega_\chi>\omega_\pi$, the derivative couplings are greater then the kinetic energy of $\pi_c$ but less then the kinetic energy of $\chi_c$. The Lagrangian in this regime looks like
\be \label{lagreg2} \mathcal{L}_{reg1}^{(2)}\simeq  \frac{1}{2}\dot{\chi}_c^2-\frac{1}{2a^2}(\partial_i\chi_c)^2-\frac{1}{2a^2}(\partial_i\pi_c)^2-\frac{1}{2}m_\chi^2(t)\chi^2-\frac{1}{2}m^2_\pi(t)\pi^2_c.\ee In this regime the reheating particles, $\chi_c$ are the effective degrees of freedom and they undergo resonance, and the presence of inflaton perturbations $\pi_c$ are completely negligable!

In the case of Hidden Preheating with $\beta_1$ there was a distinct energy range of at what scales modes of which species were effective. Here, in the presence of $\beta_2$, the effective appearance of a mode depends on its relative frequency of the mode it couples to! We show the phase space for this situation in figure \ref{fig:beta2modes}.
\begin{figure}[h]
	
	\includegraphics[width=\textwidth]{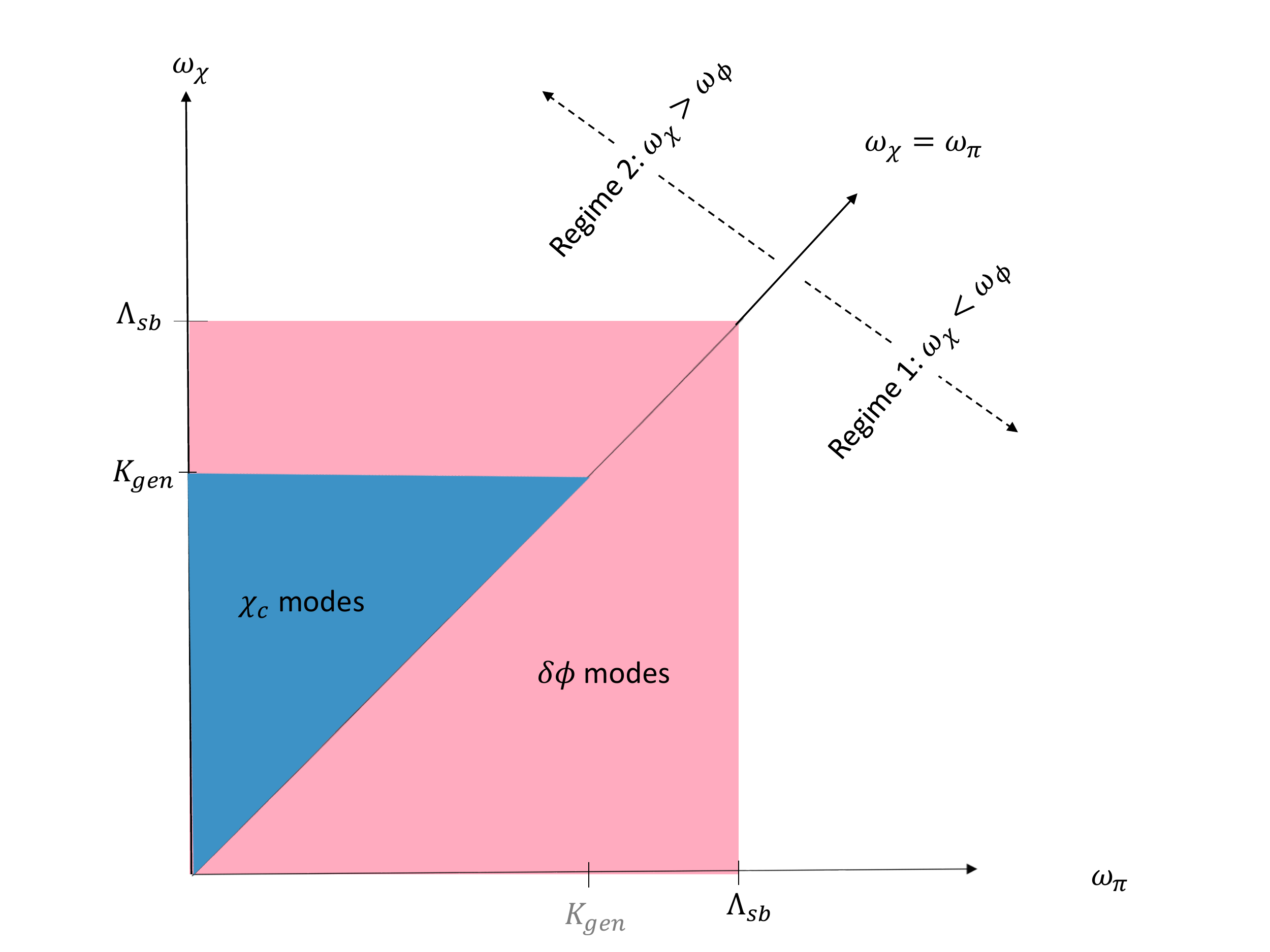}
	\caption{Here we show the regions where the inflaton modes $\delta\phi=\pi_c$ and reheating modes $\chi_c$ appear as effective degrees of freedom.}
	\label{fig:beta2modes}
\end{figure}
For the moment there are no examples we know of that fall into this category. We note again that, while the operator $\beta_1$ defined a scale $R_1$, the strength of the coupling $\rho_2$ that the operator $\beta_2$ defines, determines only the relative frequency of inflaton to reheating field. This means when a certain inflaton mode at a certain frequency is coupled to a specific mode of reheating sector with a specific frequency, out of the two modes under question the one with the greater frequency will be the mode that is the effective degree of freedom.
\subsubsection{Hidden Preheating by $\beta_3$:}

The quadratic Lagrangian in this case is
\begin{align}&\mathcal{L}^{(2)}=\frac{1}{2}\dot{\chi}^2_c+\frac{1}{2}\dot{\pi}_c^2-\frac{(\partial_i\chi_c)^2}{2a^2}-\frac{(\partial_i\pi_c)^2}{2a^2}-\frac{m_\chi^2(t)}{2}\chi_c^2-\frac{m^2_\phi(t)}{2}\pi_c^2\\
&-\frac{\dot{\beta}_3\dot{\chi}_c\pi_c}{\sqrt{-2m^2_{pl}\dot{H}}}-\frac{\ddot{H}}{\dot{H}}\frac{\beta_3\pi_c\dot{\chi}_c}{\sqrt{-2m^2_{pl}\dot{H}}}-\frac{\beta_3\dot{\pi}_c\dot{\chi}_c}{\sqrt{-2m^2_{pl}\dot{H}}}-\frac{\ddot{\beta}_3\pi_c\chi_c}{\sqrt{-2m^2_{pl}\dot{H}}}.\end{align}
This time we have different couplings at different levels,
\be \rho_3\equiv\frac{\beta_3}{\sqrt{-2m^2_{pl}\dot{H}}},~~ R_2\equiv\frac{\ddot{\beta}}{\sqrt{-2m^2_{pl}\dot{H}}},~~R_3\equiv\frac{\dot{\beta}_3}{\sqrt{-2m^2_{pl}\dot{H}}}\ee
\begin{align}\mathcal{L}^{(2)}=&\frac{1}{2}\dot{\chi}^2_c+\frac{1}{2}\dot{\pi}_c^2-\frac{(\partial_i\chi_c)^2}{2a^2}-\frac{(\partial_i\pi_c)^2}{2a^2}-\frac{m_\chi^2(t)}{2}\chi_c^2-\frac{m^2_\phi(t)}{2}\pi_c^2\\
&-R_3\dot{\chi}_c\pi_c-\frac{\ddot{H}}{\dot{H}}\rho_3\pi_c\dot{\chi}_c-\rho_3\dot{\pi}_c\dot{\chi}_c-R_2\pi_c\chi_c.
\end{align}
$\beta_3$ has mass dimension 2. This makes $\rho_3$ a dimensionless coupling strength, just like $\rho_2$. We would again expect that $\omega_\pi\sim m_\phi$ putting these two terms on equal footing.  $R_2$ and $R_3$ are the dimensionfull parameters that can set the scales here. The $R_2$ term only contributes to the over all energy. $R_3$ has dimensions of mass and works similar to $R_1$ here. So let us focus on $R_3$. This term can dominate over the kinetic energies of both $\pi_c$ and $\chi_c$ and the derivative couplings with $\rho_3$ if
\be R_3>\omega_\chi,\omega_\pi~~\text{and}~~R_3>\rho_3\omega_\pi.\ee
As $\rho_3\leq1$ the later part of the condition is automatically satisfied if the first part is already met. 
In this regime
\be\mathcal{L}^{(2)}\simeq -R_3\dot{\chi}_c\pi_c-\frac{1}{2a^2}(\partial_i\chi_c)^2-\frac{1}{2a^2}(\partial_i\pi_c)^2-\frac{1}{2}m^2_\chi(t)\chi_c^2-\frac{1}{2}m^2_\pi(t)\pi_c^2-R_2\pi_c\chi_c.\ee
This is an example where the $\chi_c$ modes are the lightest degree of freedom, and $\pi_c$ plays the role of canonical momenta
\be p_\chi\equiv\frac{\partial\mathcal{L}}{\partial\dot{\chi}_c}=-R_3\pi_c\ee
Which is similar to the case of Hidden preheating with $\beta_1$, only this time the roles of the two fields are switched around. 

The scale $R_3$ defines, the scale up to which $\chi_c$ is the single effective species, is around $\mathcal{O}\left(\frac{m_{pl}H}{m_\phi}\right)$ order of magnitudewise. In scales where we understand $\sqrt{m_{pl}H}$ to set the unit scale, the relationship between the magnitude of this scale to the $\chi$-production scale is $K_{bck}=(m_{pl}H)^{1/3}R_3^{1/3}\sim R_3^{1/3}$. As $K_{bck}$, $R_3$ are smaller than unit scale, $R^{1/3}$ is bigger then $R_3$. In conclusion at frequencies below this scales, $E<R_3$, the effective modes are reheating modes alone where as at scales $R_3<E<K_{bck}$ modes of both $\pi_c$ and $\chi_c$ are present, and above $K_bck$ there is only inflaton perturbations. We summarize these scales and the corresponding species in figure \ref{fig:beta3modes}.

\begin{figure}[h]
	\includegraphics[width=\textwidth]{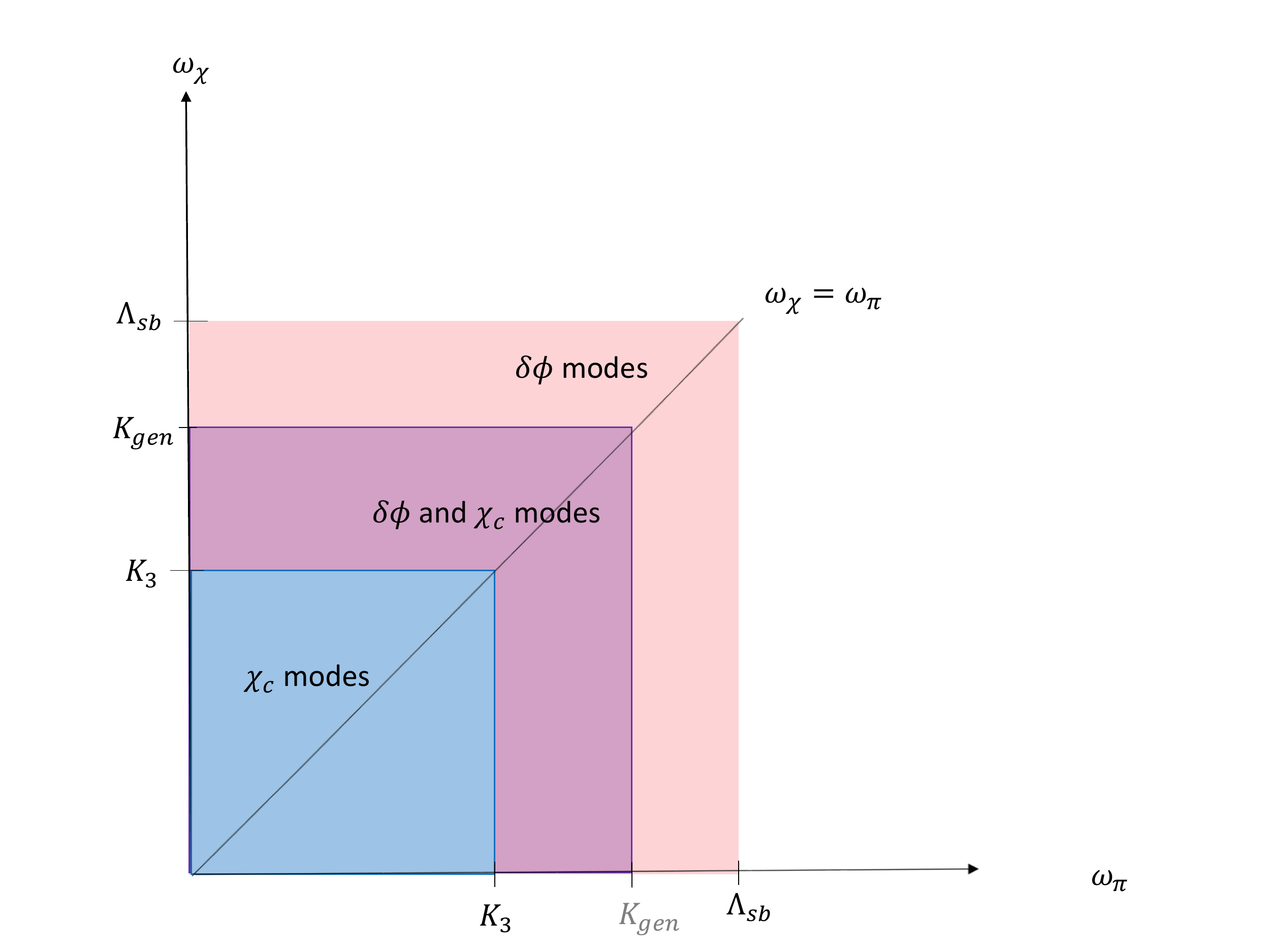}
	\caption{Here we show the regions where the inflaton modes $\delta\phi=\pi_c$ and reheating modes $\chi_c$ appear as effective degrees of freedom in the presence of $\beta_3$ interactions.}
	\label{fig:beta3modes}
\end{figure}

To summarize this section, due to the presence of derivative couplings, we asked under what conditions the system would have modes of a single species of either the inflaton or the reheating perturbations. We found that with $\beta_1$ type couplings, there is an energy scale $R_1$ below which the inflaton modes are the single species present. This means, irrespective of the coupling strength any $\beta_1$ type model will have a phase where only $\pi_c=\delta\phi$ are present. For $\beta_2$ type couplings it turned out that depending on the relative frequency of the $\chi_c$ and $\pi_c$ modes that couple with each other, only the one with greater frequency shows up to be present. In the presence of $\beta_3$ we noted that the lowest species is the reheating perturbations, up to the scale $R_3>\omega_\chi,\omega_\pi$.

Above the quadratic order, the presence of $\beta_i$ can give rise to the decay of $\chi_c$ into $\pi_c$ via the couplings $\beta_1(t)\partial_\mu\pi\partial^\mu\pi\chi$, $\beta_2(t)\partial^0\pi\partial_\nu\pi\partial^\nu\chi$ and $\beta_2(t)\partial_\mu\pi\partial^\mu\pi\partial_\nu\pi\partial^\nu\chi$. This is opposite of the intuition that the inflaton would decay into the reheating sector. 


\subsection{The Eventual Ladder of Scales}
To summarize, in the previous sections we focused on the scales associated with the production of $\chi_c$ modes, self resonance of the inflaton and strong mixing between the inflationary and reheating sector. These are the processes available during preheating below the symmetry breaking scale. By making general assumptions about the EFT coefficients $\{\alpha_i(t),\beta_i(t)\}$ we were able to obtain an overall magnitude for the strength of each of these scales. To remind ourselves, the scale of symmetry breaking, up to which this EFT formalism holds is
\be \Lambda^2_{sb}(H,m_2)=\sqrt{-2m^2_{pl}\dot{H}c_\pi}.\ee
The scale up to which $\chi_c$ modes will be set to resonance by the background is
\be K^2_{bck}(\alpha_i)=\frac{1}{c^2_\chi}\left[\left(\frac{\dot{\alpha}_3-\ddot{\alpha}_4}{2}\right)^{2/3}-\frac{\alpha_3-\dot{\alpha}_4}{\alpha_1}\right]\sim\frac{1}{c^2_\chi}\left(\frac{m^2_{pl}H^2_m}{m_\phi}\right)^{2/3}.\ee
There is also the possibility of $\chi$ production via its sound speed, which can take place up to the lower scale of
\be K_{c_\chi}=\frac{1}{2}\frac{\dot{\alpha}_1}{\alpha_1^{3/2}}\sim \frac{m^3_\phi}{m_{pl}H_m}.\ee
And we found that in the presence of $\beta_1$ there is the possibility of strong mixing between inflationary and reheating sectors below the scale 
\be R^2_1(\beta_1)= \frac{\beta^2_1}{2m^2_{pl}|\dot{H}|}\sim \frac{\Lambda^4_{sb}}{m^2_\phi} \ee
Lastly, the inflaton fluctuation can go into self resonance at
\be E^2_{\delta\phi}=\frac{1}{c_\pi}\frac{\ddot{H}}{\sqrt{-\dot{H}}}\sim \frac{1}{c_\pi}m_\phi H_m.\ee

We started our discussion by noting that to study preheating we are focusing at times where $m_\phi >H_{pre}\sim H_m$. As such, at the background level we have the hierarchy $m_{pl}> m_\phi>H_{pre}\sim H_m$, present at the epoch of interest. The background scales appear in associated scales with interactions. Hence the hierarchy at the background level implies certain hierarchies among the processes considered. This hierarchy, in the basic version of without the derivative couplings, ie $\beta_i=0$, is listed in Figure \eqref{scales_beta0}.  We have seen how the presence of field content affect the species that are present at different scales. We represent this effect among the scales in Figure \eqref{scales_withbeta1} with $\beta_1\neq0$.
\begin{figure}
	\begin{subfigure}[b]{0.5\textwidth}
		\includegraphics[width=\textwidth]{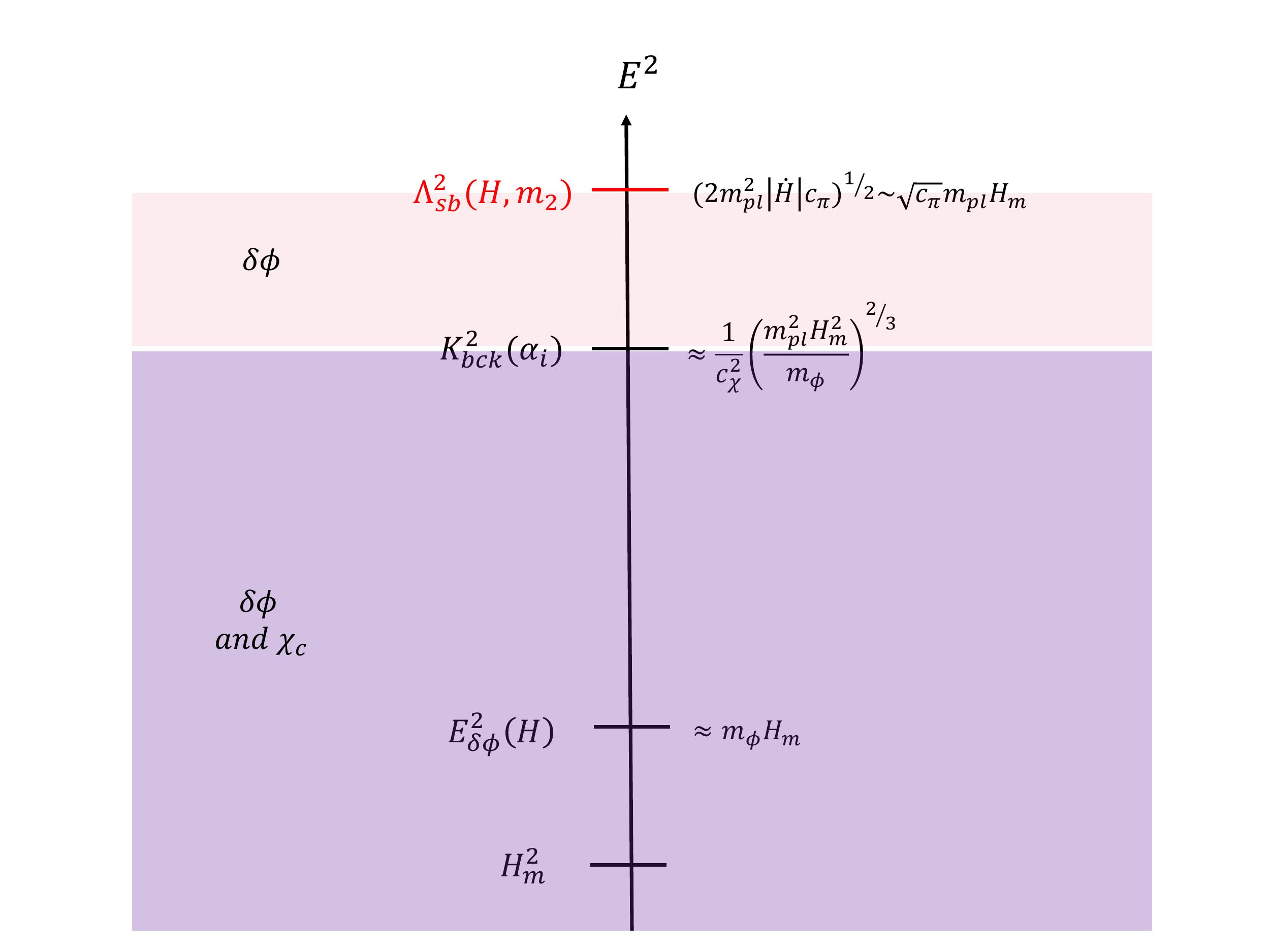}
		\caption{}
		\label{scales_beta0}
	\end{subfigure}
	~
	\begin{subfigure}[b]{0.5\textwidth}
		\includegraphics[width=\textwidth]{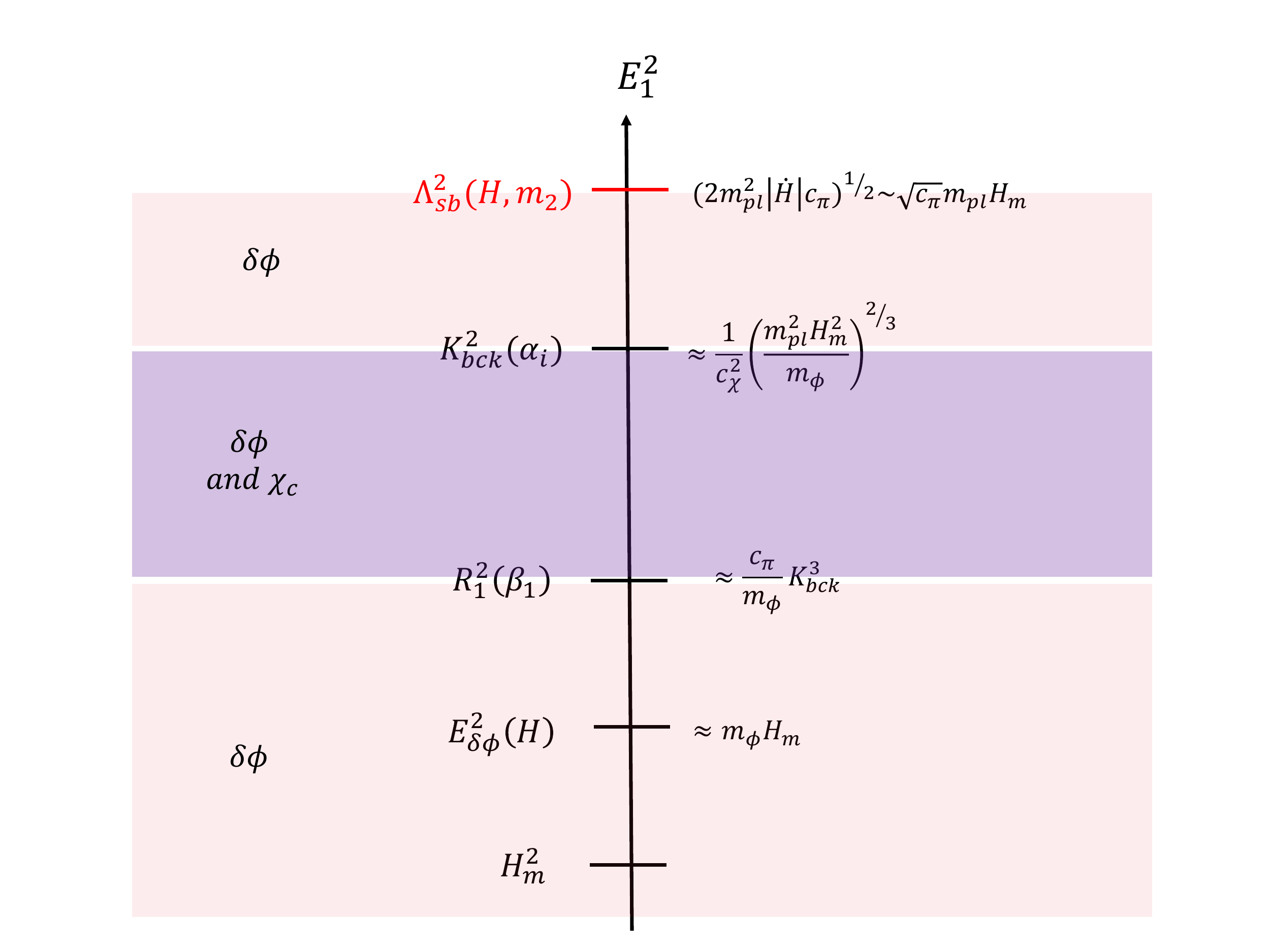}
		\caption{}
		\label{scales_withbeta1}
	\end{subfigure}
	\caption{The hierarchy between the scales and the species present in the EFT (a) with $\beta_i=0$ and (b) with $\beta_1\neq0$.}
\end{figure}




\section{What Happens to Adiabatic Modes and Who is the Goldstone Boson during Preheating?}
\label{sec:adiabaticmodes}
In section \ref{sec:gaugeinvvar} we introduced the quantities $\zeta$ and $\mathcal{R}$. Their relevance is that these are known to be conserved quantities for superhorizon inflationary modes, and hence through them one can gain knowledge of inflation from today's observations. We have mentioned their connection to metric and field perturbations
\be\label{4.5zeta} \zeta=\frac{A}{2}-H\frac{\delta\rho}{\dot{\bar{\rho}}},\ee
\be\label{4.5R}\mathcal{R}=\frac{A}{2}+H\delta u,\ee
and derived the governing equation for their  dynamics
\be\label{dzeta_X}\dot{\zeta}=\frac{\partial_i \delta{T^i}_0}{3(\bar{\rho}+\bar{p})}+\frac{\dot{\bar{\rho}}\delta p-\dot{\bar{p}}\delta\rho}{3(\bar{\rho}+\bar{p})^2},\ee
\be \label{rdot} 3\left(\bar{\rho}+\bar{p}\right)\dot{\mR}=-\delta\dot{\rho}+\partial_i\delta{T^i}_0-3H\delta\rho-H\delta{T^i}_i+3\left(\bar{\rho}+\bar{p}\right)\left[\dot{H}\delta u+H\delta\dot{u}\right].\ee
At the time of their introduction \cite{Bardeen:1980kt}, \cite{Lyth:1984gv}, the scientific community was not certain that these scalar quantities really referred to physical degrees of freedom. There was the concern that these may just be gauge modes. In \cite{Weinberg:2003sw} Weinberg came up with an argument to prove that they are physical and that on superhorizon scales they are conserved. By studying perturbed Einstein's equations at linear order in Newtonian gauge, he showed that in the limit $k\to 0$, $\zeta$ and $\mathcal{R}$ are conserved. He proved they are physical by showing that solutions of the field equations for $\zeta$ and $\mathcal{R}$ in the limit $k\to 0$ can be continued to $k\neq 0$. Here we want to check the adiabaticity of $\zeta$ during preheating. 

If certain conditions are satisfied, the reheating sector can cause growth in $\zeta$ \cite{Finelli:2000ya}. Here we focus on linear order perturbation theory. As long as a single field sets the evolution of the background, there is a difference in the introduction of the perturbations of the dominant sector and other sectors. The EFT formalism captures this difference while taking into account all possible generalizations of interactions between the perturbations. In the initial stages of preheating it is the inflaton $\phi_0(t)$ that drives the dynamics of the background $H_p(t)$. Hence the Goldstone boson is associated to scalar perturbations of the inflaton field $\delta\phi$, while the scalar field $\chi$ associated with the reheating sector is introduced as a perturbation. Goldstone modes are expected to be massless. However we saw in the earlier sections that the preheating background $H_p(t)$, can give rise to mass terms for the inflaton perturbations, $\delta\phi=\pi_c$. This raises the question of who is the Goldstone mode? By keeping track of the interactions in two different gauges, we will clarify that $\zeta$ is the variable that remains massless during preheating and can be identified as the Goldstone mode at all times rather then $\pi$ or $\pi_c$. $\zeta$ being defined at linear order, the Goldstone modes are associated only with the inflaton sector. The subtlety here is that their properties are more clear in the gauge where the fact that they are associated to the inflationary sector is not explicit.


By now we have introduced the effective field theory for perturbations for an era where the energy density is dominated by a single field species (this field being the inflaton here) in unitary gauge where constant time surfaces are aligned with the background, $\phi(x,t)=\phi_0(t)$ and $\delta\phi=0$. Within this temporal gauge the remaining spatial metric components can be fixed such that $g_{ij}=a^2(t)(1+2\zeta)\delta_{ij}$.
In our general decomposition of the metric \eqref{dec100}-\eqref{dec1ij}, the scalar contributions are
\be \label{scalarsinthemetric}ds^2=-(1+2E)dt^2+2a\partial_iFdx^idt+a^2\left[(1+2A)\delta_{ij}+2\partial_i\partial_jB\right]dx^idx^j,\ee 
and this spatial gauge choice amounts to setting  $A=\zeta$ in our decomposition of \eqref{dec1ij}, or in terms of more contemporary literature $\Psi=\zeta$. As we saw in terms of ADM decomposition, the metric components $g_{00}=-(1+2E)= -(\lambda^N)^2$ and $g_{0i}=2a(t)\partial_iF=2\lambda^i$, where we are focusing only on the scalar components, are not completely random, they need to be determined via solving the constraint equations. For us the linear order solutions are enough. However these solutions do depend on second order terms in the action. 
With the assumption $\lambda^N=1+\delta N$ where $\delta N$ is a small perturbation and hence $g_{00}\sim -(1+2\delta N)$, for $m_i=0$ in the $\zeta$ gauge \cite{Senatore:2016aui}
\be\label{admsolninzeta} \lambda^N=1+\frac{\dot{\zeta}}{H},~~\lambda_i=\partial_i\left(-\frac{1}{a^2}\frac{\dot{\zeta}}{H}-\frac{\dot{H}}{H^2}\frac{1}{\partial^2}\dot{\zeta}\right).\ee
Note that these solutions, at the perturbative level, involve terms that all come with time derivatives of $\zeta$. This fact does not change even when $m_i\neq 0$, as studied in section 3.3 of \cite{Creminelli:2006xe}. From now on we will refer to this gauge, which was historically first introduced in \cite{Maldacena:2002vr}, as the ``$\zeta$-gauge''. 
\begin{align}\nn \text{$\zeta$-gauge:}~~& \delta\phi=0,~\phi(x,t)=\phi_0(t),~g_{00}\sim-\left(1+2\frac{\dot{\zeta}}{H}\right),\\
\label{eq:zetagauge}&~g_{i0}\sim\mathcal{O}\left(\dot{\zeta}\right),~g_{ij}=a^2(t)\left(1+2\zeta\right)\delta_{ij}.\end{align}
Since there are no perturbations to the scalar matter content, at linear order for this gauge choice $\delta\rho=0$. Hence the gauge invariant quantity $\zeta$ appears explicitly as a scalar mode in $g_{ij}$.
Prior to gauge fixing, the perturbed Einstein Equations at linear order \cite{Baumann:2009ds} are
\begin{subequations}
	\begin{align}
	\label{lee1}3H\left(-\dot{A}+HE\right)+\frac{k^2}{a^2}\left[-A+H\left(a^2\dot{B}-aF\right)\right]=-\frac{\delta\rho}{2m^2_{pl}}\\
	\label{lee2}-\dot{A}+HE=-\frac{(\bar{\rho}+\bar{p})\delta u}{2m^2_{pl}}\\
	\label{lee3}-\ddot{A}-3H\dot{A}+H\dot{E}+\left(3H^2+2\dot{H}\right)E=\frac{1}{2m^2_{pl}}\left(\delta p-\frac{2}{3}k^2\sigma\right)\\
	\label{lee4}\left(\partial_t+3H\right)\left(\dot{B}-\frac{F}{a}\right)-\frac{A+E}{a^2}=\frac{\sigma}{m^2_{pl}}
	\end{align}
\end{subequations}
where $\sigma$ is the anisotropic stress defined via \eqref{dTijform}. The first two here are the constraint equations. The energy conservation equation is
\begin{align} \label{energycons}
\delta\dot{\rho}+3H\left(\delta\rho+\delta p\right)=\frac{k^2}{a^2}\delta q+\left(\bar{\rho}+\bar{p}\right)\left[-3\dot{A}+k^2\left(\dot{B}+\frac{F}{a}\right)\right]
\end{align}

Setting $E=0$, $A=\zeta$, in the case of zero anisotropic stress, via equation \eqref{lee1} the $\zeta$ gauge immediately gives $\dot{\zeta}=0$ in the limit $k\to 0$. Note that this wouldn't have happened if $\lambda^N,~\lambda^i$ had nonderivative dependence on $\zeta$. So the constancy of $\zeta$ is very much explicit and applicable to all eras in the $\zeta$ gauge. And this follows because the effective field theory of perturbations satisfies perturbed Einstein Equations. 

In the choice of the time coordinate such that the scalar perturbation presents itself in the matter sector $\delta\phi=\dot{\phi}_0\pi$, with the transformation property of $\pi\to\pi-\xi^0$ under time diffeomorphisms $t\to t+\xi^0$. 
Here the ADM variables are \cite{Baumann:2011su}
\be \label{admsolninpi}
E=\delta N=-\frac{\dot{H}}{H}\pi,~~\partial^i\lambda_i=\frac{1}{c^2_\pi}\frac{\dot{H}}{H}\dot{\pi}.\ee
From here on we will refer this gauge choice as the "$\pi$-gauge" which is, with focus on the scalar components
\be \label{eq:pigauge} \text{$\pi$-gauge:}~~\phi(x,t)=\phi_0(t)+\delta\phi,~g_{00}\sim-1+2\frac{\dot{H}}{H}\pi,~\partial^ig_{i0}=2\partial^i\left(\frac{1}{c^2_{\pi}}\frac{\dot{H}}{H}\dot{\pi}\right),~g_{ij}=a^2(t)\delta_{ij}.\ee
Because $\zeta$ is a gauge invariant variable, its properties will not change from one gauge to another. However, let us explicitly work out $\zeta$ dynamics in the $\pi$-gauge.

\subsection{$\zeta$ dynamics in the $\pi$-gauge}
In $\pi$-gauge, if we ignore complicated model dependent mixing terms, that is set the EFT parameter $\beta_i=0$, we expect the dynamics during preheating to be explainable by the matter action
\begin{align} \label{matterlag}
\nn&\mathcal{S}_m=\int d^4x\sqrt{-g}\Bigg[ -m^2_{pl}\left(3H^2(t+\pi)+\dot{H}(t+\pi)\right)\\
\nn&+m^2_{pl}\dot{H}(t+\pi)\left(g^{00}+2g^{0\mu}\partial_\mu\pi+g^{\mu\nu}\partial_\mu\pi\partial_\nu\pi\right)+\frac{1}{2}m^4_2(t+\pi)\left(\delta g^{00}+2g^{0\mu}\partial_\mu\pi+g^{\mu\nu}\partial_\mu\pi\partial_\nu\pi\right)^2\\
&-\frac{\alpha_1(t+\pi)}{2}g^{\mu\nu}\partial_\mu\chi\partial_\nu\chi+\frac{\alpha_2(t+\pi)}{2}\left(g^{0\mu}\partial_\mu\chi+g^{\mu\nu}\partial_\mu\pi\partial_\nu\chi\right)^2-\frac{\alpha_3(t+\pi)}{2}\chi^2\Bigg].\end{align}
Our aim in this section is to write the explicit dependence on $\pi$ and background dynamics for $\zeta$ and $\dot{\zeta}$ arising from \eqref{matterlag} to first order in perturbations. Any $\chi$ involvement would present itself starting from second order.

To reach $\zeta$ and $\dot{\zeta}$ we must first obtain the stress energy momentum tensor
\be T_{\alpha\beta}=-\frac{2}{\sqrt{-g}}\frac{\delta \mathcal{S}_m}{\delta g^{\alpha\beta}}.\ee
At the background level, that is neglecting any $\delta g^{\mu\nu}$, $\pi$ and $\chi$ contributions, we achieve the familiar expressions
\begin{align}
\bar{T}_{00}&=3m^2_{pl}H^2\\
\bar{T}_{0i}&=0\\
\bar{T}_{ij}&=-m^2_{pl}\bar{g}_{ij}\left(3H^2+2\dot{H}\right),\end{align}
which imply $\bar{\rho}=3m^2_{pl}H^2$ and $\bar{p}=-m^2_{pl}(3H^2+2\dot{H})$.

To first order in perturbations, by expanding the time dependent functions $f(t+\pi)$ to linear order in $\pi$ and including in $\delta g^{\mu\nu}$, after a straightforward calculation\footnote{ For example the term $\mathcal{L}_m\supset \sqrt{-g}m^2_4g^{00}g^{0\mu}\partial_\mu\pi$ leads to the term $m^4_2\bar{g}^{0\mu}\partial_\mu\pi$ in $\delta T_{00}$. And the perturbations to the inverse metric are connected to the perturbations to the metric by $\delta g^{\mu\nu}\equiv g^{\mu\nu}-\bar{g}^{\mu\nu}=-\bar{g}^{\mu\rho}\bar{g}^{\nu\sigma}\delta g_{\rho\sigma}$ so that $\delta g^{00}=1+g^{00}=-\delta g_{00}$.} we have 
\begin{align}
\label{dT00}\delta T_{00}&=m^2_{pl}\left[6H\dot{H}\pi-2\dot{H}\dot\pi-\left(3H^2+\dot H\right)\delta g_{00}\right]+m^4_2\Big[3\delta g_{00}+2\dot\pi\Big]\\
\label{dT0i}\delta T_{0i}&=-m^2_{pl}(3H^2+2\dot{H})\delta g_{0i}-2m^2_{pl}\dot{H}\partial_i\pi\\
\label{dTij}\delta T_{ij}&=-m^2_{pl}(3H^2+2\dot{H})\delta g_{ij}-m^2_{pl}\bar{g}_{ij}\left(\dot{H}\delta g_{00}+(6H\dot{H}+2\ddot{H})\pi+2\dot{H}\dot{\pi}\right).\end{align}
Equation \eqref{dT00} defines the perturbation to the energy density to be
\begin{align}
\nn\delta \rho=&\delta T_{00}+\bar{\rho}\delta g_{00}\\
\label{drho}=&m^2_{pl}\left[6H\dot{H}\pi-2\dot{H}\dot\pi-\dot H\delta g_{00}\right]+m^4_2(t)\Big[3\delta g_{00}+2\dot\pi\Big].\end{align}
The components of $\delta T_{ij}$ are related to the pressure perturbation and anisotropic stress $\sigma$ as
\be \label{dTijform} \delta T_{ij}=\bar{p}\delta g_{ij}+\bar{g}_{ij}\delta p+a^2(\partial_i\partial_j\sigma+\partial_i\sigma^j+\partial_j\sigma^j+\sigma^T_{ij})\ee
which by comparison with \eqref{dTij} implies that there is no anisotropic stress tensor involved and the pressure perturbation is
\be \label{ppert} \delta p=-m_{pl}^2\left(\dot{H}\delta g_{00}+(6H\dot{H}+2\ddot{H})\pi+2\dot{H}\dot{\pi}\right).\ee 
Thus in the $\pi$-gauge, with $A=B=0$, the gauge invariant quantity \eqref{4.5zeta} is
\be\label{ourzeta}\zeta=-H\pi+\frac{1}{3}\dot{\pi}+\frac{1}{6}\delta g_{00}-\frac{m^4_2}{m^2_{pl}\dot{H}}\Big[\frac{1}{2}\delta g_{00}+\frac{1}{3}\dot{\pi}\Big].\ee

In the long wavelength limit, where the terms with spatial derivatives in \eqref{dzeta_X} drop out, we obtain
\begin{align}\label{ourzetadot}
\dot{\zeta}_{k\to 0}=-\left(H+\frac{\ddot{H}}{6\dot{H}}\right)\left[\delta g_{00}+2\dot{\pi}\right]+\frac{m^4_2(t)}{m^2_{pl}}\left(\frac{H}{\dot{H}}+\frac{\ddot{H}}{3\dot{H}^2}\right)\left[\frac{3}{2}\delta g_{00}+\dot{\pi}\right].\end{align}
In $\pi$-gauge $\delta g_{00}=2\frac{\dot{H}}{H}\pi$, which leads to
\be \label{dzeta}\dot{\zeta}_{k\to 0}=-2\left(H+\frac{\ddot{H}}{6\dot{H}}\right)\left[\frac{\dot{H}}{H}\pi+\dot{\pi}\right]+\frac{m^4_2(t)}{m^2_{pl}}\left(\frac{H}{\dot{H}}+\frac{\ddot{H}}{3\dot{H}^2}\right)\left[\frac{3\dot{H}}{H}\pi+\dot{\pi}\right].\ee
At first sight this equation for $\zeta$ is not applicable during preheating, or any other era where an oscillating scalar field dominates, due to the presence of $\dot{H}$ in the denominator in equation \eqref{dzeta}. So we have derived the known result \cite{Finelli:2000ya} based on the equation of state $\omega$, in the $\pi$ gauge. 

At this point we have more or less made a connection between the language of cosmological perturbations in the old lore and the $\pi$-language. We have seen that $\zeta$ is conserved in the $\zeta$ gauge and we know that it is conserved in the Newtonian gauge. So it must be conserved in the $\pi$ gauge also. We need to find the variable suited for this gauge.

\subsection{The conserved variable in $\pi$-gauge}

Previously we failed in our attempt to confirm the constancy of $\zeta$ via the energy momentum conservation equation. In order to find a time-independent variable in $\pi$ gauge, let us look at the perturbed Einstein Equations in $\pi$ gauge. Setting $A=B=0$, in the case of zero anisotropic stress these are
\begin{subequations}
	\begin{align}
	\label{leep1}-3H^2E+\frac{k^2}{a^2}(aHF)=\frac{\delta\rho}{2m^2_{pl}}\\
	\label{leep2}HE=-\frac{\delta q}{2m^2_{pl}}\\
	\label{leep3}H\dot{E}+(3H^2+2\dot{H})E=\frac{\delta p}{2m^2_{pl}}\\
	\label{leep4}(\partial_t+3H)\frac{F}{a}+\frac{E}{a}=0.
	\end{align}
\end{subequations}
And the continuity equation is
\be \label{dmuTmu0} \delta\dot{\rho}+3H(\delta\rho+\delta p)=k^2\left[\frac{\delta q}{a^2}+(\bar{\rho}+\bar{p})\frac{F}{a}\right].\ee
This will eventually give us our master equation for $\dot{\zeta}$. Equations \eqref{leep1} and \eqref{leep3} give $\delta\rho+\delta p$ in terms of the metric perturbations. By inserting the expression obtained this way into \eqref{dmuTmu0}, in the long wavelength limit we obtain
\be \delta\dot{\rho}+3m^3_{pl}\left(2H^2\dot{E}+4H\dot{H}E\right)=0.\ee
This suggests that the explicitly conserved quantity at long wavelengths, in the $\pi$-gauge is 
\be Q\equiv \delta\rho+6m^2_{pl}H^2E.\ee
What is this quantity $Q$? In $\pi$-gauge, $E=-\frac{\dot{H}}{H}\pi=-\frac{1}{2}\delta g_{00}$ and we know $\delta \rho$ from equation \eqref{drho} 
\begin{align} Q=-2m^2_{pl}\dot{H}\left[\dot{\pi}+\frac{\dot{H}}{H}\pi\right]+2m^4_2(t)\left[\dot{\pi}+3\frac{\dot{H}}{H}\pi\right].
\end{align}
Our $Q$ is let alone being conserved, exactly zero for exact de Sitter as $\dot{H}\to 0$ provided there is no sound speed term. This is because there are no physical scalar modes in exact de Sitter.  More importantly it suffers no divergences during preheating. To summarize, in the $\pi$-gauge with $A=B=0$, \be\zeta=-H\frac{\delta\rho}{\dot{\bar{\rho}}}=-\frac{\delta\rho}{6m^2_{pl}\dot{H}}=-H\pi+\frac{1}{3}\dot{\pi}+\frac{\dot{H}}{3H}\pi-\frac{m^4_2}{m^2_{pl}}\left[\frac{\pi}{H}+\frac{\dot{\pi}}{3\dot{H}}\right]\ee
in $\pi$-gauge. If sound speed terms are introduced $\zeta$ itself looks divergent as expressed in $\pi$-gauge both for inflation and preheating. In terms of familiar quantities,
\be Q=6m^2_{pl}\dot{H}\left(\frac{\delta g_{00}}{2}-\zeta\right),\ee
and in a sense it keeps track of the difference between the temporal metric perturbation and the scalar component of the spatial metric perturbation. 

We can conclude that adiabatic modes do persist to be adiabatic even during preheating, yet their connection with $\zeta$ is not explicit in every gauge. Weinberg himself faces the same problem in Synchronous gauge, where the conserved quantity turns out to be \cite{Weinberg:2003sw}
\be A_W=-k^2\mathcal{R}.\ee
And he acknowledges that this discrepancy between gauges is because the limit $k\to 0$ has a different meaning in different gauges.

\subsection{$\mR$ and its dynamics}
We can use the same logic as above. To remind ourselves in terms of general gauge parameters
\be\label{introR} \mR=\frac{A}{2}+H\delta u \ee
where $\delta u$ is the velocity perturbation. It is defined via the $0i$th component of the perturbed stress energy tensor as follows
\be\delta T_{i0}=\bar{p}\delta g_{i0}-\left(\bar{\rho}+\bar{p}\right)\partial_i\delta u.\ee 
Comparing this with our $\delta T_{i0}$ from \eqref{dT0i} we see that
\be \label{deltau} \delta u=-\pi.\ee
So we can recognize $\pi$, which nonlinearly realizes time translation invariance in the Lagrangian, as the perturbation in the scalar velocity. In terms of the $\pi$-gauge
\be \label{piR} \mR=-H\pi.\ee
There is clearly a difference between $\mR$ and $\zeta$ when they are expressed in this gauge.
In literature, on superhorizon scales
\be \label{dRlit} \dot{\mathcal{R}}_{k\to 0}=\frac{\dot{\bar{\rho}}\delta p-\dot{\bar{p}}\delta\rho}{3\left(\bar{\rho}+\bar{p}\right)^2}\equiv X=\dot{\zeta}_{k\to 0}.\ee 
Let us first check if we can obtain this relation in the effective field theory set up. Using the relation between the scalar metric perturbation and $\mR$, which is $A=2\mR-2H\delta u$ we can this time turn \eqref{eq:encon} into the following equation for $\dot{\mR}$
\be \label{dotRI}\dot{\mR}=\frac{\partial_i\delta {T}^i_0}{3\left(\bar{\rho}+\bar{p}\right)^2}+\frac{\dot{\bar{\rho}}\delta p-\dot{\bar{p}}\delta\rho}{3\left(\bar{\rho}+\bar{p}\right)^2}-4m^4_2(t)\frac{\ddot H}{\dot H}\dot{\pi}-\partial_t\left(4m^4_2(t)\dot{\pi}\right).\ee
This indeed produces \eqref{dRlit} on superhorizon scales for models with canonical kinetic terms, where $m_2=0$. But it brings us back to the same problems that we faced with $\dot{\zeta}$. Which will have the same resolution.
\subsection{Who is the Goldstone Boson?}
Having reconciled the adiabaticity of $\zeta$ throughout cosmic evolution, let us now try to expose the Goldstone boson in the EFT. Because $\pi$ is introduced in a fashion so as to nonlinearly realize time translations, we have been referring to it as the Goldstone boson. Its role as the inflaton perturbation becomes apparent once it is canonically normalized. However under this process $\pi$ gains a mass term
\be m_\pi^2=-3\dot{H}c^2_\pi-\frac{1}{4}\left(\frac{\ddot{H}}{\dot{H}}-2\frac{\dot{c}_\pi}{c_\pi}\right)^2-\frac{3H}{2}\left(\frac{\ddot{H}}{\dot{H}}-2\frac{\dot{c}_\pi}{c_\pi}\right)-\frac{1}{2}\partial_t\left(\frac{\ddot{H}}{\dot{H}}-2\frac{\dot{c}_\pi}{c_\pi}\right).\ee
Even prior to canonical normalization the sound speed terms give it mass via
\be m^4_2(t+\pi)\left(\delta g^{00}\right)^2\sim m^4_2\left(\frac{\dot{H}}{H}\right)^2\pi^2.\ee
Goldstone bosons have the distinguishing property of being massless, yet here a strong time dependence of the background is making $\pi$ massive. It is at least making the inflaton perturbation in $\pi$-gauge massive.

Perhaps we face this issue because we are raising the question in a gauge which is not suitable to make conclusions, just like the problem we faced with adiabaticity. $\zeta$ being a conserved quantity, suggests that it may also be the quantity that remains massless. This would then suggest that in $\pi$ gauge, the massless Goldstone boson is actually the quantity Q. Rather then rewriting our Lagrangian in terms of Q to check this, we will consider $\zeta$ in $\zeta$-gauge 
\be S|_{\zeta-gauge}\subset\int d^4x\sqrt{-g}\Bigg[\frac{m^2_{pl}}{2}R-m^2_{pl}\left(3H^2(t)+\dot{H}(t)\right)+m^2_{pl}\dot{H}(t)g^{00}+\frac{m^4_2(t)}{2!}\left(\delta g^{00}\right)^2\Bigg].\ee
The Ricci scalar involves the kinetic terms for $\zeta$ and because $\delta g^{00}$ carries only time derivatives of $\zeta$, in the $\zeta$-gauge it is clear that $\zeta$ remains massless. 

In summary, the perturbations of the field that sets the time dependence of the background are the Goldstone bosons that nonlinearly realize time translation invariance. They are also the modes that remain adiabatic on superhorizon scales at all times. All of these properties of the Goldstone boson are easier to see in the gauge where time is fixed parallel to the time dependence of the background, in other words in the $\zeta$-gauge. In this gauge the Goldstone boson enters the system as the scalar mode in the metric. The fact that the action on the whole respects all diffeomorphisms is revealed in another gauge, the $\pi$-gauge, at the cost of making the properties of being massless and adiabatic inexplicit. However because it is the gauge where the scalar mode associated with the time dependence of the background is explicitly connected to the perturbation of the field that causes this time dependence, $\pi$-gauge is the gauge where  the scale of decoupling, from gravity and also from any other species present, is easier to see.        
\section{Will $\langle\chi\rangle$ develop time dependence?}
\label{sec:timedepof chibck}
As we conclude our discussion on preheating from an effective field theory description at the perturbative level, this is the question we leave open for future work. Making use of nonlinearly realized time diffeomorphism, we discussed how a general Lagrangian for the early stages of preheating can be captured in the effective field theory of cosmological perturbations. This description is applicable for only the early stages because we made the following two assumptions
\begin{enumerate}
	\item{We assumed that a single field dominates the energy density, and this field is the one that drove inflation.}
	\item{The field that drove inflation gives the time dependence of the background.} 
\end{enumerate}
Our first assumption implies that the reheating field, which we represented as $\chi$, is only introduced as a perturbation. However, if preheating is successful, eventually the number density of $\chi$ will grow so much that its vacuum expectation value, $\langle\chi\rangle\equiv\chi_0$, starts contributing to the background energy density. So eventually assumption 1 fails. Yet in our description all the background information is brought out by $H(t)$. No where did we make use of $\phi_0$ or $\chi_0$! The Hubble parameter keeps track of these quantities for us. Hence the EFT description of two fields will more or less remain valid. What is more crucial is assumption 2.

If the background $\chi_0$ comes out to develop time dependence, it will start setting the choice for the time diffeomorphisms. The clock will start shifting from $\phi_0(t)$ to $\chi_0(t)$. In a sense all the success of preheating and eventually reheating the universe relies on this stage. It is important to be able to move forward from an effective field theory description of the perturbations to a description that can also account for the background behavior to capture this stage. And this is what we try to address in the last section. 

As the $\chi$-background develops, and if it develops to be time dependent, what happens in terms of the effective field theory for perturbations is that the role of $\pi$ and $\chi$ in our story switch. Remember that we have handled an EFT of types of perturbations where one was the perturbation to the clock field, $\pi$. Coming from inflation the clock is set by the inflaton field and hence $\pi$ represents inflationary perturbations. The moment the time dependence in $\chi_0=\chi_0(t)$ becomes negligible, one can still make use of the same EFT setup but now $\pi$ will be denoting reheating perturbations and the inflaton perturbations will be the secondary perturbations that just happen to be present.

\chapter{Towards an Effective Field Theory of the Backgrounds}
\label{chp:towardsEFTbackground}

In chapter \ref{chp:EFTpreheating} we talked about an effective field theory at the perturbative level, for perturbations of a reheating sector coupled to Goldstone modes of the inflatonary sector. In doing so, the emphasis was put on the nonlinearly realized nature of time diffeomorphisms, and the scale at which this arises set the scale of the EFT. The advantage of following this route was that we became aware of the relationship between scales of different processes. But we were not sure about the behavior of the specific backgrounds involved. We based the EFT on the background set by the Hubble parameter $H(t)$. In general $H(t)$ has contributions coming from both the inflationary background $\phi_0(t)$ and the background of any other sector present $\chi_0(t)$. We could study the perturbations $\pi_c$ and $\chi_c$ but we were blind to what happens with $\phi_0$ and $\chi_0$. Being able to distinguish between the background contribution of the sectors involved is important when one wants to understand the species that sets the definition for the choice of time, or determine the efficiency of energy transfer from inflaton field to the preheating field. These were the questions we raised in section \ref{sec:timedepof chibck}. We now turn towards exploring formalisms which may help provide a suitable framework to investigate such questions. 

\section{Adding a Second Scalar to Weinberg's EFT of Inflation}
\label{sec:addingtoweinberg}

An important goal of EFT's is to be able to capture corrections coming from what might have been allowed at a higher energy level, perturbatively. These corrections can arise because of extra degrees of freedom, or extra symmetries like in the case we saw previously. They will appear as higher order operators suppressed by the scale of high energy physics. In natural units, the order of the operators are determined by the dimensions of energy they carry. The Ricci scalar, for example, involves two derivatives of the metric and hence is of mass dimension 2. The mass dimension of matter fields are determined from dimensional analysis. The interactions they contribute to the action with
\be S_\mathcal{O}=\int d^Dx\sqrt{-g} \mathcal{O},\ee
must guarantee that the action remains dimensionless. For example the operator $\mathcal{O}= \partial_\mu\phi\partial^\mu\phi$ contributes to the action as a dimension 4 operator in 4 dimensions. This means in four dimensions $\phi$ has mass dimension 1. In 4 dimensions an operator constructed by $\mathcal{O}_2\propto\phi^2$ needs a coupling constant with two mass units, $\mathcal{O}_2=M^2\phi^2$. If there are any heavy fields integrated out, the effects of any interactions they have with the fields that remain dynamical at the scale of interest will imply higher derivatives of these lower scale fields, $\mathcal{O}_{n,m}=\partial^n\phi^m$ and renormalize coupling parameters, such as the mass, of the lower scale fields. The heavy field will enter the lower scale only through processes that put it in the internal legs of Feynman diagrams. Hence it enters in via its propagator which effectively brings on a derivative on the low scale field $\phi$ that can be at the outer legs. Then an operator that involves $\left(\partial^\mu\phi\partial_\mu\phi\right)^2$ would have to be suppressed by some mass scale, $\mathcal{O}_8=\frac{1}{\Lambda^4}\left(\partial_\mu\phi\partial^\mu\phi\right)^2$, where that scale $\Lambda$ corresponds to the mass of the heavy field that is integrated out. Since partial derivatives of fields bring out the energy dependence this also means $\mathcal{O}_8\sim\frac{\omega^4}{\lambda^4}\phi^4$. As the energy scale of interest approaches the scale of suppression $\omega\to\Lambda$ terms like this that involve higher order derivative contributions will fail to be corrections anymore. Intuitively we reach the scales at which the heavy field also becomes dynamical and processes that involve this field at the external legs now also contribute. From the EFT point of view, what looked like an expansion in terms of derivatives, with expansion parameter $\frac{\omega}{\Lambda}$ stops being a perturbative one. If we were instead working in a 2 dimensional theory, the leading kinetic term $\mathcal{O}=\partial_\mu\phi\partial^\mu\phi$,  would have been required to be mass dimension 2, implying $\phi$ to be dimensionless in 2D.

If we count the number of derivatives and number of fields involved in an operator with parameters d and f respectively, and denote the operators as $\mathcal{O}_{d,f}$, then in D dimensions, higher order operators of form $\mathcal{O}_{d\geq D,f}$ stop being renormalizable corrections at scales close to the cutoff, and operators of form $\mathcal{O}_{d,f\geq D}$ become non-renormalizable if the amplitude of the field grows up to the strength of the cutoff.

And so one would argue that the scalar field minimally coupled to gravity, 
\be -\frac{1}{2}m^2_{pl}R-\frac{1}{2}\left(\partial\phi\right)^2-V(\phi)\ee
is only the first term in an effective field theory. In \cite{Weinberg:2008hq} the most general EFT of a scalar field in General Relativity has been developed. By making sure that the classical equations of motion are satisfied,\footnote{This requirement is actually more clear in a path integral approach to quantum perturbations, where the leading order contribution is the classical configuration. Substitution of the leading order equation of motion also guarantees avoiding the propagation of unphysical degrees at the classical level, ie avoiding the mistake of promoting auxiliary ADM variables into fields.} the leading order corrections were found to be
\be\label{Linf}\mathcal{L}_{inf}=-\frac{1}{2}m^2_{pl}R-\frac{1}{2}\left(\partial\phi\right)^2-V(\phi)+\frac{c_1}{\Lambda^4}\left(\partial\phi\right)^4+\dots\ee     
where the coefficients are in general functions of the scalar field, i.e. $c_1=c_1(\phi)$. To capture preheating at the end of inflation, we are interested in coupling a second scalar field to \eqref{Linf}.

The effective field theory of the preheating sector alone is governed by
\be\label{Lchi} \mathcal{L}_\chi=-\frac{1}{2}\left(\partial\chi\right)^2-U(\chi)+\frac{c_2}{\Lambda^4}(\partial\chi)^4+\dots\ee
The two sectors interact with each other via
\be \label{Lmix} \mathcal{L}_{mix}=-\frac{c_3}{\Lambda}(\partial\phi)^2\chi-\frac{c_4}{\Lambda^2}(\partial\phi)^2\chi^2+\dots\ee
The coefficients $c_i$ are expected to be order one constants for a UV completable EFT \cite{Adams:2006sv}, and be positive to be able to avoid pathological instabilities \cite{Easson:2016klq}. Their specific values correspond to specific models. Because we are interested in the general consequences of these interactions for the preheating epoch, rather then specific details, $c_i$ are rather like the names of the interactions, and their value is taken to be $0$ or $1$ to indicate the absence or presence of a term. 

Note the absence of operators of the form $\partial_\mu\phi\partial^\mu\chi$, $\chi\partial_\mu\phi\partial^\mu\chi$. As long as the inflaton is shift symmetric, it would intuitively be expected to couple to the reheating sectors via derivative terms like this. However, through partial integration these derivatives on $\chi$ can be shifted onto $\phi$. Then via the use of leading order $\phi$ equation of motion these terms turn out to have the form $\phi^q\chi^p$. Hence they do not respect the shift symmetry of the inflaton and are discarded \cite{Assassi:2013gxa}. And so the EFT of Inflation with an additional Reheating field is given by $\mathcal{L}_{inf}+\mathcal{L}_\chi+\mL_{mix}$, which is at leading order
\be \label{Lreheat}\mL=-\frac{1}{2}m^2_{pl}R-\frac{1}{2}f\left(\frac{\chi}{\Lambda}\right)\partial_\mu\phi\partial^\mu\phi-\frac{1}{2}\partial_\mu\chi\partial^\mu\chi-V(\phi)-U(\chi)\ee
with
\be\label{fchi}f\left(\frac{\chi}{\Lambda}\right)=1+2c_3\frac{\chi}{\Lambda}+2c_4\frac{\chi^2}{\Lambda^2}.\ee
A priori there is no specification on $\Lambda$ without having a specific UV completion in mind. 

Interactions of type \eqref{Lreheat} have first been considered as an alternative to canonical couplings of preheating in \cite{ArmendarizPicon:2007iv}, with the Natural Inflation in mind, with $\Lambda\simeq m_{pl}$ as the scale of its UV completion. Recently these interactions have also been considered within the framework of multifield inflation, under the heading \emph{geometric destabilization of inflation} \cite{Renaux-Petel:2015mga}. Both of these works set $c_3=0$ so as to ensure that at the background level the reheating field does not develop a time dependence, $\chi_0\simeq const$. Yet in both of them, effects of the $c_4$ term are important and appreciated. From an EFT point of view, the perturbative expansion can be truncated at the order of desire, yet one cannot eliminate lower order terms to focus on the higher order ones, unless there is a symmetry that prohibits them. In this respect, we cannot include the $c_4$ term without taking account of the $c_3$ term unless there is say a $Z_2$ symmetry on the reheating field, which would eliminate all interactions that involve odd powers of $\chi$ by demanding invariance under $\chi\to-\chi$. However we will see how the $c_3$ term, unavoidable with these concerns in mind, can actually become problematic for the background physics.
\subsection{Background Dynamics}
When we consider the fields as perturbed around homogeneous backgrounds
\be \phi(t,\vec{x})=\phi_0(t)+\delta\phi(t,\vec{x}),\ee
\be \chi(t,\vec{x})=\chi_0(t)+\delta\chi(t,\vec{x}),\ee
and assume that the main contribution of these fields to the energy momentum density   come from their background values the Friedmann equations for the evolution of the Hubble parameter with flat spatial sections give
\be 3m^2_{pl}H^2(t)=\frac{1}{2}f(t)\dot{\phi}_0^2+\frac{1}{2}\dot{\chi}_0^2+V(\phi_0)+U(\chi_0)=\rho(t),\ee
\be-2m^2_{pl}\dot{H}(t)=f(t)\dot{\phi}_0^2+\dot{\chi}_0^2=\rho(t)+p(t)\ee
where $\rho$ and $p$ denote the overall energy density and pressure in the universe.

The equation of motions for the background fields themselves are
\be \ddot{\phi}_+3H\dot{\phi}_0+\partial_\chi\left(ln f\right)\dot{\phi}_0\dot{\chi}_0+\frac{1}{f}\partial_\phi V=0,\ee
and
\be \ddot{\chi}_0+3H\dot{\chi}_0-\frac{1}{2}\left(\partial_\chi f\right)\dot{\phi}_0^2+\partial_\chi U=0.\ee
Notice that the presence of $\chi$ gives rise to an additional friction term $\partial_\chi\left(ln f\right)\dot{\phi}_0\dot{\chi}_0$ in the inflaton dynamics. And in return the inflaton gives rise to the following effective potential 
\be\label{Ueff} U_{eff}=U(\chi_0)-\frac{\dot{\phi}_0^2}{2}f\left(\frac{\chi_0}{\Lambda}\right)\ee
for the reheat field. These terms complicate the solution for the dynamics of the fields. However the $\chi_0$ solution will be stable if it minimizes its effective potential \eqref{Ueff}. 

Finding the minimum of the effective potential requires two conditions. The vanishing of the first derivative of the potential
\be\label{condition1} \text{condition 1}:~~\partial_\chi U_{eff}|_{\chi^*_0}=0,\ee
gives the field value which extremizes the  the potential. This solution $\chi_0^*$ will be the minimum provided that
\be\label{condition2} \text{condition 2:}~~\partial^2_\chi U_{eff}|_{\chi_0^*}>0.\ee
Assuming a Taylor expansion for the bare potential $U(\chi_0)\sim\frac{1}{2}m^2_\chi\chi_0^2$, the background solution that minimizes the effective potential is
\be\label{chi0} \chi_0(t)=c_3\frac{\dot{\phi}_0^2}{\Lambda m^2_\chi}\left[1-2c_4\frac{\dot{\phi}_0^2}{\Lambda^2m^2_\chi}\right]^{-1}\text{provided}~ \frac{m^2_\chi}{m^2_\phi}>\frac{m^2_{pl}}{\Lambda^2}.\ee
Notice that the first condition \eqref{condition1} entirely  fixes the functional form. The second condition \eqref{condition2} gives 
\be 
m^2_\chi>2c_4\frac{\dot{\phi}_0^2}{\Lambda^2}.\ee 
At the end of inflation the inflaton field is expected to have a profile of $\phi_0(t)=\Phi sin(m_\phi t)$ where $\Phi\sim \mathcal{O}(0.1)m_{pl}$. Through this profile, condition \eqref{condition2} implies the relation between the masses of the fields involved and the cutoff. More importantly notice that the time dependence of $\chi_0$ arises completely because of the $c_3$ term, which was set to zero by hand in the discussion of derivative couplings during reheating and geometric destabilization of inflation. Through a numerical analysis, we saw in \cite{Giblin:2017qjp} that the solution \eqref{chi0} sets quite a good estimate for the solution of background equations of motion and remains in agreement with an overall matter dominated phase. 

\subsection{Consistency of the EFT}
Looking at the terms in the preheating Lagrangian \eqref{Lreheat}, 
\be f\left(\frac{\chi}{\Lambda}\right)=1+2c_3\frac{\chi}{\Lambda}+2c_4\frac{\chi^2}{\Lambda^2}\ee 
this has the form of a perturbative expansion of a function in $\frac{\chi}{\Lambda}$. This expansion will hold as long as $\chi\ll \Lambda$ and appears as the effective version at low energies of some interaction potential for the $\chi$ field. We can argue that the series was terminated because higher order terms give lesser and lesser contributions.

We must keep in mind that from an EFT point of view, we cannot have the $c_4$ term of interest without also having the $c_3$ term, unless the $\chi$ sector has some further symmetries to forbid it. Yet we saw that the $c_3$ term will cause the background field $\chi_0$ to grow. It can happen that in a model whose parameters have the sufficient values for setting resonance in $\delta\chi$, the background grows to values above the cutoff, $\chi_0\geq\Lambda$. Such a model fails because of EFT concerns, the interactions $f\left(\frac{\chi_0}{\Lambda}\right)$ can no longer be written in the form we have been working with, one must instead consider the fully summed function that $f$ is. Yet we do not know the functional form of $f$. 

From solution \eqref{chi0} we see that \be \frac{\chi_0}{\Lambda}\simeq \frac{\dot{\phi}^2_0}{m^2_\chi \Lambda^2}.\ee
The stability condition of the solution \eqref{condition2} is in favor of making sure that $\chi_0$ stays under control \be\frac{\chi_0}{\Lambda}=\frac{\dot{\phi}^2_0}{m^2_\chi \Lambda^2}\ll 1.\ee Keeping in mind that $\dot{\phi}_0\sim \mathcal{O}(0.1 m_\phi m_{Pl})$, this suggests that we can access a large range for $\Lambda$ as long as we adjust the ratio of masses accordingly 
\be\label{analyticcontrol}\frac{\chi_0}{\Lambda}=\mathcal{O}\left(\frac{m_\phi}{m_\chi}\right)^2\left(\frac{m_{Pl}}{\Lambda}\right)^2<1.\ee
With small masses of reheat particles, $m_\chi\ll m_\phi$ we should require large values of the cutoff $\Lambda\gg m_{Pl}$, where as with large masses of reheat particles $m_\chi\gg m_\phi$, the cuttoff needs to be sub-Planckian $\Lambda\ll m_{Pl}$.

Now that we have an idea about the range of validty of our EFT from $\chi$ point of view, let us end this section by commenting on the validity of the EFT from the inflaton's point of view. In equation \eqref{Linf} there is the term $\frac{(\partial\phi)^4}{\Lambda^4}$ and we need to check that this term is smaller then $(\partial\phi)^2$. So we require that $\dot{\phi}^2_0<\Lambda^4$ for the validity of this perturbative EFT on the inflationary side.  Assuming that the amplitude of $\dot{\phi}_0$ is of order $m_\phi m_{Pl}$ the condition on behalf of $\phi_0$ is 
\be\label{phivalidity}\left(\frac{m_\phi}{m_{Pl}}\right)^2\left(\frac{m_{Pl}}{\Lambda}\right)^4<1,\ee
with $m_\phi \simeq  10^{-6}m_{Pl}$ this implies $\Lambda>10^{-3}m_{Pl}$. To compare, in the derivatively coupled preheating model, the UV complexion choice to chaotic inflation requires $F \gsim m_{pl}$.

These restrictions on mass and cutoff parameters for the self consistency of the EFT seem to push models out of the regimes in parameter space where resonant $\delta\chi$ production and hence preheating is likely to occur \cite{Giblin:2017qjp}.

\section{Towards a Hamiltonian Formulation}
\label{sec:towardshamiltonian}
\subsection{Canonical Transformation from field variables to energy density variables}
The effective field theory (EFT) of inflation is formulated on the observation that inflation ends at one point in time. This implies time translation symmetry is not a symmetry of the situation at hand. Inflation ends when the time dependence of the background, that is $H(t)$ the Hubble parameter becomes as important as its amplitude, $|\dot{H}|\sim H^2$. From this point onwards $H(t)$ and accordingly $\phi_0(t)$, the background inflaton  field, evolve with time. However this evolution is such that the amplitude of these fields decreases monotonically in time, hence giving a function that defines a certain direction in time. While in EFT of inflation the time dependence always enters via the time dependent EFT coefficients, in the interpretation of these coefficients one mostly has the monotonically decreasing $\phi_0(t)$ as the clock at the back of one's mind. 

During both preheating and reheating  it would be natural to expect that the dynamics of the field $\phi_0(t)$ is coupled to the dynamics of a secondary field $\chi_0(t)$. The hope is that the inflaton transfers its energy to the $\chi$ field on the whole, and this transfer involves resonant excitations of $\chi_k$ perturbations. If the efficiency of energy transfer increases over time, one can no longer treat $\phi_0$ as an isolated field. Choosing $\phi_0(t)$ as the clock of the system is no longer advisable since $\phi_0$ alone is not enough to represent the whole system. As a candidate clock we would now like to study the behavior of the system in terms of energy densities as the variable. 

\subsection{Energy density and a Hamiltonian Formulation}
As a first step in understanding how to work with energy densities, we build on the canonical model introduced for preheating with the interaction $g^2\phi^2\chi^2$ \cite{Kofman:1994rk}
\begin{align} \nn S_m=&\int d^4x \sqrt{-g} \mL \\
\label{str1}=&\int d^3x a^3\left[-\frac{1}{2}\partial^\mu\phi\partial_\mu\phi-\frac{1}{2}\partial^\mu\chi\partial_\mu\chi-\frac{1}{2}m^2_\phi\phi^2-\frac{1}{2}m^2_\chi\chi^2-\frac{1}{2}g^2\phi^2\chi^2\right].\end{align}
For the moment we will only focus on the background fields $\phi_0(t)$ and $\chi_0(t)$
\be\label{str0} S_m=\int d^4x a^3\left[\frac{1}{2}\dot\phi^2_0+\frac{1}{2}\dot\chi^2_0-\frac{1}{2}m^2_\phi\phi^2_0-\frac{1}{2}m^2_\chi\chi^2_0-\frac{1}{2}g^2\phi^2_0\chi^2_0\right].\ee
Given this matter action we can calculate the stress energy momentum tensor
\be T_{\mu\nu}=-\frac{2}{\sqrt{-g}}\frac{\delta S_m}{\delta g^{\mu\nu}},\ee
to obtain the total energy density  for the system as 
\be \rho=T_{00}=\int d^4x\frac{1}{2}\left[\dot\phi^2_0+\dot\chi^2_0+m^2_\phi\phi^2_0+m^2_\chi\chi^2_0+g^2\phi^2_0\chi^2_0\right].\ee

In literature one is used to working with the field variables $(\phi_0,\dot\phi_0,\chi_0,\dot\chi_0)$ in a Lagrangian formulation. Yet one can switch to a Hamiltonian formulation by recognizing that the conjugate momenta for the action \eqref{str0} are,
\begin{align}
\Pi_\phi&\equiv \frac{\partial\mL}{\partial\dot\phi_0}=a^3\dot\phi_0\\
\Pi_\chi&\equiv \frac{\partial\mL}{\partial\dot\chi_0}=a^3\dot\chi_0.
\end{align}
In terms of variables $(\phi_0, \Pi_\phi, \chi_0, \Pi_\chi)$ the energy density takes on the form
\be \rho (\phi_0, \Pi_\phi, \chi_0, \Pi_\chi)=\frac{1}{2}\left[\frac{\Pi^2_\phi}{a^6}+\frac{\Pi^2_\chi}{a^6}+m^2_\phi\phi^2_0+m^2_\chi\chi^2_0+g^2\phi^2_0\chi^2_0\right].\ee
The Hamiltonian which can be calculated by
\be H(\phi_0, \Pi_\phi, \chi_0, \Pi_\chi)=\int d^3x\left[\Pi_\phi\dot\phi_0+\Pi_\chi\dot\chi_0-\mL\right]=\int d^3x \mH,\ee
gives
\be \label{hamil}H(\phi_0, \Pi_\phi, \chi_0, \Pi_\chi)=\int d^3x \left[\frac{1}{2a^3}\Pi^2_\phi+\frac{1}{2a^3}\Pi^2_\chi+\frac{a^3}{2}m^2_\phi\phi^2_0+\frac{a^3}{2}m^2_\chi\chi^2_0+\frac{a^3}{2}g^2\phi^2_0\chi^2_0\right] \ee

Notice that
\be \label{rho&H} H(\phi_0, \Pi_\phi, \chi_0, \Pi_\chi)=\int d^3x a^3 \rho (\phi_0, \Pi_\phi, \chi_0, \Pi_\chi).\ee

The time evolution of the energy density generated by the Hamiltonian obeys
\be \dot\rho\equiv\{\rho, H\}+\frac{\partial\rho}{\partial t}.\ee
Looking at our observation in equation \eqref{rho&H} the Poisson Bracket $\{\rho,H\}$ vanishes. Switching to Hamiltonian formulation is showing us that the total energy might change in this standard example only because of the expansion of the universe, which is not the effect we want to study. Hence in the following we will set $a=1$. Being constant during reheating, energy density seems to be a good candidate to be the reference frame for defining the clock with respect to. 

\subsection{Canonical Transformation to Energy variables}

In the previous section we noticed that the Hamiltonian density of the system, with the expansion of the universe neglected, is equivalent to the total energy density, meaning the total energy density is conserved. So instead of the fields themselves, we would like to reformulate our theory in terms of energy densities. That is from the set of variables $(q,p):(\phi_0,\Pi_\phi; \chi_0, \Pi_\chi)$, we would like to go to a set $(Q,P)$ where one of the variables represents energy density per species. Before going on to the canonical transformation from the set $(q,p)$  to $(Q,P)$, let us take another look at the Hamiltonian \eqref{hamil} with $a=1$

\be H(\phi_0, \Pi_\phi, \chi_0, \Pi_\chi)=\int d^3x \left[\frac{1}{2}\Pi^2_\phi+\frac{1}{2}\Pi^2_\chi+\frac{1}{2}m^2_\phi\phi^2_0+\frac{1}{2}m^2_\chi\chi^2_0+\frac{1}{2}g^2\phi^2_0\chi^2_0\right] .\ee

The last term here, $g^2\phi_0^2\chi_0^2$, that is responsible for preheating is of fourth order. It represents the interaction by which we model the phenomena. We will treat this term as an interaction part added to the free Hamiltonian. That is, we will consider the system as 
\be \mH=\mH_0+\mH_{int},\ee
where
\begin{subequations}
	\begin{align}
	\mH_0&=\frac{1}{2}\Pi^2_\phi+\frac{1}{2}\Pi^2_\chi+\frac{1}{2}m^2_\phi\phi^2_0+\frac{1}{2}m^2_\chi\chi^2_0\\
	\mH_{int}&=\frac{1}{2}g^2\phi^2_0\chi^2_0.
	\end{align}
\end{subequations}
We can apply the same logic to the energy density and split it as
\be \rho=\rho_\phi+\rho_\chi+\rho_{int}\ee
where
\begin{subequations}
	\begin{align}
	\rho_\phi&=\frac{1}{2}\Pi^2_\phi+\frac{1}{2}m^2_\phi\phi^2_0\\
	\rho_\chi&=\frac{1}{2}\Pi^2_\chi +\frac{1}{2}m^2_\chi\chi^2_0\\
	\rho_{int}&=\frac{1}{2}g^2\phi^2_0\chi^2_0
	\end{align}
\end{subequations}

The first observation is that it is convenient to pick energy densities $\rho_\phi$ and $\rho_\chi$ as the new conjugate momenta $P_i$ with respect to the free Hamiltonian\footnote{In the in-in formalism for calculating cosmological correlators, the quadratic Hamiltonian, which we referred here as the free Hamiltonian, generates the time evolution of the fields \cite{Weinberg:2005vy}. Therefore we think it is appropriate to define the operators with respect to $\mL_0$ without the interactions added. }. In other words, we would like the new canonical variables to be 
\begin{subequations}
	\begin{align}
	P_\phi &\equiv \rho_\phi\equiv \frac{\partial\mL_0}{\partial\dot{Q_\phi}}\\
	P_\chi &\equiv \rho_\phi\equiv \frac{\partial\mL_0}{\partial\dot{Q_\chi}}
	\end{align}
\end{subequations}
where in terms of the old variables 
\be \mL_0=\left[\frac{1}{2}\dot\phi^2_0+\frac{1}{2}\dot\chi^2_0-\frac{1}{2}m^2_\phi\phi^2_0-\frac{1}{2}m^2_\chi\chi^2_0\right]. \ee

At this point we need to figure out what the new canonical fields $Q_i$ are. What we know is by definition the canonical fields must satisfy the classical Poisson bracket relations
\be [Q_i,P_j]=\delta_{ij}.\ee
Remembering classical harmonic oscillators and making use of the Poisson bracket condition in terms of the old variables by
\be [Q_i,P_j]=\Sigma_{k}\left[\frac{\partial Q_i}{\partial q_k}\frac{\partial P_j}{\partial p_k}-\frac{\partial Q_i}{\partial p_k}\frac{\partial P_j}{\partial q_k}\right],\ee
one can conclude the new variables to be $Q_i=\frac{1}{m_i}tan^{-1}\left(m_i\frac{q_i}{p_i}\right)$. This set of $(Q_i, P_i)$ are also known as the action-angle variables \cite{morse1953methods}.

To summarize, the phase space where the energy densities appear among the canonical variables consists of
\begin{subequations} \label{Qrho}
	\begin{align}
	Q_\phi=\frac{1}{m_\phi}tan^{-1}\left(m_\phi\frac{\phi_0}{\Pi_\phi}\right) ~~&\&~~ \rho_\phi=\frac{1}{2}\Pi^2_\phi+\frac{1}{2}m^2_\phi\phi^2_0\\
	Q_\chi=\frac{1}{m_\chi}tan^{-1}\left(m_\chi\frac{\chi_0}{\Pi_\chi}\right) ~~&\&~~ \rho_\chi=\frac{1}{2}\Pi^2_\chi+\frac{1}{2}m^2_\chi\chi^2_0
	\end{align}
\end{subequations}
In obtaining the Hamiltonian $\tilde H (Q,P,t)$ one needs to first solve for the functional that generates the transformation, $F$, according to 
\begin{subequations}
	\begin{align}
	p_i=&\frac{\partial F}{\partial q_i}\\
	P_i=-&\frac{\partial F}{\partial Q_i}.
	\end{align}
\end{subequations}
The generating functional $F$ keeps track of any explicit time dependence in the transformation. The new Hamiltonian $\tilde H (Q,P,t)$ is given by
\be \tilde H (Q,P,t)= H(Q,P,t)+\frac{\partial F}{\partial t}.\ee
By neglecting the expansion of the universe, we got rid of any explicit time dependence and for practical purposes we don't need to calculate the generating functional. In terms of the new variables our Hamiltonian is
\be \tilde{\mH} (Q_\phi, \rho_\phi, Q_\chi, \rho_\chi)= \rho_\phi +\rho_\chi.\ee

Notice that the energy density of each species is conserved as can be seen from Hamilton's equations of motion. Of course this will not be the case when we add the interaction term.

\subsection {Including Interactions}
Now that we have familiarized ourselves with the free system, it is time to study our preheating exemplary interaction 
\be \mH_{int}(\phi_o, \Pi_\phi,\chi_0,\Pi_\chi)=\frac{1}{2}g^2\phi_0^2\chi_0^2.\ee
In terms of the variables \eqref{Qrho} this interaction is
\be \label{int} \tilde{\mH}_{int}= \frac{2g^2}{m^2_\phi m^2_\chi}sin^2\left(m_\phi Q_\phi\right)sin^2\left(m_\chi Q_\chi\right)\rho_\phi\rho_\chi.\ee
Now the canonical energies $\rho_\phi$ and $\rho_\chi$ evolve in time via the interaction term. Their behavior is determined by the following equations coupled to each other
\be \dot{\rho}_\phi=-\frac{\partial \tilde{H}}{\partial Q_\phi}=-\frac{\partial \tilde{H}_{int}}{\partial Q_\phi}\ee
\be \dot{\rho}_\chi=-\frac{\partial \tilde{H}}{\partial Q_\chi}=-\frac{\partial \tilde{H}_{int}}{\partial Q_\chi}\ee
and initial condition $\rho_\phi(0)=\rho$, $\rho_\chi(0)=0$ such that initially the background energy density is dominated by the inflaton.

The concern we wanted to address was that if the clock is set by $\phi_0(t)$, it will become inappropriate at later times, by when the other sector will have developed a time dependent background $\chi_0(t)$. Although $H(t)$ can keep track of this change in the definition of time, we do not know how to interpret it in terms of species. Rather then focusing on $\phi_0(t)$, it may be advantageous to define the clock, such that surfaces of constant time correspond to surfaces of constant $\rho_{int}$. This in essence amounts to reinterpreting $H(t)$ as $\rho_{int}$.  As $\rho_{int}$ keeps track of the energy transfer between the two sectors, it is a good variable to base things off of for studying turning points in the species that dominates the overall energy density. Hence it can also be a good variable to define things with respect to in raising questions related with the efficiency of the energy transfer, such as  in asking if ever the reheating field can take over to be the dominant source at the background level. 

\chapter{Concluding Remarks}

We started this thesis by acknowledging the presence of diffeomorphism invariance in general relativity and hence the formalism that sets the grounds for studying questions related with the past and present states of the universe. Our aim was to be able to study cosmological perturbations that carry traces from these various epochs. On the big picture, the diffeomorphism invariance of gravity causes difficulties in understanding how to quantize it. Our natural understanding of quantization involves a clear distinction between spatial coordinates and time, but with the possibility to redefine coordinates via Lorentz invariance, this distinction is lost. However we do have more or less comfortable means to quantize perturbations that can handle diffeomorphism invariance accordingly. At the end of the day, within this thesis we saw ways and virtues of gauge fixing on cosmological backgrounds.

The freedom to choose coordinates has the crucial consequence that physical results must not depend on the choice made. Although one can invent gauge invariant variables to overcome this subtlety, construction or the use of such variables may not always be easy. The BRST formalism we introduced in chapter \ref{chp:quantization methods} is one way of how to handle gauge fixing for cosmological correlators. Within this formalism we studied the quantization for perturbations of a system of scalar field and metric that respects diffeomorphism invariance, by exploring the connection between diffeomorphism invariance and constrained systems. Here we explored the structure and implications of the global BRST symmetry that remains after gauge fixing, on quantization of cosmological perturbations and evaluation of expectation values. This meant the introduction of unphysical grassmann variables, which we referred to as ghosts and antighosts, that kept track of the constraints of the system and the gauge fixing employed. Along the way, we saw that only one scalar and two tensor modes are present in the cohomology of the BRST charge and constitute all of the physical degrees of freedom. In addition, this formalism gave rise to some new gauge choices, and led us to notice that the simplest quantization procedure that was developed by Dirac amounts to the choice of synchorous gauge and does not involve ghosts. 

BRST symmetry brings to light the importance of ghosts. Because these are not physical degrees of freedom they cannot contribute to probability amplitudes as the onshell end states. This means they will not contribute at the classical, tree, level but will be important at higher order contributions. Higher order contributions in perturbation theory involve contributions from high energy physics. Hence the BRST ghosts play an important role in calculations that will also address quantum gravity effects.

A powerful view on making the gauge choice is the observation that the cosmological backgrounds are backgrounds with time dependence. They imply a choice for fixing time diffeomorphisms that make it more convenient to handle scalar perturbations. A choice in which the scalar degree of freedom associated appears as a metric perturbation. An effective field theory formalism for studying perturbations on de Sitter have been developed by virtue of such a gauge choice for the time coordinate. The motivation behind the introduction of this formalism have initially been to generalize the possible interactions that can exist for perturbations, and hence generalize the possibilities for inflationary phenomenology. In retrospect, this formalism also gave rise to an organized way of handling the scales of relevance for the possible types of interactions, and also gauge choices which emphasize adiabatic nature of superhorizon scalar perturbations.

Starting from chapter \ref{chp:EFTpreheating} we looked at preheating within this formalism. We established the hierarchies between the scales of $\chi$ production and infalton self resonance, and discovered the possibility of particle production via the sound speed of the preheating sector. Moreover we established that $\zeta$, which is the scalar perturbation as it appears eaten by the metric, is the Goldstone boson that remains massless and conserved throughout preheating as well as all other epochs. We were also able to  discover regimes in which the reheating degree of freedom $\chi$ appears hidden which type of models can give rise to this and raise questions about strong coupling. 

The main focus on preheating literature have been on the effects that the inflaton background $\phi_0(t)$ can cause on a reheating field. For example, one common treatment is to look for the $\phi_0^*$ that will maximize the range of resonant momenta $k^*$. Yet during our treatment we grew aware that if one wants to understand the passage from inflaton domination to something else, one will eventually need to investigate the point at which the reheating sector takes over at the level of the background. We saw that a consistent EFT attempt to capture background dynamics restricted the system to be inefficient for preheating. But we did not see any restrictions for resonant production in studying an EFT of perturbations. The main lesson we learned from the EFT treatment of cosmological perturbations was the importance of focusing on how to set the time diffeomorphisms. To end with, hopefully we managed to point towards a direction that may be more convenient to question the efficiency of energy transfer off of the inflaton, inspired by how guiding the definition of time coordinate can be.  

%


\end{keeppage}

\addcontentsline{toc}{chapter}{Bibliography}
\bibliography{references}

\newpage
\pagestyle{empty}

{\Large{ \bf{ Gizem Şengör}}}\\Syracuse University, Physics Department\\
gizemsengor@gmail.com\\
\HRule
\vspace*{0.1cm}
\textbf{Personal Details:}\\
Born in February 11th, 1989 Istanbul. \\
Citizen of Republic of Turkey 
\vspace*{0.1cm}\\
\HHline\\
\vspace*{0.1cm}
\textbf{Languages:}\\
Turkish (Native language)\\
English (Proficient)\\
French (Basic)
\vspace*{0.1cm}\\                
\HHline\\
\vspace*{0.1cm}
\textbf{Education:}
\begin{itemize}
	\item{Syracuse University, 2013-2018 Syracuse NY, USA\\
		PhD in Physics, June 2018\\
		Advisor: Associate Professor Scott Watson\\
		Thesis title: ``Cosmological Perturbations in the Early Universe''}
	
	\item{University of Amsterdam, Fall 2017 Amsterdam, Netherlands \\
		Delta Institute of Theoretical Physics Visiting Fellow\\
		Supervisor: Associate Professor Jan Pieter van der Schaar}
	
	\item{Boğaziçi University, 2011-2013 Istanbul, Turkey\\
		M.S. in Physics\\
		Advisor: Professor Metin Ar\i k\\
		Thesis title: ``From Five Dimensional Flat Spacetime to Our Four Dimensional Braneworld via Kaluza-Klein''}
	
	\item{Boğaziçi University, 2007-2011 Istanbul, Turkey\\
		B.S. in Physics, graduated with Honors}
\end{itemize} 
\vspace*{0.1cm}
\HHline\\
\vspace*{0.2cm}
\textbf{Honors and Awards:}
\bi
\item{2017 Henry Levinstein Fellowship for Outstanding Senior Graduate Student\\
Fall 2017 Syracuse University\\
On the academic excellence in research}
\item{WISE-FPP cohort (Future Professionals Programe)\\
2015-2017 Syracuse University}
\item{2014 Henry Levinstein Fellowship Award\\
Summer 2014 Syracuse University\\
In recognition of excellent work in first year graduate courses and promise of excellence in research.}
\ei
\HHline
\vspace*{0.3cm}
\textbf{Publications:}
\begin{enumerate}
	\item{``Toward an Effective Field Theory Approach to Reheating''\\
		Ogan \"Ozsoy, John T. Giblin, Eva Nesbit, Gizem Şengör, Scott Watson\\
		Phys.Rev. D96 (2017) no.12, 123524, arxiv:1701.01455[hep-th]}
	
	\item{``BRST Quantization for Cosmological Perturbations''\\
		Cristian Armendariz-Picon, Gizem Şengör\\
		JCAP 11 (2016) 016, arxiv:1606.03823 [hep-th]
	}
	
	\item{``Non-thermal WIMPs and Primordial Black Holes''\\
		Julian Georg, Gizem Şengör, Scott Watson\\
		Phys.Rev.D 93, 123523 (2016), arXiv:1603.00023 [hep-ph]}
	
	\item{``Is the Effective Field Theory of Dark Energy Effective?''\\
		Eric Linder, Gizem Şengör, Scott Watson\\
		JCAP 1605 (2016) 053, arXiv:1512.06180 [astro-ph]}
	
	\item{``A Model Independent Approach to (p)Reheating''\\
		Ogan \"Ozsoy, Gizem Şengör, Kuver Sinha, Scott Watson\\
		arXiv:1507.06651 [hep-th]}
	
	\item{``A five dimensional model with a fifth dimension
		as fundamental as time in terms of a cosmological approach''\\
		Gizem Şengör, Metin Ar\i k\\
		Mod. Phys. Lett., A28, (2013),arXiv:1302.0947 [gr-qc]}
	\end{enumerate}
	\vspace*{0.2cm}
	\HHline\\
	\vspace*{0.3cm}
	\textbf{Talks at Workshops and Invited Seminars:}
	\vspace*{0.3cm}\\
	``A Promenade in de Sitter''\\
	Syracuse University, High Energy Theory Group Seminars, April 2018\\
	based on ongoing work and discussions with Jan Pieter van der Schaar and Dionysios Anninos
	\vspace*{0.2cm}\\
	``An Effective Field Theory Approach to Preheating''\\
	University of Amsterdam, Cosmology Journal Club Seminar, November 2017\\
	based on work with Ogan \"Ozsoy, John T. Giblin, Eva Nesbit, Scott Watson
	\vspace*{0.2cm}\\
	``BRST Quantization for Cosmological Perturbations''\\
	Istanbul Technical University, Physics Department Seminars, July 2016\\
	based on work with Cristian Armendariz-Picon
	\vspace*{0.2cm}\\
	``Is the Effective Field Theory of Dark Energy Effective?''\\
	Syracuse University, High Energy Theory Group Seminars, March 2016\\
	based on work with Eric Linder, Scott Watson
	\vspace*{0.2cm}\\
	``Is the Effective Field Theory of Dark Energy Effective?''\\
	ICTP-SAIFR, Sao Paulo, 2016,\\
	School on Effective Field Theory Across Length Scales
	\vspace*{0.2cm}\\
	``EFT Methods for Preheating''\\
	University at Buffalo 2015, Rust Belt Meeting\\
	based on work with Ogan \"Ozsoy, Kuver Sinha, Scott Watson
	\vspace*{0.2cm}\\
	``Black Hole Constraints on Moduli Cosmology''\\
	PennState 2015, Neighbourhood Workshop on Astrophysics and Cosmology\\
	based on work with Julian Georg, Scott Watson
	\vspace*{0.2cm}\\
	``Black-hole Constraints on the post-Inflationary Epoch''\\
	University of Chicago COSMO 2014 (Poster)
	\vspace*{0.2cm}\\
	``How far away is Proxima Centauri?”\\
	XXVth ICPS 2010, Graz International Conference for Physics Students
	\vspace*{0.1cm}\\
	This was work done together with fellow students Medine Tuna Pesen, Yemliha Bilal Kalyoncu, on stellar spectra, absolute and apparent luminosity of stars and the effects of   these in determining astronomic distances.                   
	\vspace*{0.2cm}\\
	\HHline
	\vspace*{0.3cm}\\
	\textbf{Conferences, Schools and Workshops Attended}
	\vspace*{0.3cm}\\
	UCLA Dark Matter Conference\\
	UCLA, Santa Barbara February 21-23, 2018
	\vspace*{0.2cm}\\
	UCLA Dark Matter Advanced Training Institute\\
	UCLA, Santa Barbara February 18-20 2018
	\vspace*{0.2cm}\\
	CERN Winter School on Supergravity, Strings and Gauge Theory 2018\\
	CERN, Geneva 12-16 February 2018
	\vspace*{0.2cm}\\
	School on Effective Field Theory Across Length Scales\\
	ICTP-SAIFR, Sao Paulo February 22- March 4, 2016
	\vspace*{0.2cm}\\	 
	Rust Belt Meeting\\
	University at Buffalo November 7-8,2015
	\vspace*{0.2cm}\\ 
	Neighbourhood Workshop on Astrophysics and Cosmology\\
	PennState March 26-27, 2015
	COSMO 2014\\
	University of Chicago August 25-29, 2014
	\vspace*{0.2cm}\\
	Neighbourhood Workshop on Astrophysics and Cosmology\\
	PennState April 3-4, 2014
	\vspace*{0.2cm}\\
	Summer School on Cosmology\\
	ICTP-Trieste July 16-27, 2012
	\vspace*{0.2cm}\\
	International School on Strings and Fundamental Physics\\
	DESY-Hamburg July 1-12, 2012
	\vspace*{0.2cm}\\
	ULUYEF (Uludağ High Energy Physics Winter School)\\
	Bursa, Turkey February 5-11,2012
	\vspace*{0.2cm}\\
	XXVIth ICPS (International Conference of Physics Students)\\
	Budapest, Hungary August 11-18, 2011
	\vspace*{0.2cm}\\
	XXVth ICPS\\
	Graz, Austria August 17-23, 2010
	\vspace*{0.3cm}\\
	IAPS2CERN (International Association of Physics Students trip to CERN)\\
	Geneva, Switzerland February 17-20, 2010 
	\vspace*{0.2cm}\\
	XXIVth ICPS\\
	Split, Croatia 10-18 August,2009
	\vspace*{0.2cm}\\
	XXIIIth ICPS\\
	Krakow, Poland August 6-13, 2009
	\vspace*{0.2cm}\\
	\HHline
	\vspace*{0.2cm}
	\textbf{Interdisciplinary Presentations:}
	\vspace*{0.2cm}\\
	“Transposition in Music via Linear Algebra”\\
	XXVIth ICPS Budapest, 2011
	\vspace*{0.2cm}\\
	This work, carried out together with fellow students Cem Er\"oncel and Medine Tuna Pesen, was about describing musical notes as basis vectors and defining a transposition matrix among different musical scales which can be used to transpose a piece of  music in a given scale to another scale, keeping the piece’s structure invariant.
	\vspace*{0.2cm}\\
	“Dancing with Physics”\\
	XXIVth ICPS Split, 2009
	\vspace*{0.2cm}\\
	In ballet moves, the motion of arms effect the technical and artistic quality of a move due to energy and momentum transfer among limbs. In literature one can find studies that especially focus on jumps and turns. The subject of this work was to investigate the momentum transfer from one move to another in a sequence of ballet steps. It is intuitive for dancers that a jump is easier to perform when it is part of a sequence. Here we wanted to give the physical explanation behind the intuition and carry on a quantitative study on the jump pas de chat. This work was completed under the supervision of Prof. Ibrahim Semiz.\\
	\vspace*{0.2cm}    
	\HHline
	\vspace*{0.2cm}\\
	\textbf{Teaching Experience:}
	\vspace*{0.2cm}\\
	Astronomy 101 Our Corner of the Universe\\ 
	Fall 2013 \& 2014, Summer 2014, lab assistant
	\vspace*{0.2cm}\\
	Astronomy 104 Stars, Galaxies \& the Universe\\
	Spring 2014 lab assistant
	\vspace*{0.2cm}\\
	PHY 661 Graduate Quantum Mechanics\\
	Spring 2016, 2017 \& 2018  grader for weekly homeworks
	\vspace*{0.2cm}\\
	PHY 731 Graduate Thermodynamics \& Statistical Mechanics\\
	Spring 2016 grader for weekly homeworks
	\vspace*{0.2cm}\\
	PHY 567 Introduction to Quantum Mechanics I\\
	Spring 2017 grader for weekly homeworks
	\vspace*{0.2cm}\\
	PHY 641 Advanced Electromagnetic Theory\\
	Spring 2018  grader for weekly homeworks\\
\HRule

\end{document}